\documentclass{article}
\usepackage[utf8]{inputenc}

\usepackage[T1]{fontenc}
\usepackage[english,german]{babel}
\usepackage[babel]{csquotes}
\usepackage{subfigure}
\usepackage{float}
\usepackage{graphicx}
\usepackage{amsmath}
\usepackage{amssymb}
\usepackage{amsfonts}
\usepackage{color}
\usepackage{braket}
\usepackage[bottom,hang,flushmargin]{footmisc} 
\usepackage{verbatim}
\usepackage{tensor}
\usepackage{bbold}
\usepackage{changepage}
\usepackage[
sorting=none
]{biblatex}
\usepackage[a4paper,left=3cm,right=2.5cm,top=2.5cm,bottom=2.5cm,twoside]{geometry}
\usepackage{scrextend}
\usepackage{pdfpages}
\usepackage{hhline}
\usepackage{pdflscape}
\usepackage{rotating}
\usepackage{tablefootnote}
\usepackage{tensor}
\usepackage{setspace}

\usepackage[hidelinks]{hyperref}

\newcommand\at[2]{\left.#1\right|_{#2}}

\newcommand{\bras}[1]{\tensor[_s]{\bra{#1}}{}}
\newcommand{\kets}[1]{\tensor[]{\ket{#1}}{_s}}
\newcommand{\brakets}[1]{\tensor[_s]{\braket{#1}}{_s}}

\newcommand{\ketsp}[1]{\tensor[]{\ket{#1}}{_{s'}}}
\newcommand{\braketsp}[1]{\tensor[_s]{\braket{#1}}{_{s'}}}

\makeatletter
\pretocmd{\section}{\addtocontents{toc}{\protect\addvspace{-35\p@}}}{}{}
\pretocmd{\subsection}{\addtocontents{toc}{\protect\addvspace{-11.3\p@}}}{}{}
\pretocmd{\subsubsection}{\addtocontents{toc}{\protect\addvspace{-12\p@}}}{}{}
\makeatother

\title{Quasi-Coherent States on Deformed Quantum Geometries}
\author{Laurin Felder}
\date{\today}

\begin{document}

\includepdf[pages=-]{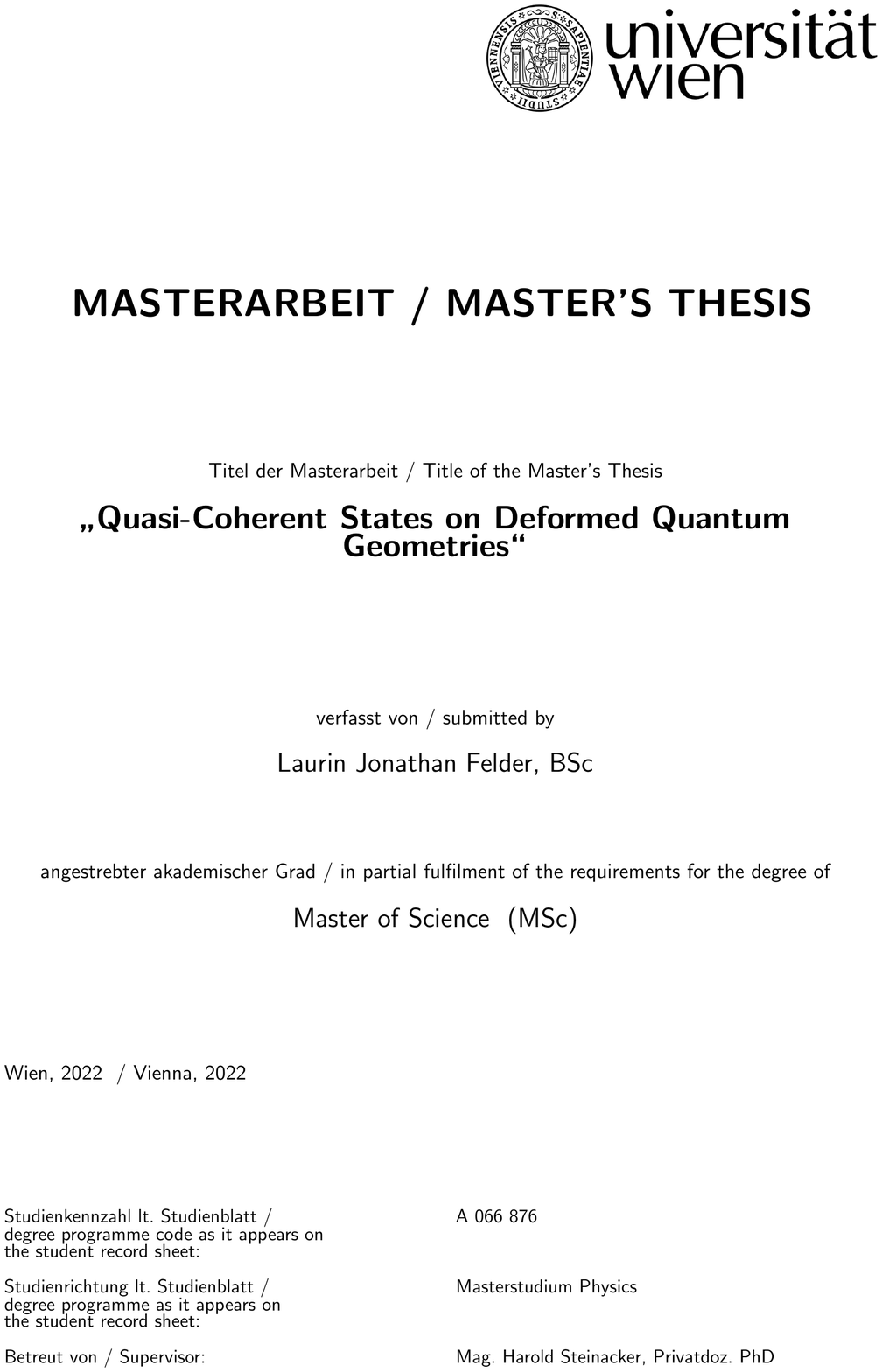}

\newpage\null\thispagestyle{empty}\newpage

\newpage
\thispagestyle{empty}

\selectlanguage{english}

\textbf{Acknowledgment}

I would like to express my deepest gratitude to my supervisor Mag. Harold Steinacker, Privatdoz. PhD for his guidance and inspiration during the writing of this thesis.
\\
Special thanks to Nora Dehmke, Vera Fulterer and Lorelies Ortner for their support.

\begin{abstract}
    Matrix configurations coming from matrix models comprise many important aspects of modern physics. They represent special quantum spaces and are thus strongly related to noncommutative geometry.
    In order to establish a semiclassical limit that allows to extract their intuitive geometrical content, 
    this thesis analyzes and refines an approach that associates a classical geometry to a given matrix configuration, based on quasi-coherent states.
    
    While, so far, the approach is only well understood for very specific cases, in this work it is reviewed and implemented on a computer, allowing the numerical investigation of deformations of these cases.
    It is proven that the classical space can be made into a smooth manifold immersed into complex projective space. Further, the necessity for the consideration of foliations thereof is shown in order to deal with the observed and subsequently described phenomenon called oxidation.
    The developed numerical methods allow the visualization of the semiclassical limit as well as quantitative calculations. Explicit examples suggest the stability under perturbations of the refined approach and highlight the physical interpretation of the construction.
    All this supports a better understanding of the geometrical content of arbitrary matrix configurations as well as their classical interpretation and establishes the determination of important quantities.
    
\end{abstract}

\selectlanguage{german}

\begin{abstract}
    Matrix-Konfigurationen als Lösungen von Matrixmodellen beinhalten viele wichtige Aspekte der modernen Physik. Sie repräsentieren spezielle Quantenräume und sind daher eng mit der nichtkommutativen Geometrie verbunden.
    Um einen semiklassischen Grenzfall ebendieser zu finden und um die darin beschriebene Geometrie zu extrahieren, wird in dieser Arbeit eine spezielle Konstruktion analysiert und verfeinert, die, basierend auf quasi-kohärenten Zuständen, einer gegebenen Matrix-Konfiguration eine klassische Geometrie zuordnet.
    
    Die Konstruktion ist nur in speziellen Fällen wirklich verstanden, daher wird diese hier diskutiert und auf einem Computer implementiert, was die numerische Untersuchung von Deformationen dieser Fälle erlaubt.
    Es wird bewiesen, dass aus der klassischen Geometrie eine glatte Mannigfaltigkeit, die in den komplexen projektiven Raum immersiert ist, gemacht werden kann. Weiters wird die Notwendigkeit zur Betrachtung von Blätterungen dieser Mannigfaltigkeit gezeigt, um das beobachtete und beschriebene Phänomen der Oxidierung handhaben zu können.
    Die entwickelten numerischen Methoden erlauben es, den semiklassischen Grenzfall zu visualisieren und quantitative Berechnungen durchzuführen. Explizite Beispiele belegen die Stabilität des verfeinerten Zugangs unter Störungen und verdeutlichen die physikalische Interpretation der Konstruktion. All dies unterstützt ein besseres Verständnis der beschriebenen Geometrie sowie der klassischen Interpretation und erlaubt die Berechnung wichtiger Größen.
\end{abstract}

\selectlanguage{english}

\newpage\null\thispagestyle{empty}\newpage

\newpage

\tableofcontents

\newpage
\section{Introduction}

Matrix configurations -- refined quantum geometries --  are extremely simple objects from a technical point of view. As solutions of matrix models, they can carry a lot of physical information, while their geometrical content is not directly accessible. They are inherently described as quantum theories, which is clearly favorable from a conceptual perspective, but coming to the price that a classical understanding is difficult. Therefore it is especially desirable to have the ability to construct a semiclassical limit, represented by a classical manifold with additional structure.
\\
One possibility to construct such a limit is inspired by ordinary quantum mechanics. There, coherent states allow one to define quantization maps which build the bridge between classical manifolds and quantum geometries. Paralleling this mechanism, one associates so called quasi-coherent states to arbitrary finite dimensional matrix configurations. Their collections can be used to define associated manifolds.
\\
This construction is well understood for (quantum) geometries associated to compact semisimple Lie groups with the fuzzy sphere as a prototypical example, but in the general case very little is known.

This motivates the scope of this thesis, that is, to refine the construction of a semiclassical limit via quasi-coherent states -- allowing one to extract the geometrical content of a matrix configuration --
and to study this limit for deformed quantum geometries, being perturbations to well known matrix configurations.\\
An important aspect is the analytic study and the local and global visualization of the associated manifolds as well as the evaluation of the approach based on quasi-coherent states away from oversimplified examples. This further includes the investigation on the stability of the framework under perturbations as well as quantitative verification (here, the focus lies on properties of the quantization map proposed in \cite{Steinacker_2021}).

This thesis is organized in three parts, where the first part focuses on the theoretical background and analytic results, while the second part is dedicated to the implementation of the framework on a computer. The third part then discusses actual numerical results.
\\
The first part (section \ref{QMG}) assumes a working knowledge in Lie theory and differential geometry, thus well established results are expected to be known. On the other hand, the more involved discussions can mostly be skipped, without foreclosing the comprehension of the argumentation. 
It is divided in four sections, where the first (section \ref{QG}) features an introductory view on quantum spaces and quantization itself, including important examples like the Moyal-Weyl quantum plane and the fuzzy sphere. Then section \ref{QMGsDescription} focuses on quantum spaces coming from matrix configurations and especially on the construction of their semiclassical limit in terms of a so called \textit{quantum manifold}. Some examples are discussed in section \ref{qmgexamples}, making the connection between earlier examples of quantum geometries and the numerical investigation in the subsequent part. Section \ref{fol} introduces different approaches to foliations of the quantum manifold, refining the latter in order to maintain stability under perturbations.\\
The second part (section \ref{implementation}) discusses the algorithms that are used for actual calculations in Mathematica, while the focus rather lies on a conceptual understanding than on the actual code, thus not much knowledge in programming is necessary.
In section \ref{BasicQuantities} the quasi-coherent states and other basic properties are calculated, followed by the visualization of the quantum manifold in section \ref{Visual}. Section \ref{Foliat} deals with calculations in the leaves (coming from the foliations from section \ref{fol}), including the integration over the latter.\\
The final part (section \ref{RESULTS}) is reserved for the discussion of actual results for important matrix configurations.
These examples are the squashed fuzzy sphere (section \ref{sfs_results}), the (with random matrices) perturbed fuzzy sphere (section \ref{fsr}),
the squashed fuzzy $\mathbb{C}P^2$ (section \ref{sfc}), the related completely squashed fuzzy $\mathbb{C}P^2$ (section \ref{csfc}) and the fuzzy torus (section \ref{ft}).\\
Further, this thesis includes three appendices, where appendix \ref{AppendixA} contains the more technical aspects related to the quasi-coherent states for which there was no place in section \ref{QMGsDescription}. Appendix \ref{AppendixB} collects the most crucial facts about the irreducible representations of the Lie algebras $\mathfrak{su}(2)$ and $\mathfrak{su}(3)$ as well as the clock and shift matrices, where the first two play an important role throughout this thesis. The last (appendix \ref{Appendix:perturbativeappr}) shows the explicit computations related to the perturbative calculations for the squashed fuzzy sphere.

The most important paper for this thesis is \cite{Steinacker_2021}, where a particular construction of the quasi-coherent states and (based on that) of a semiclassical limit has been introduced that will be used throughout this work. Thus section \ref{QMGsDescription} recapitulates some of the framework and the results -- stipulated with new findings and additional considerations.
\\
An introductory account to the background and theory of quantum spaces and matrix models can be found in the lecture notes \cite{Steinacker_2016II}, while a comprehensive introduction in form of a book by Harold C. Steinacker is in progress (some preliminary material has strongly influenced section \ref{QG} and especially both discussions of coadjoint orbits \ref{CoadjointOrbits0} and \ref{CoadjointOrbits} as well as the introduction to section \ref{QMG}).
\\
In \cite{Schneiderbauer_2015, Schneiderbauer_2016} a different method to construct quasi-coherent states, based on a Laplacian or Dirac operator, is discussed. This carries the possibility to compare some results.
\\
The origins of the generalization of coherent states to semisimple Lie algebras lie in \cite{Perelomov_1986}, while the fuzzy sphere was first described in \cite{Madore_1992}. Some completely squashed quantum geometries have already been studied in \cite{Steinacker_2015,Schneiderbauer_2015, Schneiderbauer_2016}.
\\
General results on matrix models that highlight their physical importance can for example be found in
\cite{Steinacker_2011,Steinacker_2016,Steinacker_2020,Grosse_2008,Douglas_2001,Minwalla_2000}, reaching from emergent gravity to noncommutative quantum field theory. An important matrix model is given by the so called IKKT model, introduced in \cite{Ishibashi_1997}. In accordance with its strong relation to type IIB string theory here a semiclassical limit acquires the interpretation as a brane in target space \cite{Steinacker_2011,Steinacker_2021,Ishibashi_1997}.

In this thesis, an accessible semiclassical limit for arbitrary matrix configurations is described as a refinement of the construction from \cite{Steinacker_2021}, which allows for a better understanding of the geometry encoded in these.
\newpage
\section{Quantum Matrix Geometries}
\label{QMG}

The \textit{matrix configurations} that we will consider in this section are defined in terms of extraordinarily simple objects -- then again it is far from obvious how to extract any geometrical information from them.\\
They are formulated in so called \textit{quantum spaces} (which are here described as \textit{noncommutative geometries} via noncommutative algebras) which are usually thought of as \textit{quantizations} of \textit{classical spaces} (i.e. Poisson or symplectic manifolds with their commutative algebras of functions).\\
At first, one attempts to construct quantum spaces associated to classical spaces together with a \textit{correspondence principle} implemented as \textit{quantization maps} (one then says that one \textit{quantized} the classical spaces).
But if one wants to consider matrix configurations coming from matrix models as fundamental, the task is to find classical spaces together with quantization maps as semiclassical limits in such a way that the quantum spaces are the quantizations of the classical spaces. Defining such a construction is the purpose of this section.

Since a basic understanding of the quantization of Poisson or symplectic manifolds is necessary, the here used procedures are reviewed in section \ref{QG}, together with some of the basic examples that will frequently reappear throughout this work, based on \cite{Steinacker_2021,Steinacker_2016II,Schneiderbauer_2015,Schneiderbauer_2016,Steinacker_2011,Steinacker_2020}.
\\
Having that background, in section \ref{QMGsDescription} matrix configurations are introduced as special tuples of Hermitian matrices (whereas matrix models are actions upon matrix configurations) without going too deep into their origins. Following \cite{Steinacker_2021}, a procedure for finding a corresponding classical manifold (together with additional geometrical structure) -- relying on so called \textit{quasi-coherent states} -- is introduced.\\
In section \ref{fol} a main problem of the previous procedure -- the latter constructed manifolds are often too high dimensional -- is tackled via different foliation prescriptions, mostly relying on the different geometrical structures.\\
Finally, some of the developed concepts are applied to basic examples of matrix configurations (reproducing some of the examples discussed in section \ref{QG}).

\subsection{Quantization of Classical Spaces and Quantum Geometries}
\label{QG}

There are various approaches to noncommutative geometry and the quantization of \textit{Poisson} or \textit{symplectic manifolds} (like the spectral triple \cite{Connes_1995} or deformation quantization based on formal star products \cite{Waldmann_2007}), yet here we follow a different (although related) approach.

We first look at the classical space, defined in terms of a Poisson manifold $(\mathcal{M},\{,\})$.
This means $\mathcal{M}$ is a smooth manifold, coming with its commutative algebra of smooth functions $\mathcal{C}^\infty(\mathcal{M})$ (under point wise multiplication), where we will drop the $\infty$ from now on.
The manifold is equipped with a \textit{Poisson bracket} -- an antisymmetric and bilinear map $\{,\}:\mathcal{C}(\mathcal{M})\times\mathcal{C}(\mathcal{M})\to\mathcal{C}(\mathcal{M})$, satisfying the Leibniz rule and the Jacobi identity\footnote{$\{fg,h\}=\{f,h\}g+f\{g,h\}$ respectively $\{f,\{g,h\}\}+\{g,\{h,f\}\}+\{h,\{f,g\}\}=0$ for all $f,g,h\in\mathcal{C}(\mathcal{M})$.}, making $\mathcal{C}(\mathcal{M})$ into a Lie algebra \cite{Michor_2008}.
We can simply extend the algebra of functions by allowing the latter to attain complex values. If we demand the bracket to be complex bilinear it also extends naturally. From now on, this extension is assumed and $\mathcal{C}(\mathcal{M})$ stands for the complex algebra of smooth (yet not holomorphic) complex valued functions on $\mathcal{M}$.
\\
Alternatively, we can start with a symplectic Manifold $(\mathcal{M},\omega)$ where $\omega\in \Omega^2(\mathcal{M})$ is a nondegenerate two form that is closed $d\omega=0$, called \textit{symplectic form}. Then, there is a naturally induced Poisson bracket $\{f,g\}:=-\omega(df^\sharp,dg^\sharp)\, \forall f,g\in\mathcal{C}(\mathcal{M})$ on $\mathcal{M}$, where $\sharp$ indicates the induced isomorphism between $\Omega^1(\mathcal{M})$ and $\mathfrak{X}(\mathcal{M})$, defined via $\omega(\alpha^\sharp,\xi)=\alpha(\xi)\,\forall \alpha\in\Omega^1(\mathcal{M}),\xi\in\mathfrak{X}(\mathcal{M})$. Conversely, any Poisson manifold decomposes into a foliation of symplectic leaves, thus symplectic manifolds are special cases of Poisson manifolds \cite{Michor_2008}.

While the latter algebra is commutative, we take as a model of a quantum space an (in general) noncommutative endomorphism algebra $\operatorname{End}(\mathcal{H})$ of a Hilbert space $\mathcal{H}$, coming with the commutator as a natural Lie bracket $[,]:\operatorname{End}(\mathcal{H})\times \operatorname{End}(\mathcal{H})\to\operatorname{End}(\mathcal{H})$.
This map is antisymmetric, complex bilinear and satisfies a relation parallel to the Leibniz rule and the Jacobi identity\footnote{\label{ComProps}$[FG,H]=[F,H]G+F[G,H]$ respectively $[F,[G,H]]+[G,[H,F]]+[H,[F,G]]=0$ for all $F,G,H\in\operatorname{End}(\mathcal{H})$.}.

The function algebras equipped with Poisson brackets and the endomorphism algebras provide many comparable operations and objects with physical interpretation. Table \ref{compareCQ} provides a list of related structures. For example on the classical side we can construct the $L^2$ inner product if the manifold is compact, whereas on the quantum side the Hilbert-Schmidt inner product is naturally available for finite dimensional Hilbert spaces.
Similarly, real functions respectively Hermitian operators may be interpreted as observables. If they are also positive and normalized in the respective induced norms, they may be regarded as mixed states \cite{Steinacker_2021,Schneiderbauer_2015}.

\begin{table}[H]
\begin{tabular}{c|c|c}
structure & classical space (comm.) & quantum space (noncomm.) \\ \hline
algebra & $\mathcal{C}(\mathcal{M})$ & $\operatorname{End}(\mathcal{H})$ \\
addition \& multiplication & pointwise operations & matrix operations \\
(Lie) bracket & $(f,g)\mapsto i\{f,g\}$ & $(F,G)\mapsto[F,G]$ \\
conjugation & $f\mapsto f^*$ & $F\mapsto F^\dagger$ \\
inner product\tablefootnote{$\operatorname{dim}(\mathcal{M})=:2n$ and $\Omega:=\frac{1}{n!}\omega^{\wedge n}=\sqrt{\operatorname{det}(\omega_{ab})}dx^1\wedge\dots\wedge dx^{2n}\in\Omega^{2n}(\mathcal{M})$ is the volume form coming from the symplectic form $\omega$ that is potentially induced by $\{,\}$.} (if well def.) & $(f,g)\mapsto\braket{f\vert g}_2:=\frac{1}{(2\pi)^n}\int_\mathcal{M}\Omega\, f^*g$ & $(F,G)\mapsto\braket{F\vert G}_{HS}:=\operatorname{tr}(F^\dagger G)$ \\
observable & $f^*=f$ & $F^\dagger=F$ \\
mixed state & $f\geq0$ \& $\Vert f\Vert_2=1$ & $F\geq0$ \& $\Vert F\Vert_{HS}=1$\\
\end{tabular}
\centering
\caption{Comparison of related structures on Poisson manifolds and endomorphism algebras of Hilbert spaces (for $f,g\in\mathcal{C}(\mathcal{M})$ and $F,G\in \operatorname{End}(\mathcal{H})$). Adapted from \cite{Schneiderbauer_2015} and \cite{Steinacker_2021}}
\label{compareCQ}
\end{table}

Now, \textit{quantizing} a Poisson manifold means to first find an appropriate Hilbert space and then relating algebra elements and especially classical observables to quantum observables in such a way that at least some features are preserved. Here, this is done by a \textit{quantization map}.\\
A quantization map is defined as a (complex) linear map
\begin{align}
    Q: \mathcal{C}(\mathcal{M})\to\operatorname{End}(\mathcal{H})
\end{align}
that further depends on a parameter $\theta$ called \textit{quantization parameter}\footnote{This parameter may be discrete or continuous while also $\{,\}$ and the Hilbert space itself may depend on it. It should be thought of as a formalization of the usual use of the reduced Planck constant $\hbar$, for example when saying \textit{to take $\hbar$ to zero}.} and satisfies the following axioms:
\begin{enumerate}
    \item $Q(1_\mathcal{M})=\mathbb{1}_\mathcal{H}$ (completeness relation)
    \item $Q(f^*)=Q(f)^\dagger$ (compatibility of con- and adjungation)
    \item $\lim_{\theta\to0} (Q(f\cdot g)-Q(f)\cdot Q(g))=0\quad\text{and}\quad\lim_{\theta\to0}\frac{1}{\theta} (Q(\{f, g\}))-\frac{1}{i} [Q(f),Q(g)])=0$ (asymptotic compatibility of algebra structure)
    \item $[Q(\mathcal{C}(\mathcal{M})),F]=0\implies F\propto\mathbb{1}_\mathcal{H}$ (irreducibility)
\end{enumerate}
The first condition ensures that totally mixed states are mapped to totally mixed states (up to normalization), while the second condition guarantees that observables are mapped to observables.
The third condition ensures compatibility of multiplications and brackets in the two algebras at least in the \textit{semiclassical limit} $\theta\to 0$. The last condition ensures that there are no sectors of $\operatorname{End}(\mathcal{H})$ that are left invariant under the adjoint action of the quantization of $\mathcal{C}(\mathcal{M})$ and therefore $\mathcal{H}$ is only \textit{as large as necessary}.
\\
There are further axioms that may be imposed. Naturally, we might demand
\begin{enumerate}
\setcounter{enumi}{4}
    \item $\braket{Q(f)\vert Q(g)}_{HS}=\braket{f\vert g}_2$ (isometry)
\end{enumerate}
meaning that $Q$ is an \textit{isometry} under the natural inner products.
If both $\mathcal{C}(\mathcal{M})$ and $\operatorname{End}(\mathcal{H})$ are representations of some Lie algebra $\mathfrak{g}$ with both actions denoted by $\cdot$, we can further impose
\begin{enumerate}
\setcounter{enumi}{5}
    \item $Q(x\cdot f)=x\cdot Q(f)$ (intertwiner of action)
\end{enumerate}
meaning that $Q$ is an \textit{intertwiner} of the action \cite{Steinacker_2016II,Schneiderbauer_2016,Waldmann_2007}.

Let us formulate two remarks. Since the \textit{dynamics} of classical or quantum systems are governed by the Hamilton equations in terms of the brackets with a Hamiltonian that itself is an observable, $Q$ can be said to represent the \textit{correspondence principle}.\\
If $Q$ is an isometry, completeness implies $\frac{1}{(2\pi)^n}\operatorname{vol}_\Omega(\mathcal{M})=\Vert 1_\mathcal{M}\Vert_2^2\overset{!}{=}\Vert  \mathbb{1}_\mathcal{H}\Vert_{HS}^2=\operatorname{dim}(\mathcal{H})$ what is better known as \textit{Bohr-Sommerfeld quantization condition} of the symplectic volume \cite{Steinacker_2016II}.

\subsubsection{The Moyal-Weyl Quantum Plane}

As a first example, we briefly review the Moyal-Weyl quantum plane, that reformulates results from ordinary quantum mechanics.

Let $\mathcal{M}=\mathbb{R}^{2n}$, together with the Poisson bracket defined via
\begin{align}
    \label{com1}
    \{x^a,x^b\}=\theta^{ab},
\end{align}
where the $x^a$ are the Cartesian coordinate functions\footnote{We might view $\mathbb{R}^{2n}$ as a phase space and view the first $n$ coordinates as spatial coordinates and the second $n$ coordinates as momentum coordinates.} and $\theta^{ab}$ is the constant matrix given by
\begin{align*}
(\theta^{ab})=\theta
    \begin{pmatrix}
0 & \mathbb{1}_n \\
-\mathbb{1}_n& 0 
\end{pmatrix},
\end{align*}
using the quantization parameter $\theta$ \cite{Steinacker_2016II,Steinacker_2011}.

Recalling the Stone-von Neumann theorem \cite{vonNeumann_1931}, we put $\mathcal{H}=L^2(\mathbb{R}^n)$ and define operators\footnote{In analogy, we might interpret the first $n$ operators as position operators and the second $n$ operators as momentum operators.} $X^a$ via
\begin{align}
    \label{moyalweylops}
    \left(X^i\phi\right)(q)=q^i\phi(q),\quad \left(X^j\phi\right)(q)=-i\theta \partial_j \phi(q)
\end{align}
for $i=1,\dots,n$ and $j=n+1,\dots, 2n$ and for any $\phi\in L^2(\mathbb{R}^n)$ and $q\in\mathbb{R}^n$.\\
This implies the commutation relations
\begin{align}
\label{UsualCorrPr}
[X^a,X^b]=i\theta^{ab}\mathbb{1}_\mathcal{H},
\end{align}
that already nicely compare to equation (\ref{com1}) \cite{Steinacker_2016II,Steinacker_2011}.

Since $\mathcal{M}$ itself is a Lie group (the translation group $\mathbb{R}^{2n}$ with a natural action on itself given by $t\cdot x:=x-t$), we have the induced action of the translation group $\mathbb{R}^{2n}$ on $\mathcal{C}(\mathcal{M})$
\begin{align}
    (t\cdot f)(x)=f(x+t).
\end{align}
On the other hand, $\mathcal{H}$ is a \textit{projective representation} of the translation group (respectively a representation of a central extension thereof) given by $t\cdot\ket{v}=U_t\ket{v}$ for $U_t:=\exp(i\sum_{a,b}t^a\theta^{-1}_{ab}X^a)$ and we get an induced ordinary representation on $\operatorname{End}(\mathcal{H})$ via
\begin{align}
    t\cdot F=U_t^{-1} F U_t,
\end{align}
considering the \textit{Baker-Campbell-Hausdorff formula}.\\
Under these actions, we find $t\cdot x^a=x^a+t^a$ (the function $x^a$ not a point in $\mathbb{R}^{2n}$) as well as $t\cdot X^a=X^a+t^a\mathbb{1}_\mathcal{H}$ so both observables transform in a comparable way under translations, giving them a related interpretation.
\\
We note further that this implies that the Poisson bracket is invariant under translations.

Now, we define the so called \textit{plain waves} $v_k\in\mathcal{C}(\mathcal{M})$ via
\begin{align}
    v_k(x):=\exp(i\sum_a k_a x^a)
\end{align}
and the corresponding quantum version $V_k\in\operatorname{End}(\mathcal{H})$ via 
\begin{align}
    V_k:=\exp(i\sum_a k_a X^a).
\end{align}
Then, we directly find $t\cdot v_k=\exp(i\sum_a k_a t^a)v_k$ and again via the Baker-Campbell-Hausdorff formula $t\cdot V_k=\exp(i\sum_a k_a t^a)V_k$, thus both transform identically under the translation group as common eigenmodes of all translations. It is further obvious that the $v_k$ form a basis of $\mathcal{C}(\mathcal{M})$.\\
With this at hand, we define the quantization map to be the unique linear map satisfying
\begin{align}
    Q(v_k)=V_k,
\end{align}
for all $k\in\mathbb{R}^{2n}$.
It is not hard to verify that all axioms (1-6) are satisfied and $Q$ is a quantization map -- in fact the additional axioms (5) and (6) even make $Q$ unique.

Further this implies
\begin{align}
    Q(x^{a_1}\cdot\dots\cdot {x^a})=X^{(a_1}\cdot\dots\cdot X^{a_n)},
\end{align}
what is well known as the \textit{Weyl ordering prescription}. Especially this leaves us with $X^a$ as quantization of $x^a$, giving equation (\ref{UsualCorrPr}) the interpretation of the \textit{canonical commutation relations} \cite{Steinacker_2016II,Steinacker_2011}.

\subsubsection{The Fuzzy Sphere}
\label{FuzzySphere0}

Now, we turn to a quantum analogue of the sphere $S^2$, the \textit{fuzzy sphere}. This represents the prototypical compact quantum space. 

Of course, we pick $\mathcal{M}=\mathcal{S}^2\hookrightarrow\mathbb{R}^3$, where we have the Cartesian embedding functions $(x^a):\mathcal{S}^2\to\mathbb{R}^3$, satisfying $\sum_a x^ax^a=1_{\mathcal{M}}$. Further, we define a Poisson bracket via
\begin{align}
    \label{com2}
    \{x^a,x^b\}:=\theta^{ab}(x):=\frac{2}{N}\sum_c\epsilon^{abc}x^c
\end{align}
(the use of the factor $N=2,3,4,\dots$ will become apparent later), where we view $\frac{1}{N}$ as our quantization parameter $\theta$ \cite{Steinacker_2016II, Steinacker_2011}.

On the other hand, we set $\mathcal{H}=\mathbb{C}^N$, observing the scheme: A compact (noncompact) classical space fits to a finite (infinite) quantum space. Such a link is highly plausible in the light of the combination of axiom (1) and (5).\\
At this point a little ad hoc\footnote{Alternatively, we could somehow parallel the use of Cartesian embedding functions and construct operators via the so called \textit{Jordan-Schwinger representation} or \textit{oscillator construction}, based on the operators of the Moyal-Weyl quantum plane \cite{Steinacker_2020}.}, we introduce the operators $X^a:=\frac{1}{C_N}J_N^a$, where $C_N^2:=(N^2-1)/4$ and the $J_N^a$ are the orthogonal generators of the Lie algebra $\mathfrak{su}(2)$ in the $N$ dimensional irreducible representation $\mathbb{C}^N$ (discussed in appendix \ref{su2rep}), satisfying $\sum_aX^aX^a=\mathbb{1}_\mathcal{H}$ and
\begin{align}
    [X^a,X^b]=\frac{i}{C_N}\sum_c\epsilon^{abc}X^c.
\end{align}

The appearance of $SU(2)$ and consequently $SO(3)\cong SU(2)/\{\pm 1\}$ in the last definition is not accidental: While $\mathcal{S}^2$ is exactly an $SO(3)$ orbit of the natural action on $\mathbb{R}^3$, we get the induced action on $\mathcal{C}(\mathcal{M})$ via
\begin{align}
    (R\cdot f)(x)=f(R^{-1}\cdot x).
\end{align}
Let now $R=\exp(i\sum_a r^a J^a)$ (where the $J^a$ are the Lie algebra generators of $SO(3)$), then we get from the $\mathfrak{su}(2)$ representation on $\mathcal{H}$ the induced projective $SO(3)$ representation $R\cdot \ket{v}=U_R\ket{v}$ where $U_R:=\exp(i\sum_a r^a J_N^a)$, further inducing an $SO(3)$ representation on $\operatorname{End}(\mathcal{H})$ via
\begin{align}
    R\cdot F=U_R^{-1} F U_R.
\end{align}
We then find the identical transformation behaviours $R\cdot x^a=\sum_b R^{ab}x^b$ and $R\cdot X^a=\sum_b R^{ab}X^b$.\\
Also here, we note that the Poisson bracket has been chosen such that it is invariant under rotations.

As for the Moyal-Weyl quantum plane, we consider the common eigenmodes of the rotation operators in the respective representations: For $\mathcal{C}(\mathcal{M})$, these are the well known \textit{spherical harmonics}
\begin{align}
    Y^l_m=\sum_{a_{-l},\dots,a_l}Y^{(m)}_{(a_1\dots a_l)}x^{a_1}\dots x^{a_l}
\end{align}
for $l=0,1,2,\dots$ and $m=-l,\dots,l$ and for some known coefficients $Y^{(m)}_{(a_1\dots a_l)}$, coming with also known but here irrelevant eigenvalues.
For $\operatorname{End}(\mathcal{H})$, we find the eigenmodes
\begin{align}
    \hat{Y}^l_m=\sum_{a_{-l},\dots,a_l}c^lY^{(m)}_{(a_1\dots a_l)}X^{a_1}\dots X^{a_l}
\end{align}
for some real normalization constants $c^l$ with the \textit{same} eigenvalues. However, here we have $l=0,1,2,\dots,N-1$ -- providing us with a natural \textit{ultraviolet cutoff}.\\
Now, we can define a quantization map in the same manner as for the Moyal-Weyl quantum plane as the unique linear map that satisfies
\begin{align}
    Q(Y^l_m)= \left\{
\begin{array}{ll}
\hat{Y}^l_m & l \leq N-1 \\
0 & \, \textrm{else} \\
\end{array}
\right.
\end{align}
One then verifies that the axioms (1-6) hold (while (5) only holds in the limit $\theta\to 0$ respectively $N\to\infty$). Again, the axioms (5) and (6) make $Q$ unique.
\\
An interesting consequence is the natural quantization of the prefactor $\frac{1}{N}$ in (\ref{com2}), quantizing the symplectic volume of $S^2$.

Here, we also have
\begin{align}
    Q(x^a)= X^a,
\end{align}
while for general polynomials in $x^a$ the result is (only a little bit) more involved \cite{Steinacker_2016II,Madore_1992,Steinacker_2011,Hoppe_1982}.

\subsubsection{Quantized Coadjoint Orbits}
\label{CoadjointOrbits0}

The idea behind the fuzzy sphere can be generalized from $SU(2)$ to an arbitrary compact semisimple Lie group $G$ of dimension $D$ (coming with its associated Lie algebra $\mathfrak{g}$), also providing us with a less ad hoc conception of the former. While this section is rather technical, it can in principle be skipped leaving most of the remaining comprehensible.
Let $\lambda\in\mathfrak{g}^*$ be a dominant integral element\footnote{We shall assume that a maximal Cartan subalgebra $\mathfrak{h}\subset\mathfrak{g}$ and a set of positive roots has been chosen.}.

We consider the coadjoint action of $G$ on the dual of the Lie algebra $\mathfrak{g}^*$, providing us with the (coadjoint) orbit $\mathcal{O}_\lambda:=\operatorname{Ad}^*(G)(\lambda)$ that is naturally a smooth manifold (in fact a so called \textit{homogeneous space}) and isomorphic to the quotient manifold $G/G_\lambda$, where $G_\lambda$ is the \textit{stabilizer} of $\lambda$.
As long as $\lambda$ does not lie on the border of a fundamental Weyl chamber, the stabilizer is simply isomorphic to the maximal Torus $T:=\exp(\mathfrak{h})$ (otherwise $T$ is strictly contained in $G_\lambda$) and we define
\begin{align}
    \mathcal{M}:=\mathcal{O}_\lambda\cong G/G_\lambda.
\end{align}
If we identify\footnote{Since $G$ is compact, $\mathfrak{g}$ has a natural $G$ invariant inner product (the Killing form) providing us with an inner product on $\mathfrak{g}^*$ and we should identify isometrically with respect to the standard inner product on $\mathbb{R}^D$.} $\mathfrak{g}^*\cong \mathbb{R}^D$, we find natural Cartesian embedding functions for $\mathcal{M}$.\\
The restriction of the $G$ action to $\mathcal{M}$ is by definition transitive, while the corresponding infinitesimal action induces a map, mapping Lie algebra elements $X\in\mathfrak{g}$ to vector fields $V_X\in\mathfrak{X}(\mathcal{M})$. Due to the transitivity, the image of a basis of $\mathfrak{g}$ spans the whole tangent bundle, allowing us to uniquely define the 2-form
\begin{align}
    \omega_\lambda(V_X,V_Y):=-\lambda([X,Y])
\end{align}
on $\mathcal{M}$, using the Lie bracket in $\mathfrak{g}$. It turns out that this is a symplectic form that is further invariant under the $G$ action, known as the Kirillov-Kostant-Souriau symplectic form, inducing a $G$ invariant Poisson structure on $\mathcal{O}_\lambda$.

On the other hand, by the theorem of highest weight, we have a unique finite dimensional irreducible representation $\mathcal{H}_\lambda$ of $G$ (coming with a $G$ invariant inner product) with highest weight $\lambda$, which we use as our Hilbert space
\begin{align}
    \mathcal{H}:=\mathcal{H}_\lambda.
\end{align} 

Since both $\mathcal{M}$ and $\mathcal{H}$ are equipped with $G$ actions and invariant inner products, we find induced actions (and invariant inner products) on $\mathcal{C}(\mathcal{M})$ respectively $\operatorname{End}(\mathcal{H})$, and we may decompose each into irreducible representations of $G$.

It then turns out that beneath some cutoff (comparing the Dynkin indices of the respective irreducible representations to the Dynkin indices of $\lambda$) the two algebras are isomorphic and we may construct a quantization map $Q$ that is a $G$ intertwiner and an isometry below this cutoff, where we further define a parameter $\theta$ from the Dynkin indices of $\lambda$ \cite{Steinacker_2016II,Schneiderbauer_2016,Perelomov_1986, Michor_2008, Bernatska_2012}.

Coming back to $SU(2)$ -- the universal covering group of $SO(3)$ -- maximal tori are isomorphic to $U(1)\subset SU(2)$ and all dominant integral elements $\lambda$ are labeled by a single Dynkin index $N=1,2,3,4,\dots$.\\
The stabilizer $G_\mu$ is simply given by $U(1)$, implying $\mathcal{M}\cong SU(2)/U(1)\cong \mathcal{S}^2$. $\mathfrak{su}(3)$ is three dimensional, so the above embedding exactly reproduces the Cartesian embedding $\mathcal{S}^2\hookrightarrow \mathbb{R}^3$ (again up to a scalar depending on $N$). Further the induced Poisson structure simply reproduces equation (\ref{com2}) (up to a scalar depending on $N$).
\\
Further, we find $\mathcal{H}=\mathcal{H}_\lambda\cong\mathbb{C}^N$.
\\
The irreducible representations in the decomposition of $\mathcal{C}(\mathcal{M})$ are spanned by the spherical harmonics $Y^l_m$ of fixed $l$, while in the decomposition of $\operatorname{End}(\mathcal{H})$ they are spanned by the corresponding eigenmodes $\hat{Y}^l_m$ of fixed $l$. The cutoff is then simply given by $N-1$ and we have $\theta=\frac{1}{N}$, while also the definition of $Q$ matches perfectly. Thus we conclude that the fuzzy sphere is a quantized coadjoint orbit of $SU(2)$.

Finally, we find some parallels in the construction of the Moyal-Weyl quantum plane. However the analogy works only to some extent.\\
We might start with the $2n+1$ dimensional Heisenberg group $H_{2n+1}$, being a central extension of the translation group $\mathbb{R}^{2n}$. It is rather easy to see that all coadjoint orbits are isomorphic to $\mathbb{R}^{2n}$ (carrying a natural invariant inner product), since the coadjoint action of the central charge is trivial\footnote{This is compatible with the trivial fact that $\mathbb{R}^{2n}$ is a representation of itself.}. The induced Poisson bracket is then simply given by equation (\ref{com1}) (up to a scalar).
\\
On the other hand, by a more modern version of the Stone-von Neumann theorem \cite{Rieffel_1972}, the Hilbert space $L^2(\mathbb{R}^n)$ is the only unitary irreducible representation of $H_{2n+1}$, while the action of the central charge on $\operatorname{End}(L^2(\mathbb{R}^N))$ is again trivial.\\
$Q$ is then simply chosen to be an intertwiner of the $\mathbb{R}^{2n}$ actions and an isometry, while $\theta$ is only a scale here.

\subsubsection{Coherent States on the Moyal-Weyl Quantum Plane}
\label{MoyalWeylCoherent}

As a good preparation for what is yet to come, we look at a second quantization map for the Moyal-Weyl quantum plane that is based on \textit{coherent} states -- (normalized) states in $\mathcal{H}$ that are \textit{optimally localized}.
Also their generalizations, the \textit{quasi-coherent} states, will be of great importance in the following.

As the \textit{location} of a state $\ket{v}$, we define the expectation value of the $X^a$ (defined in equation (\ref{moyalweylops})) given by $\bra{v}X^a\ket{v}$. The square of the norm of the average deviation is then given by $\sum_a(\bra{v}X^aX^a\ket{v}-(\bra{v}X^a\ket{v})^2)$. (By the Heisenberg uncertainty principle we know that this value is \textit{strictly} positive.)\\
The normalized vector $\ket{v}$ is then said to be \textit{coherent} or \textit{optimally localized} (at $\bra{v}X^a\ket{v}$) if and only if the latter expression is minimal over all normalized states in $\mathcal{H}$.

One then finds a coherent state $\ket{0}$, that is optimally localized at the origin\footnote{This actually makes $\ket{0}$ unique up to a phase.} $x=0$, implying that it is at the same time a (the) state minimizing the simpler expression $\sum_a\frac{1}{2}\bra{v}X^aX^a\ket{v}$ (where the factor $\frac{1}{2}$ is purely conventional), representing the Hamiltonian of an $n$ dimensional harmonic oscillator at the origin.
Now, taking into account the action of the translation group (recall $U_x^{-1} X^a U_x=X^a+x^a\mathbb{1}_\mathcal{H}$), this immediately implies that $\ket{x}:=U_x\ket{0}$ is a coherent state located at $x$ and minimizing the expression $\bra{v}\frac{1}{2}\sum_a(X^a-x^a)(X^a-x^a)\ket{v}$ -- the Hamiltonian of a shifted harmonic oscillator, located at $x$.

For a later use, this motivates the definition of (quasi-)coherent states $\ket{x}$ to be normalized lowest eigenvectors of the operator
\begin{align}
    \label{ham1}
    H_x=\frac{1}{2}\sum_a(X^a-x^a)(X^a-x^a).
\end{align}

Having these states at hand, we can define a new quantization map for the Moyal-Weyl quantum plane, given by
\begin{align}
    Q(f):=\frac{1}{(2\pi)^n}\int_\mathcal{M} \Omega\, f(x)\ket{x}\bra{x},
\end{align}
going under the name of \textit{coherent state quantization}, where $\Omega=\theta^{-2n} dx^1 \wedge \dots\wedge dx^{2n}$.

This quantization map satisfies the axioms (1-4) and (6), while it is not an isometry.
Axiom (1) is fittingly called \textit{completeness relation} as it is equivalent to
\begin{align}
    \frac{1}{(2\pi)^n}\int_\mathcal{M} \Omega \ket{x}\bra{x}=\mathbb{1}_\mathcal{H},
\end{align}
while the intertwiner property (axiom 6) is equivalent to
\begin{align}
    Q(v_k)=c_k V_k,
\end{align}
for some proportionality constants $c_k$ that actually tends to $1$ for $\theta\to 0$, thus the Weyl quantization and the coherent state quantization tend to each other in this limit.

Further, the quasi coherent states provide us with a map going in the other direction -- a \textit{dequantization map}: We define the so called \textit{symbol map} $\operatorname{Sym}:\operatorname{End}(\mathcal{H})\to \mathcal{C}(\mathcal{M})$ via
\begin{align}
    (\operatorname{Sym}(F))(x):=\bra{x}F\ket{x}.
\end{align}
However, at this point it should not be expected that $\operatorname{Sym}$ is inverse to $Q$ \cite{Steinacker_2021,Steinacker_2020}.

This construction generalizes to arbitrary matrix configurations, as we will see in section \ref{QMGsDescription}, although the quasi-coherent states are then no longer strictly optimally localized.

\subsection{Quantum Matrix Geometries and Quasi-Coherent States}
\label{QMGsDescription}

The starting point for our discussion is a so called \textit{matrix configuration}, an ordered set of $D$ Hermitian endomorphisms respectively matrices\footnote{We always identify $\mathcal{H}\cong\mathbb{C}^N$ and $\operatorname{End}(\mathcal{H})\cong\operatorname{Mat}(\mathbb{C}^N)$ in finite dimensions.} $X^a$ acting on a Hilbert space $\mathcal{H}$ of dimension $\operatorname{dim}(\mathcal{H})=N>1$ in an irreducible way\footnote{
This means that the natural action of the matrices on $\operatorname{End}(\mathcal{H})$ (given by $[X^a,\cdot]$) satisfies $[X^a,F]=0\forall a\implies F\propto\mathbb{1}_\mathcal{H}$ in analogy to irreducible Lie algebra actions.}. In the following (except explicitly stated otherwise), $N$ will be finite.\\
(A \textit{matrix model} is then given by an action principle for matrix configurations for fixed $D$ and $N$.)

We may view such configurations as quantum spaces which are equipped with a metric (in the light of section \ref{QG}, the quantum space itself is then given by $\operatorname{End}(\mathcal{H})\cong \operatorname{Mat}_\mathbb{C}(N)$). The heuristic explanation why the $X^a$ represent a metric comes from interpreting the $X^a$ as quantized Cartesian embedding functions for $\mathbb{R}^D$, while one can always pull back the Euclidean metric to manifolds embedded into $\mathbb{R}^D$. Yet, there is an intrinsic and more well defined explanation, based on generalized differential operators.\\
Considering a manifold, a linear differential operator $\partial$ on the algebra of smooth functions is characterized by \textit{linearity} and the satisfaction of the \textit{Leibniz rule} $\partial(fg)=\partial(f)g+f\partial(g)$ for all $f,g$ in the algebra.
Having a Poisson structure, especially every $i\{h,\cdot\}$ for $h\in\mathcal{C}(\mathcal{M})$ is a differential operator.
Thus maps of the form $\hat{\partial}=[H,\cdot]$ for $H\in\operatorname{End}(\mathcal{H})$ provide a nice generalization to quantum linear differential operators\footnote{In fact, as we already interpreted the trace as an integral, the relation $\operatorname{tr}([H,F]G)=-\operatorname{tr}(F[H,G])$ can be interpreted as partial integration where we observe no boundary terms.} since they are linear and satisfy the generalized Leibniz rule $\hat{\partial}(FG)=\hat{\partial}(F)G+F\hat{\partial}(G)$ (see especially footnote \ref{ComProps} on page \pageref{ComProps}). Having this at hand, we can define the so called \textit{matrix Laplacian}
\begin{align}
    \square:=\sum_{a,b} \delta_{ab}[X^a,[X^b,\cdot]],
\end{align}
encoding the quantum version of a metric\footnote{The use of $\square$ instead of $\Delta$ is purely conventional. In principle, $\delta_{ab}$ can be replaced with $\eta_{ab}$ if one wants to work with the Minkowski signature.}.\\
However, this point of view will not play a big role in the following \cite{Steinacker_2021,Steinacker_2016II,Schneiderbauer_2016,Barrett_2022}.

One advantage of matrix configurations and matrix models is that with them one starts directly \textit{on the quantum side} (and one does not need begin with a classical theory that is subject to quantization) what is clearly favorable from a conceptional point of view as we expect nature to be inherently a quantum theory.\\
Further, (finite dimensional) matrix configurations enjoy the generic advantages of quantum spaces as we have seen them in section \ref{QG}: There is a natural (high energy) cutoff in the observables with many consequences. Heuristically\footnote{This topic can be addressed with more rigor in terms of \textit{characters}, see for example \cite{Connes_1995,Barrett_2022}}, the high energy modes are needed to build observables that allow one to resolve small distances, hence the cutoff causes \textit{coarse-graining} which is also reflected in uncertainty relations \cite{Steinacker_2021,Douglas_2001,Barrett_2022}.

Also, one can do (quantum) field theory on matrix configurations, which goes under the name of \textit{noncommutative (quantum) field theory}. For example, we could consider a so called $\Phi^4$ Lagrangian for the field $\Phi\in\operatorname{End}(\mathcal{H})$
\begin{align}
    S[\Phi]=\operatorname{tr}\left(\frac{1}{2}\Phi\square \Phi+\frac{1}{2}m^2 \Phi ^2+\frac{\lambda}{4!}\Phi^4\right),
\end{align}
where $m$ is the \textit{mass} of the field and $\lambda$ is the \textit{coupling constant}.
This can be seen to be the quantization of the well known ordinary $\phi^4$ theory over a classical space.\\
Yet there are great consequences of the noncommutativity: One can quantize the classical dynamics described by the action via \textit{path integrals}. Especially one finds the partition function\footnote{Depending on the signature, one might have to replace $-\mapsto i$ in front of the action, causing the need for additional regularization.}
\begin{align}
    Z[J]:=\int \mathcal{D}\Phi\; e^{-S[\Phi]+\operatorname{tr}[\Phi J]}
\end{align}
for an external current $J$, while the integration runs over all matrices $\Phi\in\operatorname{End}(\mathcal{H})$. However, the measure $\mathcal{D}\Phi$ is here well defined since one simply integrates over a finite dimensional vector space.\\
Due to the natural cutoff, we might hope that no ultraviolet divergences occur, however we are disappointed here: We get a new kind of divergences, where ultraviolet and infrared contributions mix (so called UV/IR mixing) which is not acceptable in view of what we know from ordinary quantum field theory.\\
A heuristic explanation of these effects lies once again in the uncertainty relations: For example for the Moyal-Weyl quantum plane, we have seen that the average deviation of the location is strictly greater than zero, meaning that if we look at short scale effects in one direction, automatically long scale effects come into play in other directions to compensate for the total uncertainty. Evidently, such a theory then turns out to be strongly nonlocal \cite{Steinacker_2016II,Douglas_2001,Minwalla_2000}.

Although this looks catastrophic at first, a possible cure lies just in an important question that is due to be posed: Which matrix configurations should we actually consider?
\\
One answer is the so called IKKT model, a supersymmetric dynamical matrix model that describes ten Hermitian matrices and spinors (what we neglect here for simplicity) via the (simplified) $SU(N)$ invariant action
\begin{align}
S[(X^a)]=\operatorname{tr}\left(\sum_{a,b,a',b'}\eta_{aa'}\eta_{bb'}[X^a,X^b][X^{a'},X^{b'}]\right)
\end{align}
(where $\eta_{ab}$ is the Minkowski metric in $9+1$ dimensions),
preferring \textit{almost commutative} matrix configurations.
Also here, we can quantize the dynamics via a well defined path integral.
The appearance of the number ten is no coincidence as the model is actually strongly related to string theory and may be a nonperturbative formulation of type IIB string theory.
\\
The reason why this model is so interesting is that there are strong suggestions that noncommutative quantum field theories on solutions of the IKKT model do not show UV/IR mixing (this is due to the supersymmetry that relates bosons and fermions) and are supposed to be UV finite \cite{Steinacker_2011,Steinacker_2016,Minwalla_2000,Ishibashi_1997}.

Also, it has to be mentioned that matrix versions of Yang-Mills gauge theories naturally emerge if one considers the dynamics of fluctuations $\mathcal{A}^a$ of a background solution $\Bar{X}^a$ of the IKKT model
\begin{align}
    X^a=\Bar{X}^a+\mathcal{A}^a.
\end{align}
Further, the dynamics of the $X^a$ carries a dynamical behaviour of the implicitly described metric\footnote{The semiclassical limit of the metric is usually called \textit{effective metric} and its inverse is found to be $G^{ab}=\sum_{cd}\theta^{ac}\theta^{bd}g_{cd}$, where the objects on the $(rhs)$ will be introduced in the equations (\ref{hermitianform}) and (\ref{theta}).} leading to so called \textit{emergent gravity} or \textit{emergent geometry}, while there is evidence that through quantum effects, the Einstein-Hilbert action can be recovered in a semiclassical limit \cite{Steinacker_2016II, Steinacker_2015, Steinacker_2016, Steinacker_2020, Grosse_2008}.

Although all the mentioned prospects are highly interesting, here we focus on the construction of a semiclassical limit -- especially a classical manifold $\mathcal{M}$ with additional structure, assuming that we already chose a specific matrix configuration. This construction is based on quasi coherent states and has been introduced in \cite{Steinacker_2021}.
\\
In the optimal case, this limit should be a symplectic manifold $\mathcal{M}\hookrightarrow\mathbb{R}^D$ that is embedded in Euclidean space together with a quantization map $Q$, s.t.
\begin{align}
\label{quantEmb}
    Q(x^a)=X^a
\end{align}
for the Cartesian embedding functions $x^a$ -- however, to which content this is exactly achievable in general shall be discussed in the following.

\subsubsection{Quasi-Coherent States on Quantum Matrix Geometries}
\label{IntroHamiltonian}

The first step to construction of a semiclassical limit for a given matrix configuration $(X^a)$ is to define \textit{quasi-coherent states}.\\
Recalling equation (\ref{ham1}) for the Moyal-Weyl quantum plane, we define the so called \textit{Hamiltonian}
\begin{align}
    \label{Hamiltonian}
     H:\,&\mathbb{R}^D\to\operatorname{End}(\mathcal{H})\\\nonumber
    &\left(x^a\right)\mapsto H_x:=\frac{1}{2}\sum_a\left(X^a-x^a\mathbb{1}\right)^2=\frac{1}{2}\sum_{a,b}\delta_{ab}\left(X^a-x^a\mathbb{1}\right)\left(X^b-x^b\mathbb{1}\right),
\end{align}
where we call $\mathbb{R}^D$ \textit{target space} in analogy to string theory. The expression on the $(rhs)$ makes the use of the Euclidean metric explicit.

One easily verifies that for a given $x$, $H_x$ is positive definite, using the irreducibility\footnote{While positivity is obvious, definiteness follows from this argument: Assume $H_x\ket{\psi}=0$. This implies $(X^a-x^a)\ket{\psi}=0$, but then $[X^a,\ket{\psi}\bra{\psi}]=0$ for all $a$, thus $\ket{\psi}\bra{\psi}\propto \mathbb{1}_\mathcal{H}$. Now, the ranks can only match for $\ket{\psi}=0$ \cite{Steinacker_2021}.}.
This implies the existence of a minimal eigenvalue $\lambda(x)$ with corresponding eigenspace $E_x$.
In the following we often restrict to points in\footnote{In appendix \ref{Appendix:principalbundle} it is shown that $\Tilde{\mathbb{R}}^D$ is open in $\mathbb{R}^D$.} $\Tilde{\mathbb{R}}^D=\{x\in\mathbb{R}^D:\operatorname{dim(E_x)=1}\}$, where we find a corresponding normalized eigenvector $\ket{x}\in E_x\subset\mathcal{H}$ that is unique up to a complex phase. (At this point this may seem restrictive, but actually this is a core feature that is necessary to reproduce the appropriate topology as we will see for the example of the fuzzy sphere in section \ref{FuzzySphere}.)
This state we call \textit{quasi-coherent state} (at $x$).
Loosely speaking, we have defined a map $\ket{\cdot}:\Tilde{\mathbb{R}}^D\to \mathcal{H}$, yet in general we can not find a smooth global phase convention \cite{Steinacker_2021}.

Sometimes, we are interested in the full eigensystem of $H_x$, for what we introduce the notation $H_x\ket{k,x}=\lambda^k(x)\ket{k,x}=\lambda_{k,x}\ket{k,x}$ (depending on the purpose), assuming $\lambda^k(x)\leq\lambda^l(x)$ for $k<l$.

\subsubsection{The Bundle Perspective}
\label{bp}

In a more abstract language, we have defined a \textit{principal fiber bundle}
\begin{align}
    p:\mathcal{B}\subset\Tilde{\mathbb{R}}^D\times\mathcal{H}\to\Tilde{\mathbb{R}}^D\subset\mathbb{R}^D
\end{align}
with standard fiber $U(1)$ over the open $\Tilde{\mathbb{R}}^D\subset\mathbb{R}^D$. For a detailed discussion, see the appendices \ref{Appendix:smoothdependence} and \ref{Appendix:principalbundle}.
In this picture, we should regard $x\mapsto\ket{x}$ as a \textit{smooth local section} of the bundle, existing locally around any point in $\Tilde{\mathbb{R}}^D$. (If we want to stress the fact that we consider the local section and not a single quasi-coherent state, we write $\kets{\cdot}$ respectively $\kets{x}$ if we evaluated it at some $x$.)\\
The potentially from $\Tilde{\mathbb{R}}^D$ missing points (defining the set $\mathcal{K}:=\mathbb{R}^D\backslash\Tilde{\mathbb{R}}^D$) may prevent the bundle from being trivial or equivalently from admitting global smooth sections.

Now, we define the natural \textit{connection 1-form}
\begin{align}
    \label{connection}
    iA=\bras{x}d\kets{x},
\end{align}
where $A$ is real. This provides us with the \textit{gauge covariant derivative operator}
\begin{align}
    D:=d-iA,
\end{align}
acting on sections.
Again, for a detailed discussion see appendix \ref{Appendix:principalbundle}.\\
Under a \textit{gauge transformation}
\begin{align}
    \kets{x}\mapsto e^{i\phi(x)}\kets{x}
\end{align}
for some local smooth real-valued function $\phi$, we find the transformation behaviour of the connection
\begin{align}
    iA\mapsto i(A+d \phi),
\end{align}
while the gauge covariant derivative transforms as its name suggests:
\begin{align}
    D \kets{x}\mapsto e^{i\phi(x)}D \kets{x}.
\end{align}
Further, we find the \textit{field strength}\footnote{The factor $\frac{1}{2}$ is rather an accident here and is only introduced for consistency. Later results suggest that the choice $\omega=-dA$ would be optimal. One could then redefine $h_{ab}=\frac{1}{2}(g_{ab}-i\omega_{ab})$ in equation (\ref{hermitianform}) to circumvent all factors and adapt for the changes in the following.}
\begin{align}
    \omega=\frac{1}{2} dA.
\end{align}

Form here on, it is more practicable to switch to index notation.\\
Using the gauge covariant derivative, we find the gauge invariant Hermitian form
\begin{align}
    \label{hermitianform}
    h_{ab}:=((\partial_a-iA_a)\ket{x})^\dagger(\partial_b-iA_b)\ket{x}=:g_{ab}+i\omega_{ab}
\end{align}
(turning out to be the pullback of a canonical $U(1)$ invariant bundle metric on $T\mathcal{B}$)
that decomposes into the real and symmetric (possibly degenerate) \textit{quantum metric}
\begin{align}
    g_{ab}=\frac{1}{2}\left((\partial_a\bra{x})\partial_b\ket{x}+(\partial_b\bra{x})\partial_a\ket{x}-2A_a A_b\right)
\end{align}
and the real and antisymmetric \textit{would-be symplectic form}
\begin{align*}
    \omega_{ab}=\frac{1}{2i}\left((\partial_a\bra{x})\partial_b\ket{x}-(\partial_b\bra{x})\partial_a\ket{x}\right)=\frac{1}{2}\left(\partial_a A_b-\partial_b A_a\right)=\frac{1}{2}(dA)_{ab}
\end{align*}
\cite{Steinacker_2021}. For a detailed discussion, look at appendix \ref{Appendix:mom}.

\subsubsection{The Hermitian Form from Algebraic Considerations}
\label{algTricks}

The Hamiltonian (\ref{Hamiltonian}) is of extremely simple form, allowing us to compute $h_{ab}$ entirely algebraically without performing any explicit derivation. This will be very useful for the implementation on a computer in section \ref{implementation}.\\
Let $x\in\Tilde{\mathbb{R}}^D$. As a beginning we find\footnote{Here, we do not distinguish between upper and lower indices since we work with the Euclidean metric.} $\partial_a H_x=-(X^a-x^a\mathbb{1})$. Thus
\begin{align}
    (H_x-\lambda(x))\partial_a\ket{x}=\partial_a((H_x-\lambda(x))\ket{x})-\partial_a(H_x-\lambda(x))\ket{x}=\left(X^a-x^a+\partial_a\lambda(x)\right)\ket{x}, \label{derofh}
\end{align}
where we used the eigenvalue equation in the second step and inserted the derivative
of $H_x$ in the third step.
The $(lhs)$ is (again by the eigenvalue equation) orthogonal to $\ket{x}$, thus we get
\begin{align}
    \label{constraint}
    0=\bra{x}X^a\ket{x}-x^a+\partial_a\lambda(x).
\end{align}
Since the multiplicity of the eigenvalue $\lambda(x)$ equals one, we find by the spectral theorem
\begin{align}
    H_x-\lambda(x)=\sum_{k=2}^N  (\lambda^k(x)-\lambda(x))\ket{k,x}\bra{k,x}.
\end{align}
Now, we define the \textit{pseudoinverse} of $H_x-\lambda(x)$
\begin{align}
    (H_x-\lambda(x))^{-1'}:=\sum_{k=2}^N  \frac{\ket{k,x}\bra{k,x}}{\lambda^k(x)-\lambda(x)},
\end{align}
satisfying
\begin{align}
    (H_x-\lambda(x))^{-1'}(H_x-\lambda(x))=(H_x-\lambda(x))(H_x-\lambda(x))^{-1'}=\mathbb{1}-\ket{x}\bra{x}.
\end{align}
Applying this operator to (\ref{derofh}), we find
\begin{align}
    (\partial_a-iA_a)\ket{x}&=(\mathbb{1}-\ket{x}\bra{x})\partial_a\ket{x}\\\nonumber
    &=(H_x-\lambda(x))^{-1'}\left(X^a-x^a+\partial_a\lambda(x)\right)\ket{x},
\end{align}
where we used the definition of the connection 1-form \cite{Steinacker_2021}.
Since by definition $(H_x-\lambda(x))^{-1'}\ket{x}=0$, this simplifies even more and we arrive at
\begin{align}
    \label{AlgTrick}
    (\partial_a-iA_a)\ket{x}=(H_x-\lambda(x))^{-1'}X^a\ket{x}=:\mathfrak{X}^a_x\ket{x},
\end{align}
with the newly introduced operator $\mathfrak{X}^a_x:=(H_x-\lambda(x))^{-1'}X^a$ that is completely independent of any derivatives.

This allows us to calculate $h_{ab}$ complete algebraically as
\begin{align}
    h_{ab}(x)=\bra{x}(\mathfrak{X}^a_x)^\dagger \mathfrak{X}^b_x\ket{x},
\end{align}
and consequently $g_{ab}$ and $\omega_{ab}$.

\subsubsection{Algebraic Constraints on the Quasi-Coherent States}
\label{NewAlgConstr}

Again, due to the simple form of the Hamiltonian (\ref{Hamiltonian}), we can formulate nontrivial relations between its spectrum at different points.\\
We start by defining the equivalence relation $x\sim y:\iff E_x=E_y$ for all points $x,y\in\Tilde{\mathbb{R}}^D$ and label the corresponding equivalence classes by $\mathcal{N}_x:=[x]=\{y\in\Tilde{\mathbb{R}}^D\vert E_x=E_y\}$ that we call \textit{null spaces}.

A direct calculations shows
\begin{align}
    H_x=H_y+\frac{1}{2}\left(\vert x\vert^2-\vert y\vert^2\right)\mathbb{1}-\sum_a(x^a-y^a)X^a. \label{HPerturbation}
\end{align}
Assume now that $x\sim y$, then this implies $\sum_a (x-y)^a X^a\ket{x}\propto \ket{x}$ (since $\ket{x}$ is an eigenvector of all other terms) and applying $(H_x-\lambda(x))^{-1'}$  from the left, we find $\sum_a(x-y)^a\mathfrak{X}^a\ket{x}=0$ \cite{Steinacker_2021}.

Similarly, we find the relation
\begin{align}
    \label{HStraightLines}
    H_{(1-\alpha)x+\alpha y}=(1-\alpha)H_x+\alpha H_y+\frac{\alpha^2-\alpha}{2}\vert x-y\vert^2\mathbb{1}\quad \forall\alpha\in\mathbb{R}.
\end{align}
Considering again the case $x\sim y$, this shows that also all points on the straight line segment (with respect to the Euclidean metric) between $x$ and $y$ lie in the equivalence class $\mathcal{N}_x$: Since $x\sim y$ we have $E_x=E_y$, but then at $(1-\alpha)x+\alpha y$ for $\alpha\in[0,1]$ the subspace $E_x$ remains the lowest eigenspace.\\
Following this line further $\ket{x}$ remains an eigenvector, so either the line hits $\mathcal{K}$ and beyond another lowest eigenvector takes the place or $\ket{x}$ remains the lowest eigenvector until infinity.\\
This yields two important results:
\begin{enumerate}
    \item $\mathcal{N}_x$ is convex,
    \item $\mathcal{N}_x$ is closed in $\Tilde{\mathbb{R}}^D$.
\end{enumerate}

Since this implies that $\mathcal{N}_x$ is a submanifold of $\Tilde{\mathbb{R}}^D$, it makes sense to consider $T_x\mathcal{N}_x\subset T_x\Tilde{\mathbb{R}}^D$.\\
Further, any $v^a\in T_x\mathcal{N}_x$ is proportional to $(x-y)^a$ for some $y\in\mathcal{N}_x$, thus
\begin{align}
    \sum_a v^a \mathfrak{X}^a\ket{x}=0\quad\forall v\in T_x\mathcal{N}_x.
\end{align}
Our considerations from section \ref{algTricks} then imply that $T_x\mathcal{N}_x$ lies in the kernel of $h_{ab}(x)$ and consequently of $g_{ab}(x)$ and $\omega_{ab}(x)$ \cite{Steinacker_2021}.

\subsubsection{The Manifold Perspective}
\label{mfp}

In section \ref{bp}, we have seen that the quasi-coherent states define a fiber bundle over $\Tilde{\mathbb{R}}^D$. An alternative viewpoint is to consider the set of all quasi-coherent states of the matrix configuration (identifying states that discern only in a phase) as a subset of complex projective space
\begin{align}
   \mathcal{M}':=\cup_{x\in\Tilde{\mathbb{R}}^D}U(1)\ket{x}/U(1)
   \cong\left\{E_x\vert x\in\Tilde{\mathbb{R}}^D\right\}\subset\mathbb{C}P^{N-1},
\end{align}
under the identification $\mathcal{H}\cong \mathbb{C}^N$.\\
Locally, a smooth section $\kets{\cdot}$ of $\mathcal{B}$ defines a smooth map\footnote{Actually, we should consider the natural smooth projection $p:\mathbb{C}^N\to\mathbb{C}P^{N-1}$ and consequently $q_s:=p\circ
\kets{\cdot}$.} $q_s:=U(1)\kets{\cdot}:U\subset\Tilde{\mathbb{R}}^D\to\mathcal{M}'$. Since all sections only deviate in a $U(1)$ phase, all $q_s$ assemble to a global smooth and surjective map $q:\Tilde{\mathbb{R}}^D\to\mathcal{M}'$.\\
Since all our definitions fit nicely together, $q$ descends to a bijection $\underline{q}:\Tilde{\mathbb{R}}^D/\sim\to\mathcal{M}'$ \cite{Steinacker_2021}.

We now want to know whether $\mathcal{M}'$ has the structure of a smooth manifold.
It is tempting to use the map $q$ to construct local coordinates, however this only has a chance if $q$ has constant rank. 
As we will see later, this is not the case in general (look for example at section \ref{SFuzzySphere}), so instead we have to look at the subset of $\Tilde{\mathbb{R}}^D$ where the rank of $q$ is maximal.\\
We define $k=\max_{x\in\Tilde{\mathbb{R}}^D}\operatorname{rank}(T_xq)$. This allows us to define
$\hat{\mathbb{R}}^D:=\{x\in\Tilde{\mathbb{R}}^D\vert \operatorname{rank}(T_xq)=k\}$.
One can easily show that $\hat{\mathbb{R}}^D$ is open\footnote{See for example the discussion of definition 2.1 in \cite{Michor_2008}.}.
\\
Consequently, we define $\mathcal{M}:=q(\hat{\mathbb{R}}^D)$ and use the same letter $q$ for the restriction to $\hat{\mathbb{R}}^D$.

Then we find (using the constant rank theorem and the results from section \ref{NewAlgConstr}) that $\mathcal{M}$ is a smooth immersed submanifold of $\mathbb{C}P^{N-1}$ of dimension $k$ what we will call \textit{quantum manifold} or \textit{abstract quantum space}, being a candidate for the semiclassical limit of the given matrix configuration. For a detailed discussion see appendix \ref{Appendix:manifold}.

Especially, we have 
\begin{align}
    \operatorname{ker}(T_xq)=T_x\mathcal{N}_x,
\end{align}
and more importantly
\begin{align}
    \label{Tanq}
    T_x q \cdot \partial_a \cong D_a\kets{x}
\end{align}
and consequently
\begin{align}
    T_{q(x)}\mathcal{M}\cong \langle D_a\kets{x} \rangle_\mathbb{R}.
\end{align}

Further, we can pull back the Fubini–Study metric and the Kirillov-Kostant-Souriau symplectic form along the immersion $\mathcal{M}\hookrightarrow\mathbb{C}P^{N-1}$ what we (up to a scale) call $g_\mathcal{M}$ (a Riemannian metric) respectively $\omega_\mathcal{M}$ (a closed 2-form), reproducing exactly the $g_{ab}$ and the $\omega_{ab}$ if further pulled back along $q$ to $\Tilde{\mathbb{R}}^D$ \cite{Steinacker_2021}.\\
From that, we know that the kernel of $g_{ab}(x)$ coincides with the kernel of $T_xq$ (since $g_\mathcal{M}$ is nondegenerate), but we only know that the kernel of $T_xq$ lies within the kernel of $\omega_{ab}(x)$, while there might be even more degeneracy. For a detailed discussion once more see appendix \ref{Appendix:manifold}.

Finally, we define the set $\Tilde{\mathcal{M}}:=\{(\bra{x}X^a\ket{x})\vert U(1)\ket{x}\in\mathcal{M}'\}\subset\mathbb{R}^D$ and call it \textit{embedded quantum space}. In the context of the IKKT model which is strongly related to type IIB string theory, this has the interpretation as a \textit{brane} in target space. In general, this will not be a manifold\footnote{It may have self intersections or other peculiarities as we will see in section \ref{csfc}.} yet it is more accessible than $\mathcal{M}$ from an intuitive point of view, being a subset of $\mathbb{R}^D$. This embedded quantum space $\Tilde{\mathcal{M}}\hookrightarrow \mathbb{R}^D$ exactly represents the candidate for the space for which we wish equation (\ref{quantEmb}) to hold \cite{Steinacker_2021}.

\subsubsection{Properties of the Quasi Coherent States}

In the previous, we have seen a few constructions based on quasi-coherent states but have not discussed their actual meaning, except for the coherent states of the Moyal-Weyl quantum plane.

First, we consider the expectation values of the $X^a$ in the quasi-coherent state at $x\in\Tilde{\mathbb{R}}^D$
\begin{align}
    \label{xBold}
    \mathbf{x}^a(x):=\bra{x}X^a\ket{x} \quad \text{resp.}\quad \mathbf{x}^a(U(1)\ket{x}):=\bra{x}X^a\ket{x}
\end{align}
(depending on the context),
providing us with a point in $\Tilde{\mathcal{M}}$ for some given point in $\Tilde{\mathbb{R}}^D$ respectively $\mathcal{M}'$. This should be thought of as the \textit{location} associated to the quasi-coherent state $\ket{x}$.\\
Especially we might think of the map $U(1)\ket{x}\mapsto \mathbf{x}^a(x)$ as the Cartesian embedding functions\footnote{Although in general this map will not be a topological embedding or even an immersion.}  that map $\mathcal{M}$ into $\mathbb{R}^D$, subject to quantization as in equation (\ref{quantEmb}).
\\
For later use, we note
\begin{align}
\label{partialEmbedd}
    \partial_a\mathbf{x}^b(x)&=\bra{x}X^b(D_a+iA_a)\ket{x}+(\bra{x}X^b(D_a+iA_a)\ket{x})^*=\bra{x}X^b\mathfrak{X}^a\ket{x}+(\bra{x}X^b\mathfrak{X}^a\ket{x})^*=\\\nonumber
    &=2\bra{x}X^{(b}(H_x-\lambda(x))^{-1'}X^{a)}\ket{x}=\bra{x}X^b\mathfrak{X}^a\ket{x}+\bra{x}X^a\mathfrak{X}^b\ket{x},
\end{align}
where in the second step the contributions proportional to $A_a$ cancel and we used equation (\ref{AlgTrick}) \cite{Steinacker_2021}.

Based on this interpretation, it makes sense to consider the following two \textit{quality measures},
namely we define the \textit{displacement}
\begin{align}
    d^2(x):=\sum_a(\mathbf{x}^a(x)-x^a)^2
\end{align}
and the \textit{dispersion}
\begin{align}
    \delta^2(x):=\sum_a(\Delta X^a)^2,
\end{align}
for
\begin{align}
    (\Delta X^a)^2:=\bra{x}(X^a-\mathbf{x}^a(x))^2\ket{x}=\bra{x}X^aX^a\ket{x}-\mathbf{x}^a(x)\mathbf{x}^a(x).
\end{align}

Then the displacement measures how far the location of $\ket{x}$ is away from $x$ itself, while the dispersion measures how well the state $\ket{x}$ is localized at $\mathbf{x}^a(x)$ via the standard deviation. So, the \textit{coherency} of a state is the better the smaller both the displacement and the dispersion are (one directly generalizes the measures to arbitrary states of norm one).

Now, we find
\begin{align}
    \label{measuressum}
    \delta^2(x)+d^2(x)&=\sum_a\left(\bra{x}X^aX^a\ket{x}-\mathbf{x}^a(x)\mathbf{x}^a(x)+(\mathbf{x}^a(x)-x^a)^2\right)\nonumber\\
    &=2\bra{x}\frac{1}{2}\left(\sum_a(X^a-x^a)^2\right)\ket{x}=2\bra{x}H_x\ket{x}=2\lambda(x),
\end{align}
but this means that if $\lambda(x)$ is small also the dispersion and the displacement are small.\\
Choosing the lowest eigenstate of $H_x$ by definition means to minimize the sum of the quality measures, thus choosing the state of most optimally coherence -- the quasi-coherent state \cite{Steinacker_2021}. Finally, we note that if $\operatorname{dim}(\mathcal{N}_x)>0$ we automatically find by equation (\ref{measuressum}) that $\lambda$ can not be constant on $\mathcal{N}_x$ (since the displacement changes while the dispersion remains constant) thus in general we should not expect that $\lambda$ is small for all $x\in\Tilde{\mathbb{R}^D}$. In fact we will see that for the fuzzy sphere $\lambda_x\mapsto\infty$ for $\vert x \vert\to \infty$.

\subsubsection{Quantization and Dequantization}
\label{QuantMap}

Now, we turn to a very important aspect of the construction. Our intention was to find a semiclassical geometry, corresponding to a given matrix configuration for what we defined the quantum manifold $\mathcal{M}$ as a candidate.\\
Therefore, we would like to define a quantization map $Q$ from $\mathcal{C}(\mathcal{M})$ to $\operatorname{End}(\mathcal{H})$. However, so far it is necessary to assume that $\omega_\mathcal{M}$ is nondegenerate\footnote{This directly implies that $k=\operatorname{dim}(\mathcal{M})$ is even.} (thus symplectic, inducing a Poisson structure on $\mathcal{M}$) and $\mathcal{M}$ is compact, allowing us to integrate over the latter with the induced volume form $\Omega_\mathcal{M}:=\frac{1}{(k/2)!}\omega_\mathcal{M}^{\wedge k/2}$.
We will deal with the fact that this is not the case in general in section \ref{fol}.

We now define\footnote{This is meant in the sense $(\ket{\cdot}\bra{\cdot})(U(1)\ket{x}):=\ket{x}\bra{x}$.} the \textit{would-be quantization map} (paralleling the idea of the coherent state quantization for the Moyal-Weyl quantum plane in section \ref{MoyalWeylCoherent})
\begin{align}
    \label{quantmapCS}
    \mathcal{Q}:\; &\mathcal{C}(\mathcal{M})\to\operatorname{End}(\mathcal{H})\nonumber\\
    &\phi\mapsto \frac{\alpha}{(2\pi)^{k/2}}\int_\mathcal{M}\Omega_\mathcal{M} \,\phi\ket{\cdot}\bra{\cdot},
\end{align}
where $\alpha$ is chosen such that $\frac{\alpha}{(2\pi)^{k/2}}V_\omega=N$, introducing the \textit{symplectic volume} $V_\omega:=\int_\mathcal{M}\Omega_\mathcal{M}$ of $\mathcal{M}$.

On the other hand, we can define a dequantization map\footnote{The map is obviously independent of the chosen $U(1)$ phase and actually smooth as we can locally express it via smooth local sections $\kets{\cdot}$.} that we call \textit{symbol map} via
\begin{align}
    \operatorname{Symb}:\; &\operatorname{End}(\mathcal{H})\to\mathcal{C}(\mathcal{M})\nonumber\\
    &\Phi\mapsto \bra{\cdot}\Phi\ket{\cdot},
\end{align}
noting that for example $\mathbf{x}^a$ is the symbol of $X^a$ \cite{Steinacker_2021}.

Then there are a few things that we might conjecture or hope to find, at least approximately (especially that $Q$ satisfies the axioms of a quantization map in section \ref{QG}).
\newpage
\begin{enumerate}
    \item We would appreciate to find $Q\overset{?}{=}\operatorname{Symb}^{-1}$ (in the weak sense $\operatorname{Symb}\circ Q=id_{\mathcal{C}(\mathcal{M})}$).
    \item This would be plausible if $\vert\braket{x\vert y}\vert^2$ behaved somehow like a delta distribution or at least as a strongly peaked Gaussian.
    \item Considering the introduction of this section, we might conjecture equation (\ref{quantEmb}) to be true in the sense
    \begin{align}
        \label{quant}
        X^a\overset{?}{=}Q(\mathbf{x}^a)\propto \int_\mathcal{M}\Omega_\mathcal{M} \,\mathbf{x}^a\ket{\cdot}\bra{\cdot}.
    \end{align}
    (We have already seen that we should think of the $\mathbf{x}^a$ as would-be Cartesian embedding functions embedding $\mathcal{M}$ into $\mathbb{R}^D$.)
    \item In order to satisfy the first axiom, we need the \textit{completeness relation}
    \begin{align}
        \label{compl}
        \mathbb{1}_\mathcal{H}\overset{?}{=}\mathcal{Q}(1_\mathcal{M})\propto \int_\mathcal{M}\Omega_\mathcal{M} \,\ket{\cdot}\bra{\cdot}
    \end{align}
    to hold. Then the condition on $\alpha$ ensures that this equation is consistent when taking its trace. 
    \item Luckily, axiom two as well as linearity are satisfied by definition.
    \item The first part of axiom three is again plausible if $\braket{x\vert y}$ were similar to the delta distribution.
    \item The second part of axiom three becomes plausible if additionally $\{\mathbf{x}^a,\mathbf{x}^b\}\overset{?}{=}\frac{1}{i}\bra{x}[X^a,X^b]\ket{x}=:\theta^{ab}(U(1)\ket{x})$ for the Poisson structure induced by $\omega_{\mathcal{M}}$, assuming we can establish $Q\overset{?}{=}\operatorname{Symb}^{-1}$ as well as $X^a\overset{?}{=}Q(\mathbf{x}^a)$.\\
    (In any local coordinates for $\mathcal{M}$, we can calculate the Poisson bracket as\\ $\sum_{\mu,\nu}(\omega_{\mathcal{M}}^{-1})^{\mu\nu}\partial_\mu\mathbf{x}^a \partial_\nu\mathbf{x}^b=\{\mathbf{x}^a,\mathbf{x}^b\}\overset{?}{=}\theta^{ab}$.)
    \item Now, if all $X^a$ lie in the image of $Q$, axiom four is satisfied by definition.
\end{enumerate}

The first conjecture cannot hold true in the strong sense in general, since already in simple examples, $Q$ has an \textit{ultraviolet cutoff} (strongly oscillating functions are mapped to zero) and thus is not injective. However, in a certain regime in $\mathcal{C}(\mathcal{M})$ respectively $\operatorname{End}(\mathcal{H})$ the maps $Q$ and $\operatorname{Symb}$ can be shown to be approximately mutually inverse. So, we would hope that the $X^a$ themselves lie in that regime. On the other hand, the result\footnote{$\vert v\vert_g^2:=\sum_{a,b}v^av^b g_{ab}$.} $\vert \braket{x\vert y}\vert^2\approx e^{-\vert x-y\vert_g^2}$ supports the second conjecture.
\\
It further turns out that if $X^a$ itself lies in the regime, $-2\sum_c\omega_{ac}\theta^{cb} \approx p_a^b\approx \partial_a \mathbf{x}^b$ (this is the first evidence for the suggested redefinition $\omega_{ab}\mapsto -\frac{1}{2}\omega_{ab}$), where $p$ is a rank $k$ projector with kernel $T_x\mathcal{N}_x$, under the assumptions that the matrix configuration is \textit{almost commutative} (especially that the commutators $[X^a,X^b]$ are small compared to the $X^a$ in some norm\footnote{Such configurations are especially favored by the IKKT model.}, which can often be achieved by choosing $N$ large) and that the $\partial_a\lambda$ are small\footnote{This especially means that $\mathbf{x}^a\approx x^a$.} \cite{Steinacker_2021}.

We will thus focus in sections \ref{implementation} and \ref{RESULTS} on verifying the completeness relation, the quantization of the embedding functions and the compatibility of the Poisson structure induced by $\omega_\mathcal{M}$ with the would-be Poisson structure induced by $-2\theta^{ab}$
\begin{align}
    \label{CompPoisson}
    \{\mathbf{x}^a,\mathbf{x}^b\}\overset{?}{=}-2\theta^{ab}
\end{align}
via numerical computations.

\subsubsection{Comparison of the Different Structures}
\label{comparison}

We have already discussed the objects $g_{ab}$ and $\omega_{ab}$, but there are further structures available. We defined the target space together with the Euclidean metric $\delta_{ab}$ and in the last section we introduced the real antisymmetric object\footnote{Note that the components depend smoothly on $x$ and that the definition is independent of the choice of a local section.}
\begin{align}
    \label{theta}
    \theta^{ab}(x):=\frac{1}{i}\bra{x}[X^a,X^b]\ket{x} \quad \text{resp.} \quad  \theta^{ab}(U(1)\ket{x}):=\frac{1}{i}\bra{x}[X^a,X^b]\ket{x},
\end{align}
what we might call \textit{(semiclassical) Poisson tensor} since (in the light of the previous section we want to think of it as $\{\mathbf{x}^a,\mathbf{x}^b\}$ respectively as the dequantization of $\frac{1}{i}[X^a,X^a]$.

However, there is some caveat: In some sense, $g_{ab}$ and $\omega_{ab}$ are tied to the quantum manifold, while $\delta^{ab}$ and $\theta^{ab}$ are supposed to be viewed on target space.\\ 
To understand this, we note that the first two are well defined on $\mathcal{M}$ since they are pullbacks of $g_\mathcal{M}$ and $\omega_\mathcal{M}$ (as discussed in section \ref{mfp}) while they need not have to be constant on $\mathcal{N}_x$ (equation (\ref{explicitg}) shows this explicitly for the fuzzy sphere) -- what is somehow peculiar and inconsistent viewed on the target space. On the other hand, $\delta_{ab}$ and $\theta^{ab}$ are constant on $\mathcal{N}_x$, but cannot be pushed forward to $\mathcal{M}$ consistently -- in general $T_x\mathcal{N}_x$ does not even necessarily lie within the kernel of $\theta^{ab}$.
However, the component functions can be pushed forward to $\mathcal{M}$ which is evident from equation (\ref{theta}).

There is another structure on $\mathcal{M}$ that we might consider -- the $(1,1)$-tensor field $J_\mathcal{M}$ that is defined via $g_\mathcal{M}(\eta,J_\mathcal{M}(\xi))=\omega_\mathcal{M}(\eta,\xi)$ for all vector fields $\xi,\eta$. If now $J_\mathcal{M}^2=-P_\mathcal{M}$ (for some projector $P_\mathcal{M}=P_\mathcal{M}^2$) is satisfied, we call $\mathcal{M}$ \textit{almost Kähler}\footnote{Note that this is a nonstandard notion. If $\omega_\mathcal{M}$ is additionally nondegenerate, $\mathcal{M}$ gets truly Kähler.}. Namely, one then has a metric, a would-be symplectic structure and a would-be complex structure that are \textit{compatible} \cite{Steinacker_2021}.

\subsection{Examples of Quantum Matrix Geometries and Analytic Results}
\label{qmgexamples}

Having discussed quite a few constructions built on quasi-coherent states, it is time for examples that actually relate directly to the examples of section \ref{QG}, now viewed as matrix configurations.\\
We begin with the Moyal-Weyl quantum plane from section \ref{moyalweyl2}, followed by the fuzzy sphere from section \ref{FuzzySphere}.
An important new example is the \textit{squashed fuzzy sphere} discussed in section \ref{SFuzzySphere}, a perturbed version of the ordinary (round) fuzzy sphere.
Then, in section \ref{CoadjointOrbits} we look at coadjoint orbits, and as an example thereof the fuzzy $\mathbb{C}P^2$ in section \ref{fuzzycp2} -- the $SU(3)$ equivalent of the fuzzy sphere.
In section \ref{random} we end with random matrix configurations.

\subsubsection{The Moyal-Weyl Quantum Plane as a Matrix Configuration}
\label{moyalweyl2}

The Moyal-Weyl quantum plane can be reformulated as a matrix configuration. Since it guided us to the definition and use of quasi-coherent states via the Hamiltonian (\ref{Hamiltonian}) this is not surprising.
We define the matrix configuration
\begin{align}
    \mathbb{R}^{2n}_\theta:=(X^1,\dots,X^{2n})
\end{align}
via the operators from equation (\ref{moyalweylops}), meaning our target space is $\mathbb{R}^D$ for $D=2n$. This is the one and only time that we actually deal with an infinite dimensional matrix configuration ($N=\infty$), meaning that not all arguments from the last section remain valid. Still, we can reproduce most results due to the simple structure of the geometry.

In section \ref{MoyalWeylCoherent}, we have already identified the (quasi-)coherent states $\ket{x}=U_x\ket{0}$, where we used the representation of $H_{2n+1}$ on $\mathcal{H}$ and thus find $\Tilde{\mathbb{R}}^D=\mathbb{R}^D$. It is further obvious that the $\mathcal{N}_x$ are zero dimensional. After calculating $\partial_a\ket{x}$ via the Baker-Campbell-Hausdorff formula one concludes that $T_xq$ has full rank for all $x\in\mathbb{R}^D$. Thus we have $\hat{\mathbb{R}}^D=\mathbb{R}^D$ and $\mathcal{M}=H_{2n+1}\ket{0}/U(1)\cong \mathbb{R}^{2n}$.

Calculating\footnote{Actually, we should not confuse the semiclassical Poisson tensor with the Poisson tensor $\theta^{ab}$ that we introduced in section \ref{MoyalWeylCoherent}. However, it turns out that both coincide.} $\theta^{ab}$, $A^a$, $g_{ab}$ and $\omega_{ab}$ (the latter reproduces the Poisson structure from equation (\ref{com1}) up to  a scale) is a fairly simple task but not very illuminating, so we stop at this point (after noting that $Q$ simply reproduces the coherent state quantization map) and continue with the more interesting fuzzy sphere \cite{Steinacker_2021}.

\subsubsection{The Fuzzy Sphere as a Matrix Configuration}
\label{FuzzySphere}

We start with the three Hermitian matrices $J^a_N$ for $a=1,2,3$ and $N\geq 2$ -- the usual orthonormal generators of the $N$ dimensional irreducible representation of $\mathfrak{su}(2)$ -- that satisfy $[J_N^a,J_N^b]=\sum_ci\epsilon^{abc}J_N^c$ and $\sum_aJ_N^aJ_N^a=\frac{N^2-1}{4}\mathbb{1}_N$. The explicit construction can be found in appendix \ref{su2rep}, together with a quick overview on the relevant related quantities.\\
Naively, the last equation can be viewed as \textit{fixing the radius}, motivating us to normalize the matrices according to
\begin{align}
    X^a:=\frac{1}{C_N}J_N^a\;\forall a,\quad C_N:=\sqrt{\frac{N^2-1}{4}}=\sqrt{j(j+1)},
\end{align}
leaving us with the new relations
\begin{align}
    \label{theXmatrices}
    [X^a,X^b]=\sum_c\frac{i}{C_N}\epsilon^{abc}X^c,\quad \sum_aX^aX^a=\mathbb{1},
\end{align}
while we note that we have already seen these matrices in section \ref{FuzzySphere0}.

Now, the matrix configuration of the fuzzy sphere of degree $N$ is simply defined as the ordered set
\begin{align}
    \label{matricesfs}
    S^2_N:=\left(X^1,X^2,X^3\right).
\end{align}
This means, our target space is the $\mathbb{R}^3$ and our Hilbert space is $N$ dimensional, identified with $\mathbb{C}^N$.

Due to equation (\ref{theXmatrices}), the Hamiltonian has the simple form
\begin{align}
    H_x=\frac{1}{2}\left(1+\vert x\vert^2\right)\mathbb{1}-\sum_{a=1}^3x^aX^a,
\end{align}
so the quasi-coherent state at $x$ is given by the maximal eigenstate of $\sum_ax^aX^a$.
As a first consequence, we note that $0\in\mathcal{K}$, since there all eigenvalues of $H_x$ equal one half. Also, a positive rescaling of $x$ does not change the eigenspaces and the ordering of the eigenvalues, thus $\ket{x}=\ket{\lambda x}$ for $\lambda>0$ and especially $\lambda x\in\mathcal{N}_x$.

Now, using the adjoint representation of $SU(2)$ (namely $SO(3)$) allows us to calculate $\ket{x}$ elegantly.
Let $x=\vert x\vert R_x^{-1}\cdot \hat{e}_3$ for some appropriate rotation\footnote{Evidently, this can be done for all $x\in\mathbb{R}^3$ but there is no unique choice.} $R_x\in SO(3)$. Then we find $R_x=\operatorname{Ad}(U_x)$ for some\footnote{Actually, for each $R_x$ there are two $U_x$ deviating only in a sign since $SU(2)$ is the double cover of $SO(3)$.} $U_x\in SU(2)$.\\
By the orthogonality of $R_x$ we find
\begin{align}
    \sum_a x^aX^a&=\sum_a\left(\vert x\vert R_x^{-1}\cdot\hat{e}_3\right)^aX^a=\sum_a\left(\vert x\vert \hat{e}_3\right)^a\left(R_x\cdot X^a\right)\\\nonumber
    &=\sum_a \vert x\vert\delta^{3a} \operatorname{Ad}(U_x)(X^a)=\vert x\vert U_x X^3U_x^\dagger.
\end{align}
Since by construction $X^3$ has the simple eigensystem $X^3\ket{k}=\frac{1}{C_N}k\ket{k}$ for $k=-j,\dots,j$, we find
\begin{align}
    \label{fsSTATES}
    \ket{x}=U_x\Ket{\frac{N-1}{2}}=U_x\ket{j},\quad \lambda(x)=\frac{1}{2}\left(1+\vert x\vert^2\right)-\vert x\vert\frac{N-1}{2C_N}=\frac{1}{2}\left(1+\vert x\vert^2\right)-\vert x\vert \sqrt{\frac{N-1}{N+1}}.
\end{align}

But that directly implies $k=2$, $\Tilde{\mathbb{R}}^3=\hat{\mathbb{R}}^3=\mathbb{R}^3\setminus\{0\}$, $\mathcal{N}_x=\mathbb{R}^+x$ and $\mathcal{M}=SU(2)(U(1)\ket{\frac{N-1}{2}})$ (where we view $U(1)\ket{\frac{N-1}{2}}$ as a point in $\mathbb{C}P^{N-1}$).
From Lie group theory, we know that we can identify any Lie group orbit with the homogeneous space we get from the Lie group modulo the stabilizer of one point in the orbit. Here, the stabilizer of $U(1)\ket{\frac{N-1}{2}}$ is simply $U(1)\subset SU(2)$ (coming from the one parameter subgroup generated by $J^3$), thus we find
\begin{align}
    \mathcal{M}\cong SU(2)/U(1)\cong S^2,
\end{align}
so the quantum manifold of the fuzzy sphere is diffeomorphic to the classical sphere $S^2$.\\
In the same manner we find
\begin{align}
    \label{FuzzySphereExp}
    \bra{x}X^a\ket{x}&=\Bra{\frac{N-1}{2}}U_x^\dagger X^aU_x\Ket{\frac{N-1}{2}}=\sum_b(R_x^{-1})^{ab}\Bra{\frac{N-1}{2}}X^b\Ket{\frac{N-1}{2}}\\\nonumber
    &=\sqrt{\frac{N-1}{N+1}}(R_x^{-1}\cdot\hat{e}_3)^a=\sqrt{\frac{N-1}{N+1}}\frac{x^a}{\vert x\vert},
\end{align}
using $\bra{\frac{N-1}{2}}J_N^a\ket{\frac{N-1}{2}}=\delta^{a3}\frac{N-1}{2}$. Thus also $\Tilde{\mathcal{M}}$ assembles to a sphere, however with radius $\sqrt{\frac{N-1}{N+1}}$.\\
This radius is exactly at the global minimum of $\lambda(x)$, attaining the value $\frac{1}{N+1}$. The displacement is given by $d^2(x)=(\vert x\vert -\sqrt{\frac{N-1}{N+1}})^2$, while we find the dispersion $\delta^2(x)=\frac{2}{N+1}$. Thus the quality of the quasi-coherent states gets better the larger $N$ is and the closer $\vert x\vert^2$ is to $\frac{N-1}{N+1}$.\\ Yet, $\lambda(x)$ is not bounded and goes to $\infty$ as $\vert x \vert\to\infty$ \cite{Steinacker_2021}.

For the fuzzy sphere, the quantum metric and the would-be symplectic form can be calculated explicitly (this is thoroughly done in appendix \ref{Appendix:perturbativeappr}) with the result 
\begin{align}
    \label{explicitg}
    (g_{ab})=\frac{j}{2\vert x\vert^4}
        \begin{pmatrix}
    (x^2)^2+(x^3)^2 & -x^1x^2 & -x^1x^3 \\
  -x^1x^2 & (x^1)^2+(x^3)^2 & -x^2x^3 \\
   -x^1x^3 & -x^2x^3 & (x^1)^2+(x^2)^2 \\
    \end{pmatrix}=\frac{N-1}{2\vert x\vert^2}\left(\delta^{ab}-\frac{x^a}{\vert x\vert}\frac{x^b}{\vert x\vert}\right)
\end{align}
and
\begin{align}
    \label{explicitomega}
     (\omega_{ab})=\frac{j}{2\vert x\vert^3}
        \begin{pmatrix}
   0 & x^3 & -x^2\\
  -x^3 & 0 & x^1 \\
   x^2 & -x^1 & 0 \\
    \end{pmatrix}=\frac{N-1}{4\vert x\vert^2}\left(\sum_c\epsilon^{abc}\frac{x^c}{\vert x\vert}\right),
\end{align}
where $j=\frac{N-1}{2}$.
Here, we can explicitly see that $h_{ab}$ is not constant on the $\mathcal{N}_x$.

Further, we find 
\begin{align}
    \theta^{ab}=\frac{1}{i}\bra{x}[X^a,X^b]\ket{x}=\sum_c\frac{1}{C_N}\epsilon^{abc}\bra{x}X^c\ket{x}=\sum_c\frac{2}{N+1}\epsilon^{abc}\frac{x^c}{\vert x\vert}
\end{align}
and subsequently
\begin{align}
    \omega_{ab}\theta^{bc}=-\frac{1}{2}\frac{1}{\vert x\vert^2}\frac{N-1}{N+1}\left(\delta^{ab}-\frac{x^a}{\vert x \vert}\frac{x^b}{\vert x \vert}\right).
\end{align}

Also, we have
\begin{align}
    \partial_a\mathbf{x}^b=\sqrt{\frac{N-1}{N+1}}\partial_a\left(\frac{x^b}{\vert x\vert}\right)=\sqrt{\frac{N-1}{N+1}}\frac{1}{\vert x\vert}\left(\delta^{ab}-\frac{x^a}{\vert x \vert}\frac{x^b}{\vert x \vert}\right),
\end{align}
noting that $p^a_b:=(\delta^{ab}-\frac{x^a}{\vert x \vert}\frac{x^b}{\vert x \vert})$ is the projector on the tangent space of $T_x S^2\subset \mathbb{R}^3$ for the sphere of radius $\vert x \vert$.\\
Thus, $-2\sum_c\omega_{ac}\theta^{cb} \approx p_a^b\approx \partial_a \mathbf{x}^b$ holds for $\vert x\vert\to \sqrt{\frac{N-1}{N+1}}$ (the points where $\lambda$ is minimal) and for large $N$ (when the $X^a$ become almost commutative, looking at equation (\ref{theXmatrices})).

In this sense, we find the pseudoinverse of the would-be symplectic form $(\omega^{-1'})^{ab}=-\frac{4\vert x\vert^2}{N-1}\epsilon^{abc}\frac{x^c}{\vert x\vert}$, satisfying $(\omega^{-1'})^{ab}\omega_{bc}=p^a_c$ and consequently
\begin{align}
    \{\mathbf{x}^c,\mathbf{x}^d\}=\sum_{a,b}(\omega^{-1'})^{ab}\partial_a\mathbf{x}^c\partial_b\mathbf{x}^d=-\frac{4}{N+1}\epsilon^{cde}\frac{x^e}{\vert x\vert}=-2\theta^{cd},
\end{align}
so both Poisson structures coincide exactly up to the factor $-2$ that can be absorbed into the definition of $\omega_{ab}$. Further, $-2\omega_{ab}$ reproduces the Poisson structure introduced in section \ref{FuzzySphere0} up to a scale factor that goes to one for increasing $N$.

The implied $SU(2)$ invariance of $\omega_{ab}$ together with equation (\ref{fsSTATES}) shows that $Q$ is an intertwiner of the $SU(2)$ action, implying that the completeness relation holds and $Q(\mathbf{x}^a)\propto X^a$, yet $Q$ is no isometry \cite{Steinacker_2021}. This will be discussed more generally in section \ref{CoadjointOrbits}.

\newpage
\subsubsection{The Squashed Fuzzy Sphere}
\label{SFuzzySphere}

We now turn to a more complicated geometry coming with less symmetry than the round fuzzy sphere from the last section. We rename the corresponding matrices $X^a\mapsto \Bar{X}^a$ and define the \textit{squashed fuzzy sphere} of degree $N$
\begin{align}
    S^2_{N,\alpha}:=\left(\Bar{X}^1,\Bar{X}^2,\alpha \Bar{X}^3\right)=:\left(X^1,X^2,X^3\right),
\end{align}
where $\alpha\geq0$ is called the \textit{squashing parameter}.

We directly see that although the matrices still span the same Lie algebra $\mathfrak{su}(2)$ in the same representation, the Hamiltonian looses its simple form and symmetry and is given by
\begin{align}
    H_x=\frac{1}{2}\left(1+\vert x \vert^2\right)-\sum_{a=1}^3 x^a \Bar{X}^a-\left(1-\alpha^2\right)\frac{1}{2}\Bar{X}^3\Bar{X}^3+\left(1-\alpha\right)x^3\Bar{X}^3.
\end{align}
Thus we cannot expect to find an easy way to calculate $\lambda(x)$ and $\ket{x}$ explicitly using group theory.

Yet, there are a few statements that can be read off directly.\\
The quasi-coherent states at points that are related by a rotation around the $z$-axis are still related by the corresponding unitary matrix since the $SU(2)$-symmetry is only partially broken.

We consider the \textit{asymptotic Hamiltonian} for $x\neq0$
\begin{align}
    \label{HamiltonianAsympt}
    H^\infty_x:=\frac{1}{2\vert x \vert}\sum_a X^aX^a-\sum_a\frac{x^a}{\vert x \vert}X^a,
\end{align}
having the same eigenvectors as $H_x$, while the eigenvalues are shifted by a strictly monotonic function $\lambda^k_x\mapsto \frac{\lambda^k_x}{\vert x\vert}-\frac{1}{2}\vert x\vert$. Here, we have the asymptotic behaviour $H^\infty_{lx}=-\sum_a \frac{x^a}{\vert x\vert}X^a+\mathcal{O}(l^{-1})$ for $l\to\infty$.
For our current matrix configuration this means
\begin{align}
    H_x^\infty=-\frac{1}{\vert x\vert}\left(x^1\Bar{X}^1+x^2\Bar{X}^2+\alpha x^3\Bar{X}^3)\right)+\mathcal{O}\left(\frac{1}{\vert x\vert}\right).
\end{align}
Considering an $x\in \Tilde{\mathbb{R}}^3$, by continuity $\lim_{l\to\infty}\ket{l x}$ is the lowest eigenstate of $H_x^{\infty}$, neglecting the $\mathcal{O}\left(\frac{1}{\vert x\vert}\right)$ terms. But this means that in the limit $l\mapsto \infty$, we recover the quasi coherent state of the round fuzzy sphere at the point $(x^1,x^2,\alpha x^3)$.\\
Further, for $\alpha<1$ and $x=\pm\vert x\vert \hat{e}_3$, we find $H_x=\frac{1}{2}(1+\vert x\vert)\mp\alpha\vert x\vert \Bar{X}^3-(1-\alpha^2)\Bar{X}^3\Bar{X}^3$, still having $\ket{\pm \frac{N-1}{2}}$ as lowest eigenvector for $\ket{x}\neq0$, implying $\mathcal{N}_x=\pm\mathbb{R}^+\hat{e}_3$.\\
Similarly, we see $H_0=\frac{1}{2}(1+\vert x\vert)-(1-\alpha^2)\Bar{X}^3\Bar{X}^3$, implying that $E_0$ is given by the span of $\ket{\frac{N-1}{2}}$ and $\ket{-\frac{N-1}{2}}$, so $0\in\mathcal{K}$.

The result that $\ket{\pm\frac{N-1}{2}}$ is the lowest eigenvector on the whole positive respectively negative part of the $z$-axis has profound consequences: It tells us that the rank of $q$ is at most two there. So if the rank of $q$ is three at any other point, this means that the whole $z$-axis has to be excluded from $\hat{\mathbb{R}}^3$. This we will show in a moment for $N>2$ and $\alpha=1-\epsilon$ for a small $\epsilon>0$.

It is rather obvious that the quasi coherent states can not be calculated explicitly for arbitrary points.
Still, the setup is particularly well suited for perturbation theory if we are only interested in the vicinity of $\alpha=1$, thus $\alpha=1-\epsilon$.
It turns out that a lot of calculation is needed, therefore we refer to appendix \ref{Appendix:perturbativeappr} for the derivation, featuring an explicit expression for the first order correction of $h_{ab}$.

These results explicitly give the terms in the expansion $g_{ab}=g_{0,ab}+\epsilon g'_{ab}+\mathcal{O}(\epsilon^2)$, where $g_{0,ab}$ is the quantum metric for the round fuzzy sphere given in equation (\ref{explicitg}).\\
$g_{0,ab}$ has the eigenvector $x^a/\vert x\vert$ with eigenvalue $0$, so its determinant vanishes
\begin{align}
    \det{(g_{0,ab})}=0,
\end{align}
just as we would expect.\\
On the other hand we have
\begin{align}
    \det{(g'_{ab})}=&\frac{C_2^2 C_3^5 x_3^2 \left(\vert x\vert^2-x_3^2\right) }{ 128 C_1^3 \vert x\vert^{15} }\cdot\\\nonumber 
    &\cdot\left(C_2 x_3^2 (4 C_3 (C_2- C_1\vert x\vert)+C_2)+4 (C_2-1) C_3 x_3 \vert x\vert
   ( C_1\vert x\vert-C_2)- C_2^2\vert x\vert^2\right)
\end{align}
for coefficients $C_i$ that depend only on $N$ with their definition given in equation (\ref{Ccoeffs}).
\\
The reason why we look at $\det{(g'_{ab})}$ is the following: Assume that the latter is nonvanishing for some $x$, then this tells us that $g'_{ab}$ has full rank for these $x$, hence rank three.
But then, making $\epsilon$ small enough, also $g_{ab}(x)$ will have rank three (just as $T_xq$) and thus the dimension of $\mathcal{M}$ is three (at least for small $\epsilon$). (We note that in turn vanishing $\det{(g'_{ab})}$ does not imply that the rank of $g_{ab}$ is smaller than three.)
\\
We assume $N>2$ (for $N=2$ we have $C_2=0$ and thus $\det{(g'_{ab})}=0$).
Now, there are two obvious zeros, given by either $x^1=x^2=0$ (what we would expect from the discussion above) and $x^3=0$. Mathematica finds a third one that is of rather complicated form but only describes a zero set in $\mathbb{R}^3$ -- just as the first two.\\
Now, this leads us to the conclusion that the rank of $T_xq$ is three almost everywhere at least for small $\epsilon$ and thus we have $k=\operatorname{Dim}(\mathcal{M})=3$ rather than two.\\
Numerical results suggest that only along the $z$-axis the rank of $q$ is reduced.

Any further analysis of the squashed fuzzy sphere shall be postponed to section \ref{sfs_results} until we are ready to do numerical calculations.

\subsubsection{Quantized Coadjoint Orbits as Matrix Configurations}
\label{CoadjointOrbits}

We can immediately generalize the description of the fuzzy sphere analogously to the discussion in section \ref{QG}. Also this section is rather technical and can in principle be skipped, leaving most of the remaining comprehensible.\\
Let $G$ be a compact semisimple Lie group of dimension $D$ with its associated Lie algebra $\mathfrak{g}$. Assume we have chosen a maximal Cartan subalgebra $\mathfrak{h} \subset\mathfrak{g}$ and a set of positive roots.
Let $\lambda\in \mathfrak{h}^*$ be a dominant integral element, providing us with the unique irreducible representation $\mathcal{H}=\mathcal{H}_\lambda$ with highest weight $\lambda\in \mathfrak{h}^*$. This we use as our Hilbert space of dimension $N=\operatorname{dim}(\mathcal{H})$. In order not to confuse $\lambda\in \mathfrak{h}^*$ with the lowest eigenvalue of $H_x$ we write $\lambda'(x)$ for the latter in this section.

Since $G$ is compact, both $\mathfrak{g}$ and $\mathcal{H}$ carry a natural $G$ invariant inner product -- for $\mathfrak{g}$ this is the Killing form.\\
Thus we can chose orthonormal bases $T^1,\dots,T^D$ of $\mathfrak{g}$ and $b^1,\dots,b^N$ of $\mathcal{H}$. The $T^a$ then act as Hermitian matrices $T^a_\lambda$ in the basis $b^i$.
\\
Since $G$ is semisimple, in any irreducible representation, the quadratic Casimir operator $T^aT^a$ acts as a scalar $C_\lambda^2$ and we define the matrix configuration as 
\begin{align}
    G_\lambda:=(X^1,\dots,X^D)
\end{align}
for $X^a:=\frac{1}{C_\lambda}T^a_\lambda$.
\\
But this allows us to write the Hamiltonian (in analogy to the fuzzy sphere) in the very simple form\footnote{Especially this means $U(1)\kets{lx}=U(1)\kets{x}$ for $l\in\mathbb{R}^+$.}
\begin{align}
    H_x=\frac{1}{2}\left(1+\vert x\vert^2 \right)\mathbb{1}-\sum_a x^aX^a.
\end{align}
We then may consider $(x^a)$ as $x\in\mathfrak{g}^*$ via $x:=\sum_a x^a (T^a)^*$ where $(T^a)^*$ is the dual of $T^a$ with respect to the Killing form.

Let $x\in g^*$ and $\ket{\psi}\in\mathcal{H}$, then we define the coadjoint orbit $\mathcal{O}_x:=\operatorname{Ad}^*(G)(x)$ through $x$ respectively the orbit $\mathcal{O}^{\ket{\psi}}:=G\cdot (U(1)\ket{\psi})$ through $U(1)\ket{\psi}$ (where we view $U(1)\ket{\psi}$ as a point in $\mathbb{C}P^{N-1}$, thus we have already factorized out the $U(1)$ phase).\\
For any $g\in G$ we have
\begin{align}
 \sum_ax^aX^a&=\sum_ag^{-1}\cdot g\cdot (x^aX^a)\cdot g^{-1}\cdot g=\sum_ag^{-1}\cdot \operatorname{Ad}(g)(x^aX^a)\cdot g\\\nonumber
 &=\sum_ag^{-1}\cdot (\operatorname{Ad}^*(g)(x)^aX^a)\cdot g,
\end{align}
thus for $x\in \Tilde{\mathbb{R}}^D$ we find $\operatorname{Ad}^*(g)(x)\in\Tilde{\mathbb{R}}^D$ and $U(1)\ket{\operatorname{Ad}^*(g)(x)}=g\cdot U(1)\ket{x}$. In terms of orbits this means $y\in \mathcal{O}_x\implies\ket{y}\in\mathcal{O}^{\ket{x}}$.

Now, every $\mathcal{O}_x$ contains at least one $\Tilde{x}_0\in\mathfrak{h}^*$ and exactly one $x_0$ in the closure of the fundamental Weyl chamber within $\mathfrak{h}^*$.\\
Then we can write $x_0=\sum_i x_{0,i} \alpha_i$ for some coefficients $x_{0,i}\geq 0$ where the $\alpha_i$ are the positive simple roots.

On the other hand, we can consider the weight basis $\ket{\lambda^\mu}$ of $\mathcal{H}$, where the $\lambda^\mu$ are the weights of $\mathcal{H}$ and especially $\lambda^0=\lambda$. Since $\lambda$ is the highest weight, all $\lambda^\mu$ can be written as $\lambda^\mu=\lambda-\sum_i k^\mu_i\alpha_i$ for some coefficients $k^\mu_i\geq 0$ (where \textit{all} coefficients vanish if and only if $\mu=0$).

In this basis (recalling $\sum_ax_0^aX^a\in\mathfrak{h}$) we find
\begin{align}
    H_{x_0}\ket{\lambda^\mu}=\left(\frac{1}{2}\left(1+\vert x_0\vert^2\right)-\lambda^\mu\left(\sum_ax_0^aX^a\right)\right)\ket{\lambda^\mu}=\left(\frac{1}{2}(1+\vert x_0\vert^2)-\frac{1}{C_\lambda}\langle \lambda^\mu, x_0\rangle\right)\ket{\lambda^\mu}
\end{align}
where $\langle\cdot,\cdot\rangle$ is the dual of the Killing form.\\
Now, we find
\begin{align}
    \langle \lambda^\mu, x_0\rangle=\langle \lambda,x_0 \rangle-\sum_{i,j}k^\mu_i x_{0,j}\langle \alpha_i,\alpha_j\rangle=\langle \lambda,x_0 \rangle-\sum_{i}k^\mu_i x_{0,i}\leq\langle \lambda,x_0 \rangle.
\end{align}
But this exactly means that we have the smallest eigenvalue
\begin{align}
    \lambda'(x_0)=\frac{1}{2}(1+\vert x_0\vert^2)-\frac{1}{C_\lambda}\langle \lambda, x_0\rangle
\end{align}
together with a lowest eigenstate
\begin{align}
    \ket{x_0}=\ket{\lambda}.
\end{align}
If now $x_0$ lies on the border of the fundamental Weyl chamber (this exactly means that at least one $x_{0,i}=0$), there may or may not be other lowest eigenstates and consequently $x_0$ is or is not in $\mathcal{K}$, depending on whether there is a $\mu$ such that $k^\mu_i\neq 0$ and $k^\mu_j= 0$ for $j\neq i$ or equivalently if $\lambda-\alpha_i$ is a weight of the representation.

Thus, we conclude that we can always write $x=\operatorname{Ad}^*(g)(x_0)$ for a $g\in G$ a point $x_0$ in the closure of the fundamental Weyl chamber. If $x_0$ lies within its border, there is the chance that $x\in\mathcal{K}$, otherwise $\ket{x}=g\ket{\lambda}$ with $\lambda'(x)=\frac{1}{2}(1+\vert x\vert^2)-\frac{1}{C_\lambda}\langle \lambda ,x_0\rangle$.\\ Especially,
\begin{align}
    \mathcal{M}=G\cdot (U(1)\ket{\lambda})=\mathcal{O}^{\ket{\lambda}}\cong \mathcal{O}_\lambda,
\end{align}
where the last isomorphism is due to the fact that the corresponding stabilizers agree $G_\lambda=G_{U(1)\ket{\lambda}}$ for a highest weight $\lambda$.
\\
Further, we find that this implies a constraint on $G_x$ if $x\notin \mathcal{K}$: Since the set of quasi coherent-states on the coadjoint orbit through $x$ is $\mathcal{M}$, this implies that $G_x\subset G_\lambda$ (otherwise there would be more quasi-coherent states coming from $\mathcal{O}_x$ than there are points in the orbit).
If $\lambda$ is from the inner of the fundamental Weyl chamber, this implies that all orbits through its border belong to $\mathcal{K}$ (since there the stabilizer is strictly larger then $G_\lambda$, where the latter is given by the maximal torus $T:=\exp{(\mathfrak{h})}$), while these are the only ones. Similar statements can be made if $\lambda$ itself is part of the border.

By our construction, $\omega_\mathcal{M}$ coincides up to a factor with the $G$ invariant Kirillov-Kostant-Souriau symplectic form from section \ref{CoadjointOrbits0}.\\
But this means that $Q$ exactly intertwines the natural group actions on $\mathcal{C}(\mathcal{M})$ respectively $\operatorname{End}(\mathcal{H})$. Thus irreducible representations map to irreducible representations. This directly implies that $1_\mathcal{M}$ is mapped to $\mathbb{1}_\mathcal{H}$ (since both transform in the trivial representation) and consequently the completeness relation is satisfied. Since both the $\mathbf{x}^a$ and the $X^a$ transform in the adjoint representation, this also implies $Q(\mathbf{x}^a)\propto X^a$.
\\
It turns out that this $Q$ satisfies all axioms of a quantization map but is not an isometry -- thus although $(\mathcal{M},\omega_\mathcal{M})$ and $\mathcal{H}$ agree (up to a scale of the symplectic form $\omega_\mathcal{M}$) with our construction in section \ref{CoadjointOrbits0}, the quantization maps discern -- just as the Weyl quantization and the coherent state quantization were different for the Moyal-Weyl quantum plane.\\
We also mention that $\mathcal{M}$ here even is a Kähler manifold \cite{Steinacker_2021,Bernatska_2012,Neeb_1994,Kostant_1982}.

Finally, we note that this construction explicitly recovers our findings for the round fuzzy sphere.

\subsubsection{The Fuzzy \texorpdfstring{$\mathbb{C}P^2$}{CP2}}
\label{fuzzycp2}

After considering the general case, we look at a specific geometry, the generalization of the fuzzy sphere from $SU(2)$ to $SU(3)$ called the fuzzy $\mathbb{C}P^2$. It will be our simplest example with $D>3$ in our numerical considerations in section \ref{RESULTS}.\\
Let $G=SU(3)$ (then $D=8$) and consider the representation $(n,0)$ of dimension $N=\frac{(n+1)(n+2)}{2}$.\\
Then we find the matrices $T_N^a$ (generalized Gell-Mann matrices) and the quadratic Casimir $C_n^2=\frac{1}{3}(n^2+3n)$ as discussed in appendix \ref{su3rep} (with an explicit implementation for Mathematica introduced in \cite{Shurtleff_2009}).

Then we can apply the general construction and obtain the matrix configuration
\begin{align}
    \mathbb{C}P^{2}_n:=\left(X^1,\dots,X^D\right),
\end{align}
where the $X^D$ are normalized to $\sum_aX^aX^a=\mathbb{1}$.

Let us first discuss why we only consider the $(n,0)$ representations. $SU(2)$ has the two Cartan generators $T^3$ and $T^8$, thus the dimension of the maximal torus is $\operatorname{dim}(T)=2$, meaning that $\operatorname{dim}(\mathcal{M})\leq 6$ where equality exactly holds if the highest weight lies within the fundamental Weyl chamber. On the other hand, $\operatorname{dim}(\mathbb{C}P^2)=4$, thus we have to consider representations with highest weight in the border of the fundamental Weyl chamber.\\
It turns out that (except the trivial representation) the only such representations are given by $(n,0)$ with the corresponding stabilizer $SU(2)\times U(1)$ of dimension $\operatorname{dim}(SU(2)\times U(1))=3+1=4$. Here, we have
\begin{align}
    \cong SU(3)/(SU(2)\times U(1))\cong \mathbb{C}P^2,
\end{align}
explaining the name of the matrix configuration \cite{Steinacker_2021,Bernatska_2012}.

Obviously, this means that we find $\mathcal{M}\cong \mathbb{C}P^2$, yet it is instructive to sort out how $\mathcal{K}$ looks like. We know that $\mathcal{K}$ is built from coadjoint orbits through the border of the fundamental Weyl chamber.
Identifying $\mathbb{R}^8\cong\mathfrak{su}(3)^*$, the border is given by the rays $\mathbb{R}^+_0\hat{e}_8\subset \mathbb{R}^8$ and $\mathbb{R}^+_0(\frac{1}{\sqrt{3}}\hat{e}_3+\hat{e}_8)\subset \mathbb{R}^8$ for $c\geq 0$.\\
Let us now consider the fundamental representation $(1,0)$ with the Gell-Mann matrices
\begin{align}
    T^3=\left( \begin{matrix} 1 & 0 & 0 \\ 0 & -1 & 0 \\ 0 & 0 & 0 \end{matrix} \right),\quad \frac{1}{\sqrt{3}} T^8=\left( \begin{matrix} 1 & 0 & 0 \\ 0 & 1 & 0 \\ 0 & 0 & -2 \end{matrix} \right).
\end{align}

We start by considering the point $\hat{e}_8\in \mathbb{R}^8$, then the Hamiltonian is of the form
\begin{align}
    c_1\mathbb{1}-c_2 T^8
\end{align}
for positive constants $c_1, c_2$,
having the two lowest eigenstates
\begin{align}
    \left( \begin{matrix} 1 \\ 0 \\ 0 \end{matrix} \right),\quad \left( \begin{matrix} 0 \\ 1 \\ 0 \end{matrix} \right)
\end{align}
and we conclude that $\mathcal{O}_{c\hat{e}_8}\subset\mathcal{K}$ for any $c\in \mathbb{R}^+$.\\
(This fits to the observations that here we have a positive simple root $\alpha_i$ orthogonal to $\hat{e}_8$ such that $\lambda-\alpha_i$ is also a weight of the representation, where $\lambda$ is the highest weight.)
 
We continue with the point $\frac{1}{\sqrt{3}}\hat{e}_3+\hat{e}_8\in \mathbb{R}^8$, then the Hamiltonian is of the form
\begin{align}
    c_3\mathbb{1}-c_4 (\frac{1}{\sqrt{3}}T^3+ T^8)
\end{align}
for positive constants $c_3, c_4$,
having the single lowest eigenstate
\begin{align}
    \left( \begin{matrix} 1 \\ 0 \\ 0 \end{matrix} \right)
\end{align}
and we conclude that $\mathcal{O}_{c(\frac{1}{\sqrt{3}}\hat{e}_3+\hat{e}_8)}\subset\Hat{\mathbb{R}}^D$ for $c\in \mathbb{R}^+$.\\
(This again perfectly fits to the observation that for the positive simple root $\alpha_j$ orthogonal to $\frac{1}{\sqrt{3}}\hat{e}_3+\hat{e}_8$, $\lambda-\alpha_j$ is \textit{not} a weight of the representation.)

Finally, we look at $0\in \mathbb{R}^8$. Here, the Hamiltonian is of the form
\begin{align}
    c_5 \mathbb{1}
\end{align}
for a positive constant $c_5$,
having a completely degenerate spectrum, letting us conclude $\mathcal{O}_{0}=\{0\}\subset\mathcal{K}$.\\
(This corresponds to the fact, that all positive simple roots $\alpha_k$ are orthogonal to $0$ and for two of them $\lambda-\alpha_k$ is a weight) \cite{Bernatska_2012}.

\subsubsection{Random Matrix Configurations}
\label{random}

Let us assume that we know nothing about a given matrix configuration $(X^a)$, respectively that it is constituted from $D$ random $N\times N$ Hermitian matrices.\\ Then there are two constraints on the dimension $k$ of $\mathcal{M}$: By our construction $k\leq D$ and $k\leq 2N-2$. The first constraint is due to the limitation of the quasi-coherent states due to their dependence on target space $\mathbb{R}^D$, while the second constraint is due to the immersion of $\mathcal{M}$ into $\mathbb{C}P^{N-1}$.

Thus, we conclude
\begin{align}
    k=\operatorname{dim}(\mathcal{M})\leq \min\{D,2N-2\}.
\end{align}
Just as random matrices usually have maximum rank (the set of matrices with reduced rank is always a zero set), we should expect for random matrix configurations that $\mathcal{M}$ assumes its maximum dimension $k=\min\{D,2N-2\}$.

If the squashed fuzzy sphere behaved randomly, this would tell us that it had $k=3$ for $N>2$ respectively $k=2$ for $N=2$, agreeing with our results from section \ref{SFuzzySphere}.

\subsection{Foliations of the Quantum Manifold}
\label{fol}

In this section, (approximate) \textit{foliations} of $\mathcal{M}$ are discussed, allowing us to look at (approximate) \textit{leaves} in $\mathcal{M}$ (for which we write $\mathcal{L}\subset\mathcal{M}$). We will describe these through \textit{distributions} in the tangent bundle of $\mathcal{M}$ respectively of $\Tilde{\mathbb{R}}^D$ (in the latter actual calculations are easier, while using $q$ we can push forward to $\mathcal{M}$ either the distribution itself or the resulting leaves).

Recalling the squashed fuzzy sphere from section \ref{SFuzzySphere}, we have seen that there are points $x$ in $\Tilde{\mathbb{R}}^3$ where the rank of $T_xq$ equals two, while for most others it equals three (implying an awkward discrepancy between $\Tilde{\mathbb{R}}^D$ and $\hat{\mathbb{R}}^D$). Yet in the round case we globally had rank two. We will later see that in a sense one of the three directions is strongly suppressed with respect to the quantum metric in comparison to the others.\\
On the other hand, since $\mathcal{M}$ is three dimensional, it can not be symplectic with any chance -- for example prohibiting to integrate over $\mathcal{M}$.\\
Similar phenomena arise for other (deformed) quantum spaces.

If we had a degenerate Poisson structure on $\mathcal{M}$, one way to proceed would be to look at the naturally defined symplectic leaves. Yet a manifold with a degenerate symplectic form does not necessarily decay into a foliation of symplectic leaves.

However, we still can try to find (approximate) \textit{foliations} of $\mathcal{M}$.
As (exact) foliations are in one to one correspondence with smooth involutive distributions via the Frobenius theorem (see for example \cite{Michor_2008}), we define approximate foliations of $\mathcal{M}$ via general distributions in the tangent bundle $T\mathcal{M}$ that can then be \textit{integrated} numerically as we will discuss in section \ref{implementation}.
\\
For these leaves we wish the following \textit{heuristic} conditions to be fulfilled:
\begin{enumerate}
    \item The dimension of the leaves respectively the \textit{rank} of the distribution (for what we write $l\leq k$) should be even and in cases of deformed quantum spaces it should agree with the dimension of the unperturbed $\mathcal{M}$. We would also hope that $l$ is smaller or equal than the minimum of $\operatorname{rank}(T_xq)$ over $\Tilde{\mathbb{R}}^D$ -- potentially allowing us to include all points in $\Tilde{\mathbb{R}}^D$.
    \item The leaves should contain the directions that are not suppressed with respect to the quantum metric, while the suppressed directions should \textit{label} the leaves. In the latter directions the quasi-coherent states should hardly change. This means $l$ should agree with the \textit{effective dimension} of $\mathcal{M}$, where the exact definition of the latter dependece on the specific case.
    \item The restriction of $\omega_\mathcal{M}$ to the leaves (for what we write $\omega_\mathcal{L}$) should be nondegenerate and thus symplectic.
\end{enumerate}

The directions in $\mathbb{R}^D$ orthogonal to the leaves (after precomposition with $T_xq$) will then be considered as approximate generalizations of the null spaces $\mathcal{N}_x$ (we might write $\mathcal{N}^\mathcal{L}_x\subset \Tilde{\mathbb{R}}^D$ for them) on which we expect to find almost the same quasi-coherent states.
\\
Once having a foliation of $\mathcal{M}$, the question remains which leaf $\mathcal{L}$ to choose -- this problem also has to be addressed.
\\
If we made a specific choice, we can then formally refine the quantization map $Q$ from section \ref{QuantMap} by replacing all $\mathcal{M}\mapsto\mathcal{L}$.

The first approach that we look at is the most obvious way to construct symplectic leaves of maximum dimension which is explained in section \ref{symp} (we thus talk of the \textit{symplectic leaf} if we mean the construction). This approach in general turns out not to satisfy the second condition, leading us to the \textit{Pfaffian leaf} described in section \ref{pfaff} that is based on a quantum version of the Pfaffian of $\theta^{ab}$. This method turns out to be rather time consuming in calculation, directing us to a simplified version what we call \textit{hybrid leaf} discussed in section \ref{hybrid}. A final approach is based on the idea to consider leaves that are in a sense optimally Kähler\footnote{This is based on the finding  that for quantized coadjoint orbits $\mathcal{M}$ is a Kähler manifold.} (the \textit{Kähler leaf}) what is presented in section \ref{kaehler}.
From now on, we use the Einstein sum convention when appropriate.

\subsubsection{The Symplectic Leaf}
\label{symp}

Let us start with the so called symplectic leaf.
We assume that $\omega_\mathcal{M}$ is degenerate as it is in the example of the squashed fuzzy sphere. Especially all cases where the dimension $k$ of $\mathcal{M}$ is odd fall in this category.\\
Our approach is to define a distribution of rank $l=\operatorname{rank}(\omega_\mathcal{M})$ that is orthogonal to the kernel of $\omega_\mathcal{M}$ with respect to $g_\mathcal{M}$, resulting in \textit{symplectic leaves} of the maximally possible dimension.

The construction works as follows: Let $p\in \mathcal{M}$.
On $T_p\mathcal{M}$ we naturally define the degenerate subspace $T_pZ_\mathcal{M}:=\operatorname{ker}(\omega_\mathcal{M}(p))=\{\xi\in T_p\mathcal{M}\vert\; \omega(\xi,\cdot)=0\}$. Actually, we are interested in some complement $T_pL_\mathcal{M}$ of $T_pZ$ (there, $\omega_\mathcal{M}(p)$ is by definition nondegenerate).\\
In general there is no natural complement, but $\mathcal{M}$ is also equipped with a nondegenerate metric $g_\mathcal{M}$, allowing us to define $T_pL_\mathcal{M}:=T_pZ_\mathcal{M}^\perp$ via the induced orthogonal complement.\\
Then we define $TL_\mathcal{M}$ as our distribution.
Thus we find the \textit{even} symplectic dimension $l=\operatorname{rank}(\omega_\mathcal{M})$ (note that we actually do not know if $l$ is constant on $\mathcal{M}$).

In practice, we want to work with $g=(\omega_{ab}(x))$ and $\omega=(g_{ab}(x))$ in target space for a given $x\in\Tilde{\mathbb{R}}^D$ with $q(x)=p$, since calculations are much easier there. 
But then, special care has to be taken, as $q$ is not inverse to a coordinate function in the usual sense, since $D-k$ coordinates are always redundant.
In appendix \ref{mfp} we have already seen that we can lift the problem locally by choosing $k$ independent coordinates and dropping the remaining ones. However, here a different approach is more comfortable.\\
We define $T_xW:=T_x\mathcal{N}_x=\operatorname{ker}(T_xq)\subset T_x\Tilde{\mathbb{R}}^D$. Assuming $k<D$, we know that both $g$ and $\omega$ act as zero on $T_xW$. Thus, there we define an inner product $g_W$ by the inclusion into $T_x\Tilde{\mathbb{R}}^D$ and extend it by zero to the induced orthogonal complement. Then $g_W+g$ is a nondegenerate inner product on the whole $T_x\Tilde{\mathbb{R}}^D$.\\
This allows us to formulate the orthogonal complement as $T_xV:=T_xW^\perp$ with respect to $g_W+g$. This subspace is what we take as a representative for $T_p\mathcal{M}$ as we note that $T_xq(T_xV)=T_p\mathcal{M}$ and that $g$ is nondegenerate on $T_xV$. However, of course this representative is not uniquely satisfying these properties.\\
Then, we define $T_xZ:=\operatorname{ker}(g_W+\omega)$ as representative of $T_pZ_\mathcal{M}$ and $T_xL:=T_xZ^\perp$ (again with respect to $g_W+g$) as representative of $T_pL_\mathcal{M}$, as we actually find $T_xq(T_xW)=\{0\}\subset T_p\mathcal{M}$ as well as $T_xq(T_xZ)=T_pZ_\mathcal{M}$ and $T_xq(T_xL)=T_pL_\mathcal{M}$.\\
In this setting, the symplectic dimension is given by $l=D-\operatorname{dim}(\operatorname{ker}(\omega))$.

One obstacle of this leaf is that $l$ will be too large in general. For example for the squashed fuzzy $\mathbb{C}P^2$ the rank of $\omega_\mathcal{M}$ will be $l=8$, while in the round case we have $k=4$ (as we will see in section \ref{sfc}), so this method will not help us to reduce the effective dimension.

\subsubsection{The Pfaffian Leaf}
\label{pfaff}

We now look at an alternative approach to foliate $\mathcal{M}$ that is not bound to the degeneracy of $\omega_\mathcal{M}$.
Further, the formulation is more affine to target space as it is not based on the quantum metric and the would-be symplectic form.

Let $x\in\Tilde{\mathbb{R}}^D$ and $0\leq s\leq D$ be an even integer and consider the \textit{Grassmannian manifold} $\operatorname{Gr}_s(T_x\Tilde{\mathbb{R}}^D)$, the manifold of linear subspaces of $T_x\Tilde{\mathbb{R}}^D$ with dimension $s$. Then we define the \textit{generalized Pfaffian}
\begin{align}
    \mathcal{P}^s_x:\; & \operatorname{Gr}_s(T_x\Tilde{\mathbb{R}}^D)\to\mathbb{C}\\
    &V\mapsto \bra{x}[X^{a_1},X^{a_2}]\dots[X^{a_{s-1}},X^{a_s}]\ket{x}\epsilon_V^{a_1\dots a_s}\nonumber,
\end{align}
where 
\begin{align}
    \epsilon_V^{a_1\dots a_s}:=O^{a_1b_1}\dots O^{a_sb_s}\epsilon^{b_1\dots b_s}
\end{align}
is the Levi-Civita symbol on $V$, using any orthogonal map $O:\mathbb{R}^s\to V\subset T_x\Tilde{\mathbb{R}}^D$ (with respect to the standard inner product on $\mathbb{R}^s$ and the standard Riemannian metric on $\Tilde{\mathbb{R}}^D\subset\mathbb{R}^D$).

$\mathcal{P}^s_x(V)$ is called generalized Pfaffian since it can be viewed as the dequantization of a quantum version of the Pfaffian\footnote{This is up to a complex phase given by $\bra{x}[X^{a_1},X^{a_2}]\ket{x}\dots\bra{x}[X^{a_{s-1}},X^{a_s}]\ket{x}\epsilon_V^{a_1\dots a_s}$.} of $\theta\vert_V$, the volume density constructed from $\theta\vert_V$ where $\theta=(\theta^{ab}(x))$.\\
So $\mathcal{P}^s_x(V)$ should be thought of as a volume density assigned to $V$ (up to a complex phase). Then maximizing this density over $\operatorname{Gr}_s(T_x\Tilde{\mathbb{R}}^D)$ heuristically means to pick the directions that are not \textit{suppressed} in the context of condition 2.

Since $\operatorname{Gr}_s(T_x\Tilde{\mathbb{R}}^D)$ is compact, the absolute square of the generalized Pfaffian attains its maximum. We define $V^s_{x,max}$ to be the (hopefully unique) subspace of dimension $s$ where the latter  attains its maximum, coming with the absolute square of the corresponding volume density $v^s_{x,max}:=\vert \mathcal{P}^s_x(V^s_{x,max})\vert^2$.\\
This fixes two problems at hand: It allows us choose the effective dimension $l$ as the maximal $s$ such that $v^l_{x,max}\gg0$ (in whatever sense) and then choose $Tq(\sqcup_{x\in\hat{\mathbb{R}}^D} V^l_{x,max})$ as distribution, defining the Pfaffian leaf\footnote{Note that $\mathcal{P}^l_x$ is constant along the $\mathcal{N}_x$ and that a large $v^l_{x,max}$ suggests that $v^{l'}_{x,max}$ is also large for $l'<l$.}.

Yet, there is an obstacle with this leaf as we do not know if $\omega_{ab}(x)$ is nondegenerate on $V^l_{x,max}$. In principle there could even be vectors in $V^l_{x,max}$ that lie in the kernel of $T_xq$.

\subsubsection{The Hybrid Leaf}
\label{hybrid}

While the symplectic leaf does not sufficiently reduce the effective dimension, the Pfaffian leaf is numerically difficult to calculate. Thus we introduce a hybrid of the two, based on $\theta^{ab}$.\\
Let $\theta:=(\theta^{ab}(x))$ for a given $x\in\Tilde{\mathbb{R}}^D$, having the eigenvectors $v_i$ with corresponding eigenvalues $\lambda_i$ (ordered such that $\vert\lambda_i\vert\leq\vert\lambda_j\vert$ if $i>j$).\\
Since $\theta$ is skew symmetric, the eigenvalues come in pairs $\lambda_{2s-1}=+i\phi_s$ and $\lambda_{2s}=-i\phi_s$ for $\phi_s\in\mathbb{R}$ and consequently $\vert \lambda_{2s-1}\vert=\vert \lambda_{2s}\vert$.
We thus choose $l$ such, that $\vert \lambda_{l+1}\vert\approx 0$, while $\vert \lambda_{l}\vert\gg 0$. (Note that this immediately implies that $l$ is even.)\\
We also define $w_{2s-1}=\operatorname{Re}(v_{2s-1})=\operatorname{Re}(v_{2s})$ and $w_{2s}=\pm\operatorname{Im}(v_{2s-1})=\mp\operatorname{Im}(v_{2s})$. Then $\theta w_{2s-1}=\pm\phi_s w_{2s}$ and $\theta w_{2s}=\mp\phi_s w_{2s-1}$, so $\langle w_{2s-1},w_{2s}\rangle$ is the subspace corresponding to $\phi_s$. We consequently define $V_x:=\langle w_1,\ldots w_l\rangle$, leading to the distribution $Tq(\sqcup_{x\in\hat{\mathbb{R}}^D}V_x)$.

Since the Pfaffian is related to the product of the eigenvalues, this result is quite parallel to the Pfaffian method if we assume that $\bra{x}[X^{a_1},X^{a_2}]\dots[X^{a_{l-1}},X^{a_l}]\ket{x}\approx \bra{x}[X^{a_1},X^{a_2}]\ket{x}\dots\bra{x}[X^{a_{l-1}},X^{a_l}]\ket{x}$ (what is plausible if the matrices $X^a$ are in the regime briefly discussed in section \ref{QuantMap} \cite{Steinacker_2021}).

Also here, we do not know if $\omega_{ab}(x)$ and $T_xq$ are nondegenerate on $V_x$.

A slightly modified version (that is more parallel to the symplectic leaf), where we use $\omega_{ab}$ instead of $\theta^{ab}$ fixes the last problem, since then $\omega_{ab}(x)$ and $T_xq$ definitely are nondegenerate on $V_x$.
In the following, we always explicitly specify if we work  with the hybrid leaf based on $\theta^{ab}$ respectively based on $\omega_{ab}$.

\subsubsection{The Kähler Leaf}
\label{kaehler}

Finally, we discuss an approach based on the fact that for quantized coadjoint orbits the manifold $\mathcal{M}$ is a Kähler manifold, corresponding to the observation that then we can choose the local sections $\kets{\cdot}$ in a holomorphic way (but then we have to accept that the $\kets{x}$ are no longer normalized).\\
This is further directly related to the following property: $i D_a \ket{x}\in \langle D_b\ket{x}\rangle_\mathbb{R}$, noting that we should think of $\langle D_b\ket{x}\rangle_\mathbb{R}\subset \mathcal{H}\cong \mathbb{C}^N$ as a representative of $T_{q(x)}\mathcal{M}$, so the representative of the tangent space is closed under the action of $i$, assuming that we have chosen an $x\in\Tilde{\mathbb{R}}^D$ \cite{Steinacker_2021}.

For general matrix configurations, we will find $i D_a \ket{x}\notin \langle D_b\ket{x}\rangle_\mathbb{R}$, but we might try to find a maximal subspace $V_x\subset \langle D_b\ket{x}\rangle_\mathbb{R}$ that is approximately invariant under the multiplication with $i$, meaning it is (approximately) a complex vector space and thus even dimensional.
\\
In the following, we are going to identify $\mathbb{C}^N\cong \mathbb{R}^{2N}$, thus it is convenient to consider multiplication with $i$ as the application of the linear operator $J\ket{v}=i\ket{v}$.

The best way to find such a $V_x$ is to construct a function on $\operatorname{Gr}_s(\langle D_b\ket{x}\rangle_\mathbb{R})\subset \operatorname{Gr}_s(\mathbb{R}^{2N})$ for even $s$ that measures how well the subspace is closed under the action of $J$.\\
This means, we need a distance function $d: \operatorname{Gr}_s(\mathbb{R}^{2N})\times \operatorname{Gr}_s(\mathbb{R}^{2N})\to \mathbb{R}$.
A common choice for such a function is given by
\begin{align}
    d(V,W):=\sup_{v\in V, \vert v\vert=1}\inf \left\{ \vert v-w\vert \vert w\in W \right\}.
\end{align}
The latter expression can be calculated in the following way: Let $P_W$ be the orthogonal projector on $W$ and let $i_V$ be the natural inclusion of $V$ into $\mathbb{R}^{2N}$, then
\begin{align}
    d(V,W)=\Vert (P_W-\mathbb{1})\circ i_V \Vert_2,
\end{align}
where $\Vert \cdot\Vert_2$ is the \textit{spectral norm}\footnote{When we actually calculate the Kähler cost, it may be convenient to replace the spectral norm with the Hilbert-Schmidt norm $\Vert\cdot\Vert_2\mapsto \Vert \cdot \Vert_{HS}$ -- although then $d$ is not necessarily a true distance function any more.} \cite{Morris_2009}.

Having this distance function at hand, we define the so called \textit{Kähler cost}
\begin{align}
    C_x^s:\, &\operatorname{Gr}_s(\langle D_b\ket{x}\rangle_\mathbb{R})\to \mathbb{R},\\\nonumber
    & V\mapsto d(V,J V).
\end{align}
Via $q$, this can be pulled back to $\operatorname{Gr}_s(T_x\Tilde{\mathbb{R}}^D)$
and we define
\begin{align}
    c_x^s:\,&\operatorname{Gr}_s(T_x\Tilde{\mathbb{R}}^D)\to \mathbb{R}, \\\nonumber
    &\Tilde{V}\mapsto d(T_xq(\Tilde{V}),J T_xq( \Tilde{V})).
\end{align}
We note, that by our construction $C_x^s=c_x^s=0$ for quantized coadjoint orbits.

Now, we define $\Tilde{V}^s_{x,min}$ to be the (hopefully unique) subspace of dimension $s$ where $c_x^s$ attains its minimum and $v^s_{x,min}:=c_x^s(\Tilde{V}^s_{x,min})$ as the corresponding Kähler cost.
Then we choose $l$ as the maximal $s$ such that $v^s_{x,min}\approx 0$ and define the distribution as $Tq(\sqcup_{x\in\hat{\mathbb{R}}^D} \Tilde{V}^l_{x,min})$.

Yet, also here we do not know if $\omega_{ab}(x)$ and $T_xq$ are nondegenerate on $\Tilde{V}^l_{x,min}$.
\\
Even if this method should not be well suited for determining foliations (as it turns out to be the case in section \ref{implementation}), the Kähler cost remains an interesting quantity.
\newpage
\section{The Implementation}
\label{implementation}

In this section the implementation on a computer of the various concepts that were introduced in section \ref{QMG} is discussed.\\
The corresponding algorithms are rather sketched than shown in full detail. Further, the focus lies on the use of \textit{Mathematica} as the latter is used for all explicit calculations.\\
Many of the algorithms are demonstrated on the squashed fuzzy sphere (from section \ref{SFuzzySphere}) for $N=4$ and $\alpha=0.1$ respectively $\alpha=0.9$ (whatever is more instructive).\\
If not stated explicitly, we will not discern between $\Tilde{\mathbb{R}}^D$ and $\hat{\mathbb{R}}^D$ in the following since the difference will not play an important role.

In section \ref{BasicQuantities} the basic quantities like the quasi-coherent states $\ket{x}$ and the quantum metric $g_{ab}$ are calculated.
The section \ref{Visual} then discusses multiple approaches to visualize the quantum manifold $\mathcal{M}$. The final section \ref{Foliat} treats the various methods for foliations, allowing us to construct curves in the leaves (what we call \textit{integrating} curves in the leaves) and consequently adapted coordinates what enables us to integrate over them globally.

\subsection{The Basic Quantities}
\label{BasicQuantities}

Assume we have chosen a point $x=(x^a)\in\mathbb{R}^D$.
As a beginning, we are interested in finding the basic quantities $\ket{x}$, $g_{ab}(x)$ and similar\footnote{We will mostly drop the argument from $g_{ab}(x)$ and similar.}. This can be achieved in the following steps.
\begin{itemize}
    \item $\lambda(x)$ and $\ket{x}$ can easily by calculated via the eigensystem of $H_x$ (given by equation (\ref{Hamiltonian})).\\
    Often it is useful to introduce a phase convention (for example $\braket{N\vert x}\geq0$ where $\ket{N}:=\hat{e}_N\in \mathbb{C}^N$), although we have to accept that there is no smooth global convention.
    \item Already a bit more involved is the calculation of $(\partial_a-iA_a)\ket{x}$. There are two available methods: Either one can calculate the difference quotient of $\ket{x}$ for a choice of some small $\epsilon$ and calculate $A_a$ using equation (\ref{connection}) or one proceeds using the eigensystem of $H_x$ following section \ref{algTricks}, especially using equation (\ref{AlgTrick}). The latter has the advantage that the calculation is pointwise and independent of some arbitrary $\epsilon$ (thus that will be the standard choice), while in the first some kind of phase fixing is particularly important in order to obtain a smooth dependence of $\ket{x}$ on $x$ (we once again note that this can only be achieved locally).
    \item Having these quantities, $h_{ab}$, $g_{ab}$ and $\omega_{ab}$ can readily be calculated using equation (\ref{hermitianform}).
    \item We can also calculate $J^a_b$, related to $J_\mathcal{M}$ from \ref{comparison}, but unless $g_{ab}$ is nondegenerate, we need some pseudoinverse\footnote{The pseudoinverse satisfies ${g'}^{ac}g_{cb}=p^a_b=g_{bc}{g'}^{ca}$ for a projector $p^a_b$ of the same rank as $g_{ab}$. Such a method is already part of Mathematica.} ${g'}^{ab}$ of $g_{ab}$. Then we find $J^a_c={g'}^{ab}\omega_{bc}$.
    \item Finally, $\theta^{ab}$ can be calculated using equation (\ref{theta}), while we find $\mathbf{x}^a$ by equation (\ref{xBold}) and $\partial_a\mathbf{x}^b$ using equation (\ref{partialEmbedd}).
\end{itemize}

Being able to calculate these important quantities, we are further interested in the calculation of simple invariants like $k=\operatorname{dim}(\mathcal{M})$ or the rank of $\omega_{ab}$. Also, we may want to check if for example the kernels of $\theta^{ab}$ and $g_{ab}$ agree.
\begin{itemize}
    \item Actually, before we do anything else, we should check whether $x$ lies within $\Tilde{\mathbb{R}}^D$ or $\mathcal{K}$. This can be done by checking the degeneracy of the lowest eigenvalue of $H_x$ via the respective eigensystem.
    \item The dimension\footnote{Strictly speaking, we have to assume that $x\in \hat{\mathbb{R}}^D$, else we only calculate the local rank of $T_xq$. Thus in order to calculate the true $k$ it is advisable to calculate the respective rank for multiple points and take the maximum. Then only the points where the maximum is attained lie within $\hat{\mathbb{R}}^D$.} $k$ of $\mathcal{M}$ can be calculated via the rank of $T_xq$.
    According to equation (\ref{Tanq}), we can calculate the rank of $T_xq$ via the real rank of $((\partial_a-iA_a)\ket{x})$, considered as a matrix.
    The real rank of a complex matrix $A$ is simply defined as
    \begin{align}
        \operatorname{rank}_\mathbb{R}(A):=\operatorname{rank}
            \left(\begin{array}{r}
            \operatorname{Re}(A) \\
	        \operatorname{Im}(A)\\
            \end{array}\right).
    \end{align}
    \item The ranks of $g_{ab}$, $\omega_{ab}$ and $\theta^{ab}$ can be calculated directly, where we already know that $\operatorname{rank}(g_{ab})\overset{!}{=}k$, while the rank of $\omega_{ab}$ is even and bounded from above by $k$. For the rank of $\theta^{ab}$ we have no such constraints in general.
    \item Also, we are interested in the kernels\footnote{Although $\theta^{ab}$ takes 1-forms as arguments, we can implicitly convert the kernel to vectors using the isomorphism induced by $\delta_{ab}$.} of $T_xq$ (which we can calculate using the above identification), $g_{ab}$, $\omega_{ab}$ and $\theta^{ab}$. Mathematica outputs for these a basis of the respective vector spaces.
    \item This allows for a sequence of checks: We know that the kernel of $T_xq$ should be contained in the kernel of $\omega_{ab}$ and agree with the kernel of $g_{ab}$. We might want to verify if that actually holds true.\\
    Since we are only in possession of bases of the respective kernels, we use the following relation\footnote{To see this, we note that the span of the $v_i$ is contained in the span of the $w_i$ if and only if adding any $v_i$ to the span of the $w_i$ does not increase the vector space. When a vector space is extended, its dimension always increases while the dimension equals the rank of the matrix built from basis vectors, thus the rank increases.}:
    \begin{align}
        \langle v_1,\ldots,v_m\rangle \subset \langle w_1,\ldots,w_n\rangle \Longleftrightarrow \operatorname{rank}(w_1,\ldots,w_n)=\operatorname{rank}(w_1,\ldots,w_n,v_1,\ldots,v_m),
    \end{align}
    where the rank is calculated for the matrices constituted from the respective column vectors.
    \item Finally, we can check the almost Kähler condition defined in \ref{comparison}: It is satisfied if and only if $(J^a_b)^2$ has eigenvalues $-1$ or $0$.
\end{itemize}

While the actual discussion of the squashed fuzzy sphere shall be postponed to section \ref{sfs_results}, we can demonstrate the output of what we achieved so far. (Here, we use $\alpha=0.9$, thus $\epsilon=0.1$.)

For the random point $x=(-0.414, -0.584, 0.161)$, we find
\begin{align}
    (g_{ab})= \begin{pmatrix}
1.014 & -0.658 & 0.192 \\
-0.658 & 0.554 & 0.271 \\
0.192 & 0.271 & 1.299 
\end{pmatrix}\approx \begin{pmatrix}
1.044 & -0.688 & 0.190 \\
-0.688 & 0.562 & 0.267 \\
0.190 & 0.267 & 1.458 
\end{pmatrix}=(g_{0,ab})+0.1(g'_{ab})
\end{align}
and
\begin{align}
    (\omega_{ab})= \begin{pmatrix}
0. & 0.356 & 1.131 \\
-0.356 & 0. & -0.803 \\
-1.131 & 0.803 & 0. 
\end{pmatrix}\approx \begin{pmatrix}
0. & 0.336 & 1.219 \\
-0.336 & 0. & -0.865 \\
-1.219 & 0.865 & 0. 
\end{pmatrix}=(\omega_{0,ab})+0.1(\omega'_{ab}),
\end{align}
where the $(rhs)$ comes from the perturbation theory in appendix \ref{Appendix:perturbativeappr}.\\
Here $T_xq$ and $g_{ab}(x)$ have the maximal rank three, while $\omega_{ab}(x)$ and $\theta^{ab}(x)$ have rank two.
Consequently, the kernel of the first two is $\{0\}$, thus they are automatically contained in all other kernels. For $\omega_{ab}(x)$ and $\theta^{ab}(x)$ we find that the respective kernels do \textit{not} agree. Further $(J^a_b)^2$ has the eigenvalues $-1.000,-1.000,0.000$, thus $\mathcal{M}$ is almost Kähler.\\
Considering the special point $x=(0,0,1)$, we find that all $T_xq$, $g_{ab}(x)$, $\omega_{ab}(x)$ and $\theta^{ab}(x)$ have rank two and that all kernels agree.\\
Thus we have $k=3$ and $(-0.414, -0.584, 0.161)\in \hat{\mathbb{R}}^3$, while $(0,0,1)\in\Tilde{\mathbb{R}}^3$ but not in $\hat{\mathbb{R}}^3$.

\subsection{The Visualization}
\label{Visual}

In order to obtain an intuitive understanding of $\mathcal{M}$, it would be advantageous to depict $\mathcal{M}$ or subsets thereof, described via points or subsets of $\Tilde{\mathbb{R}}^D$. (Often, for the sake of convenience, we will speak of points in $\Tilde{\mathbb{R}}^D$ but mean either their image under $q$ or $(\mathbf{x}^a)$, yet this should always become clear from the context.)\\
Unfortunately, there is no general elegant way to plot subsets of (respectively points in) $\mathbb{C}P^{N-1}$ and consequently of $\mathcal{M}$.
Yet, we can define a smooth map $v:\mathbb{C}P^{N-1}\setminus\mathcal{Z} \to\mathbb{C}^{N-1}\cong\mathbb{R}^{2(N-2)}$ by 
\begin{align}
    [(a_1,\ldots,a_N)]\mapsto \frac{\vert a_N\vert}{\vert(a_1,\ldots,a_N)\vert}\cdot(a_1/a_N,\ldots, a_{N-1}/a_N),
\end{align}
where\footnote{The prefactor could also be omitted, but since we always work with normalized vectors, this choice lies at hand.} $\mathcal{Z}=\{[b_1,\ldots,b_{N}]\in\mathbb{C}P^{N-1} \vert\; b_N=0\}$.
This is equivalent to choosing the unique normalized representative in the quotient with $a_N\geq 0$.\\
Although any such construction can not work globally, this map still is not too bad since the points in $\mathcal{Z}$ that we had to exclude are confined to a set of measure zero with respect to the Fubini-Study metric.
Plots generated with this method will be referred to as \textit{plots of} $\mathcal{M}$.

Alternatively, we can define the map $w:\mathbb{C}P^N\cong S^{2N-1}/U(1)\to \mathbb{R}^D$ via $U(1)\ket{v}\mapsto (\bra{v}X^a\ket{v})$ (thus by equation (\ref{xBold}) $w\vert_\mathcal{M}=(\mathbf{x}^a)$) in order to obtain points in $\mathbb{R}^D$.
In this sense depicting $\mathcal{M}$ via this constructions is equivalent to depicting $\Tilde{\mathcal{M}}$ and we will be referring to plots based on this approach as \textit{plots of} $\Tilde{\mathcal{M}}$.\\
Also, we note that in the case of the round fuzzy sphere for $N=2$ this is exactly the Hopf map up to a rescaling.
\\
Further, if we are concerned with foliations as described in section \ref{fol}, we directly obtain varieties in $\Tilde{\mathbb{R}}^D$ that we might want to depict. Consequently we will talk of \textit{plots of} $\Tilde{\mathbb{R}}^D$.

For all of the three cases, we still have to plot subsets of $\mathbb{R}^l$ for some $l$ that in general is bigger than three.\\
The first approach to this is to choose a rank three projector $P$ on $\mathbb{R}^l$ and an $SO(l)$ operator $O$ that maps the image of $P$ to $\mathbb{R}^3\times \{0\}$, resulting in a map $\mathcal{P}=O P:\mathbb{R}^l\to\mathbb{R}^3$.\\
If we want to apply $\mathcal{P}$ after $v$, our generic choice is to map
\begin{align}
    [(a_1,\ldots,a_N)]&\overset{v}{\mapsto} \frac{\vert a_N\vert}{\vert(a_1,\ldots,a_N)\vert}\cdot(a_1/a_N,\ldots, a_{N-1}/a_N)\\\nonumber
    &\overset{\mathcal{P}}{\mapsto} \frac{\vert a_N\vert}{\vert(a_1,\ldots,a_N)\vert} (\operatorname{Re}(a_1/a_N),\operatorname{Re}(a_2/a_N),\operatorname{Im}(a_1/a_N)),
\end{align}
if we want to apply it to points of $\mathbb{R}^D$, it is
\begin{align}
    (x^1,\dots,x^D)\overset{\mathcal{P}}{\mapsto}(x^1,x^2,x^3).
\end{align}
Plots based on this approach will be referred to as \textit{projective plots}.

Alternatively, we might only consider points that lie within a three dimensional hyperplane of $\mathbb{R}^l$. Let again $\mathcal{P}=OP$ and $v\in\mathbb{R}^l$, then every three dimensional hyperplane can be written as $\{x\in\mathbb{R}^l\vert\; (\mathbb{1}-P)(x-v)=0 \}$ for some rank three projector $P$ and some $v$. Thus, our method looks as such: Check if $x$ lies within the hyperplane (if yes keep the point, else drop it) and then apply $\mathcal{P}$.\\
In practice, we will not demand the constraint exactly, but rather $\vert(\mathbb{1}-P)(x-v)\vert<\epsilon$ for some small cutoff $\epsilon$ for obvious reasons.\\
If we use this approach, we will refer to \textit{sliced plots}.

\begin{figure}[H]
\centering
\begin{minipage}{.3\textwidth}
  \centering
  \includegraphics[height=.7\linewidth]{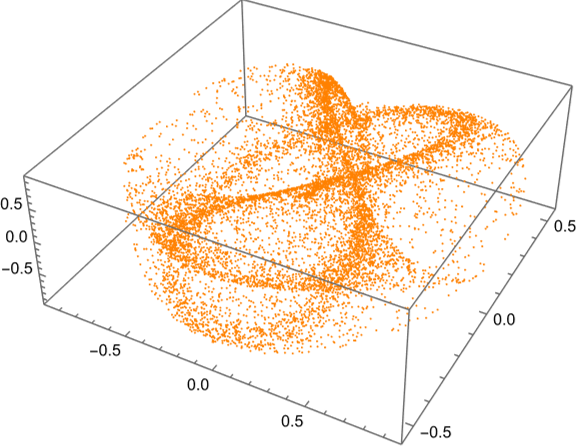}
\end{minipage}%
\begin{minipage}{.3\textwidth}
  \centering
  \includegraphics[height=.7\linewidth]{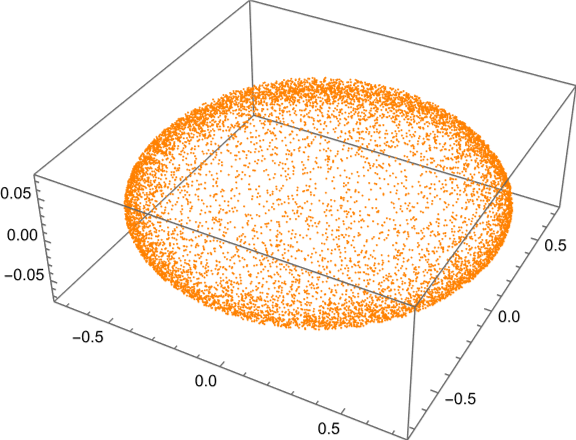}
\end{minipage}
\begin{minipage}{.3\textwidth}
  \centering
  \includegraphics[height=.7\linewidth]{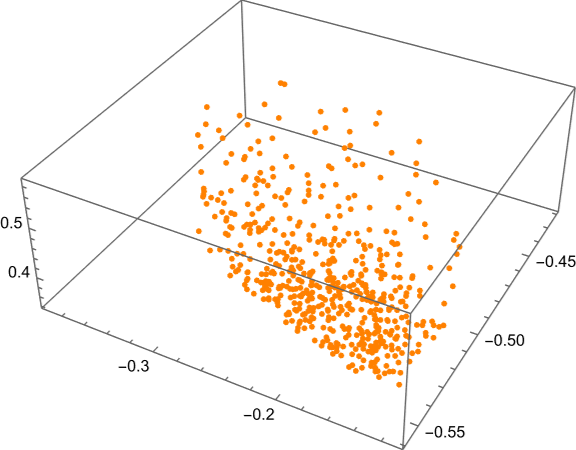}
\end{minipage}
\caption{Random points in the squashed fuzzy sphere for $N=4$ and $\alpha=0.1$. Left: projective plot of $\mathcal{M}$ for $10000$ points, middle: projective plot of $\Tilde{\mathcal{M}}$ for $10000$ points, right: sliced plot of $\mathcal{M}$ for $50000$ points}
\label{fig:Implementation/2}
\end{figure}

In figure \ref{fig:Implementation/2} we demonstrate projective plots of $\mathcal{M}$ and $\Tilde{\mathcal{M}}$ and a sliced plot of $\mathcal{M}$ for $\epsilon=0.1$. (Here, with $\epsilon$ the tolerance is meant and not $1-\alpha$.) In all cases we show random points in $\Tilde{\mathbb{R}}^3$ (strictly speaking points in $\mathbb{R}^3$), lying in the unit ball.\\
Here, we directly see a disadvantage of the sliced plots: Most points lie outside the slice, resulting in many computations that do not contribute to the picture.

\begin{figure}[H]
  \centering
  \begin{minipage}{.3\textwidth}
  \centering
  \includegraphics[height=.7\linewidth]{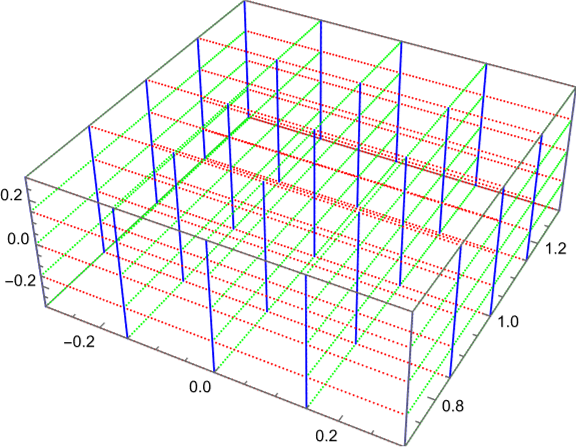}
\end{minipage}%
\begin{minipage}{.3\textwidth}
  \centering
  \includegraphics[height=.7\linewidth]{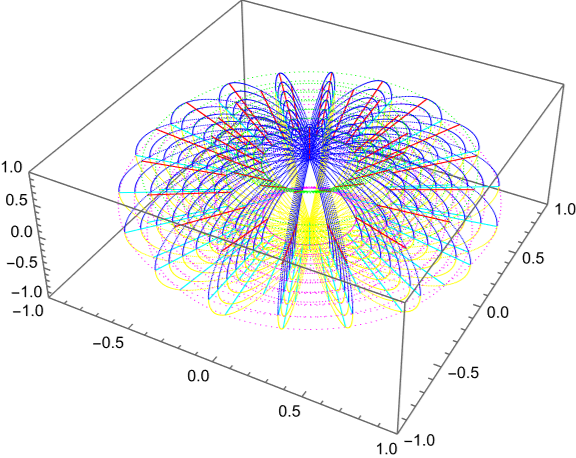}
\end{minipage}
\begin{minipage}{.3\textwidth}
  \centering
  \includegraphics[height=.7\linewidth]{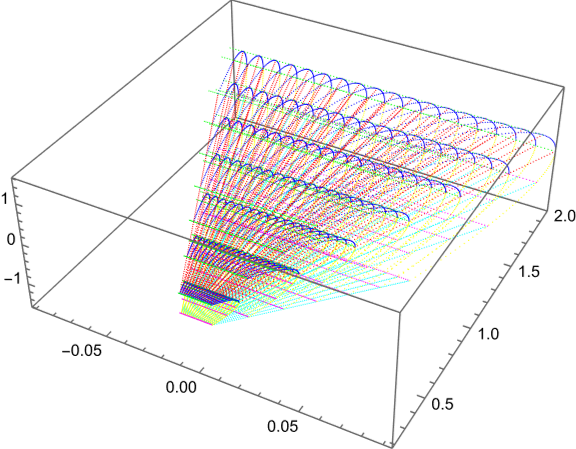}
\end{minipage}
\caption{Coordinate lines in $\mathbb{R}^3$. Left: Cartesian coordinate lines around $x=(0,1,0)$, middle: spherical coordinate lines, right: smaller sector of spherical coordinate lines at equator. The latter two have inverted colors in the southern hemisphere}
\label{fig:Implementation/3A}
\end{figure}

In order to get a more detailed understanding of how the mappings $q$ and $\mathbf{x}^a$ work and how $\mathcal{M}$ and $\Tilde{\mathcal{M}}$ look, it can be rewarding to consider coordinate lines in $\Tilde{\mathbb{R}}^D$ in favor of random points. In the simple case of $D=3$ we might for example use Cartesian or spherical coordinate lines as shown in figure \ref{fig:Implementation/3A}.

\begin{figure}[H]
\centering
\begin{minipage}{.3\textwidth}
  \centering
  \includegraphics[height=.7\linewidth]{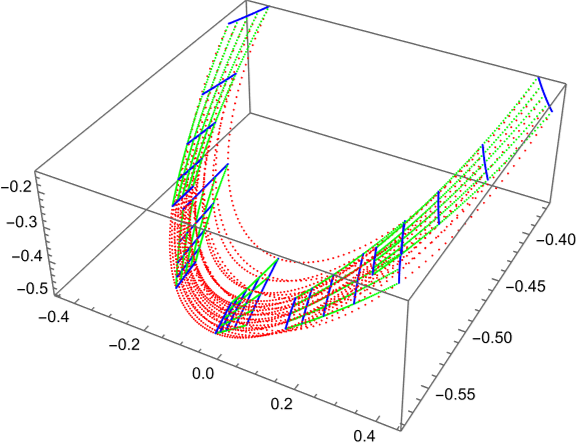}
  \includegraphics[height=.7\linewidth]{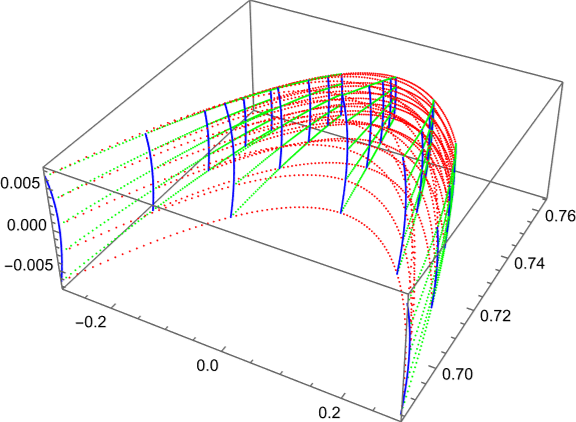}
\end{minipage}%
\begin{minipage}{.3\textwidth}
  \centering
  \includegraphics[height=.7\linewidth]{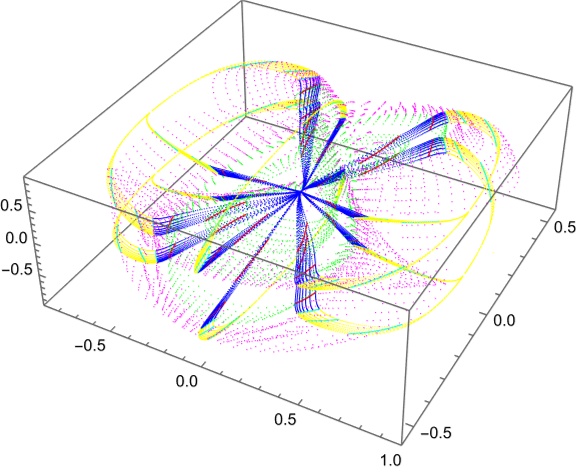}
  \includegraphics[height=.7\linewidth]{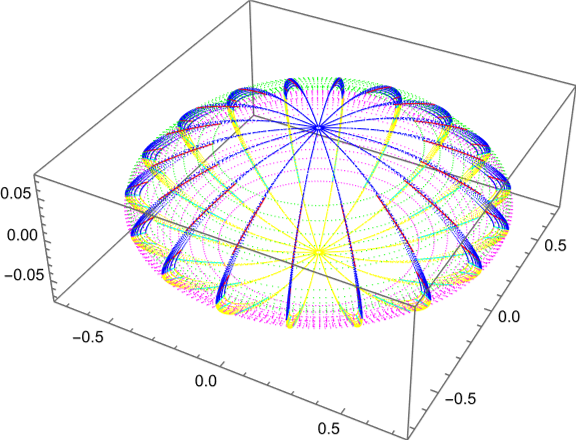}
\end{minipage}
\begin{minipage}{.3\textwidth}
  \centering
  \includegraphics[height=.7\linewidth]{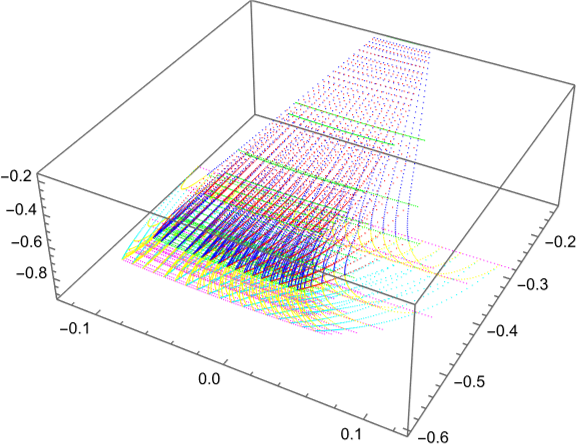}
  \includegraphics[height=.7\linewidth]{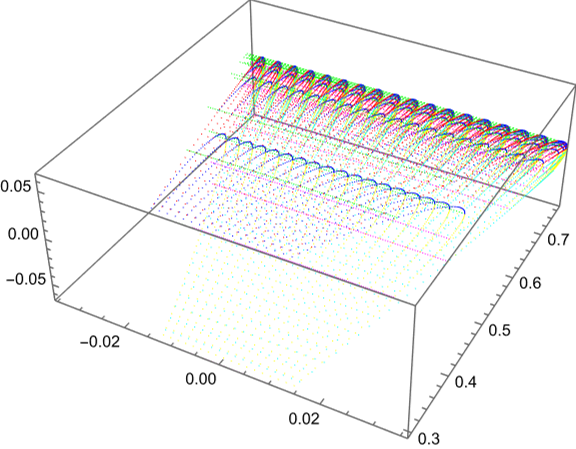}
\end{minipage}
\caption{The squashed fuzzy sphere for $N=4$ and $\alpha=0.1$. Left to right: coordinate lines from figure \ref{fig:Implementation/3A}; top: plot of $\mathcal{M}$, bottom: plot of $\Tilde{\mathcal{M}}$}
\label{fig:Implementation/3B}
\end{figure}

From these, we get plots of $\mathcal{M}$ and $\Tilde{\mathcal{M}}$ as shown in \ref{fig:Implementation/3B}. Fitting to our previous knowledge, it is plain to see that the dimension of $\mathcal{M}$ is (at least) three.

The discussion so far gave us the possibility to generate plots for points in $\Tilde{\mathbb{R}}^D$. Yet, for given $x$ we might also consider the states that we get for $\lim_{l\to\infty}\ket{lx}$, the so called asymptotic states. We can obtain them by considering 
the asymptotic Hamiltonian (\ref{HamiltonianAsympt}) where we drop the $\mathcal{O}(\vert x\vert^{-1})$ terms, allowing us to explicitly compute\footnote{Especially we can restrict to points of norm $\vert x\vert=1$.} the latter as lowest eigenstate of $-\sum_a\frac{x^a}{\vert x \vert}X^a$ (assuming that the lowest eigenspace is one dimensional). If we look at plots that show asymptotic states, we speak of \textit{asymptotic plots}.
\\
In some sense, these states form part of the closure of $\mathcal{M}$.

\begin{figure}[H]
\centering
\begin{minipage}{.3\textwidth}
  \centering
  \includegraphics[height=.7\linewidth]{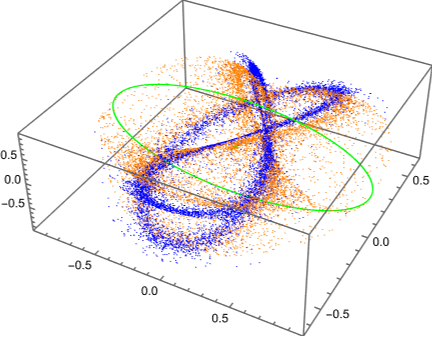}
\end{minipage}%
\begin{minipage}{.3\textwidth}
  \centering
  \includegraphics[height=.7\linewidth]{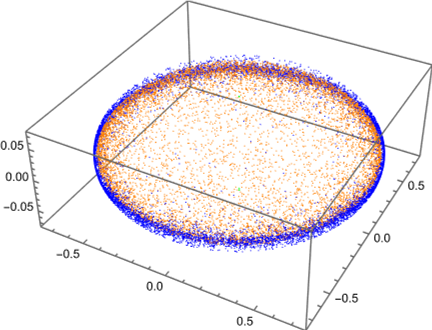}
\end{minipage}%
\caption{Random points (orange), random asymptotic points (blue) and random points on a sphere of radius $0.001$ (green) in the squashed fuzzy sphere for $N=4$ and $\alpha=0.1$. Left: projective plot of $\mathcal{M}$, right: projective plot of $\Tilde{\mathcal{M}}$}
\label{fig:Implementation/4}
\end{figure}

In figure \ref{fig:Implementation/4} we can see such plots together with random points within a ball of unit radius as in figure \ref{fig:Implementation/2} and points on a sphere of radius $0.001$, where the latter represents the opposite limit for $\lim_{l\to0}\ket{l x}$.\\ For example on the left, we see that the asymptotic states (the blue points) form a boundary of the random points within (the orange points). Note that the gap in between is a result of our restriction to the unit ball for the random points.

\newpage
\subsection{Foliations, Local Coordinates, Integration and Completeness}
\label{Foliat}

The next step is to implement the various distributions from section \ref{fol}. In section \ref{calcDist} we begin with calculating the projectors on the distribution, in turn allowing us to integrate curves in the leaves $\mathcal{L}$ in section \ref{calcCurv}.\\
With such curves we \textit{scan} $\mathcal{L}$ globally in section \ref{calcScan} and construct local coordinates in section \ref{calcCoord}.\\
Via \textit{tilings}, introduced in section \ref{calcTile}, we are ready to integrate over the leaves in section \ref{calcInt}, allowing us to check for example the completeness relation (\ref{compl}).\\
Finally, in section \ref{calcMin} an approach for choosing a special leaf is presented, based on the attempt to minimize $\lambda$ over $\mathcal{N}^\mathcal{L}_x$.

\subsubsection{Calculation of the Distributions}
\label{calcDist}

In section \ref{fol} we have defined the respective leaves via distributions in the tangent bundle in the spirit of the Frobenius theorem. Especially, we were working with $\Tilde{\mathbb{R}}^D$, where the tangent spaces are isomorphic to $\mathbb{R}^D$ and thus rather simple. Any such distribution induces a distinguished subspace $V_x$ in $T_x\Tilde{\mathbb{R}}^D\cong \mathbb{R}^D$ and consequently we get an orthogonal projector $p_x$ on this subspace with respect to any inner product.\\
In principle, the leaf is independent of the choice of an inner product, while the specific curves that we will integrate in the next section are dependent. Thus we use $\delta_{ab}$ as our generic choice, while it is conceptually more obvious to use $g_{ab}$ for the symplectic leaf and the hybrid leaf using $\omega_{ab}$. (In fact, in the latter case both versions will be implemented and it will always be specified which metric is used.)\\
Knowing these projectors for all $x\in\Tilde{\mathbb{R}}^D$ characterizes the distribution, thus our current task is to calculate the orthogonal projector $p_x$ on $V_x$ for given $x$.

Let us begin with the \textit{symplectic leaf}, for which the construction has been discussed in section \ref{symp}: We split the tangent space at $x$ in $T_x\Tilde{\mathbb{R}}^D=T_xW\oplus T_xZ\oplus T_xL$ and use $T_xL$ as our desired local subspace $V_x$.
\begin{itemize}
    \item The starting point is to calculate $g_{ab}$ and $\omega_{ab}$ as discussed in section \ref{BasicQuantities}.
    \item We find a basis of $T_xW$ by calculating the kernel of $(g_{ab})$.
    \item Then, we orthonormalize this basis with respect to $\delta_{ab}$, providing us with the normalized vectors $w_1,\ldots,w_r$ where $r=\operatorname{dim}(\operatorname{ker}(\operatorname{g_{ab}}))=D-k$.
    \item This defines a nondegenerate bilinear form on the whole $T_x\Tilde{\mathbb{R}}^D$ via $\Tilde{g}_{ab}:=\sum_{m=1}^r (w_m)^a(w_m)^b+g_{ab}$. (In section \ref{symp} this has been called $g_W$.)
    \item We would get the complement $T_xV=T_xW^\perp$ with respect to $\Tilde{g}_{ab}$ exactly via the kernel of $(\sum_{m=1}^r (w_m)^a(w_m)^b)$, but we actually do not need that.
    \item A basis of $T_xZ$ (by definition lying within $T_xV$) can then be obtained by calculation of the kernel of $\Tilde{\omega}_{ab}:=\sum_{m=1}^r (w_m)^a(w_m)^b+\omega_{ab}$. After another orthonormalization with respect to $\Tilde{g}_{ab}$, we find the basis $z_{r+1},\ldots,z_{r+s}$ of $T_xZ$, where $s=\operatorname{dim}(\operatorname{ker}(\Tilde{\omega}_{ab}))$.
    \item Then, a basis of $T_xL=(T_xW\oplus T_xZ)^\perp$ with respect to $\Tilde{g}_{ab}$ can be found via the kernel of $((w_1)^a,\ldots,(w_r)^a,(z_{r+1})^a,\ldots,(z_{r+s})^a)\cdot (\Tilde{g}_{ab})$, which after one further orthonormalization with respect to $\Tilde{g}_{ab}$ provides us with a basis $l_{r+s+1},\ldots,l_{D}$. This then also provides us with the symplectic dimension $l=D-r-s=k-s$.
    \item Finally, we get our desired projector $p_x=(\sum_{m=1}^s (l_{r+s+m})^a(l_{r+s+m})^c)\cdot(g_{cb})$.
\end{itemize}

In the case of the \textit{Pfaffian leaf}, the procedure is rather different. Here, we want to maximize the function $\mathcal{P}^s_x$ over the linear subspaces of $T_x\Tilde{\mathbb{R}}^D$ with fixed (even) dimension $s$ as discussed in section \ref{pfaff}, resulting in the optimal subspace $V_x$. The effective dimension $l$ is then determined in retrospect.
\begin{itemize}
    \item We start with a random subspace $W\subset T_x\Tilde{\mathbb{R}}^D\cong \mathbb{R}^D$ of dimension $s$, represented by an orthonormal basis $w_1,\dots,w_s$.
    \item The first task is to calculate $\mathcal{P}^s_x(W)=\bra{x}[X^{a_1},X^{a_2}]\dots[X^{a_{s-1}},X^{a_s}]\ket{x}\epsilon_W^{a_1\dots a_s}$ (with $\epsilon_W^{a_1\dots a_s}:=O^{a_1b_1}\dots O^{a_sb_s}\epsilon^{b_1\dots b_s}$ for any orthogonal $O$ mapping $\mathbb{R}^s\times \{0\}$ to $W$).\\
    Such an $O^{ab}$ (we only need the components with $b\leq s$) is simply given by the matrix $(w_1,\dots,w_s)$.\\
    The rest of the calculation is trivial but comes with large computational cost, so a clever implementation is advisable. A way in that direction is to define the matrices $\Tilde{X}^b:=\sum_a O^{ab}X^a$ for $b\leq s$, resulting in $\mathcal{P}^s_x(W)=2^{s/2}\bra{x}\Tilde{X}^{b_1}\dots \Tilde{X}^{b_s}\ket{x}\epsilon^{b_1\dots b_s}$.\\
    This can then simply be rewritten as 
    \begin{align}
        \mathcal{P}^s_x(W)=2^{s/2}\sum_{\sigma}\operatorname{sgn}(\sigma) \bra{x}\Tilde{X}^{\sigma_1}\dots \Tilde{X}^{\sigma_s}\ket{x},
    \end{align}
    where the sum runs over all permutations $\sigma$ of $(1,\dots,s)$. In this formulation, we have much fewer summands than in the original formulation.
    \item For the optimization process, it is also necessary to calculate the \textit{gradient} and the \textit{hessian} of $\mathcal{P}^s_x$ with respect to $O^{ab}$ (as the optimization will actually be done with respect to $O^{ab}$, representing $W$).\\
    Analytically, the gradient is given by
    \begin{align}
        \frac{d}{dO^{cd}}\mathcal{P}^s_x(W)=2^{s/2}\sum_{t=1}^s\sum_{a=1}^D\sum_{\sigma}\operatorname{sgn}(\sigma)\bra{x}\Tilde{X}^{\sigma_1}\dots X^{a}\mathcal{M}^{a\sigma_t}_{cd} \dots  \Tilde{X}^{\sigma_s}\ket{x}
    \end{align}
    with $\mathcal{M}^{ab}_{cd}:=\delta^{a}_c\delta^{b}_d$.\\
    Similarly, we find the hessian
    \begin{align}
        \frac{d}{dO^{ef}}\frac{d}{dO^{cd}}\mathcal{P}^s_x(W)=2^{s/2}\sum_{t\neq u=1}^s\sum_{a,a'=1}^D \sum_{\sigma}\bra{x}\Tilde{X}^{\sigma_1}\dots X^{a}\mathcal{M}^{a\sigma_t}_{cd} \dots \dots X^{a'}\mathcal{M}^{a'\sigma_u}_{cd}  \Tilde{X}^{\sigma_s}\ket{x}.
    \end{align}
    \item For further efficiency, we note the relations $\mathcal{P}^s_x(W)=\frac{1}{s}O^{cd}\frac{d}{dO^{cd}}\mathcal{P}^s_x(W)$ and $\frac{d}{dO^{cd}}\mathcal{P}^s_x(W)=\frac{1}{s-1}O^{ef}\frac{d}{dO^{ef}}\frac{d}{dO^{cd}}\mathcal{P}^s_x(W)$.
    \item Since we are actually interested in the absolute square of $\mathcal{P}^s_x$, marginal modifications have to be made, but that shall not bother us any further.
    \item For the actual maximization, we use the algorithm described in \cite{Johnsson_2012}, providing a gradient descent method and a Newton procedure for real valued functions on Stiefel manifolds adapted to Grassmannian manifolds, where the optimization is with respect to the $O^{ab}$ from above. The output is then given by a matrix $O^{ab}$, corresponding to the optimal subspace $V_x=O(\mathbb{R}^s\times\{0\})$. The orthogonal projector on $V_x$ is simply given by $(p_x)^a_c=\sum_{b=1}^sO^{ab}O^{cb}$.
    \item However, we still have to determine the correct $s$ that we use as effective dimension $l$.
    Therefore we calculate $v_x^s=\vert\mathcal{P}^s_x(V_x)^s\vert^2$ for all even $s$ and define $l$ as the largest $s$ such that $v_x^s\gg 0$ (actually defined as $v_x^s> \epsilon$ for a lower bound $\epsilon$ of our choice).\\
    Practically, it suffices to check $l$ for a few points and then always use $s=l$. 
\end{itemize}

For the \textit{hybrid leaf}, it suffices to know $\theta^{ab}(x)$ (respectively $\omega_{ab}(x)$ and $g_{ab}(x)$), while the remaining calculations are then straight forward.
\begin{itemize}
    \item All we need is the eigensystem of $\theta:=(\theta^{ab}(x))$, given in the ordering $\vert\lambda_i\vert\leq\vert\lambda_j\vert$ if $i>j$ together with the corresponding eigenvectors $v_i$.
    \item As described in section \ref{hybrid}, we choose $l$ such that $\vert \lambda_l\vert\gg 0$ while $\vert \lambda_{l+1}\vert\approx0$.
    \item Then, we define $w_{2s-1}=\operatorname{Re}(v_{2s-1})=\operatorname{Re}(v_{2s})$ and $w_{2s}=\pm\operatorname{Im}(v_{2s-1})=\mp\operatorname{Im}(v_{2s})$ for $s=1,\dots,l/2$, providing us with the basis $w_1,\dots,w_l$ of $V_x$.
    \item Finally, we calculate a corresponding orthonormal basis $\Tilde{w}_1,\dots,\Tilde{w}_l$, allowing us to define $(p_x)^a_b=\sum_{s=1}^l (\Tilde{w}_s)^a(\Tilde{w}_s)^b$.
    \item When we use $g_{ab}$ as our metric instead of $\delta_{ab}$ two simple adaptions have to be made: At first, we have to orthonormalize $w_1,\dots,w_l$ with respect to $g_{ab}$, resulting in the basis $\Tilde{w}_1,\dots,\Tilde{w}_l$. Then we get the modified projector $(p_x)^a_b=\sum_c\sum_{s=1}^l (\Tilde{w}_s)^a(\Tilde{w}_s)^c g_{cb}$.
    \item For $\omega_{ab}(x)$, the procedure is exactly the same. Having in mind the discussion in section \ref{comparison}, it is natural to use $\delta_{ab}$ when using $\theta^{ab}$ and $g_{ab}$ when using $\omega_{ab}$, but using $\delta_{ab}$ anyways is also reasonable. To keep the overview we introduce the following nomenclature: If we talk of the \textit{hybrid leaf} we use $\theta^{ab}$ and $\delta_{ab}$. For the \textit{hybrid leaf using $\omega$}, we use $\omega_{ab}$ and $\delta_{ab}$ and only if we talk of the \textit{hybrid leaf using $\omega$ and $g$} we actually work with $\omega_{ab}$ and $g_{ab}$.
\end{itemize}

Finally, we consider the \textit{Kähler leaf}. Here, the calculation is similar to the Pfaffian leaf, but our cost function looks different.
\begin{itemize}
    \item Again, we start with a random subspace $W\subset T_x\Tilde{\mathbb{R}}^D\cong \mathbb{R}^D$ of dimension $s$, represented by an orthonormal basis $w_1,\dots,w_s$.
    \item Here, we have to calculate the cost $c^s_x(W)=d(T_xq(W),J (T_xq(W)))=:d(V,V')$, where \\$d(V_1,V_2)=\Vert (P_{V_2}-\mathbb{1})\circ i_{V_1} \Vert_2$ and $J$ represents the multiplication with $i$ in $\mathbb{C}^N$.\\
    We start by calculating $T_xq(w_t)=\sum_{a=1}^D(w_t)^aD_a\ket{x}$, which we then view as a real vector in $\mathbb{R}^{2N}$.
    In general, this gives us an overcomplete basis of $T_xq(W)$ that is \textit{not} orthonormal, thus we orthonormalize it and obtain the basis $\hat{w}_1,\dots,\hat{w}_{s'}$, where $s'$ is the dimension of $T_xq(W)$. The matrix $J$ depends on our choice on how we identify $\mathbb{C}^N\cong \mathbb{R}^{2N}$, but in any case $J$ is a simple orthogonal matrix, thus $\Tilde{w}_1,\dots,\Tilde{w}_{s'}$ for $\Tilde{w}_i:=J\hat{w}_i$ is an orthonormal basis of $J(T_xq(W))$.\\
    In order to write down the operator $(P_{V'}-\mathbb{1})\circ i_{V}$ explicitly, we need to extend\footnote{This can be done as follows: We get an orthonormal basis of the complement by calculating the kernel of $(\sum_{t=1}^{s'}(\Tilde{w}_t)^a(\Tilde{w}_t)^b)$ and subsequently orthonormalizing the obtained basis.} the basis $\Tilde{w}_i$ to an orthonormal basis $\Tilde{w}_{1},\dots,\Tilde{w}_{2N}$ of $\mathbb{R}^{2N}$.
    Then, in this basis we find $((P_{V'}-\mathbb{1})\circ i_{V})^{ab}=\sum_{c=1}^{s'}(\Tilde{w}_a\cdot \hat{w}_c)(\hat{w}_c\cdot \Tilde{w}_b)-\delta_{ab}$, where $a=1,\dots,D$ and $b=1,\dots,s'$.
    The spectral norm can then be calculated using built in methods.
    
    If we replace the norm $\Vert\cdot\Vert_2\mapsto\Vert \cdot\Vert_{HS}$, we find the explicit result
    \begin{align}
        \Vert(P_{V_2}-\mathbb{1})\circ i_{V_1}\Vert_{HS}=\sqrt{s'-\sum_{a,b=1}^{s'}(\Tilde{w}_a\cdot\hat{w}_b)^2}.
    \end{align}
    \item While it is possible to calculate the gradient of $c_x^s$ explicitly, it is computationally faster to calculate it via finite differences. This is due to the fact that the gradient of the Gram-Schmidt procedure is rather complicated.
    \item The remaining part of the procedure is completely parallel to the Pfaffian leaf, only we maximize $-c_x^s$ instead of $\vert\mathcal{P}^s_x\vert^2$, thus we do not repeat the discussion.
\end{itemize}

For any of the methods above, having a basis of $V_x\subset T_x\Tilde{\mathbb{R}}^D$, it is an easy task to calculate a basis of the corresponding subspace $T_xq(V_x)=T_{q(x)}\mathcal{L}\subset T_{q(x)}\mathcal{\mathcal{M}}$ via $T_xq$, while in practice this will not be needed since most computations will take place in $\Tilde{\mathbb{R}}^D$.\\
All of the following discussion is completely agnostic to the choice of a special leaf. What is important is that we can calculate the respective $p_x$.

\subsubsection{Integrating Curves in the Leaves}
\label{calcCurv}

As we are able to calculate the orthogonal projectors $p_x$ onto the distributions (in target space) for given $x\in\Tilde{\mathbb{R}}^D$, our next task is to integrate (discretized) curves within the respective leaf through $q(x)$.\\ In the following steps we explicitly construct a discrete curve in $\Tilde{\mathbb{R}}^D$ for a given initial tangent vector $v\in T_x\Tilde{\mathbb{R}}^D\cong\mathbb{R}^D$. Using $q$, this can then be lifted to a discrete curve in the leaf $\mathcal{L}\subset\mathcal{M}$ through $q(x)$ with initial tangent vector $T_xq(p_x(v))$. (Thus slightly abusively we also say that the curve in $\Tilde{\mathbb{R}}^D$ \textit{lies within the leaf} through $x$.)
\begin{itemize}
    \item At first, we have to fix a small but finite step length $\delta$ and define $x_0=x$ as well as $v_0=v$.
    \item If we already have $x_i$ and $v_i$, we define
    \begin{align}
        x_{i+1}=x_i+\delta \frac{p_{x_i}(v_i)}{\vert p_{x_i}(v_i)\vert},
    \end{align}
    where the norm is usually taken with respect to the metric $\delta_{ab}$, while we would also be free to use $g_{ab}$ in principle.
    \item In order to proceed further it is also necessary to define $v_{i+1}$. For that we have the two choices
    \begin{align}
        v_{i+1}=v \quad \text{or} \quad v_{i+1}=p_{x_i}(v_i).
    \end{align}
    In the first case we hold the unprojected tangent vector fixed (we then speak of a \textit{fixed tangent vector}), while in the second we constantly project the previous tangent vector back into the distribution (thus we speak of an \textit{adapted tangent vector}).
    \item We continue this procedure until we have calculated the desired amount of points.
\end{itemize}
Let us briefly discuss what we have gained.
We constructed a set of points $\{x_i\}\subset\Tilde{\mathbb{R}}^D$ for $i=1,\dots,n$, which we take as an approximation to a curve $\gamma:[0,n]\to\Tilde{\mathbb{R}}^D$ with $\gamma(i)=x_i$. By our construction, we have $\Dot{\gamma}(i)\approx \delta p_{x_i}(v_i)/\vert p_{x_i}(v_i)\vert$, thus all tangent vectors of $\gamma$ approximately lie within the chosen distribution -- but this exactly means that the curve approximately lies within the corresponding leaf.\\
Further, the norm of the tangent vector is approximately confined to $\delta$, so the curve is parameterized by path length (up to $\delta$) with respect to the chosen metric and has approximately length $n\delta$. 
If we want to calculate the length with respect to another metric $g'_{ab}$ (for example if we used $\delta_{ab}$ in the construction but want to know the length with respect to $g_{ab}$), this can be done via
\begin{align}
    \vert \gamma \vert_{g'}\approx\sum_{i=1}^{n-1} \sqrt{g'_{ab}(x_i)(x_{i+1}-x_i)^a(x_{i+1}-x_i)^b}.
\end{align}
Finally, we note that if we choose to adapt the tangent vector, this means that $\gamma$ approximately is a geodesic within the leaf with respect to the restriction of the chosen metric\footnote{This is only true if the metric used here and the metric used to define $p_x$ coincide.}.

\begin{figure}[H]
\centering\begin{minipage}{.3\textwidth}
  \centering
  \includegraphics[height=.7\linewidth]{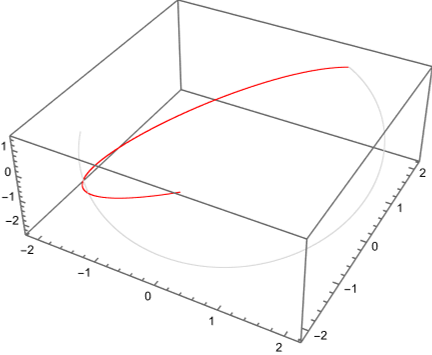}
  \includegraphics[height=.7\linewidth]{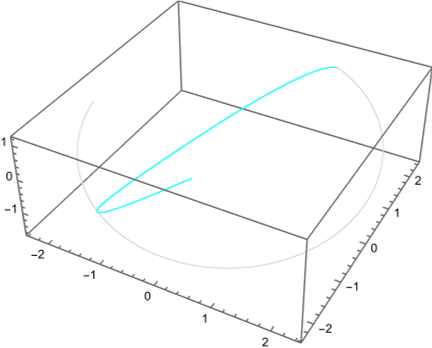}
\end{minipage}%
\begin{minipage}{.3\textwidth}
  \centering
  \includegraphics[height=.7\linewidth]{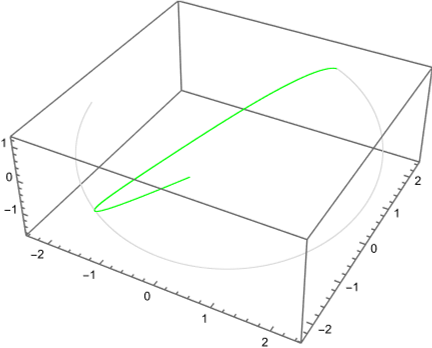}
  \includegraphics[height=.7\linewidth]{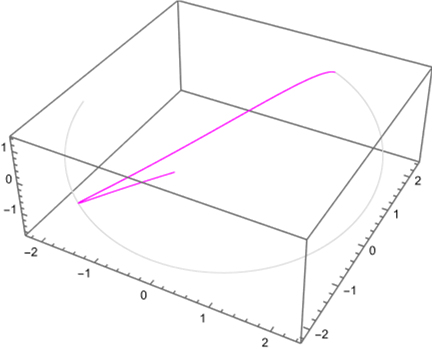}
\end{minipage}%
\begin{minipage}{.3\textwidth}
  \centering
  \includegraphics[height=.7\linewidth]{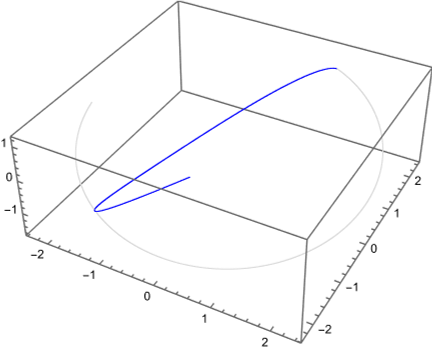}
  \includegraphics[height=.7\linewidth]{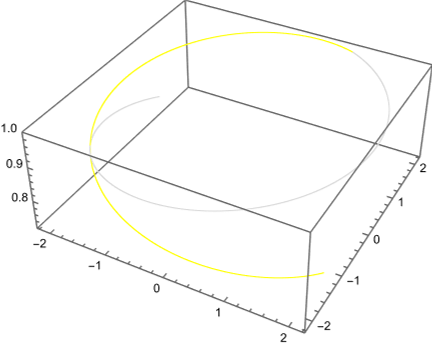}
\end{minipage}%
\\
\begin{minipage}{.3\textwidth}
  \centering
  \includegraphics[height=.7\linewidth]{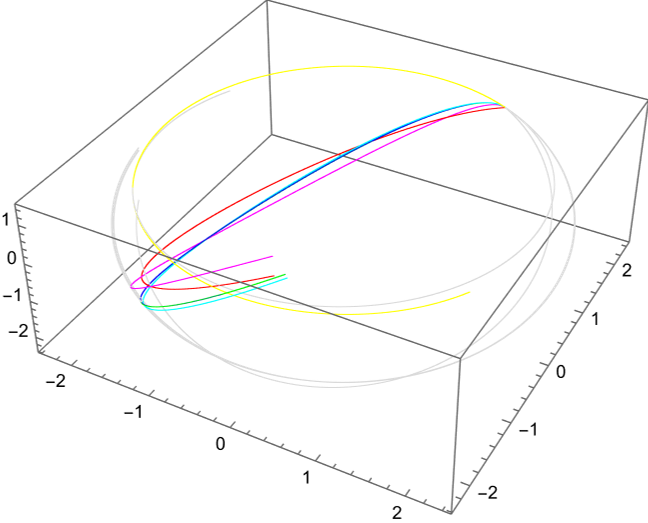}
\end{minipage}%
\begin{minipage}{.3\textwidth}
  \centering
  \includegraphics[height=.7\linewidth]{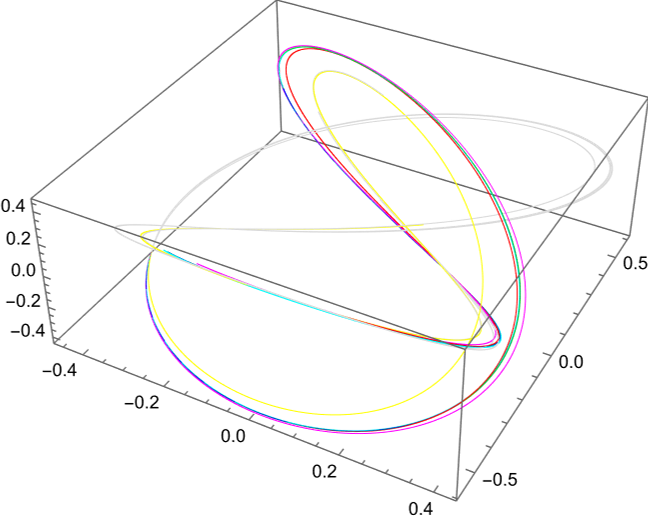}
\end{minipage}%
\caption{Curves in the respective leaves starting at $x=(1,2,1)$ with adapted tangent vectors $v_1=(0,1,0)$ and $v_2=(1,0,0)$ in the squashed fuzzy sphere for $N=4$ and $\alpha=0.1$. Symplectic leaf: red and gray, Pfaffian leaf: green and gray, hybrid leaf: blue and gray, Hybrid leaf using $\omega$: cyan and gray, hybrid leaf using $\omega$ and $g$: magenta and gray, Kähler leaf: yellow and gray; top: projective plots of $\Tilde{\mathbb{R}}^D$ for the six different leaves, bottom-left: projective plot of $\Tilde{\mathbb{R}}^D$ of all leaves, bottom-right: projective plot of $\mathcal{M}$ of all leaves. Note that the green and blue line agree \textit{exactly}}
\label{fig:Implementation/5A}
\end{figure}

In figure \ref{fig:Implementation/5A} we see curves in all the leaves with same initial $x$ and two different initial adapted tangent vectors $v_1$ and $v_2$. Looking at the single leaves, the really interesting observation is that in every leaf the two curves intersect each other away from the initial point (both in $\Tilde{\mathbb{R}}^D$ and $\mathcal{M}$), meaning the distributions are (approximately) integrable. However, we have never demanded that they are already integrable in $\Tilde{\mathbb{R}}^D$, but only in $\mathcal{M}$. Later, it turns out that this is only an artifact of the fact that we considered the squashed fuzzy sphere, while the integrability in $\mathcal{M}$ also remains intact for more general matrix configurations.\\
Comparing the different leaves, we directly note that not all are equivalent as not all colored curves intersect away from the initial point.\\
We see that only the curves in the Pfaffian leaf and the hybrid leaf (using $\theta$) strictly agree (That they agree is obvious for $D=3$. Also $l=2$ is immediate.), while at least the curves in the symplectic leaf and the hybrid leaf using $\omega$ are not too different (both in $\Tilde{\mathbb{R}}^D$ and $\mathcal{M}$).\\
That the curves in the hybrid leaf using $\omega$ respectively the hybrid leaf using $\omega$ and $g$ intersect each other away from the initial point fits to the fact that both leaves are in principle identical, while the curves themselves do not agree due to the different projections $p_x$.\\ 
However, the Kähler leaf is strongly different. Graphically, we see that there we almost have\\ $p_x(v_1)/\vert p_x(v_1)\vert\approx-p_x(v_2)/\vert p_x(v_2)\vert$.\\
Here, also the computational time is interesting: While the hybrid methods as well as the symplectic method are rather fast, the Pfaffian method is a bit slower. The Kähler method is already much slower.

\subsubsection{Scanning the Leaves}
\label{calcScan}

The next step is to analyze the leaves globally (or at least on a large scale). This can be done by a simple procedure that we call \textit{scanning the leaf} through $x\in\Tilde{\mathbb{R}}^D$.
\begin{itemize}
    \item For each $i=0,\dots,n'$ we define the points $x_{ij}$ ($j=1,\dots,n$) as the points we get from integrating the curve through $x_{i0}$ with adapted initial tangent vector $v_{i0}$.
    \item It remains to define $x_{i0}$ and $v_{i0}$. This we do iteratively.
    \item For $i=0$, we put $x_{i0}=x$ and choose $v_{i0}$ randomly.
    \item For $i>0$, we randomly pick a point from the $x_{i'j}$ for $i'<i$ that we use as $x_{i0}$ and again choose a completely random $v_{i0}$.
    \item In order to obtain reproducible results it is advisable to \textit{seed} the random number generator in advance.
\end{itemize}
This procedure never leaves the leaf through $x$ since all the curves that we integrate remain within individually. Choosing the $v_{i0}$ randomly allows us to \textit{look} into many different directions.

\begin{figure}[H]
\centering
\begin{minipage}{.3\textwidth}
  \centering
  \includegraphics[height=.7\linewidth]{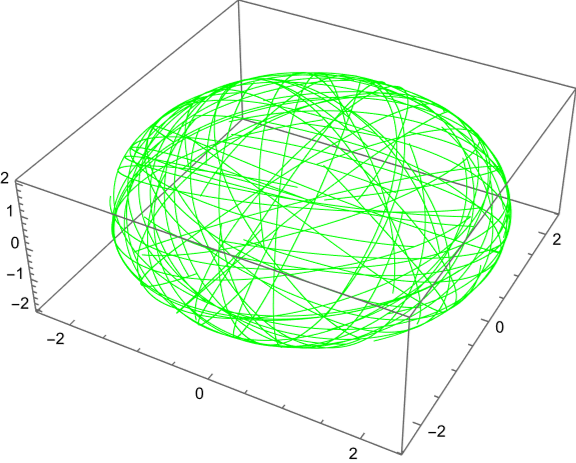}
\end{minipage}%
\begin{minipage}{.3\textwidth}
  \centering
  \includegraphics[height=.7\linewidth]{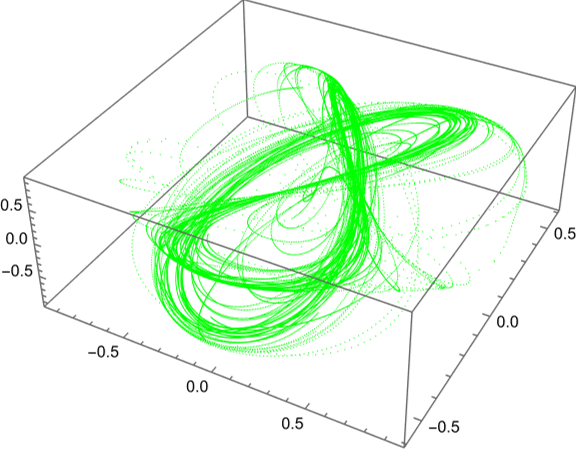}
\end{minipage}%
\begin{minipage}{.3\textwidth}
  \centering
  \includegraphics[height=.7\linewidth]{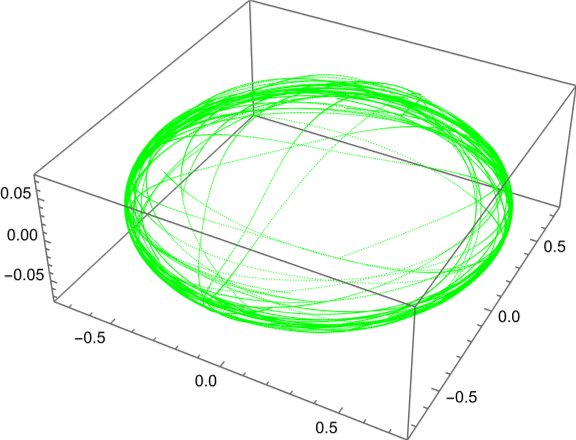}
\end{minipage}%
\caption{Scan of the hybrid leaf through $x=(1,2,1)$ in the squashed fuzzy sphere for $N=4$ and $\alpha=0.1$. Left: projective plot of $\Tilde{\mathbb{R}}^D$, middle: projective plot of $\mathcal{M}$, right: projective plot of $\Tilde{\mathcal{M}}$}
\label{fig:Implementation/6}
\end{figure}

Figure \ref{fig:Implementation/6} shows a scan of the hybrid leaf. We can see that this allows us to understand the global structure of the leaf and strengthens our trust in the integrability (of the hybrid leaf). In the plot of $\Tilde{\mathcal{M}}$ we see how strongly the squashing \textit{pushes} points towards the equator, yet still points in the polar regions remain part of $\Tilde{\mathcal{M}}$.

\subsubsection{Construction of Local Coordinates}
\label{calcCoord}

Having found a tool to analyze the global structure of the leaves, we now need to consider the local structure. Therefore we want to find well adapted \textit{local coordinates}.
In principle, there are two kinds of coordinates that we can construct.

Let us start with the \textit{coordinates of the first kind} around $x\in\Tilde{\mathbb{R}}^D$.
\begin{itemize}
    \item Our aim is to iteratively construct points $x_{i_1\dots i_l}$ with $x_{0\dots 0}=x$ and $i_j=-n,\dots,n$.
    \item We pick an orthonormal basis $v_1,\dots,v_l$ of $V_x\subset T_x\Tilde{\mathbb{R}}^D$. Usually, we do this with respect to $\delta_{ab}$, but in principle we could also use $g_{ab}$.
    \item Then, we define the points $x_{i_1\dots i_{s-1} i_s 0\dots 0}$ as the points we get from integrating the curve through $x_{i_1\dots i_{s-1} 0 0\dots 0}$ with fixed tangent vector $v_s$ if $i_s>0$ (respectively $-v_s$ if $i_s<0$) and step length $\delta$. This uniquely fixes all $x_{i_1\dots i_l}$.
    \item We also want to calculate the Jacobian $J_{i_1\dots i_l}$ for each $y:=x_{i_1\dots i_l}$.\\
    Therefore, we define $y_s^\pm$ as the point we reach by integrating one step with step length $\epsilon\delta$ with initial tangent vector $\pm v_s$.
    Then, we set $(J_{i_1\dots i_l})^a_s:=\frac{1}{2\epsilon}(y_s^+-y_s^-)^a$ for some small $\epsilon>0$ that allows us to build a difference quotient.
    \item Let us finally consider the integration of any of the involved curves, producing the point $y_i$ for $i=0,\dots,\Tilde{n}$ with step length $\Tilde{\delta}$.
    In order to improve our results it may be advantageous to reduce the step length $\Tilde{\delta}\mapsto \frac{1}{m}\Tilde{\delta}$ and consequently calculate $m \Tilde{n}$ points $y'_j$ for $j=0,\dots,m\Tilde{n}$ and finally put $y_i=y'_{mi}$ for some $m\in\mathbb{N}$. (Of course, this also applies to the calculation of the Jacobians.)\\
    If we proceed after this scheme, we say that we \textit{use} $m-1$ \textit{intermediate steps} in the curve integration, reducing the numerical errors we make by taking smaller step lengths.
\end{itemize}
Having this construction, we may think of the $x_{i_1\dots i_l}$ as a discretization of inverse coordinate functions $\psi^{-1}:(-n-\epsilon,n+\epsilon)^l\to \Tilde{\mathbb{R}}^D$ (in order to obtain true coordinate functions around $q(x)\in \mathcal{L}\subset\mathcal{M}$, we further had to apply $q$, but we will keep that implicit) such that $\psi^{-1}(i_1,\dots, i_l)=x_{i_1\dots i_l}$, where by construction all points $x_{i_1\dots i_l}$ lie within the leaf through $x$. In this picture we have $\partial_s(\psi^{-1})^a\vert_{(i_1,\dots, i_l)}\approx (J_{i_1\dots i_l})^a_s$.

For the \textit{coordinates of the second kind} around $x\in\Tilde{\mathbb{R}}^D$, we have to recall that the curves generated by adapting the tangent vector are geodesics in the leaf (with respect to the restriction of the chosen metric $\delta_{ab}$ or $g_{ab}$ to the leaf). Thus, we can simply construct an \textit{exponential map} analogous to the exponential map from Riemannian geometry.
\begin{itemize}
    \item Again, for a given initial point $x$ we choose an orthonormal basis $v_1,\dots,v_l$ of $V_x$.
    \item Let $\chi\in\mathbb{R}^l$. Then we define $v:=\delta \sum_{s=1}^l\chi^s v_s\in V_x$. We pick an $n\in\mathbb{N}$ such that $\frac{1}{n}\vert v\vert$ is of order $\delta$ (with respect to the chosen metric).
    \item Then we \textit{define} $\exp_x(\delta\chi)$ as the end point we get by integrating a curve with initial point $x$, adapted tangent vector $v$ and step length $\frac{1}{n}\vert v\vert$ for exactly $n$ steps.
    \item We can calculate the Jacobian simply as the differential quotient $\partial_s (\exp_x)^a\vert_\chi\approx \frac{1}{2\epsilon}(\exp(\chi+\epsilon v_s)-\exp(\chi-\epsilon v_s))^a$.
    \item In practice, we will evaluate $\exp_x$ on a lattice $\{- n,\dots, n\}^{\times l}$, providing us with the points $x_{i_1\dots i_l}:=\exp_x(i_1,\dots, i_l)$ and the Jacobians $(J_{i_1\dots i_l})^a_s:=\partial_s (\exp_x)^a\vert_{(i_1,\dots, i_l)}$.
\end{itemize}
Here, we should really think of $\exp_x: V_x\to \Tilde{\mathbb{R}}^D$ as the map that maps radial lines in the distribution at $x$ to geodesics in the leaf through $x$, locally defining so called \textit{normal coordinates}.

\begin{figure}[H]
\centering
\begin{minipage}{.3\textwidth}
  \centering
  \includegraphics[height=.7\linewidth]{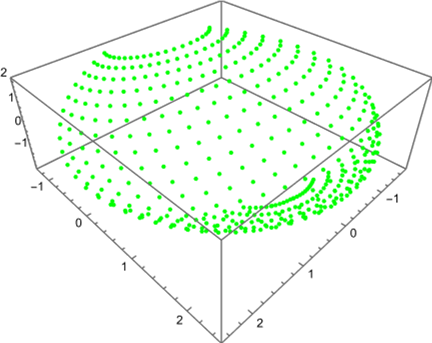}
  \includegraphics[height=.7\linewidth]{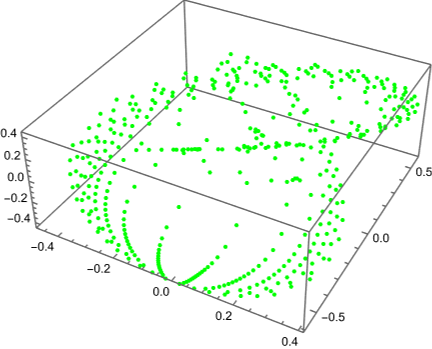}
\end{minipage}%
\begin{minipage}{.3\textwidth}
  \centering
  \includegraphics[height=.7\linewidth]{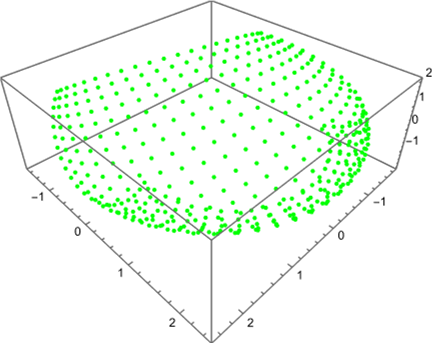}
  \includegraphics[height=.7\linewidth]{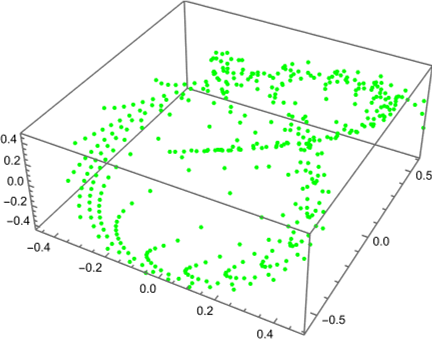}
\end{minipage}%
\caption{Local coordinates around $x=(1,2,1)$ for the hybrid leaf in the squashed fuzzy sphere for $N=4$ and $\alpha=0.1$. Top: projective plot of $\Tilde{\mathbb{R}}^D$, bottom: projective plot of $\mathcal{M}$; left: coordinates of the first kind, right: coordinates of the second kind}
\label{fig:Implementation/7}
\end{figure}

In figure \ref{fig:Implementation/7} we can see local coordinates of the first and second kind for the hybrid leaf.\\
The benefit of the coordinates of first kind is that they are fast to calculate, while for $y=x_{i_1,\dots,i_l}$ far away from $x$ it might happen that some of the $v_s$ lie within the kernel of $p_y$, meaning that the coordinates can not be continued further, while also the ordering of the $v_s$ takes a large role. \\
On the other hand the benefit of the coordinates of the second kind is that they are constructed via geodesic and completely independent of the ordering of the $v_s$. Further, they can easily be calculated for arbitrary $\chi$. Yet, the calculation is more cumbersome since it is not built on a recursion.\\
In the following, we will always use coordinates of the first kind.

\subsubsection{Tilings}
\label{calcTile}

In the following we want to integrate over the leaf $\mathcal{L}$ through some $x\in \Tilde{\mathbb{R}}^D$. Analytically, we would choose a covering with local coordinates and a subordinate partition of unity to calculate any integral.\\
Here, we use this idea to patch local coordinate charts together to a global picture. The key to that is to \textit{tile} $\Tilde{\mathbb{R}}^D$. By a tile, we mean a set of the form $T=[a_1,b_1]\times\dots\times[a_D,b_D]\subset\mathbb{R}^D$.
\begin{itemize}
    \item We begin by scanning the leaf through $x$.
    \item Then we choose tiles $T^\alpha$ in $\mathbb{R}^D$ that only overlap at their borders and satisfy the following conditions:
    They should be so small that the intersection of the leaf with a tile is diffeomorphic to $[0,1]^l$ (thus topologically trivial) but so large that the coarseness of the scan is small in comparison. Also, aside from where the leaf intersects the borders of a tile, it should not come too close to them. Finally, the unification of all tiles has to \textit{cover the scan}.
    \item Then, for each tile $T^\alpha$ that contains at least one point of the scan (a \textit{nonempty tile}) we choose a point $x^\alpha\in T^\alpha$ of the scan that is located as centered as possible. To achieve this, we might calculate the center of the tile, measure its distance (with respect to the $\Vert\cdot\Vert_{max}$ norm) to all points from the scan that lie within and then choose a point of minimal distance.
    \item For each nonempty tile $T^\alpha$, we generate local coordinates (of the first kind) around $x^\alpha$, while choosing $n$ so large that $x^\alpha_{i_1\dots,i_l}\notin T^\alpha$ if for at least one $s$ we have $i_s=\pm n$ (then we \textit{filled} the tile).
    \item Finally, for each nonempty tile we drop the coordinate points that lie outside the tile.
\end{itemize}

\begin{figure}[H]
\centering
\begin{minipage}{.3\textwidth}
  \centering
  \includegraphics[height=.7\linewidth]{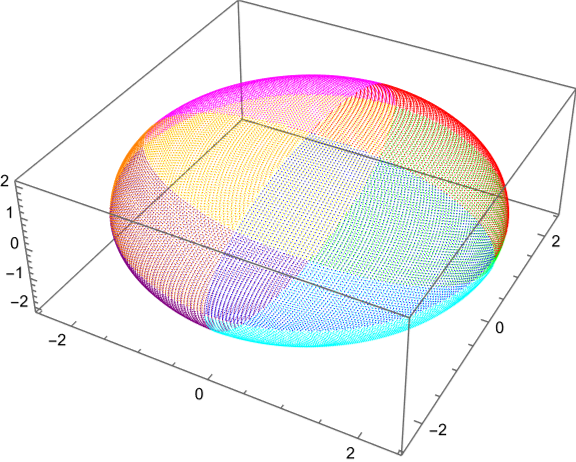}
  \includegraphics[height=.7\linewidth]{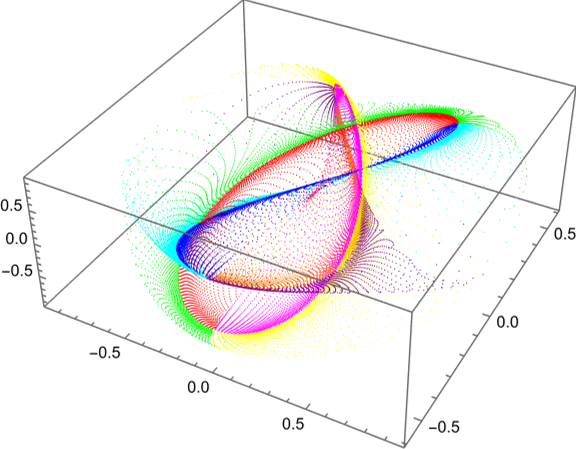}
\end{minipage}%
\begin{minipage}{.3\textwidth}
  \centering
  \includegraphics[height=.7\linewidth]{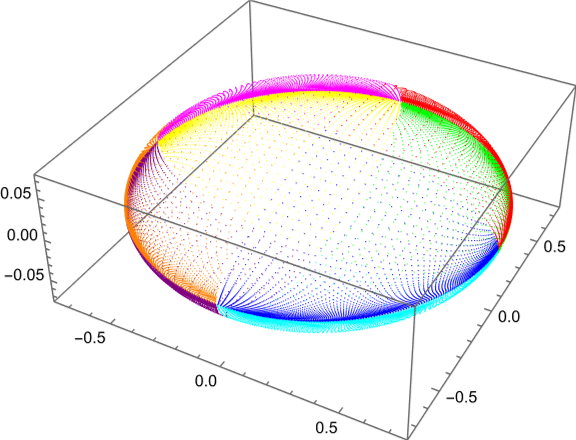}
  \includegraphics[height=.7\linewidth]{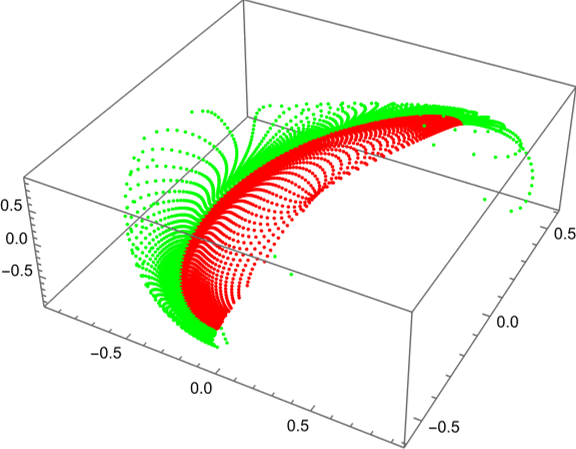}
\end{minipage}%
\caption{Tiling of the hybrid leaf through $x=(1,2,1)$ in the squashed fuzzy sphere for $N=4$ and $\alpha=0.1$ where the tiles are given by the octants in $\mathbb{R}^3$. Top-left: projective plot of $\Tilde{\mathbb{R}}^D$, top-right: projective plot of $\Tilde{\mathcal{M}}$, bottom-left: projective plot of $\mathcal{M}$, bottom-right: projective plot of $\mathcal{M}$ of two octants}
\label{fig:Implementation/8}
\end{figure}

In figure \ref{fig:Implementation/8} we can see a \textit{covering with local coordinates} patched together to give a global picture of the hybrid leaf, where we have chosen the tiles to be the octants in $\mathbb{R}^3$. We can see that we filled all tiles, no tiles are empty and obviously we covered the whole scan. While the individual coordinates do not fit together (this we can see readily from their random directedness), this construction will turn out to be very practical in the next section.

\subsubsection{Integration over the Leaves}
\label{calcInt}

Now that we covered the whole leaf with local coordinates\footnote{Here, we used the step length $0.05$. This will be the generic choice in most later examples, although weighted with the quotient of the respective $\vert x\vert$.}, it is an easy task to integrate over it. The main purpose is to verify the \textit{completeness relation} (\ref{compl}) and the \textit{quantization of the} $\mathbf{x}^a$ as in equation (\ref{quant}). Accordingly, we perform the following steps.
\begin{itemize}
    \item We cover the leaf with coordinates $x_{i_1\dots i_l}^\alpha$ as above and also calculate the corresponding Jacobians $J_{i_1\dots i_l}^\alpha$, where $\alpha$ is the index of the (nonempty) tile $T^\alpha$. We define $I(x_{i_1\dots i_l}^\alpha)$ as one if $x_{i_1\dots i_l}^\alpha\in T^\alpha$ and zero otherwise, implementing a discrete version of a partition of unity.
    \item Since our volume form is defined as $\Omega_\mathcal{L}=\frac{1}{(l/2)!}\omega_\mathcal{L}^{\wedge l/2}$, we calculate the pullback of $\omega_\mathcal{M}$ to the leaf. In our coordinates, we have
    \begin{align}
        \omega^\alpha_{st}(i_1,\dots, i_l )=\sum_{a,b}(J_{i_1\dots i_l}^\alpha)^a_s(J_{i_1\dots i_l}^\alpha)^b_t\omega_{ab}(x_{i_1\dots i_l}^\alpha)
    \end{align}
    (where at least for the symplectic leaf and the hybrid leaf based on $\omega_{ab}$ these $\omega^\alpha_{st}$ are guaranteed to be nondegenerate).
    \item Let now $f$ be a map from the leaf to a vector space. Then, we find the numerical approximation to the integral of $f$ over $\mathcal{L}$ with respect to $\Omega_\mathcal{L}$
    \begin{align}
        \int_\mathcal{L}\Omega_\mathcal{L}\; f\approx\sum_\alpha \sum_{i_1,\dots, i_l} I(x_{i_1\dots i_l}^\alpha) \sqrt{\vert\operatorname{det}(\omega^\alpha_{st}(i_1,\dots, i_l)) \vert}\; f(q(x_{i_1\dots i_l}^\alpha)),
    \end{align}
    noting that every point $(i_1,\dots,i_l)$ represents a unit volume in $\mathbb{R}^l$.
    \item If we want to use $f(q(x))=\ket{x}\bra{x}$, then all we need to do is calculate $\ket{x_{i_1\dots i_l}^\alpha}$ for all $x_{i_1\dots i_l}^\alpha$, while for $f(q(x))=\mathbf{x}^a\ket{x}\bra{x}$ we additionally need $\mathbf{x}^a(x_{i_1\dots i_l}^\alpha)$. For $f(q(x))=1$, we simply find the \textit{symplectic volume} of the leaf.
\end{itemize}

We once again consider the squashed fuzzy sphere for $N=4$ and $\alpha=0.1$.\\
We define the quantization of $1_\mathcal{L}$ (up to a proportionality factor)
\begin{align}
    \mathbb{1}':=\int_\mathcal{L}\Omega_\mathcal{L}\;\ket{\cdot}\bra{\cdot}
\end{align}
as well as the quantization of $\mathbf{x}^a$ (again up to a proportionality factor)
\begin{align}
    (X')^a:=\int_\mathcal{L}\Omega_\mathcal{L}\;\mathbf{x}^a\ket{\cdot}\bra{\cdot}
\end{align}
and the symplectic volume
\begin{align}
    V_\omega=\int_\mathcal{L}\Omega_\mathcal{L}=\operatorname{tr}(\mathbb{1}'),
\end{align}
where here $\mathcal{L}$ is the hybrid leaf through $x=(1,2,1)$.
Since we have the relation
\begin{align}
    \frac{\alpha}{(2\pi)^{l/2}}V_\omega=N
\end{align}
for the constant $\alpha$ (that is not related to the squashing parameter) in equation (\ref{QuantMap}),\\
this completely fixes the proportionality factor from above to $\frac{N}{V_\omega}$.
\\
Numerically, we find $V_\omega=10.515$ and consequently $\alpha= 2.390$.

According to the above, we suspect the  $\frac{N}{V_\omega} \mathbb{1}' \overset{?}{=}\mathbb{1}$ (assuming that the completeness relation (\ref{compl}) holds).
We find the mean of the eigenvalues of $\frac{N}{V_\omega} \mathbb{1}'$ to be $\mu_{\mathbb{1}'}=1.000$ (what is clear by construction) with a standard deviation of $\sigma_{\mathbb{1}'}=0.139$, so the result is promising. Alternatively, we find the relative deviation $d_{\mathbb{1}'}:=\Vert \frac{N}{V_\omega} \mathbb{1}'-\mathbb{1}\Vert_{HS}/\Vert \mathbb{1}\Vert_{HS}=0.121$.

Now, we come to the verification of equation (\ref{quant}) (the quantization of the $\mathbf{x}^a$): We thus suspect $\frac{N}{V_\omega} (X')^a\overset{?}{=}X^a$. Similarly as before, we consider the relative deviation\footnote{Here, by $\Vert\cdot\Vert_{HS'}$ we mean the square root of the sum over the square of all components, generalizing the Hilbert-Schmidt norm to objects with three indices.} $\Vert \frac{N}{V_\omega} X'-(X^a)\Vert_{HS'}/\Vert (X^a)\Vert_{HS'}$\\$=0.422$ which is not very encouraging. Thus, we look at $n_1:=\Vert \frac{N}{V_\omega} X'\Vert_{HS'}/(DN^2)=0.020$ and $n_2:=\Vert (X^a)\Vert_{HS'}/(DN^2)=0.034$ and consequently $n_{X'}:=\frac{n_2}{n_1}=1.725\gg 1$, so the previous result is not surprising.
So we may still hope and suspect $n_{X'}\frac{N}{V_\omega} (X')^a\overset{?}{=}X^a$. Here, we find the relative deviation $d_{X'}:=\Vert n_{X'}\frac{N}{V_\omega}  X'-(X^a)\Vert_{HS'}/\Vert (X^a)\Vert_{HS'}=0.058$ -- which looks much better.

\subsubsection{The Minimization of \texorpdfstring{$\lambda$}{lambda}}
\label{calcMin}

Yet, there is a vast choice of initial points $x\in\Tilde{\mathbb{R}}^D$, while of course not all belong to the same leaf.
Thus, having a unique rule to select a leaf via some initial point would be welcome.\\
A natural choice is to try to approximately minimize $\lambda$. This in turn means that $\partial_a\lambda$ will be small and consequently by equation (\ref{constraint}) $x^a$ will be near $\mathbf{x}^a$.

On the other hand, it is not advisable to search for the global minimum of $\lambda$. We should rather find a minimum in $\mathcal{N}_x$. Since we are working with leaves, it would be even better to actually work with $\mathcal{N}^{\mathcal{L}}_x$.\\
In our framework, it is not hard to concretize these objects introduced in section \ref{fol}: Since we have a distribution in the tangent bundle that describes the leaf $\mathcal{L}$, we can simply use the orthogonal complement $V_x^\perp$ (coming with the orthogonal projector $(\mathbb{1}-p_x)$) to define a distribution that in turn defines the generalization of $\mathcal{N}_x$ to $\mathcal{N}^{\mathcal{L}}_x$ which we call \textit{null leaf}.

So, we start from a random point $x\in\Tilde{\mathbb{R}}^D$, try to find the minimum of $\lambda$ in the null leaf at some $x'$ and use that point as our true initial point, defining the leaf $\mathcal{L}'$.\\
Then, we have to hope that for the sake of consistency $y'$ also lies in (or at least near) the leaf through $x'$ for every $y\in\Tilde{\mathbb{R}}^D$.

For any initial $x\in\Tilde{\mathbb{R}}^D$, we can use an adapted gradient descent procedure to find $x'$.
\begin{itemize}
    \item We construct a sequence $x_i$ in $\mathcal{N}^{\mathcal{L}}_x$ with $x_0=x$ that (hopefully) converges against a local minimum of $\lambda$ in the null leaf.
    \item Assume, we already know $x_i$. Then, using equation (\ref{constraint}), we calculate the gradient $\partial_a\lambda(x_i)=-(\mathbf{x}^a(x_i)-x_i^a)$. In the spirit of gradient descent methods, we put $v_i=-\partial_a\lambda(x_i)$.
    \item Then, we simply define $x_{i+1}=x_i+(\mathbb{1}-p_{x_i})(v_i)$.
    \item For every $i$ we check if $\vert(\mathbb{1}-p_{x_i})(\partial_a\lambda(x_i))\vert< \epsilon$ for some cutoff $\epsilon>0$. If this holds true, we define $x':=x_i$.
\end{itemize}

\begin{figure}[H]
\centering
\begin{minipage}{.99\textwidth}
  \centering
  \includegraphics[height=.17\linewidth]{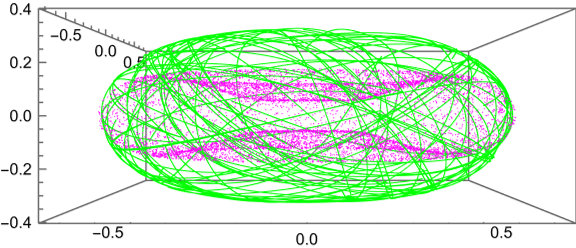}
\end{minipage}%
\caption{Hybrid leaf in the squashed fuzzy sphere for $N=4$ and $\alpha=0.1$. Green: scan of hybrid leaf through $x'$ where $x=(1,2,1)$, purple: $x'_i$ for $10000$ random $x_i$. Projective plot of $\Tilde{\mathbb{R}}^D$}
\label{fig:Implementation/10}
\end{figure}

In figure \ref{fig:Implementation/10} we see that our hope is not fulfilled and we do not find a unique leaf by this procedure. Yet, still this procedure might select a scale for us for choosing $\vert x\vert$. If this scale actually selects a leaf of good \textit{quality} for us will become clearer in section \ref{sfs_results}.

\newpage
\section{The Squashed Fuzzy Sphere and Other Results}
\label{RESULTS}

This section is dedicated to the discussion of actual numerical results for some of the matrix configuration defined in section \ref{qmgexamples} as well as a few others, based on the algorithms from section \ref{implementation}.
\\
In section \ref{sfs_results} the squashed fuzzy sphere is studied in more detail (compared to what we have already seen in section \ref{implementation}). This is followed by section \ref{fsr}, where we add random matrices to the round fuzzy sphere.\\
In section \ref{sfc} we then turn to the squashed fuzzy $\mathbb{C}P^2$, our first example with $D>3$, accompanied by the related completely squashed fuzzy $\mathbb{C}P^2$ in section \ref{csfc}.\\
The final example in section \ref{ft} is then given by the so called \textit{fuzzy torus} $T^2_N$, our first example that is not derived from a semisimple Lie group.

\subsection{The Squashed Fuzzy Sphere}
\label{sfs_results}

In section \ref{implementation} we have already seen some results for the squashed fuzzy sphere (that has been introduced in section \ref{SFuzzySphere}) for $N=4$ and $\alpha=0.1,0.9$. Now, we are going to study the latter geometry in more detail, including the dependence on $N$ and $\alpha$ and a detailed discussion of the results coming from the integration over $\mathcal{L}$.\\
This will be accompanied by several plots that allow us to understand the geometry visually.

\subsubsection{First Results and Dimensional Aspects}
\label{basciccalculations}

As a beginning, we want to determine the dimension $k$ of $\mathcal{M}$ (given by the rank of $T_xq$) as well as the ranks of $g_{ab}(x)$, $\omega_{ab}(x)$ and $\theta^{ab}(x)$. Also, we want to know which of the kernels of the latter agree and which points lie in $\Tilde{\mathbb{R}}^D$ respectively $\hat{\mathbb{R}}^D$ (note that we already know from analytic considerations that $0\in\mathcal{K}$).\\
Therefore, we perform a cascade of calculations: For the points $(0,0,1)$, $(0,1,0)$ as well as for three random points in the unit ball, for $N=2,3,4,10,100$ and for $\alpha=1,0.5,0.1,0$ we check if $x\in\hat{\mathbb{R}}^D$, calculate the latter objects and the associated ranks and compare the kernels.

We begin with the discussion of the choice $N=2$, turning out to be fundamentally different to $N>2$.\\
For the round case $\alpha=1$ we find that all considered points lie in $\Tilde{\mathbb{R}}^3$, while
we always find that the dimension $k$ as well as all ranks equal two. This in turn shows that all these points also lie in $\hat{\mathbb{R}}^3$. Further, all kernels agree and the almost Kähler condition is satisfied. These results are fitting well to our knowledge $\operatorname{dim}(\mathcal{M})=2$ and $\Tilde{\mathbb{R}}^3=\hat{\mathbb{R}}^3=\mathbb{R}^3\setminus\{0\}$ from section \ref{FuzzySphere}.\\
In the squashed cases $\alpha=0.5,0.1$ all considered points still lie in $\Tilde{\mathbb{R}}^3$, the dimension and the ranks remain at two but the kernel of $\theta^{ab}$ discerns from all the other kernels (except for $(0,0,1)$). The almost Kähler condition remains intact. So we find $k=2$ and conjecture $\Tilde{\mathbb{R}}^3=\hat{\mathbb{R}}^3=\mathbb{R}^3\setminus\{0\}$.\\
In the completely squashed case $\alpha=0$ we find $(0,0,1)\in\mathcal{K}$ while the other points remain in $\Tilde{\mathbb{R}}^3$. The dimension of $\mathcal{M}$ turns to one just as the rank of $g_{ab}$. Accordingly, $\omega_{ab}$ and $\theta^{ab}$ vanish. The kernel of $T_xq$ and $g_{ab}$ agree. Since $\omega_{ab}$ vanishes, the almost Kähler condition is meaningless. Here, we conjecture $\Tilde{\mathbb{R}}^3=\hat{\mathbb{R}}^3=\mathbb{R}^3\setminus \mathbb{R}\hat{e}_3$.

Let us now consider the more interesting choices $N>2$.\\
In the round case $\alpha=1$ nothing changes and we obtain the expected results.\\
For the squashed cases $\alpha=0.5,0.1$ we witness an interesting effect (we have already stumbled upon it in section \ref{SFuzzySphere}) that we will call \textit{oxidation} in the following. For all points except $(0,0,1)$ the dimension as well as the rank of $g_{ab}$ turn to three. The ranks of $\omega_{ab}$ and $\theta^{ab}$ remain two, while their kernels again discern.
At $(0,0,1)$ the dimensions as well as the ranks are still two and all kernels agree.
The almost Kähler condition only remains true approximately. These observations together with our analytic results from section \ref{SFuzzySphere} encourage us to conjecture $\Tilde{\mathbb{R}}^3=\mathbb{R}^3\setminus\{0\}$, while $\hat{\mathbb{R}}^3=\mathbb{R}^3\setminus \mathbb{R}\hat{e}_3$.\\
Also the case $\alpha=0$ changes: The dimension of $\mathcal{M}$ and the rank of $g_{ab}$ return to two with agreeing kernels, while $\omega_{ab}$ and $\theta^{ab}$ vanish, noting that again $(0,0,1)\in\mathcal{K}$. So we conjecture that also here $\Tilde{\mathbb{R}}^3=\hat{\mathbb{R}}^3=\mathbb{R}^3\setminus \mathbb{R}\hat{e}_3$.

These conjectures are supported by numerous further calculations. Table \ref{table1} summarizes the findings for points $x\in\mathbb{R}^3\setminus\mathbb{R}\hat{e}_3$.\\
Let us explain heuristically why these results are plausible:
In section \ref{random} we have seen that $k$ is bounded from above by $\min(D,2N-2)$, explaining why we find no oxidation for $N=2$. Again by arguments from the same section it is expected that $k=3$ for $N>2$ under the assumption that the squashed fuzzy sphere already \textit{behaves like a random matrix configuration}. In section \ref{SFuzzySphere} we have even seen analytically that $k=3$ at least for $\alpha=1-\epsilon$ for small $\epsilon>0$. Further, complete squashing effectively reduces $D$ from three to two, again supporting the validity of the findings.
\\
That the behaviour is special on the $\hat{e}_3$ axis fits to the fact that the $SO(3)$ symmetry remains unbroken for rotations around this axis and that analytically the quasi-coherent states are unperturbed there.

\begin{table}[H]
\begin{tabular}{ll|l|l|l|l}
                           &              & $\operatorname{dim}(\mathcal{M})$ & $\operatorname{rank}(g)$ & $\operatorname{rank}(\omega)$ & $\operatorname{rank}(\theta)$ \\ \hline
\multicolumn{1}{l|}{$N=2$} & $\alpha=1$   & 2                                 & 2                        & 2                             & 2                             \\ \cline{2-6} 
\multicolumn{1}{l|}{}      & $0<\alpha<1$ & 2                                 & 2                        & 2                             & 2                             \\ \cline{2-6} 
\multicolumn{1}{l|}{}      & $\alpha=0$   & 1                                 & 1                        & 0                             & 0                             \\ \hline
\multicolumn{1}{l|}{$N>2$} & $\alpha=1$   & 2                                 & 2                        & 2                             & 2                             \\ \cline{2-6} 
\multicolumn{1}{l|}{}      & $0<\alpha<1$ & 3                                 & 3                        & 2                             & 2                             \\ \cline{2-6} 
\multicolumn{1}{l|}{}      & $\alpha=0$   & 2                                 & 2                        & 0                             & 0                            
\end{tabular}
\centering
\caption{Overview of the dimensions and ranks in different scenarios for points $x\in\mathbb{R}^3\setminus\mathbb{R}\hat{e}_3$}
\label{table1}
\end{table}

We further refer to the comparison of $g_{ab}$ and $\omega_{ab}$ once calculated numerically and once via perturbation theory in section \ref{BasicQuantities}, mutually validating both approaches. This has also been checked for further scenarios.

\subsubsection{A Graphical Account}

\begin{figure}[H]
\centering
\begin{minipage}{.3\textwidth}
  \centering
  \includegraphics[height=.7\linewidth]{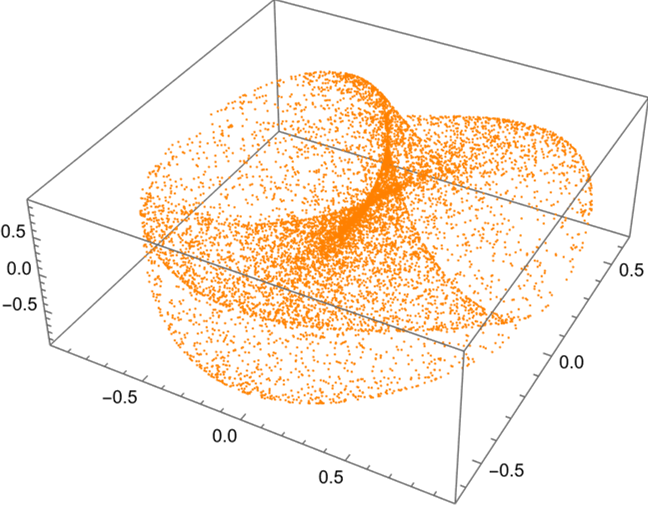}
  \includegraphics[height=.4\linewidth]{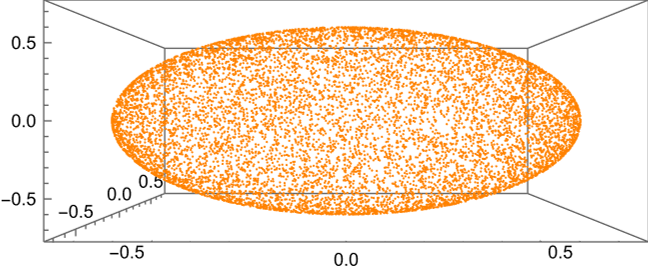}
\end{minipage}%
\begin{minipage}{.3\textwidth}
  \centering
  \includegraphics[height=.7\linewidth]{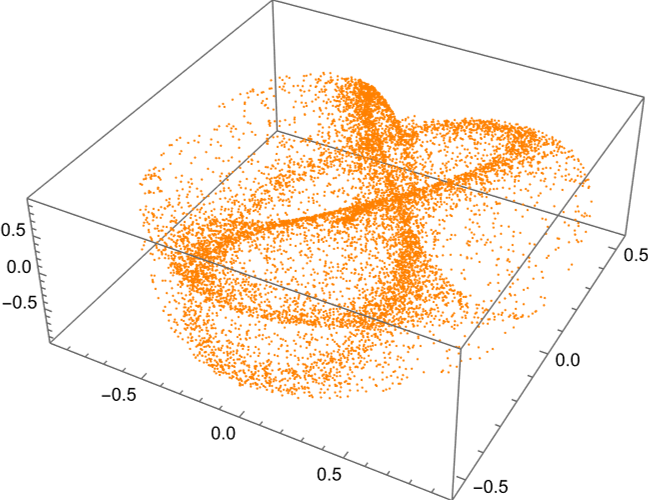}
  \includegraphics[height=.4\linewidth]{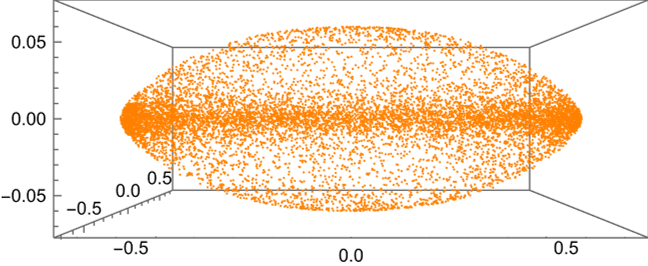}
\end{minipage}
\begin{minipage}{.3\textwidth}
  \centering
  \includegraphics[height=.7\linewidth]{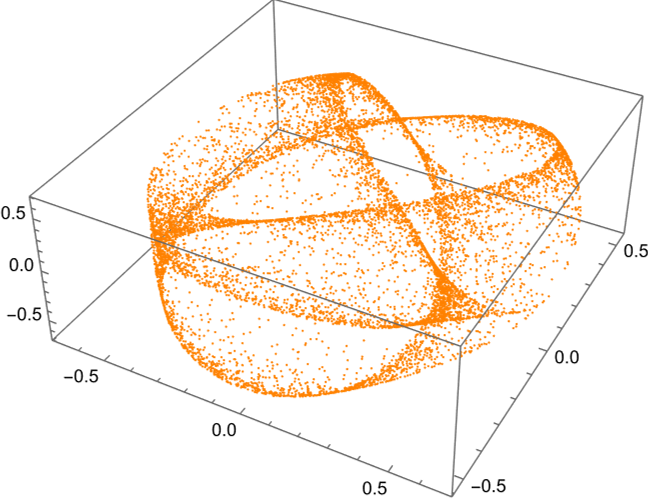}
  \includegraphics[height=.4\linewidth]{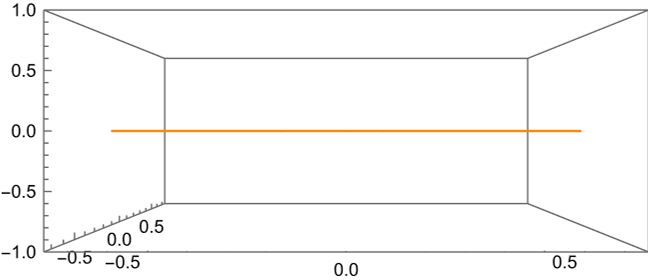}
\end{minipage}
\caption{$10000$ random points in the squashed fuzzy sphere for $N=4$. Left to right: $\alpha=1,0.1,0$; top: projective plot of $\mathcal{M}$, bottom: projective plot of $\Tilde{\mathcal{M}}$}
\label{fig:SU2/2}
\end{figure}

In figure \ref{fig:SU2/2} we see plots of $\mathcal{M}$ and $\Tilde{\mathcal{M}}$ for random points in the (squashed) fuzzy sphere for $N=4$ and for $\alpha=1,0.1,0$.\\
Looking at $\mathcal{M}$, we immediately witness self-intersections of the shape. This is not surprising since we projected from $\mathbb{R}^6$ to $\mathbb{R}^3$ in order to generate the plots. Also, the large scale shape depends strongly on the choice of a projection $P$. On the other hand we see that this large scale shape is preserved during the squashing. This is not surprising as we have seen in section \ref{SFuzzySphere} that the quasi coherent states of the round fuzzy sphere remain as asymptotic states of the squashed fuzzy sphere for $0<\alpha<1$.\\
While in the plots of $\mathcal{M}$ it is hardly possible to recognize the actual geometry, that is very present in the plots of $\Tilde{\mathcal{M}}$. In the round case, the result exactly is a sphere\footnote{Note that in these plots the $\hat{e}_3$ axis is scaled differently than the $\hat{e}_1$ and $\hat{e}_2$ axes!}, while for $\alpha=0.1$ we approximately find an ellipsoidal and in the completely squashed case a disc. The interesting result of course is the squashed case where we see that points accumulate (and in fact oxidize) at the equator while at the polar regions there are fewer points than in the round case.

\begin{figure}[H]
\centering
\begin{minipage}{.3\textwidth}
  \centering
  \includegraphics[height=.7\linewidth]{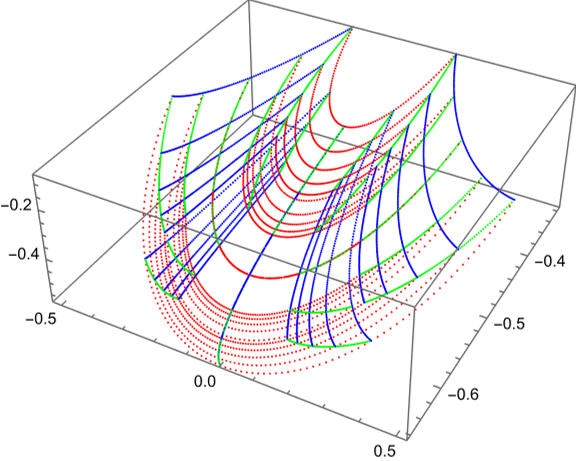}
  \includegraphics[height=.7\linewidth]{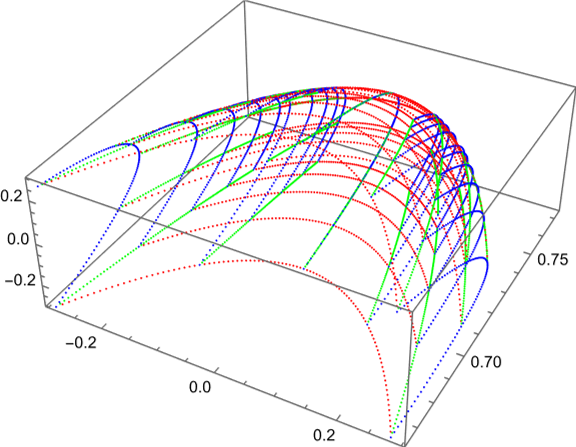}
\end{minipage}%
\begin{minipage}{.3\textwidth}
  \centering
  \includegraphics[height=.7\linewidth]{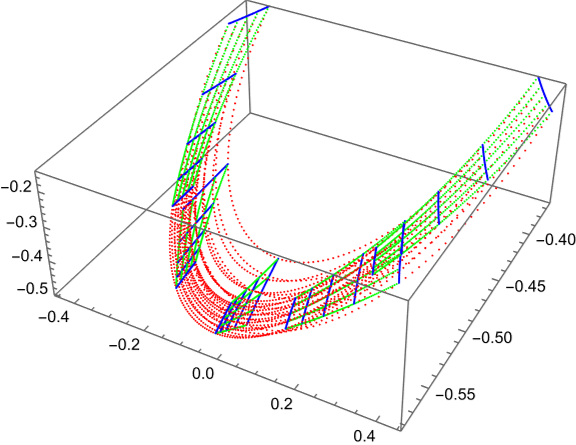}
  \includegraphics[height=.7\linewidth]{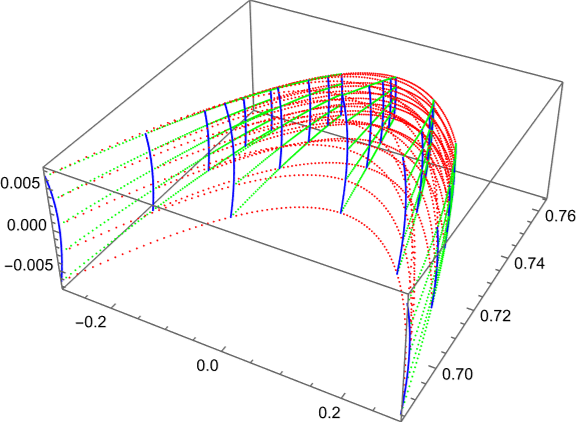}
\end{minipage}
\begin{minipage}{.3\textwidth}
  \centering
  \includegraphics[height=.7\linewidth]{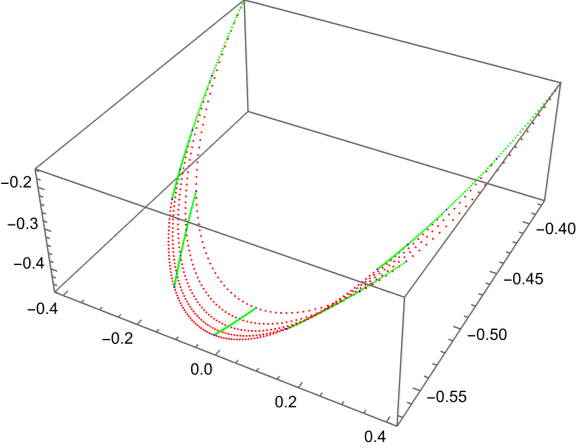}
  \includegraphics[height=.7\linewidth]{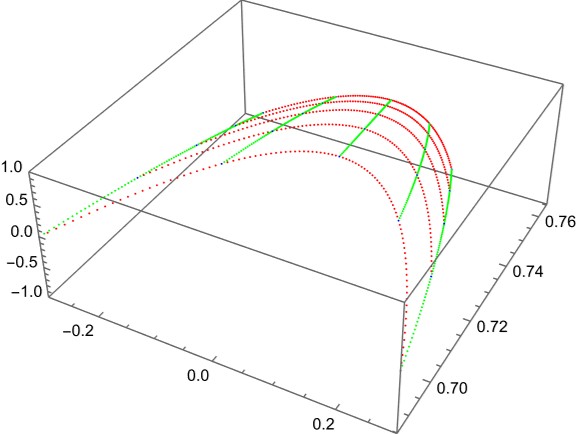}
\end{minipage}
\caption{Cartesian coordinate lines around $(0,1,0)$ in the squashed fuzzy sphere for $N=4$. Left to right: $\alpha=1,0.1,0$; top: projective plot of $\mathcal{M}$, bottom: projective plot of $\Tilde{\mathcal{M}}$}
\label{fig:SU2/3}
\end{figure}

In the latter figure it is hardly possible to determine $k$ visually. For that reason, in figure \ref{fig:SU2/3} the Cartesian coordinate lines centered at $(0,1,0)$ shown in figure \ref{fig:Implementation/3A} are plotted.\\
In the round case we directly see that one coordinate is always redundant, matching to our knowledge $k=2$ and to our discussion\footnote{There we found that we can construct true coordinates of $\mathcal{M}$ by dropping redundant coordinates.} in appendix \ref{Appendix:manifold}.
\\
In the squashed case, these coordinates are not redundant anymore and $\mathcal{M}$ visually turns out to be (at least) three dimensional. Interestingly, the same holds for $\Tilde{\mathcal{M}}$.\\
In the completely squashed case we once again see that one coordinate is redundant (especially the $\hat{e}_3$ coordinate) and $\mathcal{M}$ is therefore two dimensional.

Here we can also see how the large scale shape is preserved throughout the squashing, while the actual shape strongly changes.

\begin{figure}[H]
\centering
\begin{minipage}{.3\textwidth}
  \centering
  \includegraphics[height=.7\linewidth]{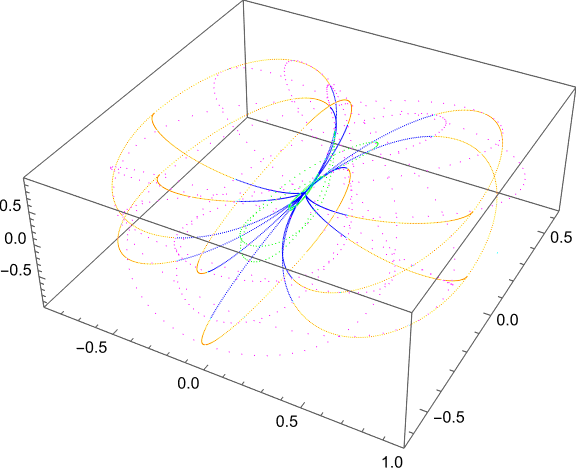}
  \includegraphics[height=.7\linewidth]{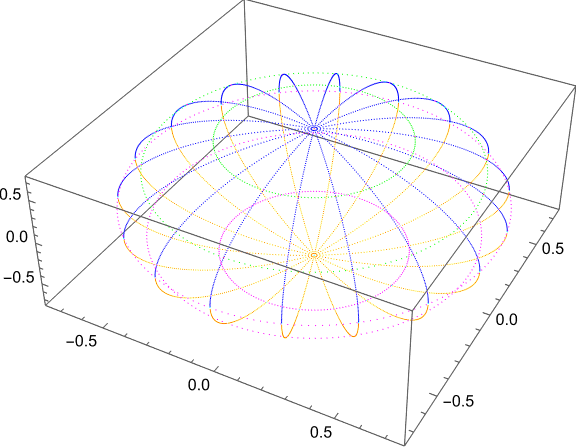}
\end{minipage}%
\begin{minipage}{.3\textwidth}
  \centering
  \includegraphics[height=.7\linewidth]{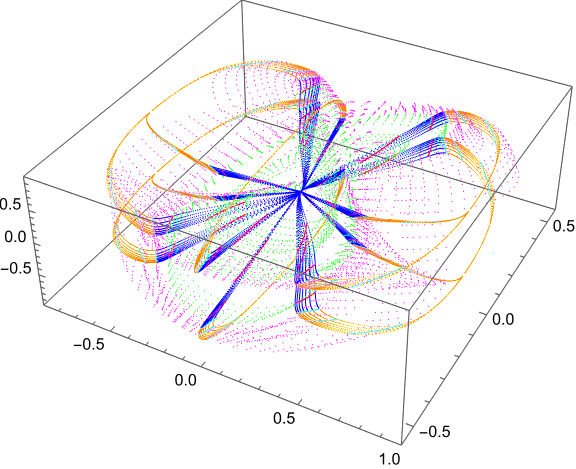}
  \includegraphics[height=.7\linewidth]{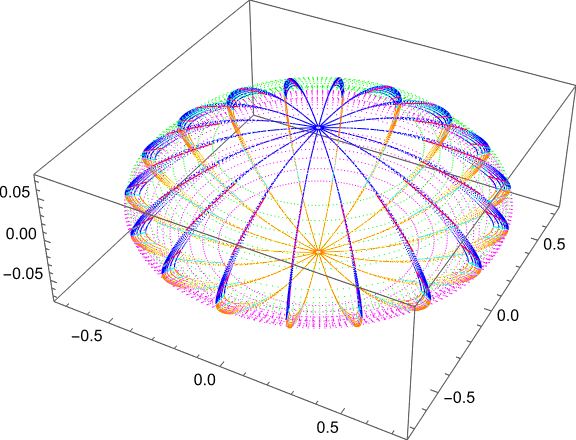}
\end{minipage}
\begin{minipage}{.3\textwidth}
  \centering
  \includegraphics[height=.7\linewidth]{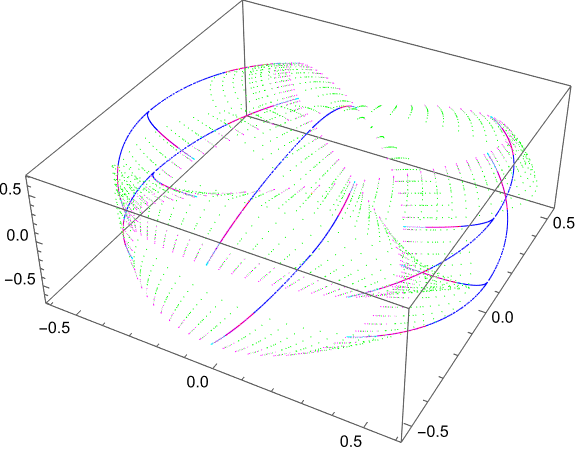}
  \includegraphics[height=.7\linewidth]{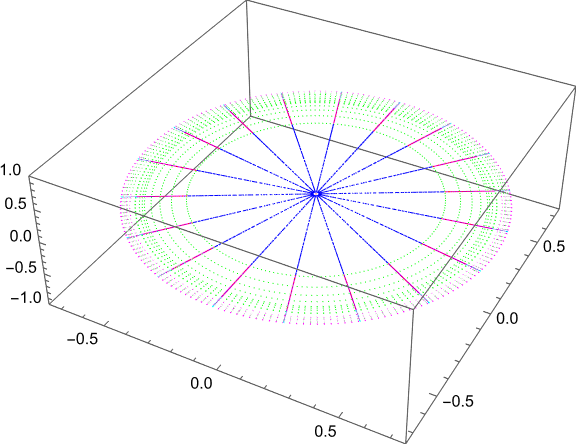}
\end{minipage}
\caption{Spherical coordinate lines in the squashed fuzzy sphere for $N=4$. Left to right: $\alpha=1,0.1,0$; top: projective plot of $\mathcal{M}$, bottom: projective plot of $\Tilde{\mathcal{M}}$}
\label{fig:SU2/3A}
\end{figure}

\newpage
For the sake of a more systematic global understanding, in figure \ref{fig:SU2/3A} the spherical coordinate lines from figure \ref{fig:Implementation/3A} are plotted\footnote{Here we replaced the color yellow with orange for a better visibility.}.\\
We directly see that these coordinates are better adapted to the geometry. In the round case it is evident that the radial coordinate is redundant. Further, we immediately see that $\mathcal{M}$ is three dimensional for $0<\alpha<1$. In the plot of $\Tilde{\mathcal{M}}$ we perceive that at the poles the radial direction is degenerate, fitting to our previous observations.

\subsubsection{The Dependence on \texorpdfstring{$N$}{N}}

\begin{figure}[H]
\centering
\begin{minipage}{.3\textwidth}
  \centering
  \includegraphics[height=.7\linewidth]{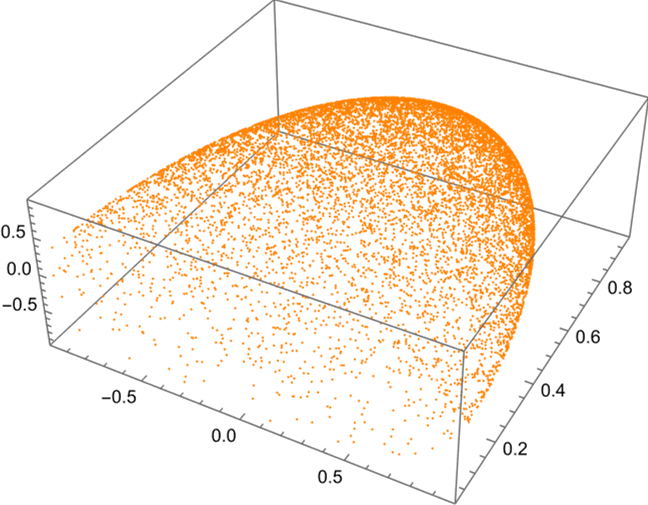}
  \includegraphics[height=.4\linewidth]{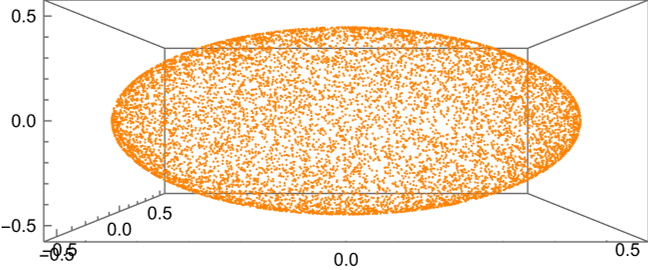}
\end{minipage}%
\begin{minipage}{.3\textwidth}
  \centering
  \includegraphics[height=.7\linewidth]{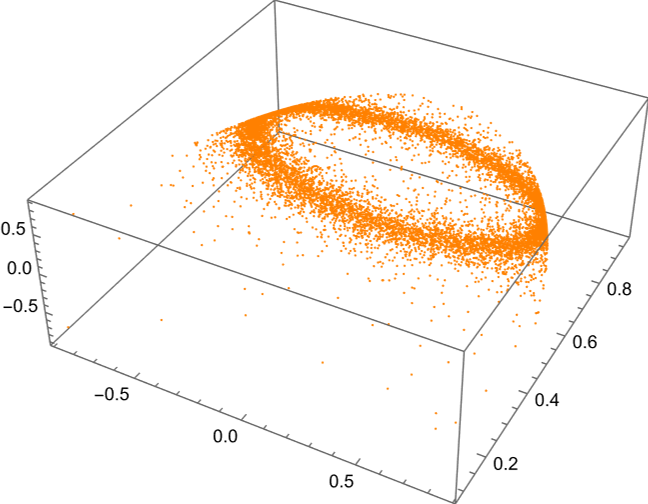}
  \includegraphics[height=.4\linewidth]{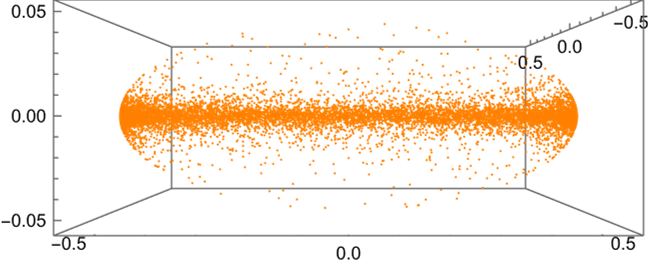}
\end{minipage}
\begin{minipage}{.3\textwidth}
  \centering
  \includegraphics[height=.7\linewidth]{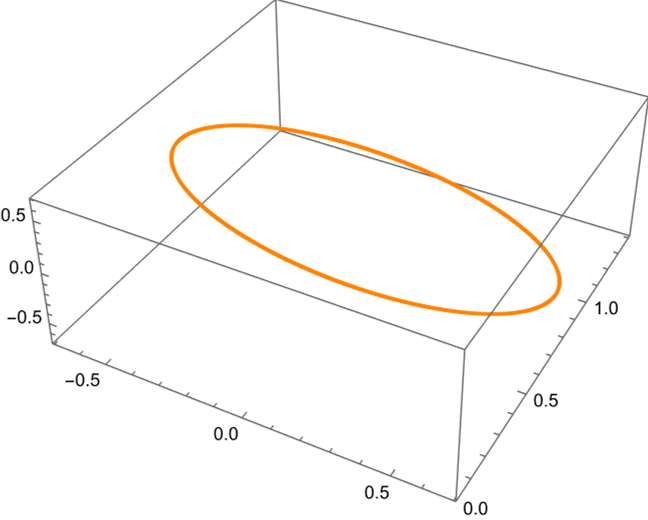}
  \includegraphics[height=.4\linewidth]{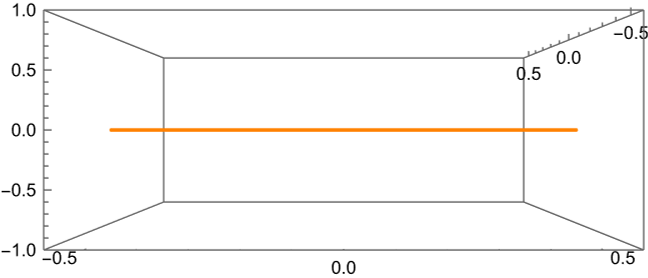}
\end{minipage}
\caption{Random points in the squashed fuzzy sphere for $N=2$. Left to right: $\alpha=1,0.1,0$; top: projective plot of $\mathcal{M}$, bottom: projective plot of $\Tilde{\mathcal{M}}$}
\label{fig:SU2/4}
\end{figure}

We have already seen that $N=2$ is a special case, so we included figure \ref{fig:SU2/4} (which is the analogue of figure \ref{fig:SU2/2}). Let us first consider the plots of $\Tilde{\mathcal{M}}$: The round case looks just as the one for $N=4$, while only the radius deviates, fitting to equation (\ref{FuzzySphereExp}). From this perspective it looks as if the same held in the completely squashed case, but this is not true. For $N=4$ we obtained a disk, while here we would only see a circle. The partially squashed case is also interesting as we again see an accumulation of points at the equator, while here the shape is much more that of an ellipsoidal than for $N=4$ as there is no oxidation.\\ 
Let us come to the plots of $\mathcal{M}$. We see that the large scale shape is much simpler than for $N=4$, but we cannot recognize a sphere in the round case. This is an artifact of our plotting procedure where we trivialized $\mathbb{C}P^{N-1}\cong S^{2N-1}/U(1)$. Since there is no global trivialization we cannot expect to get a global result. Practically this means that the border of the plotted shape is given by the states in the set $\mathcal{Z}$ defined in section \ref{Visual} where we could not choose a unique representative by our rule. That the shape looks like a half sphere is merely a coincidence, while the points on the border are glued together in a nontrivial way.
Further, here we can nicely see how the squashing takes place: The points get more and more confined to a circle, while we have a complete collapse for $\alpha=0$.

\begin{figure}[H]
\centering
\begin{minipage}{.19\textwidth}
  \centering
  \includegraphics[height=.7\linewidth]{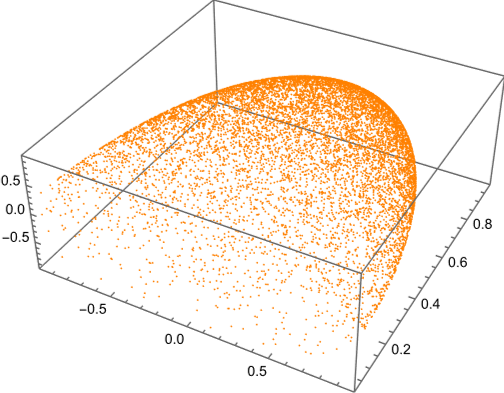}
  \includegraphics[height=.7\linewidth]{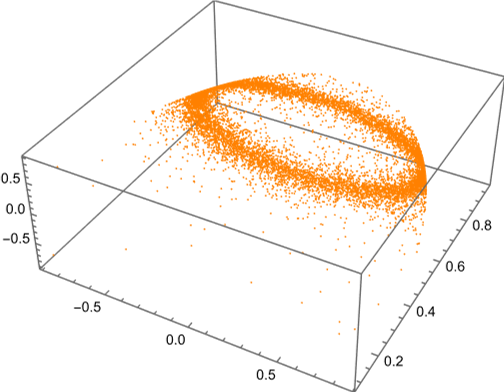}
  \includegraphics[height=.7\linewidth]{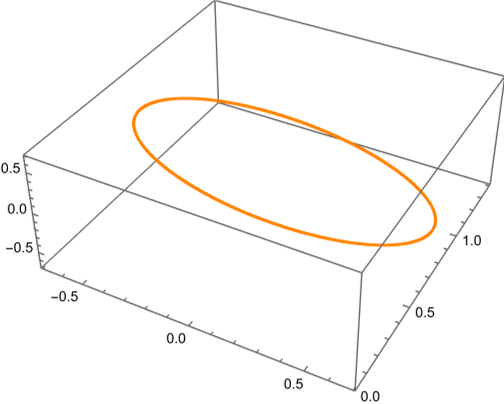}
\end{minipage}%
\begin{minipage}{.19\textwidth}
  \centering
  \includegraphics[height=.7\linewidth]{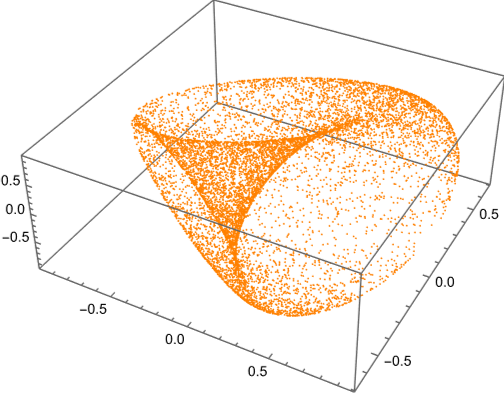}
  \includegraphics[height=.7\linewidth]{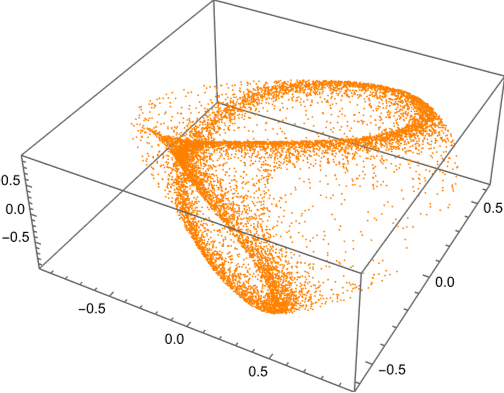}
  \includegraphics[height=.7\linewidth]{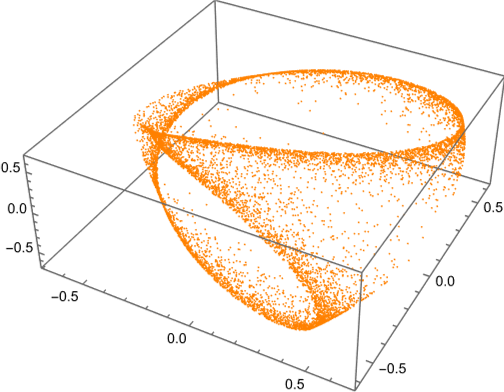}
\end{minipage}%
\begin{minipage}{.19\textwidth}
  \centering
  \includegraphics[height=.7\linewidth]{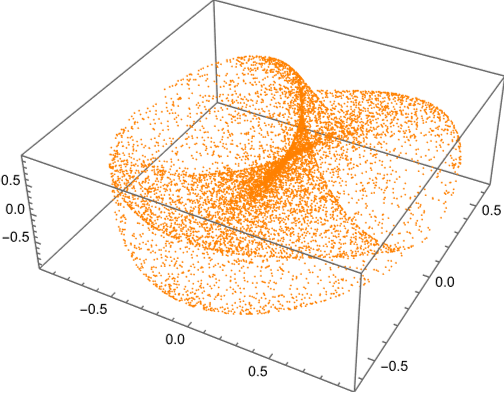}
  \includegraphics[height=.7\linewidth]{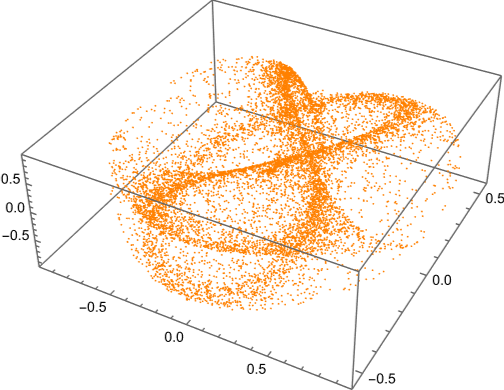}
  \includegraphics[height=.7\linewidth]{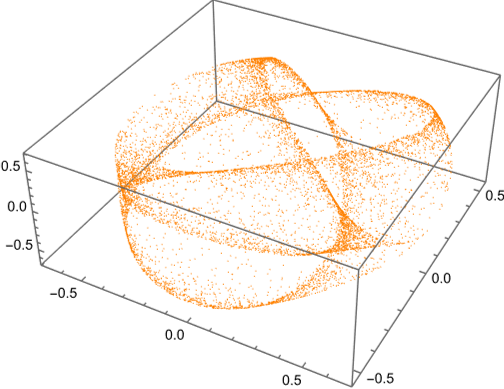}
\end{minipage}%
\begin{minipage}{.19\textwidth}
  \centering
  \includegraphics[height=.7\linewidth]{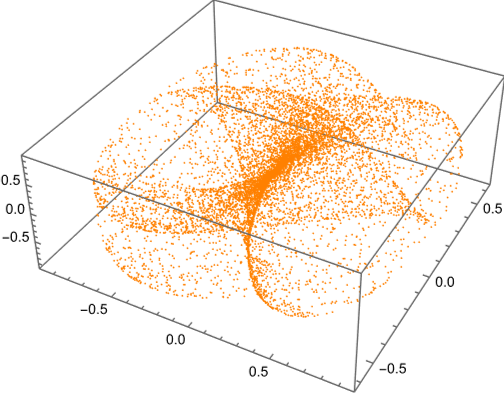}
  \includegraphics[height=.7\linewidth]{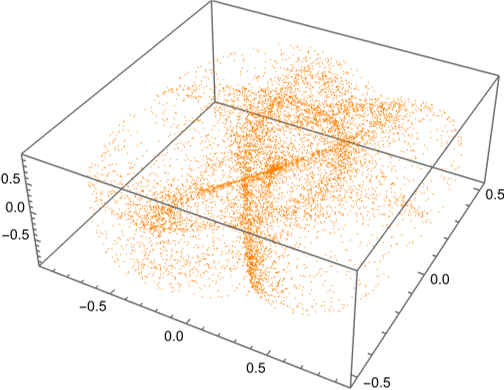}
  \includegraphics[height=.7\linewidth]{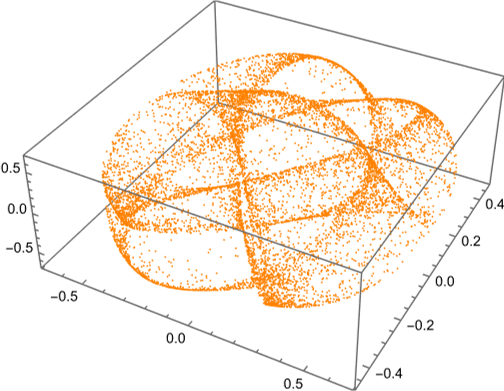}
\end{minipage}%
\begin{minipage}{.19\textwidth}
  \centering
  \includegraphics[height=.7\linewidth]{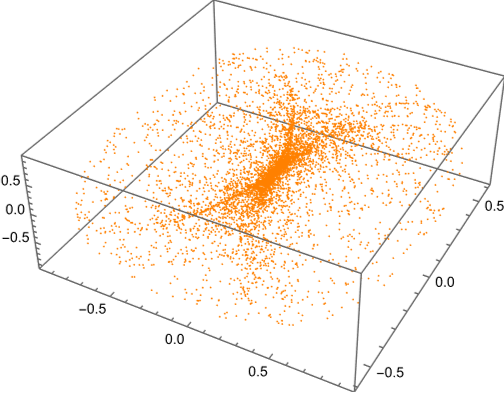}
  \includegraphics[height=.7\linewidth]{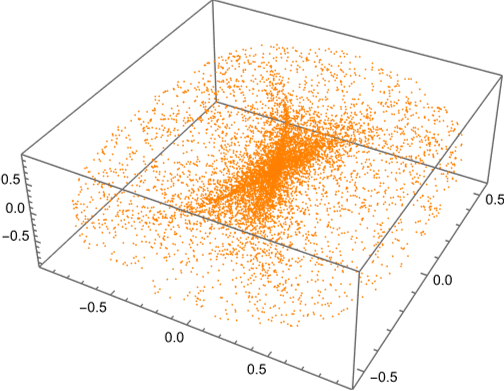}
  \includegraphics[height=.7\linewidth]{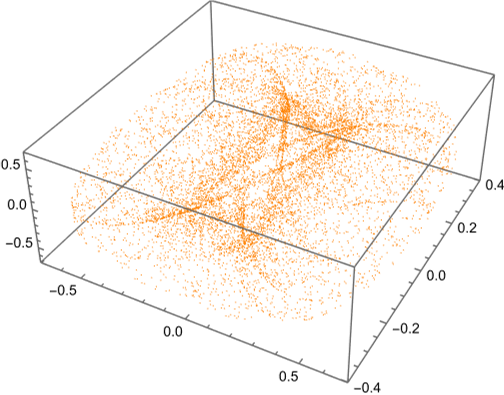}
\end{minipage}%
\caption{Random points in the squashed fuzzy sphere, projective plot of $\mathcal{M}$. Left to right: $N=2,3,4,5,10$; top to bottom: $\alpha=1,0.1,0$}
\label{fig:SU2/5}
\end{figure}

In figure \ref{fig:SU2/5} we see plots of $\mathcal{M}$ for $N=2,3,4,5,10$ and $\alpha=1,0.1,0$. We already know that in the round case $\mathcal{M}\cong S^2$ for all $N$, thus in the first line we only see projections of different immersions of $S^2\hookrightarrow \mathbb{C}P^{N-1}$. With increasing $N$ the large scale shape winds more and more around the origin, until we lose almost all information for too large $N$ since then the $\mathbb{R}^3$ gets much too low dimensional to capture all aspects of the immersion\footnote{That does not mean that it is impossible to visualize $\mathcal{M}$ locally.}.\\
In the squashed and completely squashed cases we see similar behaviour.

\newpage
\subsubsection{The Dependence on \texorpdfstring{$\alpha$}{alpha} and the Oxidation Process}
\label{DependanceAlpha}

\begin{figure}[H]
\centering
\begin{minipage}{.3\textwidth}
  \centering
  \includegraphics[height=.7\linewidth]{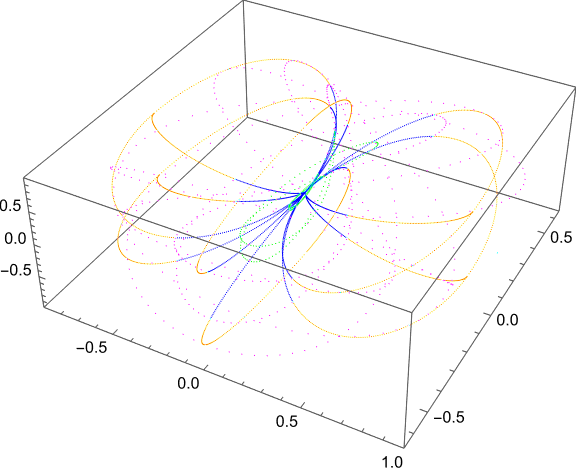}
  \includegraphics[height=.7\linewidth]{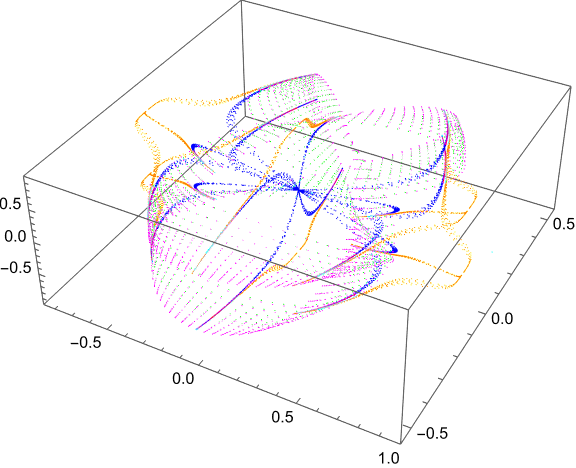}
\end{minipage}%
\begin{minipage}{.3\textwidth}
  \centering
  \includegraphics[height=.7\linewidth]{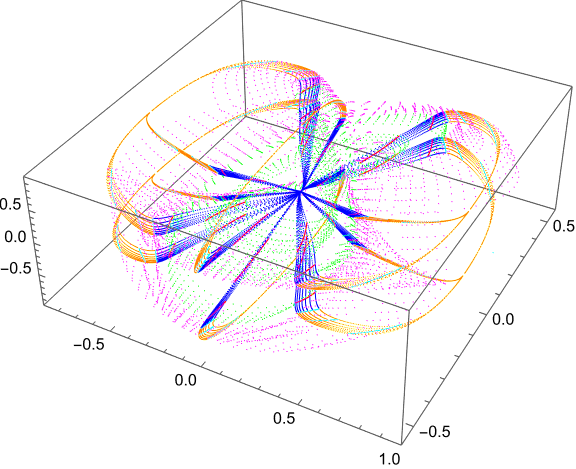}
  \includegraphics[height=.7\linewidth]{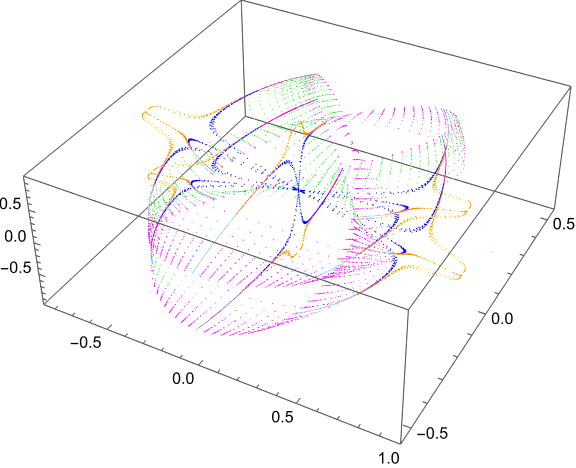}
\end{minipage}%
\begin{minipage}{.3\textwidth}
  \centering
  \includegraphics[height=.7\linewidth]{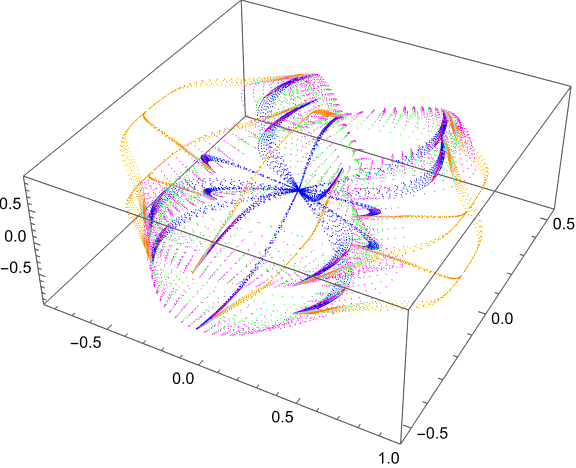}
  \includegraphics[height=.7\linewidth]{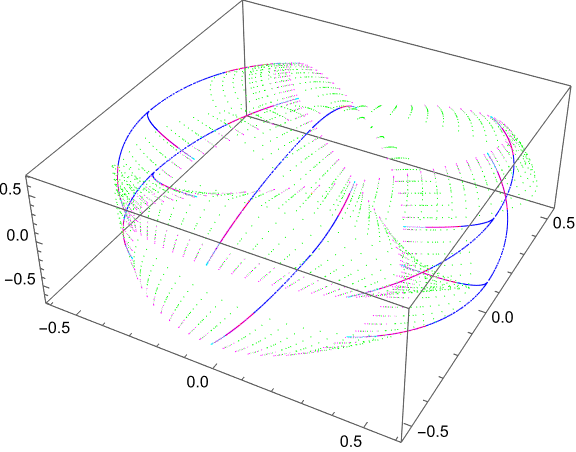}
\end{minipage}%
\caption{Spherical coordinate lines in the squashed fuzzy sphere for $N=4$. Projective plot of $\mathcal{M}$. Top: left to right: $\alpha=1,0.1,0.01$; bottom: left to right: $\alpha=0.001,0.0001,0$}
\label{fig:SU2/6}
\end{figure}

Now we want to focus on the \textit{squashing} of the fuzzy sphere. Therefore we once again specialize to $N=4$.
In figure \ref{fig:SU2/6} we plotted\footnote{Again, the color yellow has been replaced with orange.} spherical coordinate lines as shown in figure \ref{fig:Implementation/3A} for exponentially decreasing $\alpha$.\\
What we directly see is that the radial coordinate (red/cyan) does not play any role in the round case, while in the completely squashed case the radial and the polar coordinate lines (green/magenta) play the same role (this is clear from analytical considerations) and the top and bottom hemisphere produce exactly the same states (which is again obvious).\\
For the azimuthal coordinate lines (blue/orange) one sees especially well how the upper hemisphere \textit{moves closer} to the lower hemisphere, starting at the equator where they touch.\\
Note that the broken symmetry between the upper and lower hemisphere is only an artifact of the way we produced the plots via the procedure described in section \ref{Visual}.

\begin{figure}[H]
\centering
\begin{minipage}{.23\textwidth}
  \centering
  \includegraphics[height=.4\linewidth]{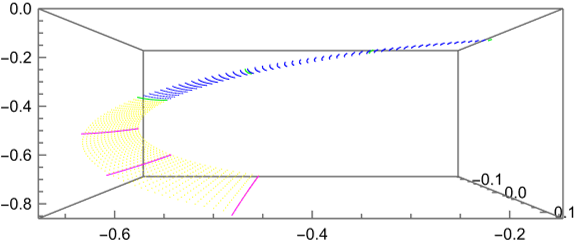}
\end{minipage}%
\begin{minipage}{.23\textwidth}
  \centering
  \includegraphics[height=.4\linewidth]{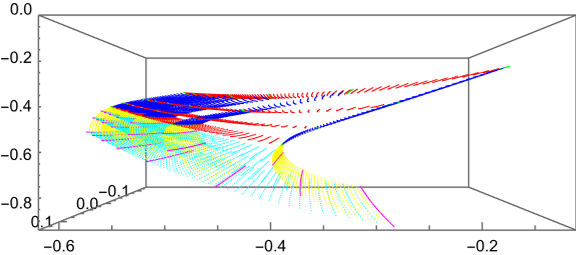}
\end{minipage}%
\begin{minipage}{.23\textwidth}
  \centering
  \includegraphics[height=.4\linewidth]{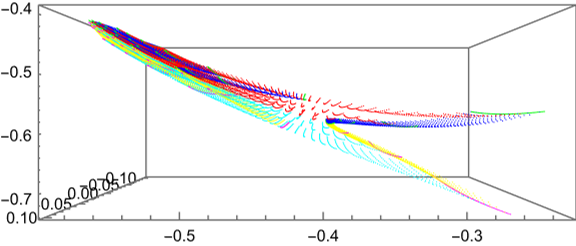}
\end{minipage}%
\begin{minipage}{.23\textwidth}
  \centering
  \includegraphics[height=.4\linewidth]{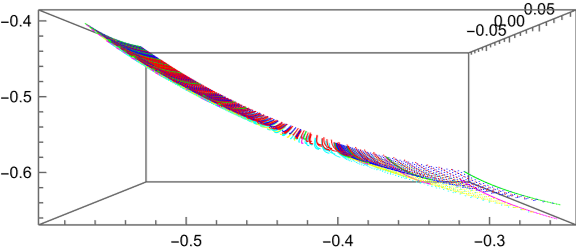}
\end{minipage}%
\caption{Sector of spherical coordinate lines (located at the equator) in the squashed fuzzy sphere for $N=4$. Projective plot of $\mathcal{M}$. Left to right: $\alpha=1,0.1,0.01,0.001$}
\label{fig:SU2/7}
\end{figure}

In figure \ref{fig:SU2/7} we plotted a sector of spherical coordinate lines centered at the equator (this sector is shown in figure \ref{fig:Implementation/3A}). Here, we can easily trace the oxidation process and we can see remarkably well how the upper and the lower hemisphere join together during squashing. Yet we should not think of an upper layer hovering above a lower layer since the smaller we make $\alpha$, the further inwards both layers heuristically \textit{touch}. (In the next section we will see that this view is anyways completely wrong.)\\
These plots also allow us to visually trace the \textit{thickness} of $\mathcal{M}$. For $\alpha=0.1$
 this oxidation is at a peak and again becomes almost irrelevant for smaller scales of $\alpha$.

\subsubsection{Asymptotic States and Topological Aspects}

In order to better understand the topology of $\mathcal{M}$, we want to know the behaviour for large and small $\vert x\vert$.\\
For large absolute values this is simple since in the limit $\vert x\vert\to\infty$, we can consider the \textit{asymptotic states} (that we have discussed in section \ref{SFuzzySphere}).\\
In the opposite limit $\vert x\vert\to0$ it is not that trivial to find the quasi-coherent states. Yet numerically, we can simulate this limit by considering random points on a \textit{tiny sphere} of radius $\epsilon$ in $\Tilde{\mathbb{R}}^D$.

\begin{figure}[H]
\centering
\begin{minipage}{.24\textwidth}
  \centering
  \includegraphics[height=.7\linewidth]{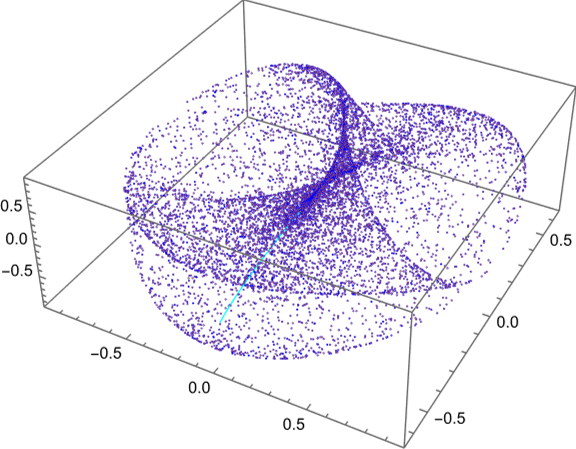}
  \includegraphics[height=.7\linewidth]{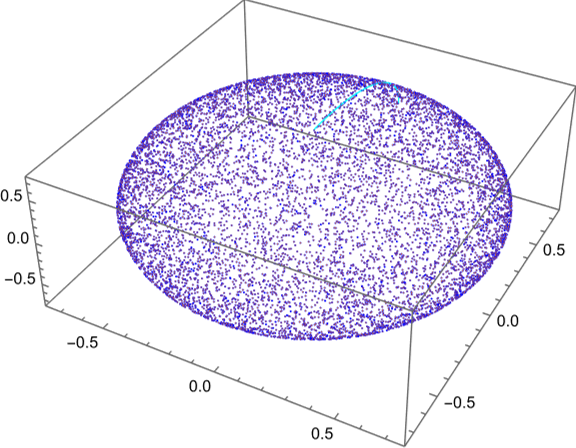}
\end{minipage}%
\begin{minipage}{.24\textwidth}
  \centering
  \includegraphics[height=.7\linewidth]{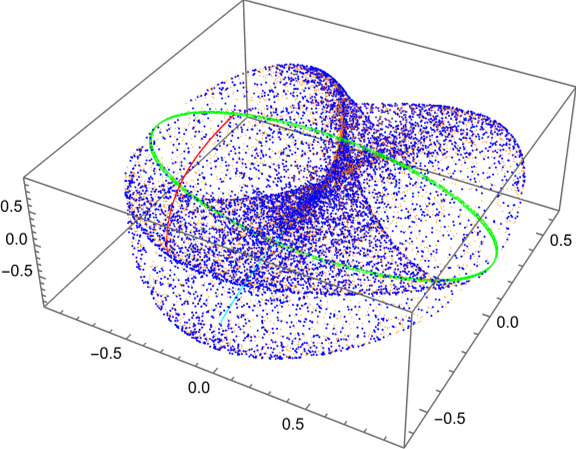}
  \includegraphics[height=.7\linewidth]{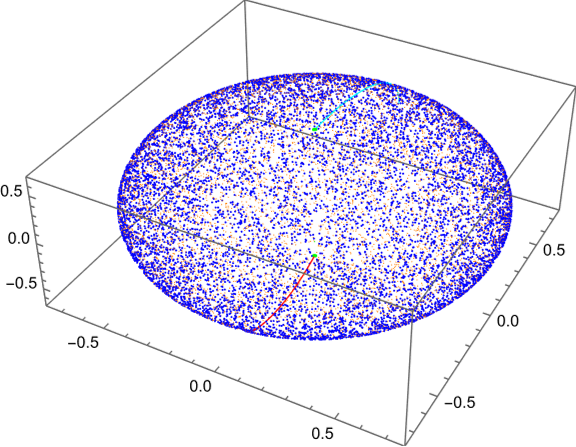}
\end{minipage}%
\begin{minipage}{.24\textwidth}
  \centering
  \includegraphics[height=.7\linewidth]{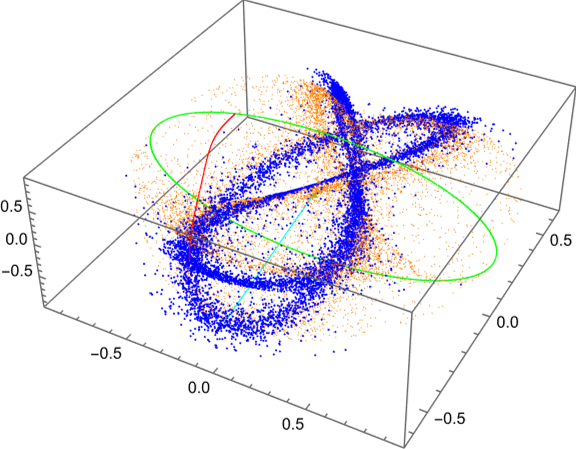}
  \includegraphics[height=.7\linewidth]{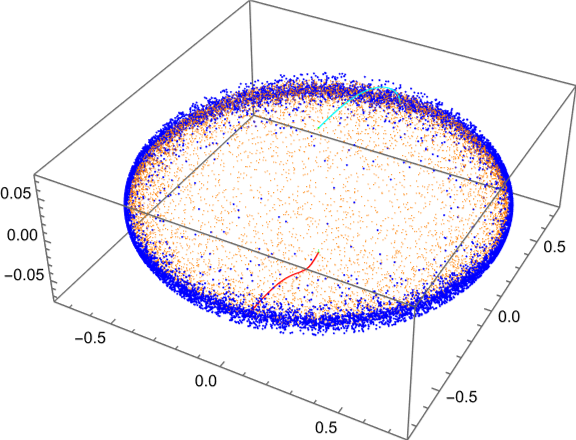}
\end{minipage}%
\begin{minipage}{.24\textwidth}
  \centering
  \includegraphics[height=.7\linewidth]{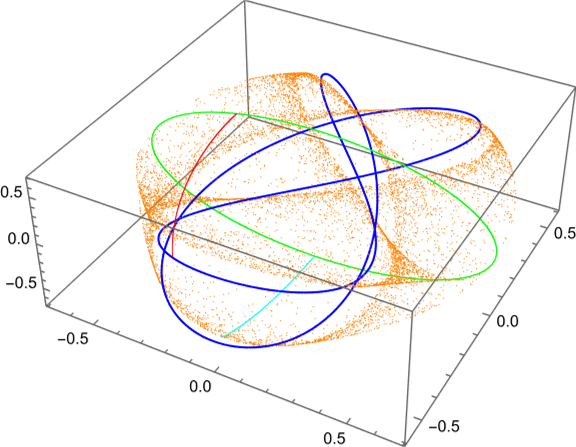}
  \includegraphics[height=.7\linewidth]{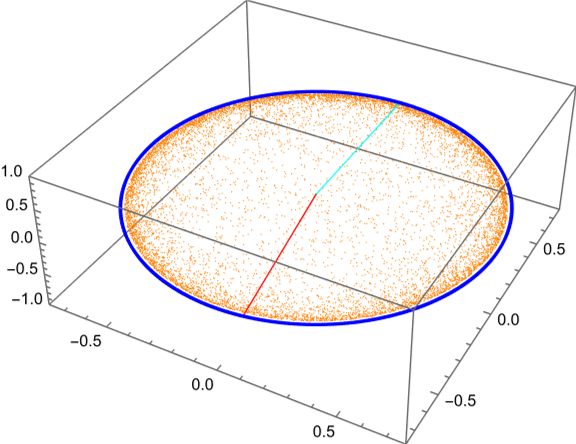}
\end{minipage}%
\caption{Random points from unit ball (orange), random asymptotic points (blue), random points from a tiny sphere of radius $0.001$ (green), radial line through random point (red) and polar line from $(0,0,1)$ to $(0,1,0)$ (cyan) in the squashed fuzzy sphere for $N=4$. Top: projective plot of $\mathcal{M}$, bottom: projective plot of $\Tilde{\mathcal{M}}$; left to right: $\alpha=1,0.9,0.1,0$}
\label{fig:SU2/8}
\end{figure}

\begin{figure}[H]
\centering
\begin{minipage}{.3\textwidth}
  \centering
  \includegraphics[height=.7\linewidth]{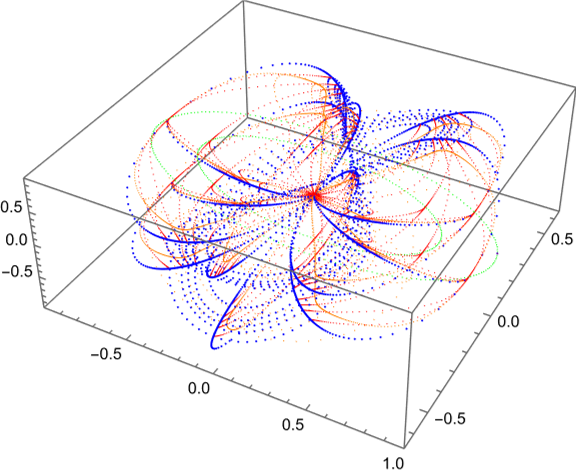}
\end{minipage}%
\begin{minipage}{.3\textwidth}
  \centering
  \includegraphics[height=.7\linewidth]{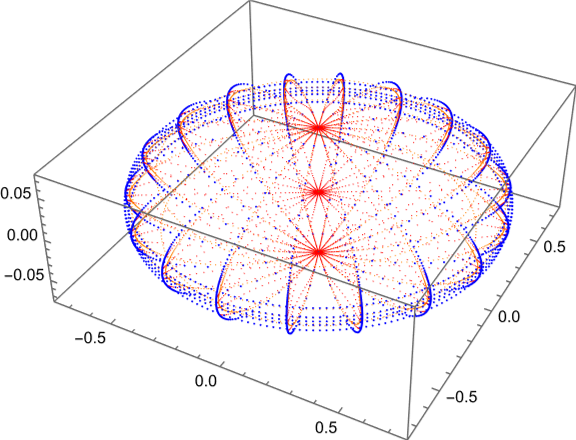}
\end{minipage}%
\caption{Spherical coordinate lines of radius $1$ (orange), asymptotic spherical coordinate lines (blue), spherical coordinate lines of radius $0.001$ (green) and radial line coordinate lines (red) in the squashed fuzzy sphere for $N=4$ and $\alpha=0.1$. Left: projective plot of $\mathcal{M}$, right: projective plot of $\Tilde{\mathcal{M}}$}
\label{fig:SU2/8A}
\end{figure}

\begin{figure}[H]
\centering
\begin{minipage}{.3\textwidth}
  \centering
  \includegraphics[height=.7\linewidth]{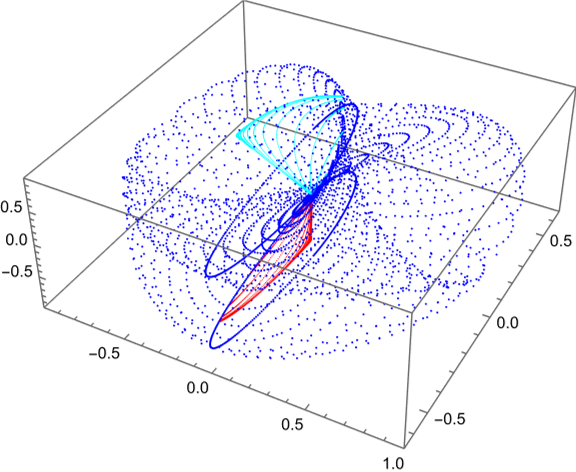}
\end{minipage}%
\begin{minipage}{.3\textwidth}
  \centering
  \includegraphics[height=.7\linewidth]{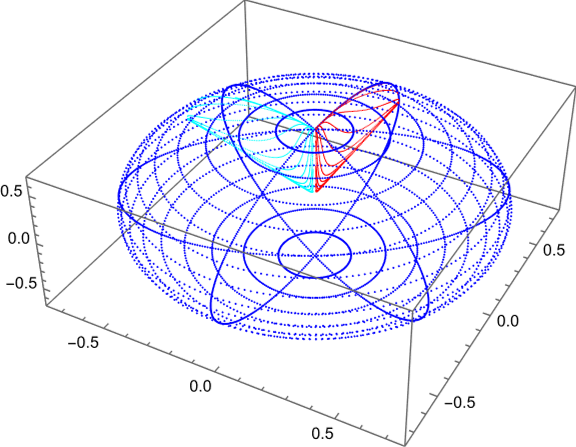}
\end{minipage}%
\caption{Spherical coordinate lines through $(0,1,\sigma)$ (red) respectively $(-1,0.3,\sigma)$ (cyan) for different $\sigma>0$ and asymptotic spherical coordinate lines (blue) in the squashed fuzzy sphere for $N=4$ and $\alpha=0.9$. Left: projective plot of $\mathcal{M}$, right: projective plot of $\Tilde{\mathcal{M}}$}
\label{fig:SU2/8B}
\end{figure}

In figure \ref{fig:SU2/8} we plotted random points from the unit ball (orange), random asymptotic points\footnote{This means, we calculated the corresponding asymptotic states in favor of the quasi-coherent states.} (blue) and random points from a tiny sphere of radius $0.001$ (green) as well as a radial curve through a random point (red) and a polar curve connecting the points\footnote{Prior knowledge tells us that the point $(0,1,0)$ is at the end of the cyan curve that is lower with respect to the print for the plots of $\mathcal{M}$ respectively higher for the plots of $\Tilde{\mathcal{M}}$.} $(0,0,1)$ and $(0,1,0)$ (cyan).\\
For a more complete understanding figure \ref{fig:SU2/8A} (instead of random points) shows corresponding coordinate lines for the case $\alpha=0.1$, while figure \ref{fig:SU2/8B} features specially chosen radial coordinate lines together with asymptotic spherical coordinate lines, visualizing the \textit{inside} of $\Tilde{\mathcal{M}}$.\\
We can make a number of observations:
\begin{itemize}
    \item In the round case we see that the asymptotic states, the random states and the states from a tiny sphere agree. The radial curve shrinks to a point. All this is fairly obvious as in the round case the quasi-coherent states do not depend on the radius.
    \item For all squashed cases $0<\alpha<1$ the asymptotic states recover the round case, but the \textit{distribution} of the points changes strongly. In the plots of $\Tilde{\mathcal{M}}$, the asymptotic states form an ellipsoidal that is squashed according to the factor $\alpha$ compared to the round case.\\
    The plots in figure \ref{fig:SU2/8A} suggest that the states at $\vert x\vert=1$ discern only slightly from the asymptotic states, while the radial and the polar coordinate lines behave similarly. On the other hand, in the right plot we see that there are radial coordinate lines that terminate at the origin.\\
    Figure \ref{fig:SU2/8B} resolves the discrepancy: The latter are exactly those with $x^3=0$, while for $\vert x \vert \sim \vert x^3\vert$, the radial coordinate lines tend to behave like polar coordinate lines. Only for $\vert x \vert \gg \vert x^3\vert $, the radial coordinate lines come near to the origin, before they drift off to a pole.\\
    From that we conclude that in $\Tilde{\mathcal{M}}$, the whole inside (with respect to the asymptotic states) is filled except from the $\hat{e}_3$ axis itself\footnote{Since $0\in\mathcal{K}$ and the rest of the axis is in $\mathcal{N}_{\pm\hat{e}_3}$, only two points lie on the $\hat{e}_3$ axis in $\Tilde{\mathcal{M}}$, distinguished by a relative sign.}, while unless $1\gg \vert x \vert \gg x^3$ the states are strongly confined to the asymptotic states.
    \item In the completely squashed case the asymptotic states form a one dimensional manifold and $\mathcal{M}$ turns two dimensional again. In the plots of $\Tilde{\mathcal{M}}$ this assembles to a disk where the center is coming from the $\hat{e}_3$ axis (omitting $0$) in $\Tilde{\mathbb{R}}^D$, while the points on the tiny sphere in $\Tilde{\mathbb{R}}^D$ lie in a tiny disk around the origin. The border of the whole disk is given by the asymptotic states. The radial and the polar coordinate lines in $\Tilde{\mathbb{R}}^D$ both behave identically as radial lines in the disk.
    \item Note that the gap in between the blue and the orange points is only an artifact of the fact that the random points are confined to the unit ball in $\mathbb{R}^3$.
\end{itemize}

The discussion of the actual \textit{thickness} of the auxiliary direction in the squashed case will be addressed in the following section.

\subsubsection{Foliations, Adapted Coordinates and Integration -- The Prototypical Result}

Now, we turn to foliations of $\mathcal{M}$ that allow us to calculate adapted coordinates and to integrate over a chosen leaf $\mathcal{L}$. That enables us to check the completeness relation and similar properties. In this section, we present the \textit{prototypical results} for the following choices: We put $N=4$, $\alpha=0.9$, $x=(1,2,1)$ (what specifies the actual leaf in the foliation) and use the hybrid leaf.\\
In the following section, we will then discuss what changes if we modify these parameters or use different leaves.

In section \ref{calcCurv} we have already seen that the leaves are (approximately) integrable for these choices, so we directly turn to the global discussion.

\begin{figure}[H]
\centering
\begin{minipage}{.3\textwidth}
  \centering
  \includegraphics[height=.7\linewidth]{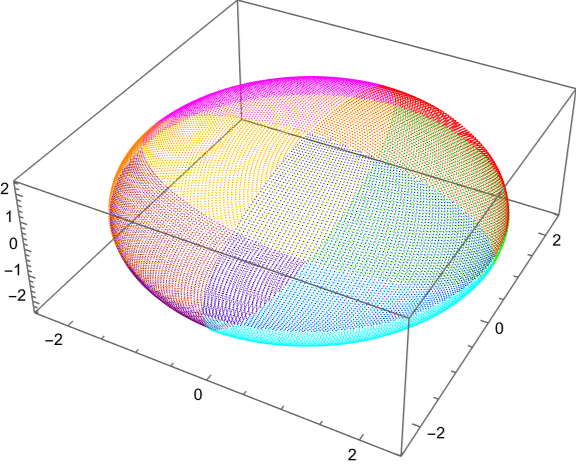}
\end{minipage}%
\begin{minipage}{.3\textwidth}
  \centering
  \includegraphics[height=.7\linewidth]{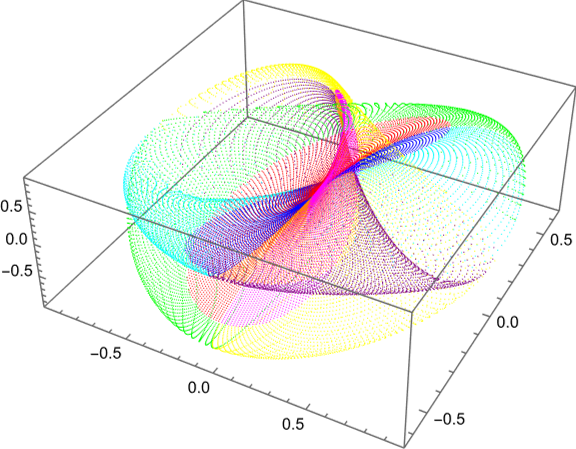}
\end{minipage}%
\begin{minipage}{.3\textwidth}
  \centering
  \includegraphics[height=.7\linewidth]{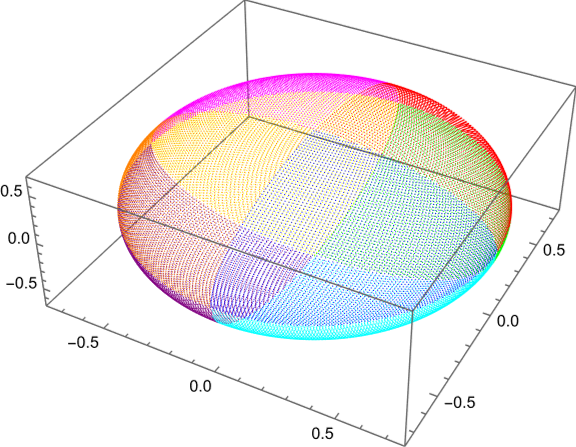}
\end{minipage}%
\caption{Tiling of the hybrid leaf through $x=(1,2,1)$ in the squashed fuzzy sphere for $N=4$ and $\alpha=0.9$. Left: projective plot of $\Tilde{\mathbb{R}}^D$, middle: projective plot of $\mathcal{M}$, right: projective plot of $\Tilde{\mathcal{M}}$}
\label{fig:SU2/9PROTO}
\end{figure}

In figure \ref{fig:SU2/9PROTO} we see a covering with coordinates of the hybrid leaf through $x$.
Once again, we note how well the leaf is integrable, especially already in $\Tilde{\mathbb{R}}^D$. In the plot of $\mathcal{M}$ we recognize the large scale shape that we know from the previous sections, while we now look at a two dimensional manifold $\mathcal{L}\subset\mathcal{M}$. Here the different colors (corresponding to different tiles) allow us to understand how the shape can be entangled and confirms that the plot only looks so complicated because the plotting procedure from section \ref{Visual} maps different regions in $\mathbb{C}P^{N-1}$ onto another.
The plot of $\Tilde{\mathcal{M}}$ already looks very much like an ellipsoidal.

Now we can leave the qualitative discussion behind in favor of quantitative results.\\
Our first interest lies on the \textit{completeness relation}.
We find that the leaf has the symplectic volume $V_\omega=10.058$.
According to section \ref{calcInt} we consider $\frac{N}{V_\omega}\mathbb{1}'$, representing $Q(1_\mathcal{M})$, which has eigenvalues $1.025,1.013,0.991,0.971$. If the completeness relation (\ref{compl}) holds, all eigenvalues should equal $1$, so the results are not far away from that.\\
To quantify the deviation, we look at the \textit{standard deviation}\footnote{By construction we have the mean $\mu_{\mathbb{1}'}=1$.} of the eigenvalues that is given by $\sigma_{\mathbb{1}'}=0.024$. Another measure is the \textit{relative deviation} of $\frac{N}{V_\omega}\mathbb{1}'$ to $\mathbb{1}$ with respect to the Hilbert-Schmidt norm, turning out to be $d_{\mathbb{1}'}=0.021$.

Let us turn to the \textit{quantization of the embedding functions} $\mathbf{x}^a$.
Here we look at $n_{X'}\frac{N}{V_\omega}X'$ from section \ref{calcInt}, representing $n_{X'}(Q(\mathbf{x}^a))$ for the proportionality constant $n_{X'}$. If equation (\ref{quantEmb}) holds, this should agree with $(X^a)$.
Explicitly, we find the relative deviation $d_{X'}=0.025$ with the \textit{correction factor} $n_{X'}=1.668$, which is pretty good.

As we have addressed two important conjectures from section \ref{QuantMap}, we also want to verify if $\{\mathbf{x}^a,\mathbf{x}^b\}\overset{?}{=}-2\theta^{ab}$ (the \textit{compatibility of the respective Poisson structures}) and 
$-2\omega_{ab}\theta^{bc}\overset{?}{=}p^c_a$ as well as $\partial_a\mathbf{x}^b\overset{?}{=}p^b_a$ for a projector $p^a_b$.\\
As a measure for the compatibility of the respective Poisson structures we define the \textit{relative deviation} $d_{\{\}}:=\Vert(\{\mathbf{x}^a,\mathbf{x}^b\}-(-2)\theta^{ab}) \Vert_{HS'}/\Vert(\{\mathbf{x}^a,\mathbf{x}^b\}) \Vert_{HS'}=0.00014$, which is remarkably good.\\
For the others it suffices to look at the eigenvalues of the $(lhs)$ (they should be given by $1$ with multiplicity $l'$ and $0$ with multiplicity $D-l'$, where $l'$ is the symplectic dimension of $\mathcal{M}$). Explicitly, the eigenvalues of $-2\omega_{ab}\theta^{bc}$ and $\partial_a\mathbf{x}^b$ are  given by $0.088,0.088,0.000$ respectively $0.319,0.277,0.000047$. So in both cases the conjectures are heavily violated. But as we are far away from the local minima of $\lambda(x)$ this is not astonishing at all and we will come back to this in the following section when we look at the dependence on $x$.

Further, it is interesting to look at the \textit{Kähler cost} of the subspace $V_L:=V_x\subset T_x\mathbb{R}^D$ coming from the distribution corresponding to the leaf.\\
As the result is only a number of generic scale we compare it to the cost of a random subspace $V_R\subset T_x\mathbb{R}^D$ and the optimal Kähler subspace $V_K\subset T_x\mathbb{R}^D$ (both of the same dimension as $V_L$). Accordingly, we find\footnote{From now on we omit the $x$ in $c^s_x$ and \textit{round} results by the following scheme: In principle we keep three digits behind the comma unless the number is too small. Then we round such that we keep two nonzero digits. If the result is smaller than $10^{-5}$ we only show its scale.} 
$c^2(V_R)=0.077$, $c^2(V_L)=0.0010$ and $c^2(V_K)=0.000062$.
This means that the hybrid leaf is rather well optimized with respect to the Kähler cost (compared to a random subspace), while it is still far away from the absolute optimum.\\
In order to arrive at the specific $V_K$ (there are plenty of local minima of $c^2$ that lie even higher than $c^2(V_L)$) it was necessary to start from twenty different random subspaces for which the gradient descent method was performed. This makes the finding of the optimal subspace for the Kähler leaf even harder (we have already mentioned that the computational effort is rather high right from the beginning) and untrustworthy. On the other hand, the results on the Kähler cost are interesting on their own.

\begin{figure}[H]
\centering
\begin{minipage}{.3\textwidth}
  \centering
  \includegraphics[height=.7\linewidth]{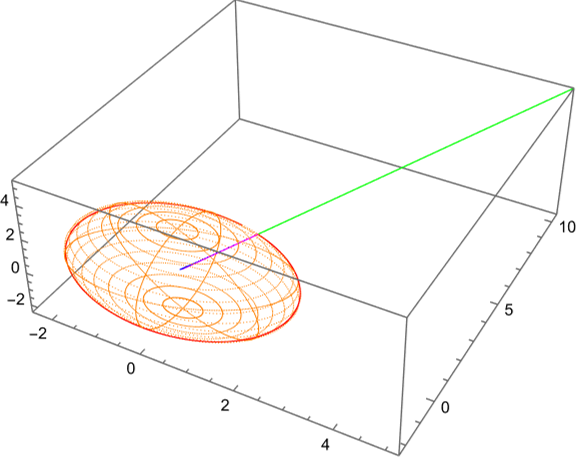}
\end{minipage}%
\begin{minipage}{.3\textwidth}
  \centering
  \includegraphics[height=.7\linewidth]{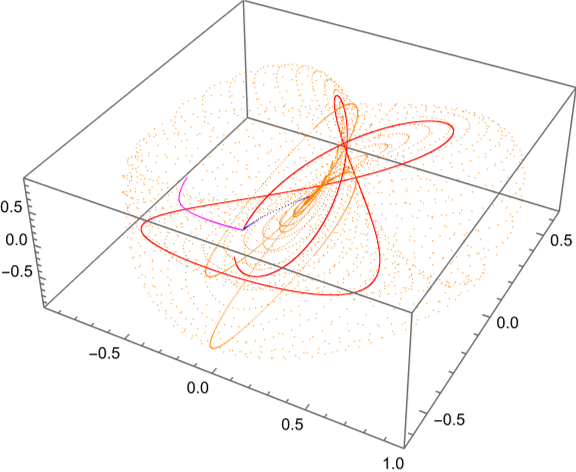}
\end{minipage}%
\begin{minipage}{.3\textwidth}
  \centering
  \includegraphics[height=.7\linewidth]{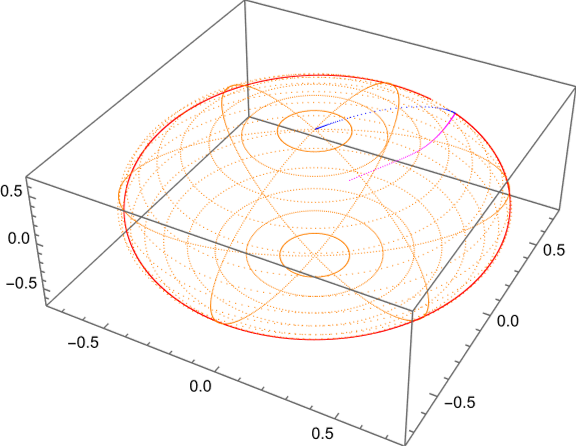}
\end{minipage}%
\caption{Determination of  the thickness in hybrid leaf through $x=(1,2,1)$. Spherical coordinate net through $x$ (orange), inwards pointing radial curve (blue), curve in leaf (red), inwards pointing curve in null leaf (magenta) and outwards pointing curve in null leaf (green) in the squashed fuzzy sphere for $N=4$ and $\alpha=0.9$. Left: projective plot of $\Tilde{\mathbb{R}}^D$, middle: projective plot of $\mathcal{M}$, right: projective plot of $\Tilde{\mathcal{M}}$}
\label{fig:SU2/9PROTOA}
\end{figure}

We then come to another interesting question, namely the \textit{thickness} of $\mathcal{M}$ in the directions orthogonal to the leaf $\mathcal{N}^\mathcal{L}_x$.\\
We address the question via the following attempt: First, we calculate the length\footnote{With respect to $g_{ab}$.} $l_L$ of a curve in the leaf starting at $x$ that approximately goes once around the origin. This gives us a reference length scale. Then we calculate the lengths $l_I$ and $l_O$ of two curves in the null leaf $\mathcal{N}^\mathcal{L}_x$ starting at $x$ -- one going inwards (we let it terminate shortly before we arrive at the origin) and one going outwards. This setup is shown in figure \ref{fig:SU2/9PROTOA}, where we additionally plotted spherical coordinate lines and a radial coordinate line through $x$ for a better orientation.\\
We find the lengths $l_L=5.323$, $l_I=0.891$ and $l_O=0.0061$. This means, that the outwards pointing curve is of negligible length, while the inwards pointing one acquires a finite length. This means we should not think of $\mathcal{M}$ as extremely thin in the direction of the null leaf,  even for $\alpha=0.9$. It turns out that the greatest contribution comes from the vicinity of the origin, fitting to the observation that in the round case $g_{ab}$ diverges for $\vert x\vert\to 0$. Thus it is hard to say how sensitive $l_I$ is to the chosen endpoint of the curve and the results should be considered cautiously.\\
While the quantitative result is not very enlightening, the qualitative results in the figure are interesting. In the left plot we see that the radial curve is almost identical with the inward pointing curve. On the other hand, considering the right plot, we note that in $\Tilde{\mathcal{M}}$ the two curves discern strongly as the radial curve terminates at the north pole, while the inwards pointing curve heads to the origin.

Finally, a short comment on the numerical quality of the results is due.\\
While the verification of the completeness relation and the quantization of the $\mathbf{x}^a$ depends on numerical integration over the constructed leaf (and thus on a finite step length, here taken to be $0.05$) the results on the compatibility of the Poisson structures and the conjectures $-2\omega_{ab}\theta^{bc}\overset{?}{=}p^c_a$ and $\partial_a\mathbf{x}^b\overset{?}{=}p^b_a$ are calculated at the point $x$. So for the latter three we should expect a much lower numerical deviation from the exact results than for the first two.\\
While the Kähler cost of $V_L$ should be of descent numerical quality, the cost of $V_R$ depends on a random choice and the determination of $V_K$ is problematic itself.\\
Concerning the thickness, we have already stated that the lengths $l_L$ and $l_O$ are rather unproblematic, while the length $l_I$ strongly depends on the exact endpoint.

\subsubsection{Foliations, Adapted Coordinates and Integration -- Dependence on Parameters}
\label{sfsfoliat}

Having seen the prototypical result for $N=4$, $\alpha=0.9$, $x=(1,2,1)$ in the hybrid leaf, we now look at the dependence of the results on these parameters. Thus we subsequently vary $\alpha$, $N$, $x$ and the leaf while holding the other parameters fixed.

As a beginning, we look at the dependence on the squashing parameter $\alpha$ and calculate the respective quantities from the last section for $\alpha=1,0.9,0.5,0.1$. The most important results are listed in the tables \ref{DepAlpha} and \ref{DepThickAlpha}.\\
Let us briefly focus on the round case. There we would expect $\sigma_{\mathbb{1}'}=d_{\mathbb{1}'}=d_{X'}=d_{\{\}}=c^2(V_R)=c^2(V_L)=c^2(V_K)=l_I=l_O=0$. This turns out to hold true up to numerical deviations. The size of these deviations perfectly fits the discussion at the end of the last section.\\
The symplectic volume $V_\omega$ is almost independent of $\alpha$, while the quality of the completeness relation and the quantization of the $\mathbf{x}^a$ is high in the round case and decreases with increasing $\alpha$. For example for $\alpha=0.5$ the results are still tolerable, while for $\alpha=0.1$ $d_{\mathbb{1}'}>10\%$.\\
The correction factor $n_{X'}$ seems to be almost independent of $\alpha$ with a slight tendency to increase with decreasing $\alpha$.\\
The compatibility of the two Poisson structures holds remarkably well until $\alpha=0.5$, but for $\alpha=0.1$ we can clearly see a violation.
\\
Now, we come to the Kähler cost: For all three subspaces, the cost increases with $\alpha$, while the discrepancy between $c^2(V_L)$ and $c^2(V_K)$ grows worse with decreasing $\alpha$. Still, in all cases the subspace $V_L$ is much better conditioned than the random subspace.
\\
Looking at the thickness, we witness an interesting behaviour: While in the round case $l_I$ and $l_O$ vanish, the values of $l_I$ jumps immediately from $0$ to approximately $1$ and then remains almost fixed. This means $\mathcal{M}$ instantly \textit{grows thick} in the auxiliary dimension when going away from $\alpha=1$. The length $l_O$ on the other hand slowly increases with decreasing $\alpha$, meaning that even for small $\alpha$ the quasi-coherent states at $x=(1,2,1)$ are not \textit{far away} from the asymptotic states.\\
We conclude that most of the appreciated properties of the round fuzzy sphere are preserved approximately during the squashing, while the quality decreases with $\alpha$. Only the thickness jumps immediately when leaving the round case. This suggests a stable behaviour of the construction under squashing.

\begin{table}[H]
\centering
\begin{tabular}{l|l|ll|ll|l|lll}
 $\alpha$ & $V_\omega$ & $\sigma_{\mathbb{1}'}$ & $d_{\mathbb{1}'}$ & $d_{X'}$ & $n_{X'}$ & $d_{\{\}}$ & $c^2(V_R)$ & $c^2(V_L)$ & $c^2(V_K)$ \\\hline
$1$ & $10.043$ & $0.0010$ & $0.0086$ & $0.015$ & $1.666$ & $\sim 10^{-9}$ & $\sim 10^{-8}$ & $\sim 10^{-8}$ & $\sim 10^{-8}$ \\
$0.9$ & $10.058$ & $0.024$ & $0.021$ & $0.025$ & $1.667$ & $0.00014$ & $0.077$ & $0.0010$ & $0.000062$ \\
$0.5$ & $10.102$ & $0.061$ & $0.053$ & $0.030$ & $1.688$ & $\sim 10^{-6}$ & $0.555$ & $0.032$ & $0.0013$\\
$0.1$ & $10.571$ & $0.139$ & $0.120$ & $0.056$ & $1.724$ & $0.190$ & $1.335$ & $0.312$ & $0.0028$
\end{tabular}
\caption{Dependence of various quantities on the squashing parameter $\alpha$}
\label{DepAlpha}
\end{table}

\begin{table}[H]
\centering
\begin{tabular}{l|lll}
 $\alpha$ & $l_L$ & $l_I$ & $l_O$ \\\hline
$1$ & $5.293$ & $\sim 10^{-7}$ & $\sim 10^{-9}$ \\
$0.9$ & $5.323$ & $0.927$ & $0.0061$ \\
$0.5$ & $4.435$ & $0.878$ & $0.029$ \\
$0.1$ & $5.497$ & $0.848$ & $0.042$
\end{tabular}
\caption{Dependence of the thickness on the squashing parameter $\alpha$}
\label{DepThickAlpha}
\end{table}

\begin{figure}[H]
\centering
\begin{minipage}{.3\textwidth}
  \centering
  \includegraphics[height=.7\linewidth]{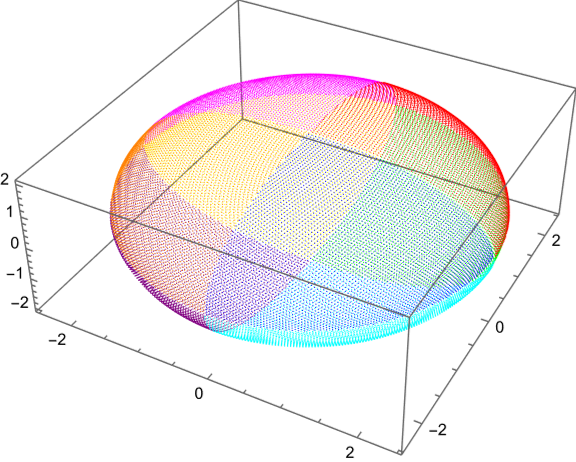}
\end{minipage}%
\begin{minipage}{.3\textwidth}
  \centering
  \includegraphics[height=.7\linewidth]{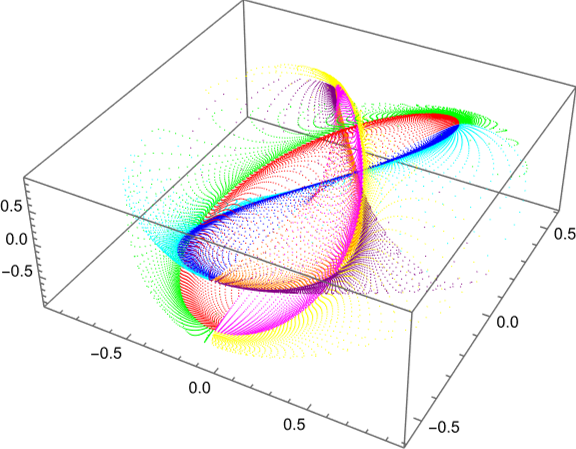}
\end{minipage}%
\begin{minipage}{.3\textwidth}
  \centering
  \includegraphics[height=.7\linewidth]{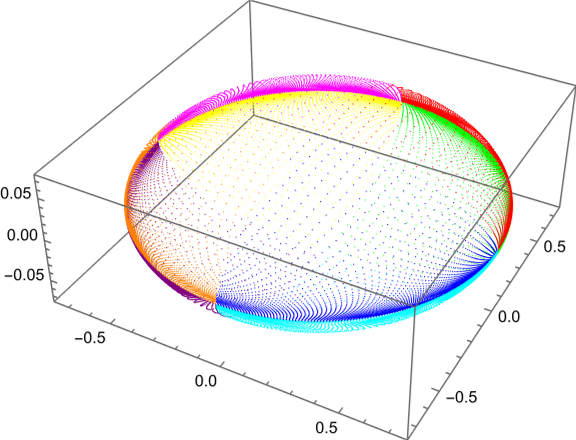}
\end{minipage}%
\caption{Tiling of the hybrid leaf through $x=(1,2,1)$ in the squashed fuzzy sphere for $N=4$ and $\alpha=0.1$. Left: projective plot of $\Tilde{\mathbb{R}}^D$, middle: projective plot of $\mathcal{M}$, right: projective plot of $\Tilde{\mathcal{M}}$}
\label{fig:SU2/9alpha}
\end{figure}

Figure \ref{fig:SU2/9alpha} shows plots for $\alpha=0.1$ that are the analogues to the ones in figure \ref{fig:SU2/9PROTO}. We see that the shape in $\Tilde{\mathbb{R}}^D$ does not change strongly, while $\mathcal{M}$ looks slightly different (comparable to the results of section \ref{DependanceAlpha}). $\Tilde{\mathcal{M}}$ looks like a squashed sphere (an ellipsoidal) as we would expect heuristically. Also these graphical results support the conjectured stability of the construction under squashing as the dimension of $\mathcal{L}$ (and thus the local structure) is preserved.

Let us discuss the dependence on $N$. In the tables \ref{DepN} and \ref{DepThickN} the most relevant quantities are displayed for $N=2,3,4,10,100$.\\
Here, $N=2$ turns out to be the special case where the same quantities vanish numerically as before for $\alpha=1$.\\
We deduce that the symplectic volume follows the rule $V_\omega\approx(N-1) \cdot 3.4$. We might explain the numerical factor by $3.4\approx \frac{1}{2}\cdot(2\pi)$, although the deviation is rather large. The additional factor $\frac{1}{2}$ is related to the fact that we should actually replace $\omega\mapsto-\frac{1}{2}\omega$ (as discussed in section \ref{bp}), while then $V_\omega \approx 2\pi N$ and $\alpha\approx1$ for large $N$ as suggested in \cite{Steinacker_2021}.\\
For the completeness relation and the quantization of $\mathbf{x}^a$ we find that the quality slightly decreases with $N$ but stagnates for large $N$. We will come back to that in a moment.\\
The correction factor follows the simple rule $n_{X'}\approx\frac{N+1}{N-1}$ (which exactly is $\vert(\mathbf{x}^a)\vert^{-2}$ in the round case).\\
The compatibility of the Poisson structures at first gets worse with increasing $N$ but for $N=100$ it is again much better.\\ 
For the Kähler costs of $V_L$ and $V_R$, we find a similar behaviour.
For $V_K$ the result is paradoxical since for $N=10,100$ the cost is \textit{higher} than for $V_L$. This is once again related to the fact that $c^2$ does not have a unique local minimum. This problem shows to increase with $N$.\\
Further, we find that all three lengths $l_L$, $l_I$ and $l_O$ increase with $N$. For $l_L$, we find the dependence $l_L\propto \sqrt{N-1}$ (this is just fine as the symplectic volume showed to scale with $N-1$). If we correct for this rescaling, both $l_I$ and $l_O$ \textit{decrease} again for large $N$.

Let us come back to the discussion of the quality of the completeness relation and the quantization of the $\mathbf{x}^a$. We have already seen that the relative deviation approximately stagnates for large $N$.\\
We now assume that $\frac{N}{V_{\omega}}\mathbb{1'}=\mathbb{1}+C$ where $C$ is the constant matrix with entries $c$. This models a \textit{constant error per component}. Under this assumption, we find the relative deviation
\begin{align}
    d_{\mathbb{1}'}=\frac{\Vert \frac{N}{V_{\omega}}\mathbb{1'}-\mathbb{1}\Vert_{HS}}{\Vert \mathbb{1}\Vert_{HS}}=\frac{\sqrt{\sum_{i,j=1}^N C_{ij}^2}}{\sqrt{\sum_{i,j=1}^N \delta_{ij}^2}}=\vert c\vert\frac{\sqrt{\sum_{i,j=1}^N 1}}{\sqrt{\sum_{i=1}^N 1}}=\vert c\vert\frac{\sqrt{N^2}}{\sqrt{N}}=\vert c\vert\sqrt{N}.
\end{align}
This result suggests that if $d_{\mathbb{1}'}$ is approximately constant with $N$, the error per component scales with $c\propto N^{-\frac{1}{2}}$ and thus the quality in fact \textit{increases} with $N$.
\\
So, we conclude that for our actual results the quality of the completeness relation improves like $N^{-\frac{1}{2}}$ in the large $N$ limit.\\
Similar arguments can be made for the quantization of the $\mathbf{x}^a$.

Under these considerations we conclude that the large $N$ limit is likely to restore the results from the round case, although for smaller $N$ the behaviour is vice versa.

\begin{table}[H]
\centering
\begin{tabular}{l|l|ll|ll|l|lll}
 $N$ & $V_\omega$ & $\sigma_{\mathbb{1}'}$ & $d_{\mathbb{1}'}$ & $d_{X'}$ & $n_{X'}$ & $d_{\{\}}$ & $c^2(V_R)$ & $c^2(V_L)$ & $c^2(V_K)$ \\\hline
 $2$ & $3.348$ & $0.0064$ & $0.0045$ & $0.0090$ & $3.001$ & $\sim 10^{-9}$ & $0$ & $\sim 10^{-8}$ & $0$ \\
 $3$ & $6.6803$ & $0.020$ & $0.0164$ & $0.018$ & $2.003$ & $0.00012$ & $0.073$ & $0.00094$ & $\sim 10^{-7}$ \\
$4$ & $10.058$ & $0.024$ & $0.021$ & $0.025$ & $1.667$ & $0.00014$ & $0.077$ & $0.0010$ & $0.000062$ \\
$10$ & $30.031$ & $0.033$ & $0.031$ & $0.034$ & $1.223$ & $0.00090$ & $0.064$ & $0.00058$ & $0.0023$ \\
$100$ & $331.062$ & $0.064$ & $0.064$ & $0.063$ & $1.019$ & $0.000011$ & $0.023$ & $0.00018$ & $0.0027$
\end{tabular}
\caption{Dependence of various quantities on $N$}
\label{DepN}
\end{table}

\begin{table}[H]
\centering
\begin{tabular}{l|lll}
 $N$ & $l_L$ & $l_I$ & $l_O$ \\\hline
 $2$ & $3.055$ & $\sim 10^{-7}$ & $\sim 10^{-9}$ \\
 $3$ & $4.338$ & $0.705$ & $0.0048$ \\
$4$ & $5.323$ & $0.927$ & $0.0061$ \\
$10$ & $9.251$ & $1.218$ & $0.0083$ \\
$100$ & $30.744$ & $3.665$ & $0.010$
\end{tabular}
\caption{Dependence of the thickness on $N$}
\label{DepThickN}
\end{table}

The next step is to study the dependence on $x$. Here we define $x_0=(1,2,1)$ and consider the cases $x=100 \cdot x_0$, $x_0$, $0.05\cdot x_0$ and $x_\lambda$, where $x_\lambda:=x_0'$ is the point in the null leaf $\mathcal{N}^{\mathcal{L}}_{x_0}$ that minimizes $\lambda$. Using the algorithm from section \ref{calcMin}, this can be calculated and one finds $x_\lambda=(0.316,0.632,0.287)\approx 0.310\cdot x_0$. Thus we find $\vert x_\lambda\vert\approx \sqrt{\frac{4-1}{4+1}}$, being consistent with the result from the round case, which is discussed in section \ref{FuzzySphere}.\\
The most relevant quantities\footnote{Here, we do not look at the thickness once again as this is not fruitful for points on the same radial line.} are provided in table \ref{DepX}.
\\
Also here, we find a special case (that is $x=100\cdot x_0$) where the same quantities almost vanish as before for $\alpha=1$.
Consequently, we therefore conjecture that for $\vert x\vert\to\infty$ (meaning we work with the asymptotic states) the round case is recovered. This would fit to the observation that the asymptotic states are exactly the quasi-coherent states from the round case. Nevertheless, $\omega_{ab}$ in principle does not have to be the same. Yet in appendix \ref{Appendix:perturbativeappr} we have seen that in the round case $\omega_{ab}\sim \vert x\vert^{-2}$, while in the squashed case the first perturbation $\omega'_{ab}\sim\vert x\vert^{-3}$. Thus, the latter is negligible compared to the unperturbed $\omega_{ab}$ in the limit, making the behaviour at least plausible.
\\
Here, we find that the symplectic volume and the correction factor $n_{X'}$ are almost independent of $x$, while the quality of the completeness relation and the quantization of $\mathbf{x}^a$ gets worse and worse with decreasing\footnote{Note that the points are \textit{not} ordered by their absolute value as $\vert x_\lambda\vert\gg 0.05\cdot x_0$.} $\vert x\vert$.\\
Also the compatibility of the different Poisson structures and all three Kähler costs grow worse with decreasing $\vert x\vert$.\\
This leads us to the conclusion that the quality is better for large $\vert x\vert$ and that the approach to take the leaf through $x_\lambda$ as default is not promising for the squashed fuzzy sphere.

Finally, we once again come back to the conjectures $-2\omega_{ab}\theta^{bc}\overset{?}{=}p^c_a$ and $\partial_a\mathbf{x}^b\overset{?}{=}p^b_a$ which are supposed to hold if $\lambda$ is minimal, so they should hold at $x_\lambda$.
Here, we find the eigenvalues of $-2\omega_{ab}\theta^{bc}$ and $\partial_a\mathbf{x}^b$, given by $1.011,0.949,0.0016$ respectively $0.960,0.960,0$. In both cases we have the approximate eigenvalues $1$ with multiplicity $2$ and $0$ with multiplicity $1$, so the conjectures hold at least approximately.

\begin{table}[H]
\centering
\begin{tabular}{l|l|ll|ll|l|lll}
 $x$ & $V_\omega$ & $\sigma_{\mathbb{1}'}$ & $d_{\mathbb{1}'}$ & $d_{X'}$ & $n_{X'}$ & $d_{\{\}}$ & $c^2(V_R)$ & $c^2(V_L)$ & $c^2(V_K)$ \\\hline
$100\cdot x_0$ & $10.069$ & $0.015$ & $0.013$ & $0.019$ & $1.668$ & $\sim 10^{-8}$ & $0.00073$ & $\sim10^{-6}$ & $0.00024$ \\
$x_0$ & $10.058$ & $0.024$ & $0.021$ & $0.025$ & $1.667$ & $0.00014$ & $0.077$ & $0.0010$ & $0.000062$ \\
$0.05\cdot x_0$ & $10.146$ & $0.199$ & $0.172$ & $0.105$ & $1.741$ & $0.079$ & $1.264$ & $0.018$ & $0.016$ \\
$x_\lambda$ & $9.999$ & $0.035$ & $0.030$ & $0.022$ & $1.669$ & $0.0015$ & $0.241$ & $0.0023$ &
$0.00065$ \end{tabular}
\caption{Dependence of various quantities on $x$}
\label{DepX}
\end{table}

\begin{figure}[H]
\centering
\begin{minipage}{.3\textwidth}
  \centering
  \includegraphics[height=.7\linewidth]{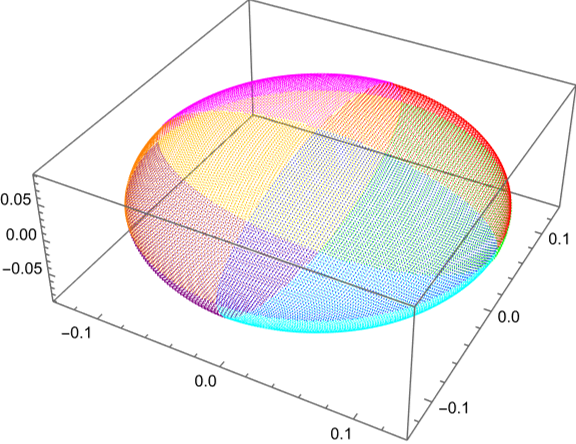}
\end{minipage}%
\begin{minipage}{.3\textwidth}
  \centering
  \includegraphics[height=.7\linewidth]{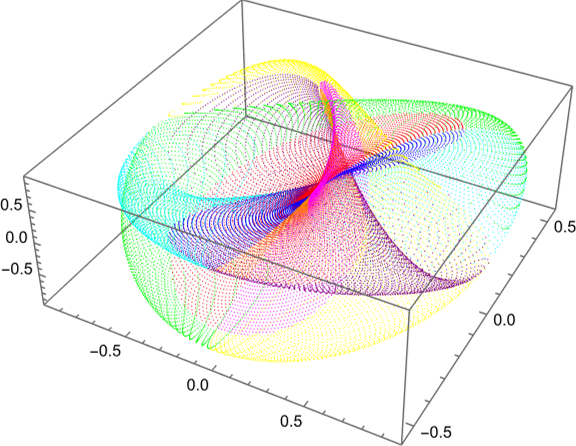}
\end{minipage}%
\begin{minipage}{.3\textwidth}
  \centering
  \includegraphics[height=.7\linewidth]{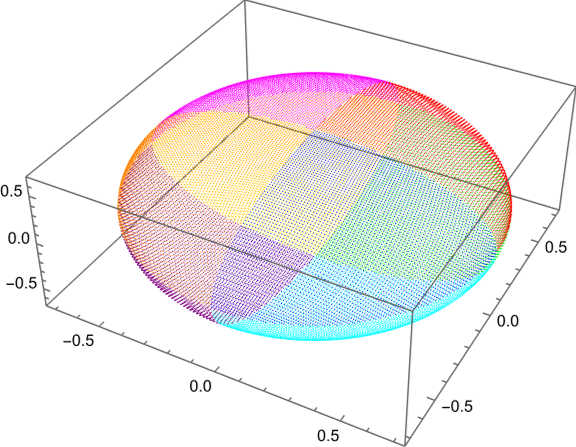}
\end{minipage}%
\caption{Tiling of the hybrid leaf through $x=0.05\cdot(1,2,1)$ in the squashed fuzzy sphere for $N=4$ and $\alpha=0.9$. Left: projective plot of $\Tilde{\mathbb{R}}^D$, middle: projective plot of $\mathcal{M}$, right: projective plot of $\Tilde{\mathcal{M}}$}
\label{fig:SU2/9x}
\end{figure}

In figure \ref{fig:SU2/9x} we see comparable plots to the ones in figure \ref{fig:SU2/9PROTO} but with $x=0.05\cdot (1,2,1)$.
Here we note that the shape in the plot of $\Tilde{\mathbb{R}}^D$ is rescaled with the factor $0.05$, while the plot of $\mathcal{M}$ is only slightly different. $\Tilde{\mathcal{M}}$ almost remains unchanged.
This supports the interpretation of the null leafs $\mathcal{N}^\mathcal{L}_y$ as the approximate generalization of the $\mathcal{N}_y$.

\begin{figure}[H]
\centering
  \centering
  \includegraphics[height=.17\linewidth]{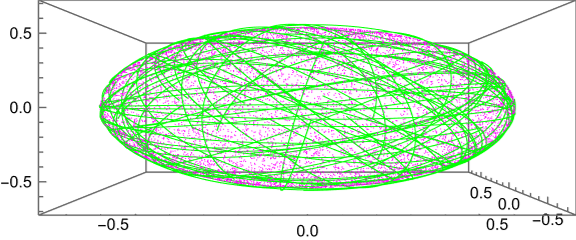}
\caption{Hybrid leaf in the squashed fuzzy sphere for $N=4$ and $\alpha=0.9$. Green: scan of hybrid leaf through $x_0'=:x_\lambda$, purple: $x'_i$ for $10000$ random $x_i$.  Projective plot of $\Tilde{\mathbb{R}}^D$}
\label{fig:SU2/9Xlambda}
\end{figure}

In figure \ref{fig:SU2/9Xlambda} we see a comparable plot to the one in figure \ref{fig:Implementation/10} but with $\alpha=0.9$ instead of $\alpha=0.1$. Here, the points that minimize $\lambda$ over the null leaves approximately lie on a single hybrid leaf. Thus here, choosing the leaf through $x_\lambda$ is consistent and independent of the initial $x_0$ while this has not been the case for $\alpha=0.1$.

Let us finally come to the dependence on the chosen leaf, an important discussion that hopefully guides us to a preferred method. We look at the following leaves\footnote{The symplectic leaf is omitted since most of the relevant procedures have not been implemented on Mathematica due to the fact that the leaf is not promising for arbitrary geometries as discussed in section \ref{fol}.}: the hybrid leaf (H), the hybrid leaf using $\omega$ (O), the hybrid leaf using $\omega$ and $g$ (G), the Pfaffian leaf (P) and the Kähler leaf (K). The tables \ref{DepLeaf} and \ref{DepThickLeaf} show the most interesting quantities while some are missing for P and K (for these some quantities have not been calculated due to the large computational effort for these two leaves).
\\
From the three implemented hybrid leaves (H, O and G) -- these are the ones which are rather less computationally demanding -- we find that the results discern only slightly, yet they are best for G. The results obtained by P are catastrophically bad. This is caused by the enormous computational expense of this leaf that made it necessary to rely on much fewer points in the integration. Since we have seen in section \ref{calcCurv} that P is almost identical to H for the squashed fuzzy sphere, we should be able to arrive at adequate results in principle. Nevertheless, the needed computational effort forces us to discard P. For K almost the same holds, although the effort is not as large. Although here we used fewer points for the integration (compared to the hybrid methods) the results are remarkably good. Due to the unreliability discussed in the previous section, we nonetheless discard K.\\
We finally conclude that all hybrid leaves generate acceptable and comparable results, but at least for the squashed fuzzy sphere G is the superior choice.

\begin{table}[H]
\centering
\begin{tabular}{l|l|ll|ll|l|lll}
 Leaf & $V_\omega$ & $\sigma_{\mathbb{1}'}$ & $d_{\mathbb{1}'}$ & $d_{X'}$ & $n_{X'}$ & $d_{\{\}}$ & $c^2(V_R)$ & $c^2(V_L)$ & $c^2(V_K)$ \\\hline
H & $10.058$ & $0.024$ & $0.021$ & $0.025$ & $1.667$ & $0.00014$ & $0.077$ & $0.0010$ & $0.000062$ \\
O & $10.008$ & $0.020$ & $0.017$ & $0.017$ & $1.668$ & $0.00014$ & $0.077$ & $0.00089$ & $0.000062$ \\
G & $10.032$ & $0.013$ & $0.011$ & $0.014$ & $1.667$ & $0.00014$ & $0.077$ & $0.00089$ & $0.000062$ \\
P & $1.244$ & $1.533$ & $1.327$ & - & - & $0.00014$ & $0.077$ & $1.414$ & $0.000062$ \\
K & $9.951$ & $0.021$ & $0.018$ & $0.031$ & $1.663$ & $0.00014$ & $0.077$ & $0.000062$ & $0.000062$
\end{tabular}
\caption{Dependence of various quantities on the leaf}
\label{DepLeaf}
\end{table}

\begin{table}[H]
\centering
\begin{tabular}{l|lll}
Leaf & $l_L$ & $l_I$ & $l_O$ \\\hline
H & $5.323$ & $0.927$ & $0.0061$ \\
O & $5.323$ & $0.941$ & $0.0061$ \\
G & $5.323$ & $0.937$ & $0.0061$
\end{tabular}
\caption{Dependence of the thickness on the leaf}
\label{DepThickLeaf}
\end{table}

\begin{figure}[H]
\centering
\begin{minipage}{.3\textwidth}
  \centering
  \includegraphics[height=.7\linewidth]{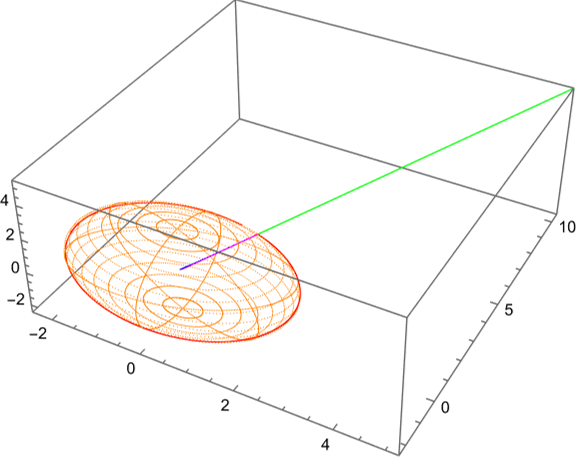}
\end{minipage}%
\begin{minipage}{.3\textwidth}
  \centering
  \includegraphics[height=.7\linewidth]{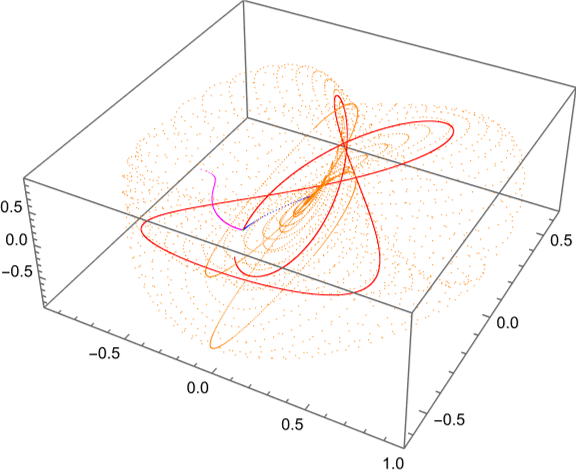}
\end{minipage}%
\begin{minipage}{.3\textwidth}
  \centering
  \includegraphics[height=.7\linewidth]{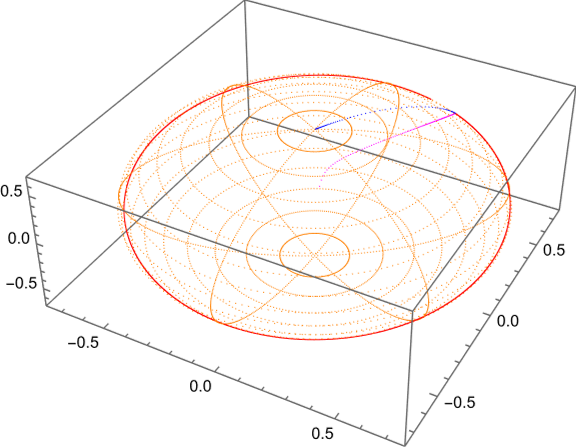}
\end{minipage}%
\caption{Determination of  the thickness in hybrid leaf using $\omega$ and $g$ through $x=(1,2,1)$. Spherical coordinate net through $x$ (orange), inwards pointing radial curve (blue), curve in leaf (red), inwards pointing curve in null leaf (magenta) and outwards pointing curve in null leaf (green) in the squashed fuzzy sphere for $N=4$ and $\alpha=0.9$. Left: projective plot of $\Tilde{\mathbb{R}}^D$, middle: projective plot of $\mathcal{M}$, right: projective plot of $\Tilde{\mathcal{M}}$}
\label{fig:SU2/9leafG}
\end{figure}

In figure \ref{fig:SU2/9leafG} we see plots comparable to figure \ref{fig:SU2/9PROTOA} but using G instead of H.
We see that although in $\Tilde{\mathbb{R}}^D$ the respective plots are almost identical, there is some deviation for small radii in $\Tilde{\mathcal{M}}$.

\begin{figure}[H]
\centering
  \centering
  \includegraphics[height=.21\linewidth]{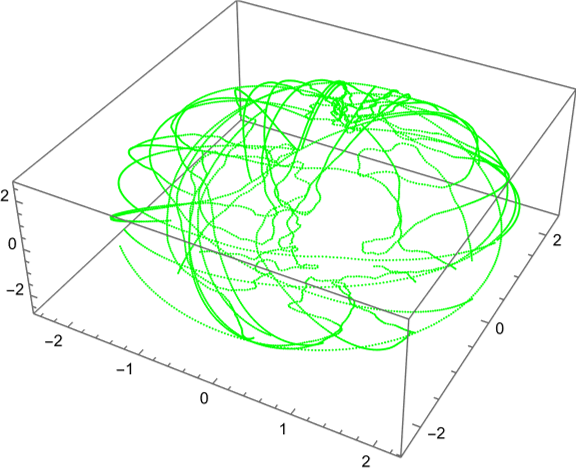}
\caption{Scan of Kähler leaf through $(1,2,1)$ in the squashed fuzzy sphere for $N=4$ and $\alpha=0.9$. Projective plot of $\Tilde{\mathbb{R}}^D$}
\label{fig:SU2/9leafK}
\end{figure}

In figure \ref{fig:SU2/9leafK} we see a scan of K that shows how unreliable the calculation is. At some locations the curves look smooth while at others the curves bend erratically.

Let us shortly summarize what we have learned in this section.\\
We concluded that the quality reduces with decreasing $\alpha$, decreasing $\vert x\vert$ and increasing $N$ -- while for large $N$ the quality improves again.\\
Further, all three hybrid leaves produce good results while the hybrid leaf using $\omega$ and $g$ turns out to be preferable.\\
In the regimes of good quality, we have found the compatibility of the Poisson structures induced by $\omega_{ab}$ and $-2\theta^{ab}$
\begin{align}
    \label{comp3}
    \{\mathbf{x}^a,\mathbf{x}^b\}\approx -2\theta^{ab},
\end{align}
the symplectic volume
\begin{align}
    \label{VN}
    V_\omega\approx (N-1)\cdot 3.4,
\end{align}
the completeness relation
\begin{align}
    \frac{N}{V_\omega} \int_\mathcal{L}\Omega_\mathcal{M} \;\ket{x}\bra{x}\approx\mathbb{1}_\mathcal{H},
\end{align}
and the quantization of the $\mathbf{x}^a$
\begin{align}
    \frac{N+1}{N-1}\frac{N}{V_\omega} \int_\mathcal{L}\Omega_\mathcal{M} \;\mathbf{x}^a \; \ket{x}\bra{x}\approx X^a.
\end{align}
This shows that many important axioms of the quantization map
\begin{align}
    Q(f):=\frac{N}{V_\omega} \int_\mathcal{L}\Omega_\mathcal{M} \; f \; \ket{x}\bra{x}
\end{align}
are satisfied at least approximately and especially
\begin{align}
    Q(1_\mathcal{M})\approx \mathbb{1}_\mathcal{H} \text{ and } Q(\mathbf{x}^a)\approx \frac{N-1}{N+1} X^a.
\end{align}
Finally, we have seen that equation (\ref{comp3}) should be viewed as a generalization of the conditions $-2\omega_{ab}\theta^{bc}\overset{?}{=}p^c_a$ and $\partial_a\mathbf{x}^b\overset{?}{=}p^b_a$ that in contrast to the latter two has a chance to hold away from points that minimize $\lambda$.

\subsection{The Fuzzy Sphere and Random Matrices}
\label{fsr}

We now define a new matrix configuration that is partially random. Here, we pick three random Hermitian $N\times N$ matrices $R^a$ such that the norm of all components is bounded by $1$. Then we define the matrix configuration\footnote{This guarantees that $S^2_{N,1,R}=S^2_N$ as we had $S^2_{N,1}=S^2_N$ for the squashed fuzzy sphere.}
\begin{align}
    S^2_{N,\beta,R}:=\left(\Bar{X}^a+(1-\beta) R^a\right)
\end{align}
where the $\Bar{X}^a$ are the matrices from the round fuzzy sphere $S^2_N$. Here, we might speak of the \textit{perturbed fuzzy sphere}.\\
Of course, also the random matrices themselves constitute a matrix configuration
\begin{align}
    R=\left(R^a\right)
\end{align}
for what we formally write $S^2_{N,-\infty,R}$ for convenience.

\subsubsection{First Results and Dimensional Aspects}

In table \ref{table1R} the dimensions of $\mathcal{M}$ and the ranks of $g_{ab}$, $\omega_{ab}$ and $\theta^{ab}$ are listed for different scenarios.
Obviously, we recover the results of the round fuzzy sphere for $\beta=1$, while in the other cases we see exactly the behaviour for random matrix configurations suggested in section \ref{random}.

\begin{table}[H]
\begin{tabular}{ll|l|l|l|l}
                           &              & $\operatorname{dim}(\mathcal{M})$ & $\operatorname{rank}(g)$ & $\operatorname{rank}(\omega)$ & $\operatorname{rank}(\theta)$ \\ \hline
\multicolumn{1}{l|}{$N=2$} & $\beta=1$   & 2                                 & 2                        & 2                             & 2                             \\ \cline{2-6} 
\multicolumn{1}{l|}{}      & $\beta<1$ & 2                                 & 2                        & 2                             & 2                             \\ \cline{2-6} 
\multicolumn{1}{l|}{}      & $\beta=-\infty$   & 2                                 & 2                        & 0                             & 0                             \\ \hline
\multicolumn{1}{l|}{$N>2$} & $\beta=1$   & 2                                 & 2                        & 2                             & 2                             \\ \cline{2-6} 
\multicolumn{1}{l|}{}      & $\beta<1$ & 3                                 & 3                        & 2                             & 2                             \\ \cline{2-6} 
\multicolumn{1}{l|}{}      & $\beta=-\infty$   & 3                                 & 3                        & 2                             & 2                            
\end{tabular}
\centering
\caption{Overview of the dimensions and ranks in different scenarios}
\label{table1R}
\end{table}

\begin{figure}[H]
\centering
\begin{minipage}{.3\textwidth}
  \centering
  \includegraphics[height=.7\linewidth]{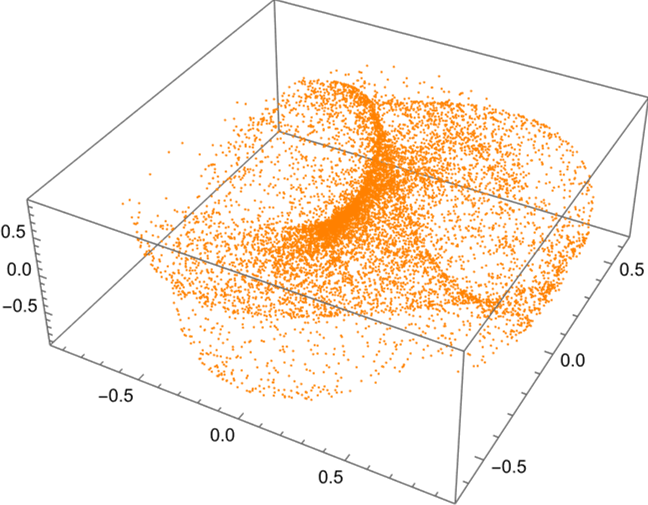}
  \includegraphics[height=.7\linewidth]{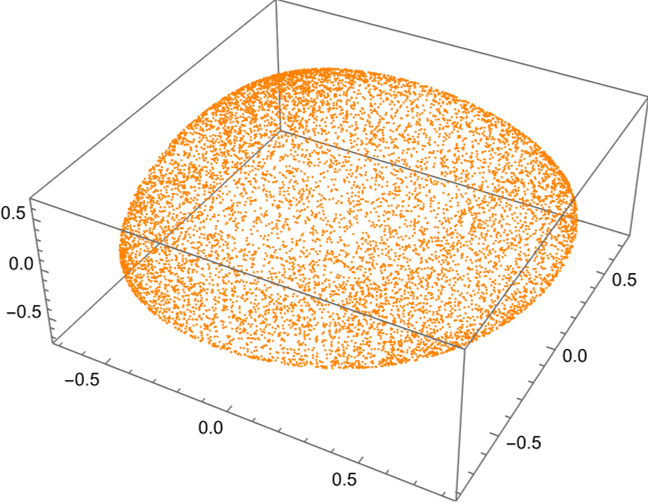}
\end{minipage}%
\begin{minipage}{.3\textwidth}
  \centering
  \includegraphics[height=.7\linewidth]{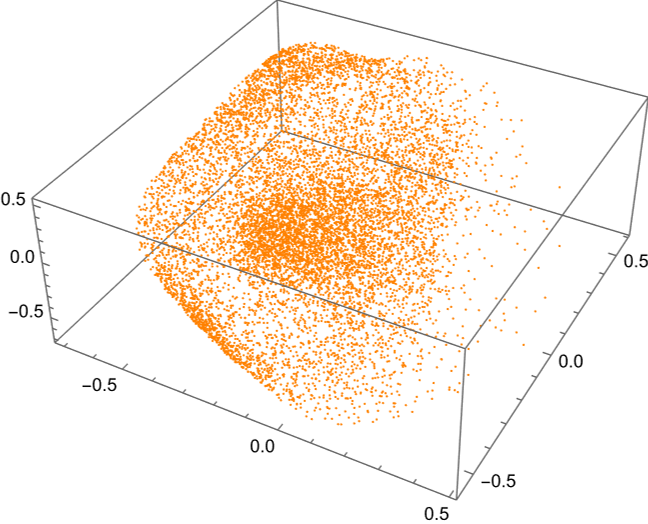}
  \includegraphics[height=.7\linewidth]{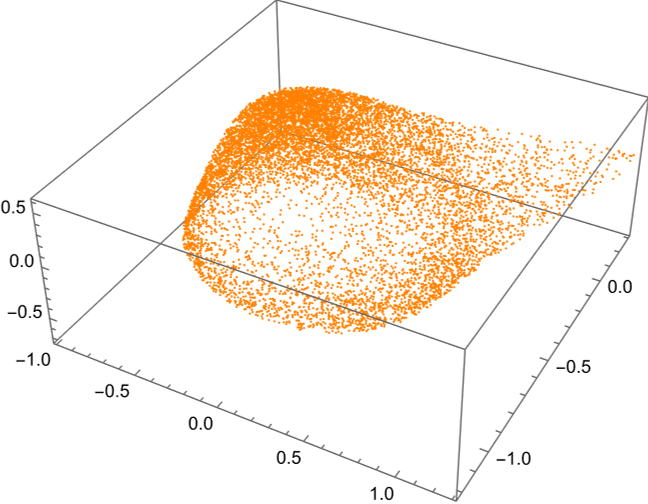}
\end{minipage}%
\caption{Random points in the perturbed fuzzy sphere. Left: $\beta=0.9$, right: $\beta=-\infty$; top: projective plot of $\mathcal{M}$, bottom: projective plot of $\Tilde{\mathcal{M}}$}
\label{fig:SU2R/2}
\end{figure}

In the random cases we further find (as the only nontrivial result) that the kernels of $\omega_{ab}$ and $\theta^{ab}$ do not agree.\\
In figure \ref{fig:SU2R/2} (that should be compared to figure \ref{fig:SU2/2} for the squashed fuzzy sphere) we see plots of random points for $\beta=0.9,-\infty$. In the case of $\beta=0.9$ we can still recognize the perturbed shape of the round sphere, while for $-\infty$ we can hardly see any geometry.

\begin{figure}[H]
\centering
\begin{minipage}{.3\textwidth}
  \centering
  \includegraphics[height=.7\linewidth]{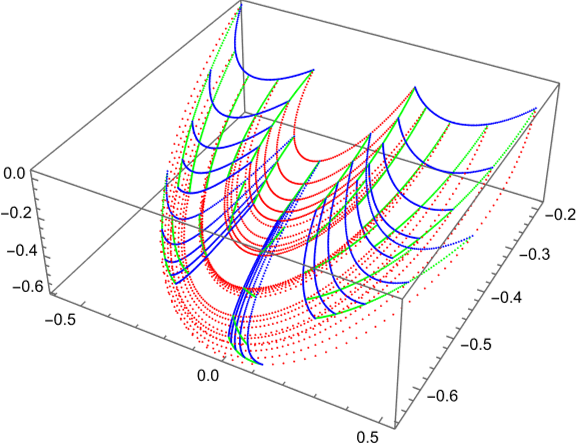}
  \includegraphics[height=.7\linewidth]{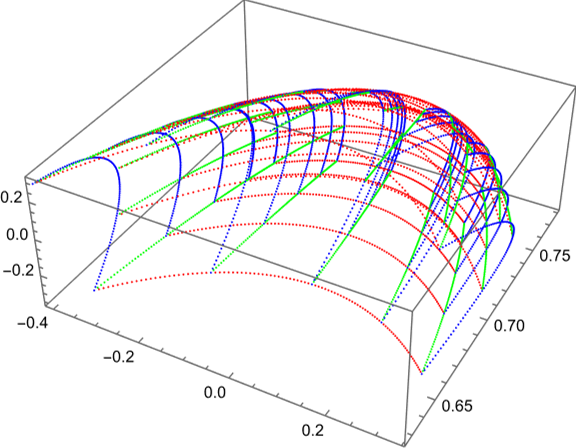}
\end{minipage}%
\begin{minipage}{.3\textwidth}
  \centering
  \includegraphics[height=.7\linewidth]{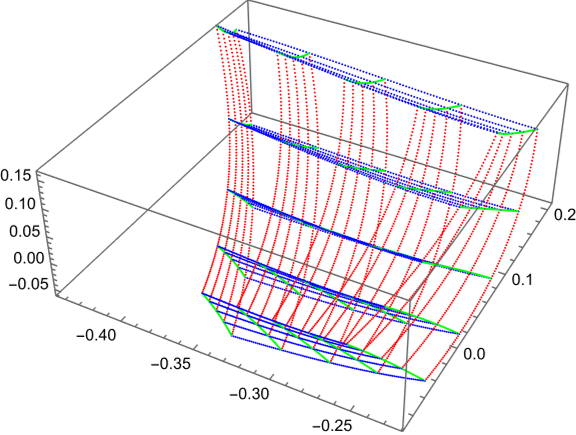}
  \includegraphics[height=.7\linewidth]{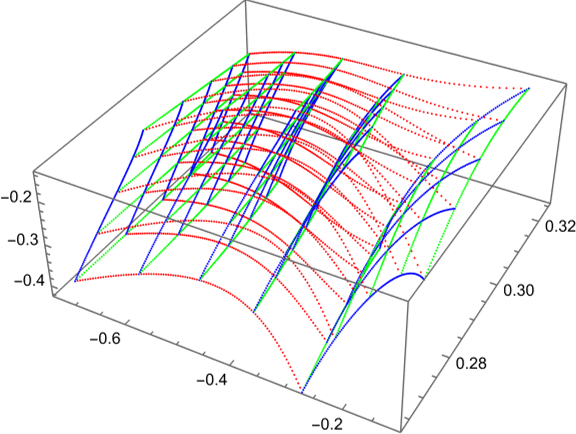}
\end{minipage}%
\caption{Cartesian coordinate lines around $(0,1,0)$ in the perturbed fuzzy sphere. Left: $\beta=0.9$, right: $\beta=-\infty$; top: projective plot of $\mathcal{M}$, bottom: projective plot of $\Tilde{\mathcal{M}}$}
\label{fig:SU2R/3}
\end{figure}

Figure \ref{fig:SU2R/3} (here, we should compare to figure \ref{fig:SU2/3}) shows Cartesian coordinate lines. We note that even in the completely random case $\beta=-\infty$, the coordinate lines look smooth, while for $\beta=0.9$ we can still recognize the shape of the round case. For both choices it is evident that $\mathcal{M}$ is three dimensional.

\subsubsection{Foliations and Integration}

\begin{figure}[H]
\centering
\begin{minipage}{.3\textwidth}
  \centering
  \includegraphics[height=.7\linewidth]{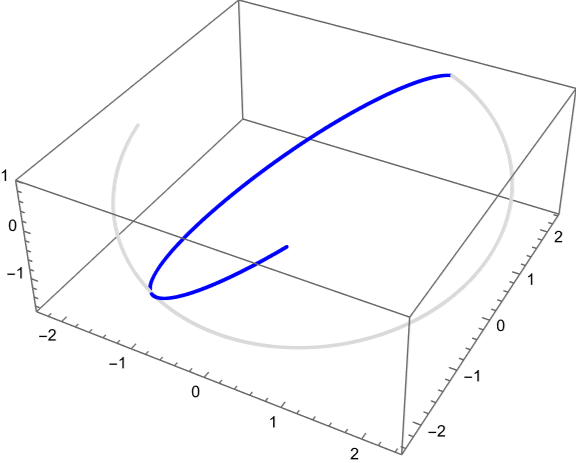}
  \includegraphics[height=.7\linewidth]{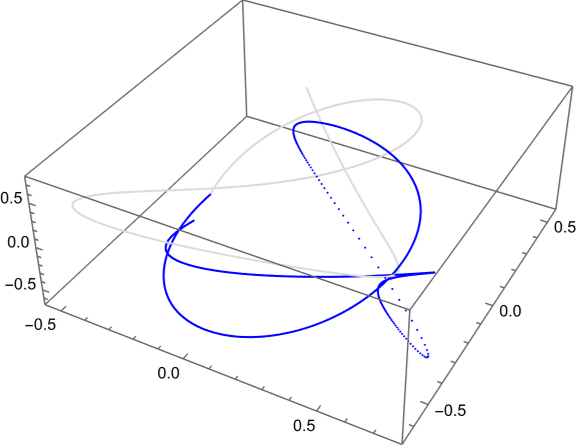}
\end{minipage}%
\begin{minipage}{.3\textwidth}
  \centering
  \includegraphics[height=.7\linewidth]{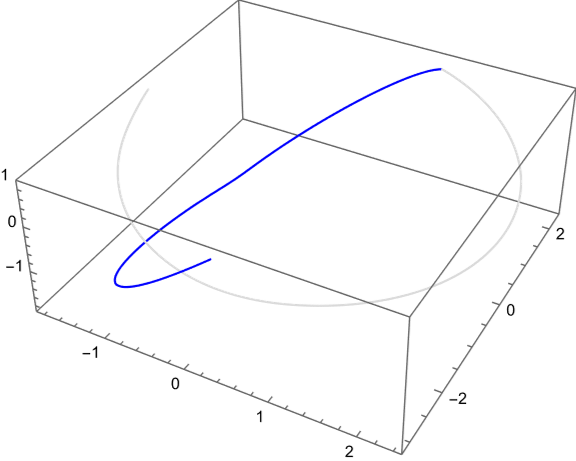}
  \includegraphics[height=.7\linewidth]{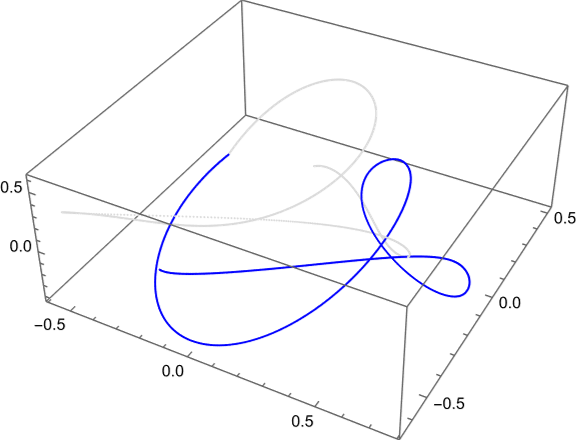}
\end{minipage}%
\caption{Curves in the hybrid leaf starting at $x=(1,2,1)$ with adapted tangent vectors $v_1=(0,1,0)$ and $v_2=(1,0,0)$ in the perturbed fuzzy sphere for $N=4$. Top: $\beta=0.9$, bottom: $\beta=0.7$; left: projective plot of $\Tilde{\mathbb{R}}^D$, right: projective plot of $\mathcal{M}$}
\label{fig:SU2R/4}
\end{figure}

In this section, we want to look at foliations of $\mathcal{M}$, while keeping the hybrid leaf as default. Our first task is to check if the leaf is still integrable.\\
In figure \ref{fig:SU2R/4} we see two different curves in the hybrid leaf starting from the same initial point. For $\beta=0.9$ we find that the curves visually intersect away from the initial point both in $\Tilde{\mathbb{R}}^D$ and $\mathcal{M}$, while for $\beta=0.7$ it is clear that the curves miss each other in $\Tilde{\mathbb{R}}^D$ but still intersect\footnote{This can be seen much better directly in Mathematica when it is possible to adjust the viewpoint.} in $\mathcal{M}$.\\
Therefore we conclude that the hybrid leaf is still integrable, however only in $\mathcal{M}$ but no longer in $\Tilde{\mathbb{R}}^D$. So the integrability in $\Tilde{\mathbb{R}}^D$ for the squashed fuzzy sphere was only a consequence of the special geometry.

\begin{figure}[H]
\centering
\begin{minipage}{.3\textwidth}
  \centering
  \includegraphics[height=.7\linewidth]{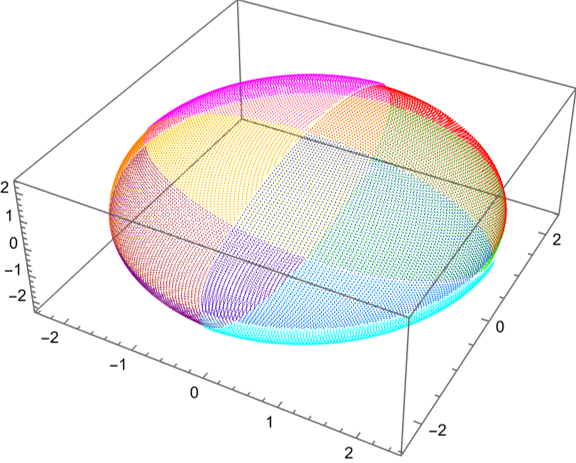}
\end{minipage}%
\begin{minipage}{.3\textwidth}
  \centering
  \includegraphics[height=.7\linewidth]{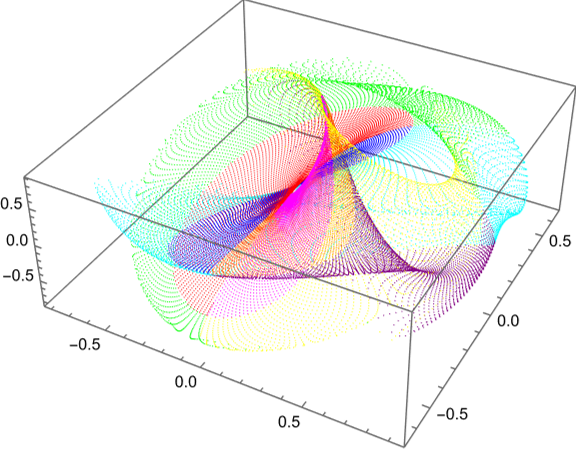}
\end{minipage}%
\begin{minipage}{.3\textwidth}
  \centering
  \includegraphics[height=.7\linewidth]{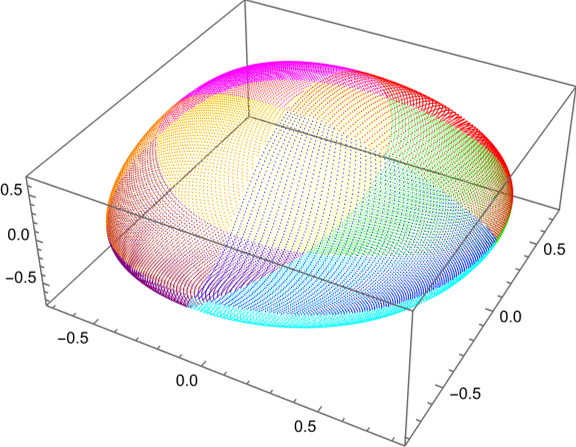}
\end{minipage}%
\caption{Tiling of the hybrid leaf through $x=(1,2,1)$ in the perturbed fuzzy sphere for $N=4$ and $\beta=0.9$. Left: projective plot of $\Tilde{\mathbb{R}}^D$, middle: projective plot of $\mathcal{M}$, right: projective plot of $\Tilde{\mathcal{M}}$}
\label{fig:SU2R/5PROTO}
\end{figure}

Knowing about the integrability of the leaf, we are ready to generate global coverings with coordinates. As our reference configuration we take $N=4$, $\beta=0.9$, $x=(1,2,1)$ and the hybrid leaf. Figure \ref{fig:SU2R/5PROTO} shows the plots in analogy to figure \ref{fig:SU2/9PROTO}. Here, we recognize the shape from figure \ref{fig:SU2R/3}.

Let us now discuss the dependence on the parameter $\beta$. Table \ref{RDepBeta} shows the most relevant quantities for $\beta=1,0.9,0.7$.\\
Of course, for $\beta=1$, we reproduce the results from section \ref{sfs_results} for $\alpha=1$, given in table \ref{DepAlpha}. Also for $\beta=0.9$, the results for the completeness relation are comparable to the respective result for $\alpha=0.9$, while the quality of the quantization of $\mathbf{x}^a$ is worse. The same holds for the compatibility of the Poisson structures and the Kähler properties. By decreasing $\beta$ further, the results impair even more.\\
In total, we conclude that the results are better for higher $\beta$, while the quantization of $\mathbf{x}^a$ works worse than for the squashed fuzzy sphere.

\begin{table}[H]
\centering
\begin{tabular}{l|l|ll|ll|l|lll}
$\beta$ & $V_\omega$ & $\sigma_{\mathbb{1}'}$ & $d_{\mathbb{1}'}$ & $d_{X'}$ & $n_{X'}$ & $d_{\{\}}$ & $c^2(V_R)$ & $c^2(V_L)$ & $c^2(V_K)$ \\\hline
$1$ & $10.043$ & $0.010$ & $0.0086$ & $0.015$ & $1.666$ & $\sim 10^{-9}$ & $\sim 10^{-8}$ & $\sim 10^{-8}$ & $\sim 10^{-8}$ \\
$0.9$ & $10.126$ & $0.019$ & $0.017$ & $0.095$ & $1.682$ & $0.084$ & $0.388$ & $0.262$ & $0.148$ \\
$0.7$ & $10.915$ & $0.072$ & $0.063$ & $0.236$ & $1.775$ & $0.487$ & $1.159$ & $0.713$ & $0.382$
\end{tabular}
\caption{Dependence of various quantities on the parameter $\beta$}
\label{RDepBeta}
\end{table}

\begin{figure}[H]
\centering
\begin{minipage}{.3\textwidth}
  \centering
  \includegraphics[height=.7\linewidth]{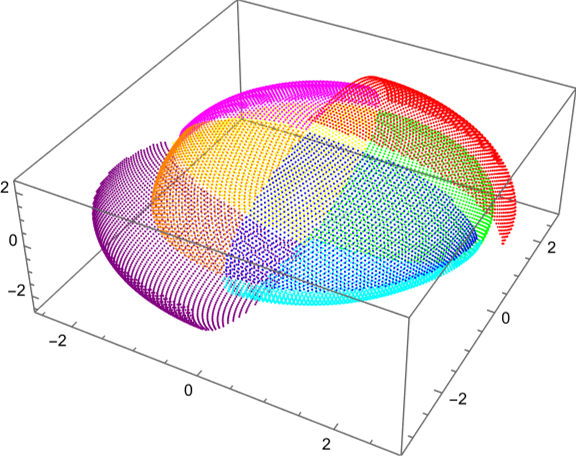}
\end{minipage}%
\begin{minipage}{.3\textwidth}
  \centering
  \includegraphics[height=.7\linewidth]{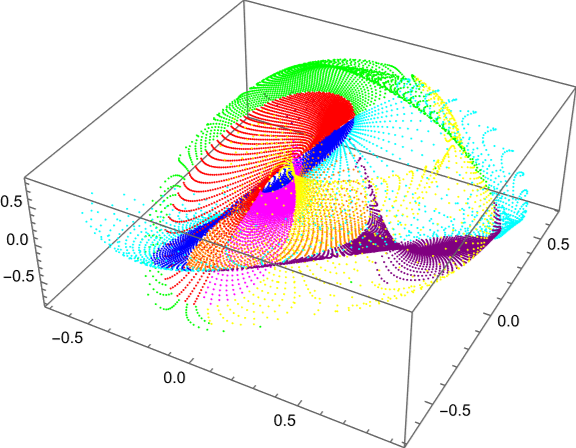}
\end{minipage}%
\begin{minipage}{.3\textwidth}
  \centering
  \includegraphics[height=.7\linewidth]{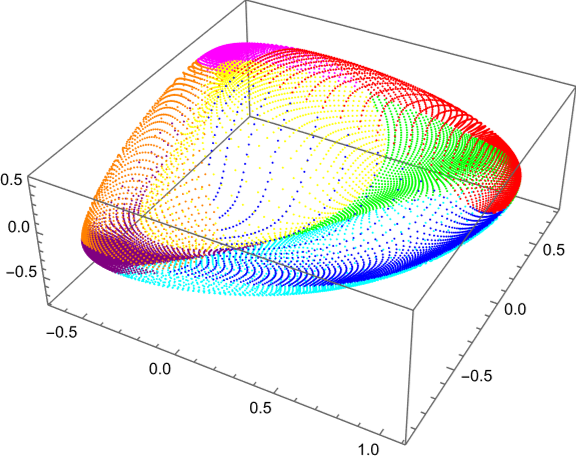}
\end{minipage}%
\caption{Tiling of the hybrid leaf through $x=(1,2,1)$ in the perturbed fuzzy sphere for $N=4$ and $\beta=0.7$. Left: projective plot of $\Tilde{\mathbb{R}}^D$, middle: projective plot of $\mathcal{M}$, right: projective plot of $\Tilde{\mathcal{M}}$}
\label{fig:SU2R/5beta}
\end{figure}

In figure \ref{fig:SU2R/5beta} we see a covering with coordinates for $\beta=0.7$. Here the deviation from the round sphere is already much stronger and it is plain to see that the leaf is not integrable in $\Tilde{\mathbb{R}}^D$. Also the shape in $\mathcal{M}$ is now far away from the well studied round case.

We now come to the dependence on $N$. In table \ref{RDepN}, the relevant quantities are listed for $N=4,10,20$.\\ 
We see that $V_\omega$ still follows the rule in equation (\ref{VN}). On the other hand, the results are getting much poorer already for $N=10$, while it is not foreseeable that they start to improve again for larger $N$.
\\
If we view the perturbations $R^a$ to the $\Bar{X}^a$ as gauge fields as proposed in \cite{Steinacker_2015}, the discussion in \cite{Steinacker_2021} suggests that we should only consider perturbations that are in some sense considered as \textit{almost local} (respectively in the regime we touched upon in section \ref{QuantMap}). For the fuzzy sphere that means to impose a cutoff on the allowed $SU(2)$ modes in the $R^a$ of order $\mathcal{O}(\sqrt{N})$ (or equivalently, only considering polynomials of order $\mathcal{O}(\sqrt{N})$ in the $\bar{X}^a$) \cite{Steinacker_2021}. Since by the discussion in section \ref{FuzzySphere0} random matrices will include all modes up to order $N-1$, it is no wonder that for large $N$ the results are far from optimal, yet it remains to check if a restriction to almost local random matrices improves the quality.

\begin{table}[H]
\centering
\begin{tabular}{l|l|ll|ll|l|lll}
$N$ & $V_\omega$ & $\sigma_{\mathbb{1}'}$ & $d_{\mathbb{1}'}$ & $d_{X'}$ & $n_{X'}$ & $d_{\{\}}$ & $c^2(V_R)$ & $c^2(V_L)$ & $c^2(V_K)$ \\\hline
$4$ & $10.126$ & $0.019$ & $0.017$ & $0.095$ & $1.682$ & $0.084$ & $0.388$ & $0.262$ & $0.148$ \\
$10$ & $30.870$ & $0.151$ & $0.143$ & $0.146$ & $1.233$ & $0.209$ & $1.147$ & $0.262$ & $0.311$ \\
$20$ & $57.395$ & $0.591$ & $0.576$ & $0.449$ & $0.988$ & $0.479$ & $0.742$ & $0.559$ & $0.509$
\end{tabular}
\caption{Dependence of various quantities on $N$}
\label{RDepN}
\end{table}

\begin{figure}[H]
\centering
\begin{minipage}{.3\textwidth}
  \centering
  \includegraphics[height=.7\linewidth]{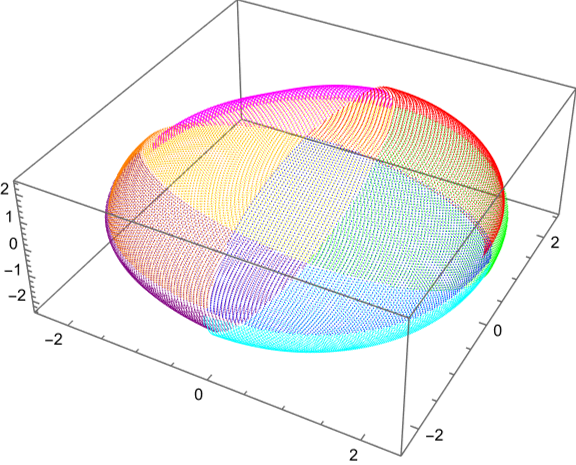}
  \includegraphics[height=.7\linewidth]{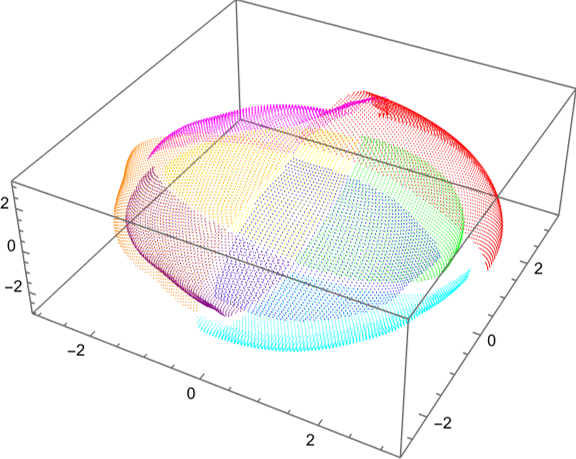}
\end{minipage}%
\begin{minipage}{.3\textwidth}
  \centering
  \includegraphics[height=.7\linewidth]{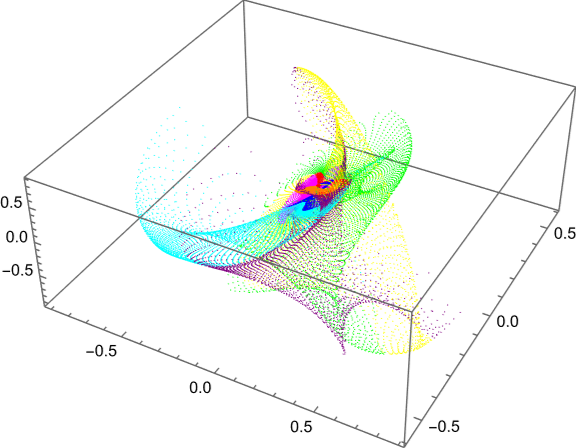}
  \includegraphics[height=.7\linewidth]{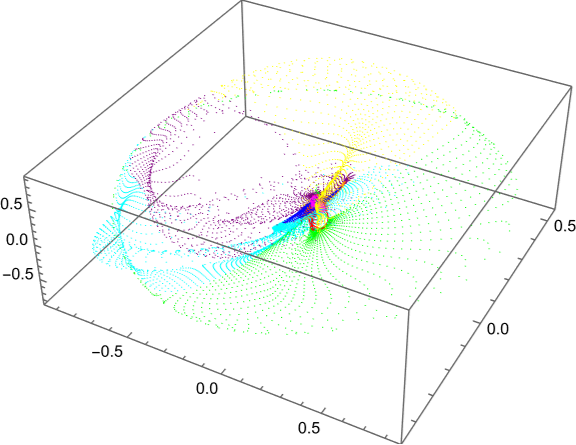}
\end{minipage}%
\begin{minipage}{.3\textwidth}
  \centering
  \includegraphics[height=.7\linewidth]{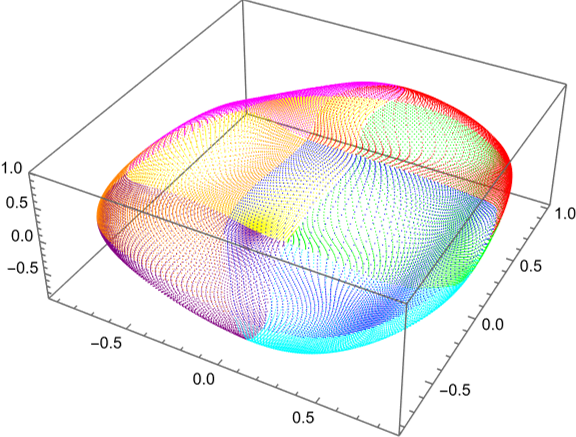}
  \includegraphics[height=.7\linewidth]{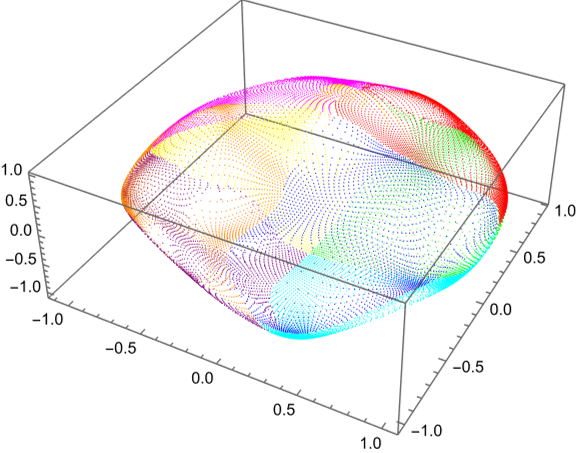}
\end{minipage}%
\caption{Tiling of the hybrid leaf through $x=(1,2,1)$ in the perturbed fuzzy sphere for $N=4$ and $\beta=0.9$. Top: $N=10$, bottom: $N=20$; left: projective plot of $\Tilde{\mathbb{R}}^D$, middle: projective plot of $\mathcal{M}$, right: projective plot of $\Tilde{\mathcal{M}}$}
\label{fig:SU2R/5N}
\end{figure}

In that context, increasing $N$ means that we add more and more modes to the gauge fields.
This behaviour can be seen very well in figure \ref{fig:SU2R/5N}. We can nicely observe how the \textit{degrees of freedom} grow with $N$, letting the shape become more and more complicated, changing on a shorter and shorter length scale.
\\
The plot for $N=20$ further makes it crystal clear that the leaf is no longer integrable in $\Tilde{\mathbb{R}}^D$. 

Now, we discuss the dependence on $x$. In table \ref{RDepX}, the relevant results are shown for $x=100\cdot x_0,x_0,0.1\cdot x_0,x_\lambda$, where $x_0=(1,2,1)$ and $x_\lambda=x'_0$ is the point in $\mathcal{N}^\mathcal{L}_{x_0}$ that minimizes $\lambda$.\\
At first we note that $x_\lambda=(0.339,0.651,0.278)\approx 0.320 \cdot x_0$ and there, the eigenvalues of $-2\omega_{ab}\theta^{bc}$ and $\partial_a\mathbf{x}^a$ are given by $1.127,1.217,0$ respectively $1.274,1.024,0.015$, which is approximately what we would expect from section \ref{QuantMap} (although the results are worse than the comparable results for the squashed fuzzy sphere).
\\
On the other hand, we observe that the results for the completeness relation and the quantization of the $\mathbf{x}^a$ are best for $x_0$, while the results for the compatibility of the different Poisson structures is better for $x_\lambda$. Yet for $0.1\cdot x_0$, the results are already very bad.\\
We conclude that the behaviour is fundamentally different from the squashed fuzzy sphere where we obtained the optimal results for larger and larger $\vert x\vert$, while here the optimal results can be found in the vicinity of $x_\lambda$ and $x_0$.

\begin{table}[H]
\centering
\begin{tabular}{l|l|ll|ll|l|lll}
$x$ & $V_\omega$ & $\sigma_{\mathbb{1}'}$ & $d_{\mathbb{1}'}$ & $d_{X'}$ & $n_{X'}$ & $d_{\{\}}$ & $c^2(V_R)$ & $c^2(V_L)$ & $c^2(V_K)$ \\\hline
$100\cdot x_0$ & $10.177$ & $0.038$ & $0.033$ & $0.087$ & $1.681$ & $0.286$ & $0.286$ & $0.286$ & $0.284$ \\
$x_0$ & $10.126$ & $0.019$ & $0.017$ & $0.095$ & $1.682$ & $0.084$ & $0.388$ & $0.262$ & $0.148$ \\
$0.1\cdot x_0$ & $9.958$ & $0.184$ & $0.202$ & $0.199$ & $1752$ & $0.126$ & $1.412$ & $0.161$ & $0.134$ \\
$x_\lambda$ & $10.091$ & $0.054$ & $0.047$ & $0.111$ & $1.689$ & $0.072$ & $0.669$ & $0.225$ & $0.145$
\end{tabular}
\caption{Dependence of various quantities on $x$}
\label{RDepX}
\end{table}

Let us finally discuss the dependence on the leaf. Table \ref{RDepLeaf} collects the most important results for the hybrid leaf (H), the hybrid leaf using $\omega$ (O) and the hybrid leaf using $\omega$ and $g$ (G).\\
We see that the results are almost identical for H and O, while they are much worse for G. This is curious since for the squashed fuzzy sphere the latter method produced the best results. Yet, the explanation is simple: For G the constructed coordinates have a much smaller range where they \textit{behave well} (i.e. there are no self-intersections or strong accumulations), thus providing a covering with global coordinates is much harder and more susceptible for numerical errors. The same happens for O when $N$ grows large, while the coordinates obtained by H are still well behaved.

\begin{table}[H]
\centering
\begin{tabular}{l|l|ll|ll|l|lll}
L & $V_\omega$ & $\sigma_{\mathbb{1}'}$ & $d_{\mathbb{1}'}$ & $d_{X'}$ & $n_{X'}$ & $d_{\{\}}$ & $c^2(V_R)$ & $c^2(V_L)$ & $c^2(V_K)$ \\\hline
H & $10.126$ & $0.019$ & $0.017$ & $0.095$ & $1.682$ & $0.084$ & $0.388$ & $0.262$ & $0.148$ \\
O & $10.100$ & $0.018$ & $0.016$ & $0.094$ & $1.682$ & $0.084$ & $0.388$ & $0.262$ & $0.148$ \\
G & $10.367$ & $0.110$ & $0.095$ & $0.179$ & $1.669$ & $0.084$ & $0.388$ & $0.262$ & $0.148$
\end{tabular}
\caption{Dependence of various quantities on the leaf}
\label{RDepLeaf}
\end{table}

\begin{figure}[H]
\centering
\begin{minipage}{.3\textwidth}
  \centering
  \includegraphics[height=.7\linewidth]{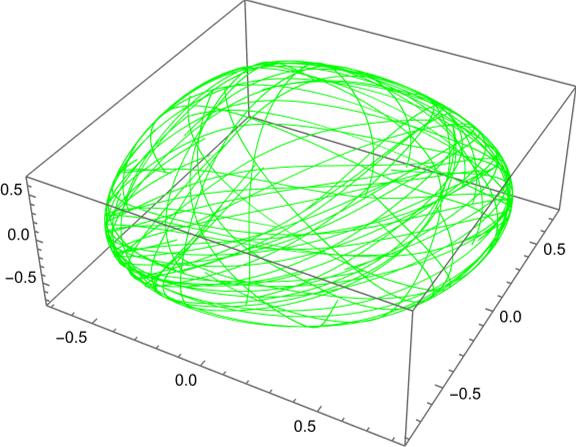}
  \includegraphics[height=.7\linewidth]{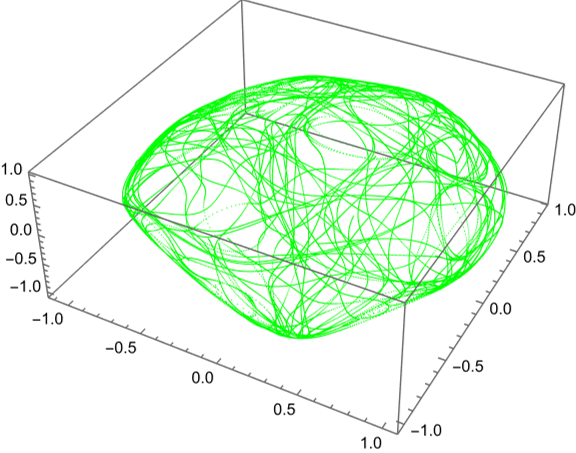}
\end{minipage}%
\begin{minipage}{.3\textwidth}
  \centering
  \includegraphics[height=.7\linewidth]{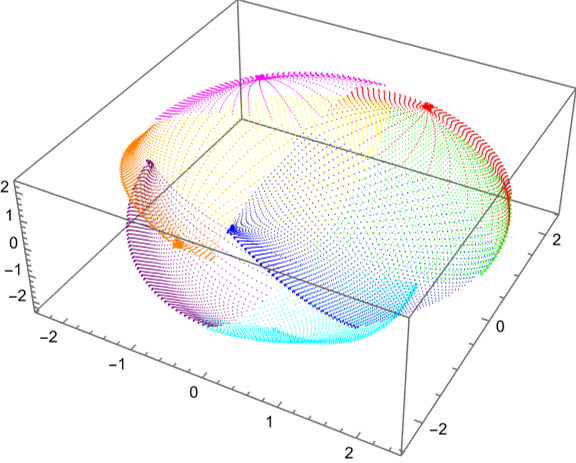}
  \includegraphics[height=.7\linewidth]{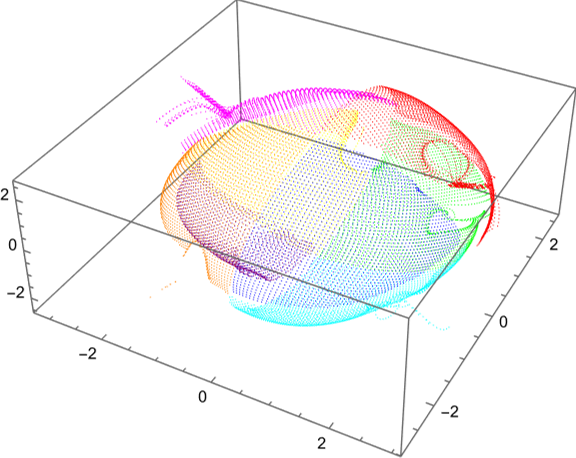}
\end{minipage}%
\caption{The perturbed fuzzy sphere for $\beta=0.9$. Top: hybrid leaf using $\omega$ and $g$ for $N=4$, bottom: hybrid leaf using $\omega$ for $N=20$; left: projective plot of $\Tilde{\mathcal{M}}$ for a scan through $x=(1,2,1)$, right: projective plot of $\Tilde{\mathbb{R}}^D$ for a tiling through $x=(1,2,1)$}
\label{fig:SU2R/5COORDS}
\end{figure}

In figure \ref{fig:SU2R/5COORDS} we see what can possibly go wrong when constructing coordinates for the leaves. While the hybrid leaf using $\omega$ is more resilient than the hybrid leaf using $\omega$ and $g$ (in the first case we have to go to $N=20$ until we find such problems, but they are present already for $N=4$ in the second case), for both leaves we encounter situations where the coordinates intersect or accumulate.\\
On the other hand, the projective plots of $\Tilde{\mathcal{M}}$ suggest that in principle it should be possible to pick a finer tiling to circumvent the problems, meaning we pick more but smaller coordinate charts.

\begin{figure}[H]
\centering
\begin{minipage}{.3\textwidth}
  \centering
  \includegraphics[height=.7\linewidth]{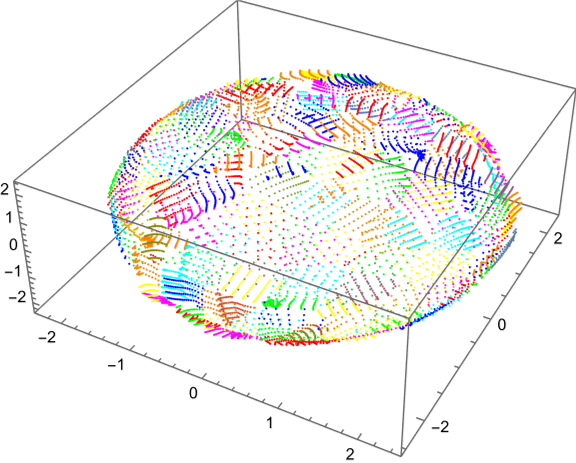}
\end{minipage}%
\begin{minipage}{.3\textwidth}
  \centering
  \includegraphics[height=.7\linewidth]{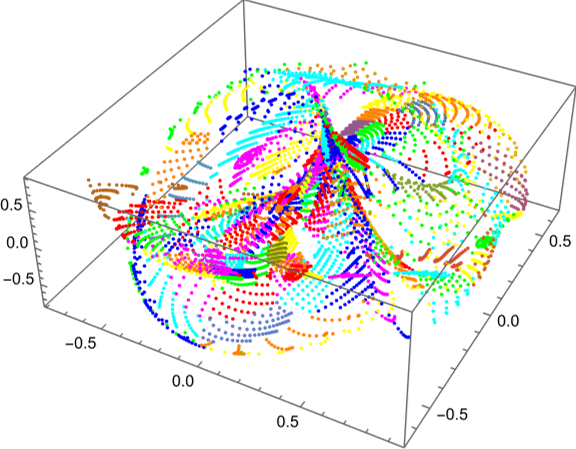}
\end{minipage}%
\begin{minipage}{.3\textwidth}
  \centering
  \includegraphics[height=.7\linewidth]{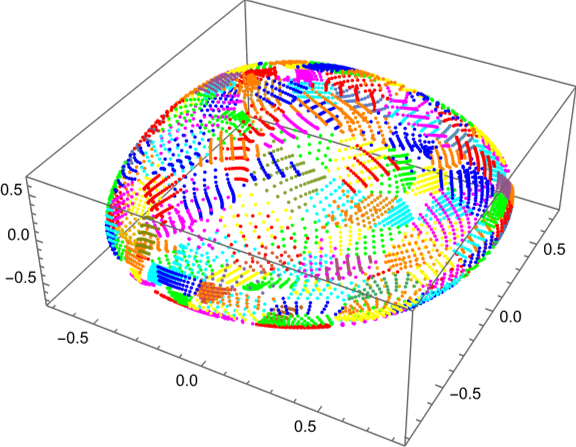}
\end{minipage}%
\caption{Tiling of the hybrid leaf using $\omega$ and $g$ through $x=(1,2,1)$ in the perturbed fuzzy sphere for $N=4$ and $\beta=0.9$, using a refined tiling. Left: projective plot of $\Tilde{\mathbb{R}}^D$, middle: projective plot of $\mathcal{M}$, right: projective plot of $\Tilde{\mathcal{M}}$}
\label{fig:SU2R/5gREFINED}
\end{figure}

Such refined charts are shown in figure \ref{fig:SU2R/5gREFINED}. These have been used to calculate the results in table \ref{RDepLeaf}. It is no wonder that there is more room for numerical errors. Yet, it is in principle possible to improve the results by constructing coordinates with shorter step lengths and more coordinate points.

\subsection{The Squashed Fuzzy \texorpdfstring{$\mathbb{C}P^2$}{CP2}}
\label{sfc}

Until now, we have only studied matrix configurations for $D=3$. Thus, it is time to look at a higher dimensional example, namely the squashed fuzzy $\mathbb{C}P^2$ with $D=8$. The round case has been discussed in section \ref{fuzzycp2}.

The general squashed fuzzy $\mathbb{C}P^2$ is then defined as the matrix configuration
\begin{align}
    \mathbb{C}P^2_{n,\alpha_a}=(\alpha_1 \bar{X}^1,\alpha_2\bar{X}^2,\alpha_3 \bar{X}^3,\alpha_4\bar{X}^4,\alpha_5\bar{X}^5,\alpha_6\bar{X}^6,\alpha_7\bar{X}^7,\alpha_8 \bar{X}^8)=:(X^a)
\end{align}
(where the $\bar{X}^a$ are the matrices from the round fuzzy $\mathbb{C}P^2$),
for arbitrary parameters $\alpha_a\geq 0$, while we mostly restrict ourselves to the special case
\begin{align}
    \mathbb{C}P^2_{n,\alpha}=(\bar{X}^1,\bar{X}^2,\alpha \bar{X}^3,\bar{X}^4,\bar{X}^5,\bar{X}^6,\bar{X}^7,\alpha \bar{X}^8),
\end{align}
for a single $\alpha\geq 0$, where we only modify the Cartan generators $\bar{X}^3$ and $\bar{X}^8$ as we have done for the squashed fuzzy sphere.

\subsubsection{First Results and Dimensional Aspects}

We start with an incomplete discussion of $\Tilde{\mathbb{R}}^D$.
In section \ref{fuzzycp2} we have seen that in the round case for $n=1$ all points in $\mathbb{R}^+_0\hat{e}_8$ lie within $\mathcal{K}$. This remains true for all $n$ and for all $0\leq\alpha\leq1$, but not for random $\alpha_a$.

In table \ref{table2}, the dimension of $\mathcal{M}$ and the ranks of $g_{ab}$, $\omega_{ab}$ and $\theta^{ab}$ are given for randomly chosen points.
In the round case, we find that the dimension as well as all ranks are four for all $n$. Further, the kernels of $T_xq$, $g_{ab}$ and $\omega_{ab}$ agree, while the kernel of $\theta^{ab}$ discerns in general.\\
When squashing, we have to treat the case $n=1$ (the fundamental representation) separately. This is analogous to the special case $N=2$ for the squashed fuzzy sphere. There, we see the same behaviour as in the round case.\\
Yet for $n>1$ the behaviour is different. Here the dimension of $\mathcal{M}$ and the ranks of $g_{ab}$ and $\omega_{ab}$ jump to eight, while the rank of $\theta^{ab}$ goes to six.

\begin{table}[H]
\begin{tabular}{ll|l|l|l|l}
                           &              & $\operatorname{dim}(\mathcal{M})$ & $\operatorname{rank}(g)$ & $\operatorname{rank}(\omega)$ & $\operatorname{rank}(\theta)$ \\ \hline
\multicolumn{1}{l|}{$n=1$} & $\alpha=1$   & 4                                 & 4                        & 4                             & 4                             \\ \cline{2-6} 
\multicolumn{1}{l|}{}      & $0<\alpha<1$ & 4                                 & 4                        & 4                             & 4                             \\ \cline{2-6} 
\multicolumn{1}{l|}{}      & $\alpha=0$   & 4                                 & 4                        & 4                             & 4                             \\ \cline{2-6}
\multicolumn{1}{l|}{}      & random $\alpha_a$   & 4                                 & 4                        & 4                             & 4                             \\ \hline
\multicolumn{1}{l|}{n>1} & $\alpha=1$   & 4                                 & 4                        & 4                             & 4                             \\ \cline{2-6} 
\multicolumn{1}{l|}{}      & $0<\alpha<1$ & 8                                 & 8                        & 8                             & 6                             \\ \cline{2-6} 
\multicolumn{1}{l|}{}      & $\alpha=0$   & 6                                 & 6                        & 6                             & 6 \\\cline{2-6} 
\multicolumn{1}{l|}{}      & random $\alpha_a$   & 8                                 & 8                        & 8                             & 6     
\end{tabular}
\centering
\caption{Overview of the dimensions and ranks in different scenarios}
\label{table2}
\end{table}

Now we consider the special point $x=(0,1,0,0,0,0,0,0)$. Table \ref{table3} shows the values analogous to the ones in table \ref{table2} for $x$. Here, we see a completely different behaviour compared to randomly chosen points: In the round case nothing changes, but for $n>1$ and $0\leq\alpha<1$ the dimension of $\mathcal{M}$, $g_{ab}$ and $\omega_{ab}$ are reduced, while for $\theta^{ab}$ this only holds true for $\alpha=0$. Considering random $\alpha_a$, only the rank of $\omega_{ab}$ is reduced.\\
Thus we conclude that $x\notin\hat{\mathbb{R}}^D$ for $0\leq \alpha \leq1$ but $x\in\hat{\mathbb{R}}^D$ for $\alpha=1$ and random $\alpha_a$. So, this behaviour very much compares to the behaviour at the special point $(0,0,1)$ for the squashed fuzzy sphere.

\begin{table}[H]
\begin{tabular}{ll|l|l|l|l}
                           &              & $\operatorname{dim}(\mathcal{M})$ & $\operatorname{rank}(g)$ & $\operatorname{rank}(\omega)$ & $\operatorname{rank}(\theta)$ \\ \hline
\multicolumn{1}{l|}{$n=1$} & $\alpha=1$   & 4                                 & 4                        & 4                             & 4                             \\ \cline{2-6} 
\multicolumn{1}{l|}{}      & $0<\alpha<1$ & 4                                 & 4                        & 4                             & 4                             \\ \cline{2-6} 
\multicolumn{1}{l|}{}      & $\alpha=0$   & 3                                 & 3                        & 2                             & 2                             \\ \cline{2-6}
\multicolumn{1}{l|}{}      & random $\alpha_a$   & 4                                 & 4                        & 4                             & 4                             \\ \hline
\multicolumn{1}{l|}{n>1} & $\alpha=1$   & 4                                 & 4                        & 4                             & 4                             \\ \cline{2-6} 
\multicolumn{1}{l|}{}      & $0<\alpha<1$ & 7                                 & 7                        & 6                             & 6                             \\ \cline{2-6} 
\multicolumn{1}{l|}{}      & $\alpha=0$   & 6                                 & 6                        & 4                             & 4 \\\cline{2-6} 
\multicolumn{1}{l|}{}      & random $\alpha_a$   & 8                                 & 8                        & 6                             & 6     
\end{tabular}
\centering
\caption{Overview of the dimensions and ranks in different scenarios for $x=(0,1,0,0,0,0,0,0)$}
\label{table3}
\end{table}

Coming back to random points in $\Tilde{\mathbb{R}}^D$, in the squashed cases the behaviour of the dimension of $\mathcal{M}$ and the ranks of $g_{ab}$ and $\omega_{ab}$ can very well be explained by the discussion of section \ref{random}, assuming that the matrix configuration behaves randomly itself, yet this does not explain why the rank of $\theta^{ab}$ is bounded by six in all cases.
To solve this riddle, we look at a group theoretical explanation.\\
By definition, we have
\begin{align}
    \theta^{ab}=\frac{1}{i}\bra{x}[\alpha_a\bar{X}^a,\alpha_b \bar{X}^b]\ket{x}=\sum_c \alpha_a\alpha_b f^{abc}\bra{x}\bar{X}^c\ket{x},
\end{align}
where the so called \textit{structure constants} $f^{abc}$ are defined via $[\bar{X}^a,\bar{X}^b]=:\sum_c\frac{1}{C_n} if^{abc}\bar{X}^c$.\\
Since all $\bar{X}^a\in\mathfrak{su}(3)$, also the vector $\xi:=\sum_c\bra{x}\bar{X}^c\ket{x}\bar{X}^c\in\mathfrak{su}(3)$.\\
Let us now consider the adjoint representation of $\xi$, given by $\Xi:=\operatorname{ad}(\xi)$. Due to $\operatorname{ad}(\bar{X}^c)(\bar{X}^b)=[\bar{X}^c,\bar{X}^b]=\sum_a\frac{1}{C_n} if^{cba}\bar{X}^a$, we find the components of $(\Xi)^{ab}$ in the basis $\bar{X}^a$
\begin{align}
    (\Xi)^{ab}=\sum_c\bra{x}X^c\ket{x}\operatorname{ad}(\bar{X}^c)^{ab}=\sum_c\bra{x}\bar{X}^c\ket{x} \frac{1}{C_n}if^{cba}.
\end{align}
On the other hand, for sure $\xi$ is an eigenvector of $\Xi$ with eigenvalue zero -- we insert into the definition $\Xi(\xi)=\operatorname{ad}(\xi)(\xi)=[\xi,\xi]=0$ -- thus the rank of $\Xi$ is bounded by seven.\\
Using the complete antisymmetry of the $f^{abc}$ for $\mathfrak{su}(3)$ and introducing the matrix $P=(\alpha_a\delta^{ab})$, we find
\begin{align}
    \theta^{ab}=-C_n (P\cdot\Xi\cdot P)^{ab}.
\end{align}
But since the rank of $\Xi$ is bounded by seven, the same holds for $\theta^{ab}$, as multiplication with $P$ from the left and right does not increase the rank. 
By antisymmetry, this means that the rank of $\theta^{ab}$ is bounded by six.

\begin{figure}[H]
\centering
\begin{minipage}{.24\textwidth}
  \centering
  \includegraphics[height=.7\linewidth]{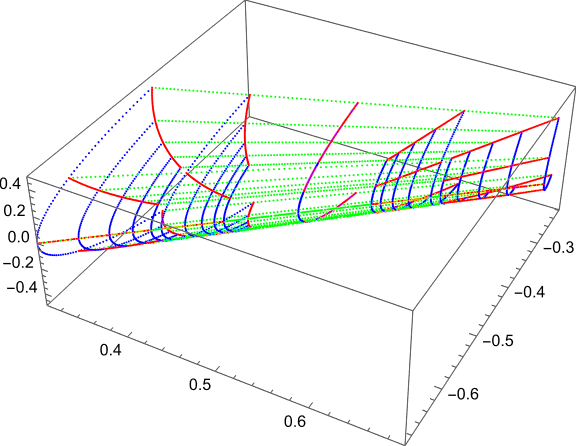}
  \includegraphics[height=.7\linewidth]{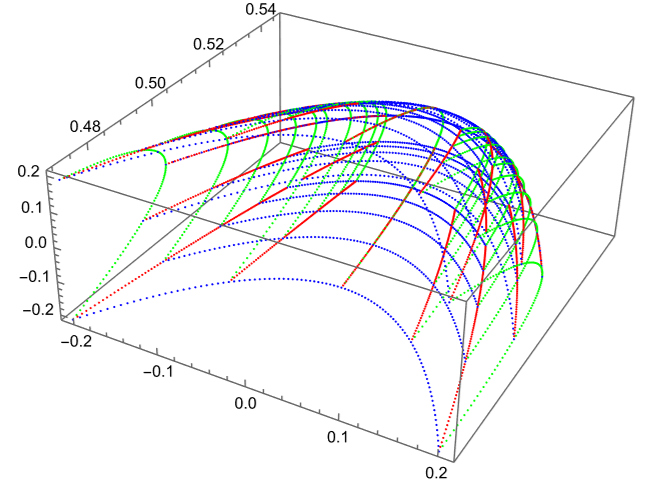}
\end{minipage}%
\begin{minipage}{.24\textwidth}
  \centering
  \includegraphics[height=.7\linewidth]{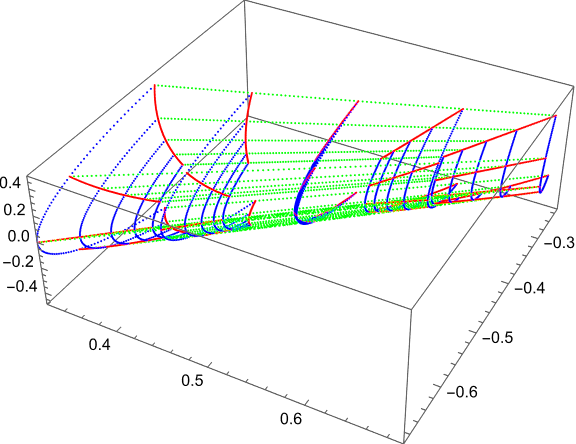}
  \includegraphics[height=.7\linewidth]{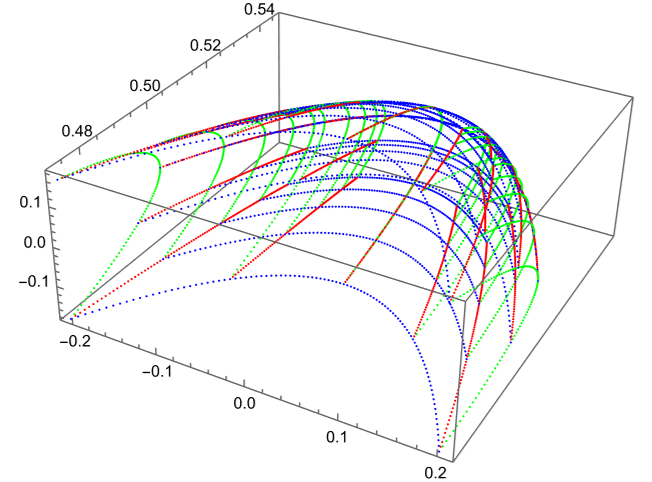}
\end{minipage}%
\begin{minipage}{.24\textwidth}
  \centering
  \includegraphics[height=.7\linewidth]{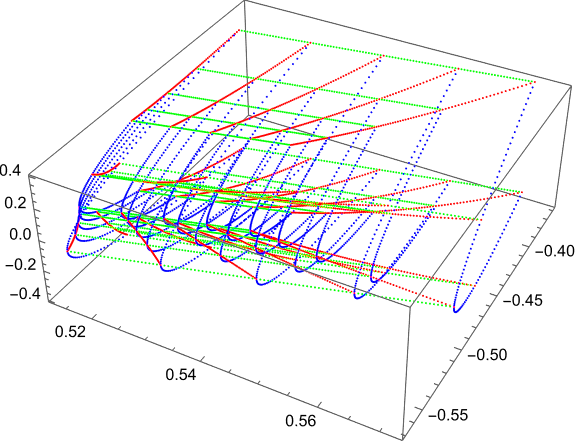}
  \includegraphics[height=.7\linewidth]{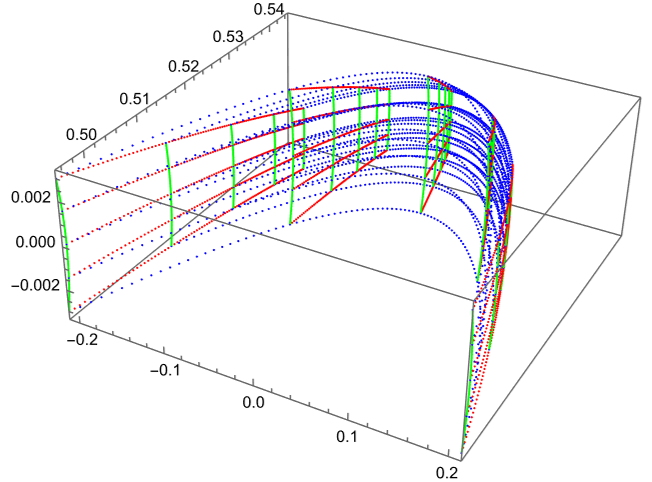}
\end{minipage}%
\begin{minipage}{.24\textwidth}
  \centering
  \includegraphics[height=.7\linewidth]{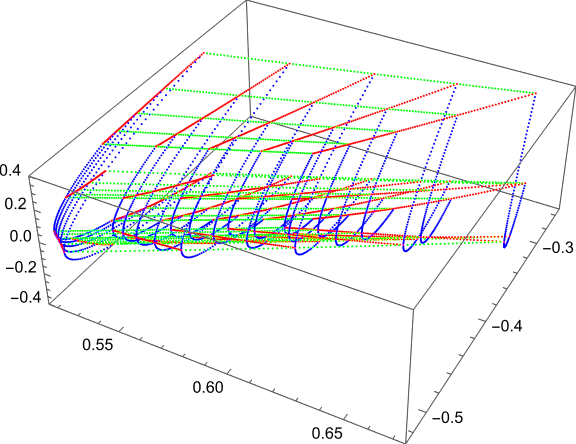}
  \includegraphics[height=.7\linewidth]{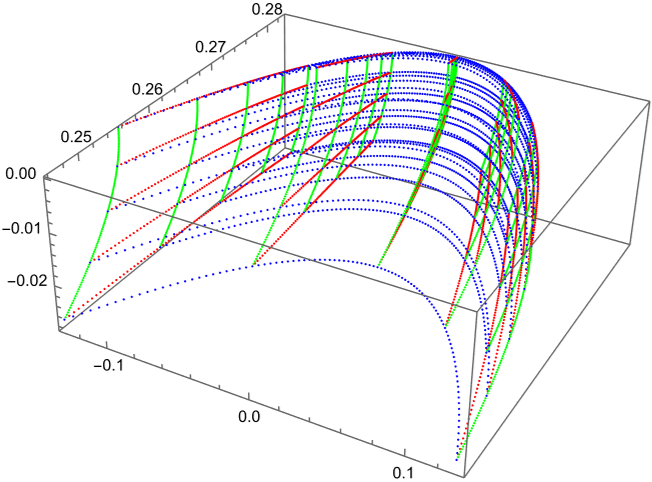}
\end{minipage}%
\caption{Cartesian coordinate lines in the directions $\hat{e}_1,\hat{e}_2,\hat{e}_3$ around $(0,1,0,0,0,0,0,0)$ in the squashed fuzzy $\mathbb{C}P^2$ for $n=2$. Left to right: $\alpha=1,\alpha=0.9,\alpha=0.1,\text{random }\alpha_i$; top: projective plot of $\mathcal{M}$, bottom: projective plot of $\Tilde{\mathcal{M}}$}
\label{fig:SU3/2}
\end{figure}

In figure \ref{fig:SU3/2} we can see plots of $\mathcal{M}$ and $\Tilde{\mathcal{M}}$ that show Cartesian coordinate lines in the directions $\hat{e}_1$, $\hat{e}_2$ and $\hat{e}_3$ in $\Tilde{\mathbb{R}}^D$ for $n=2$.
\\
If we compare the plots of $\Tilde{\mathcal{M}}$ to the corresponding plots in figure \ref{fig:SU2/3}, we can see that some of the structure of the squashed fuzzy sphere is preserved in the squashed fuzzy $\mathbb{C}P^2$.
\\
Here we can also witness that whilst squashing the dimension of $\mathcal{M}$ grows larger.

\begin{figure}[H]
\centering
\begin{minipage}{.24\textwidth}
  \centering
  \includegraphics[height=.39\linewidth]{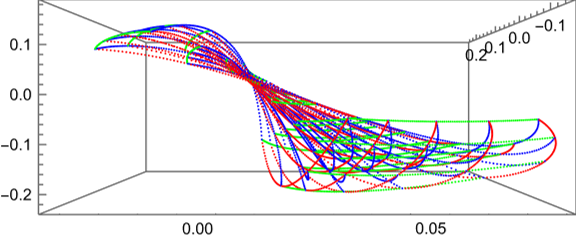}
  \includegraphics[height=.69\linewidth]{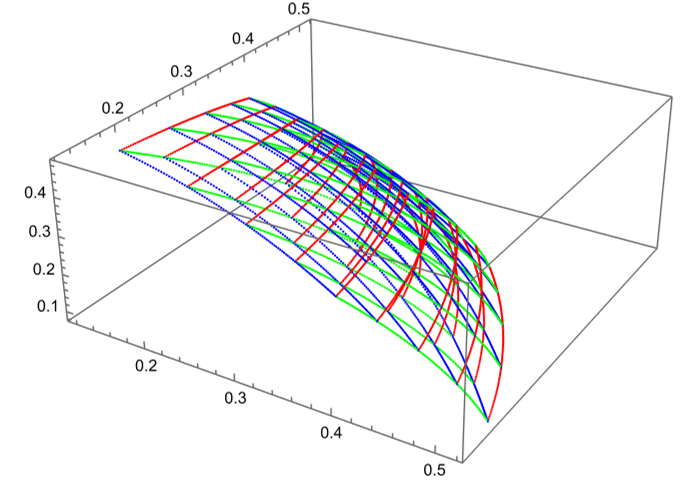}
\end{minipage}%
\begin{minipage}{.24\textwidth}
  \centering
  \includegraphics[height=.39\linewidth]{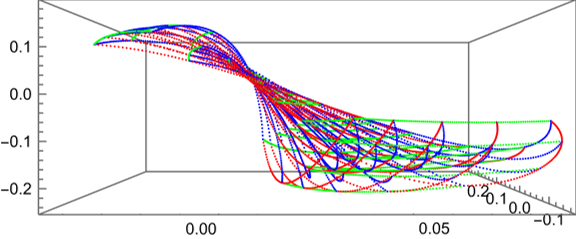}
  \includegraphics[height=.69\linewidth]{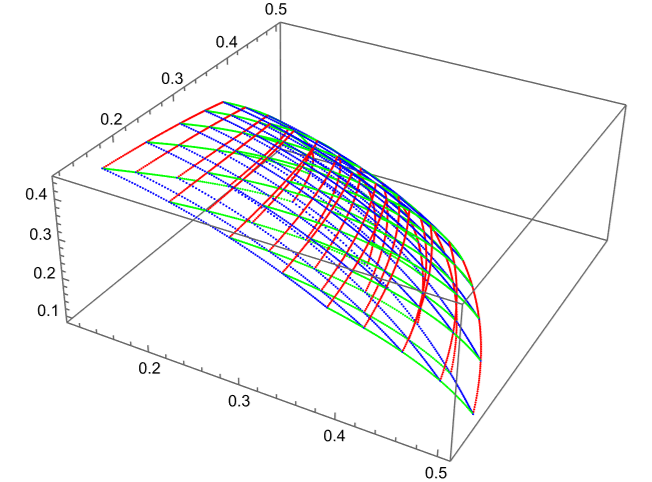}
\end{minipage}%
\begin{minipage}{.24\textwidth}
  \centering
  \includegraphics[height=.39\linewidth]{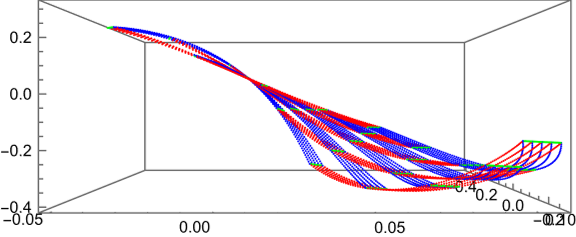}
  \includegraphics[height=.69\linewidth]{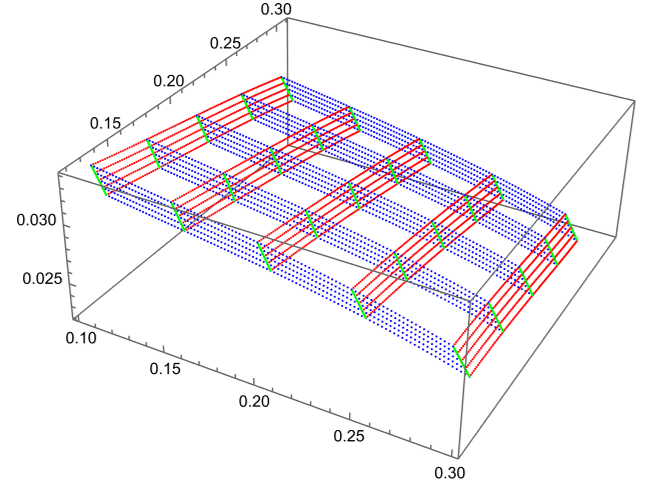}
\end{minipage}%
\begin{minipage}{.24\textwidth}
  \centering
  \includegraphics[height=.39\linewidth]{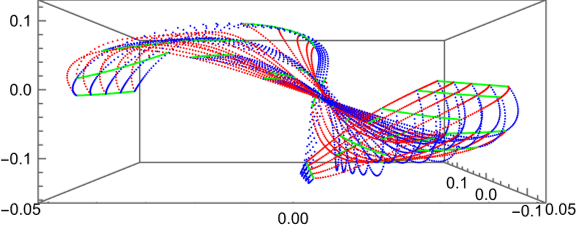}
  \includegraphics[height=.69\linewidth]{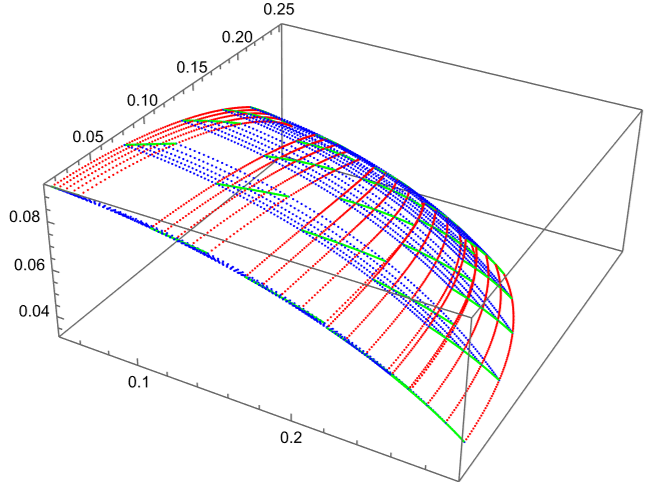}
\end{minipage}%
\caption{Cartesian coordinate lines in the directions $\hat{e}_1,\hat{e}_2,\hat{e}_3$ around a random point in the squashed fuzzy $\mathbb{C}P^2$ for $n=5$. Left to right: $\alpha=1,\alpha=0.9,\alpha=0.1,\text{random }\alpha_i$; top: projective plot of $\mathcal{M}$, bottom: projective plot of $\Tilde{\mathcal{M}}$}
\label{fig:SU3/2A}
\end{figure}

Figure \ref{fig:SU3/2A} shows a similar plot for $n=5$ and Cartesian coordinate lines centered at a random point. Here, we can see very well how the large scale structure is preserved during squashing, while in detail there are significant changes taking place.

\subsubsection{A Global View}

\begin{figure}[H]
\centering
\begin{minipage}{.24\textwidth}
  \centering
  \includegraphics[height=.7\linewidth]{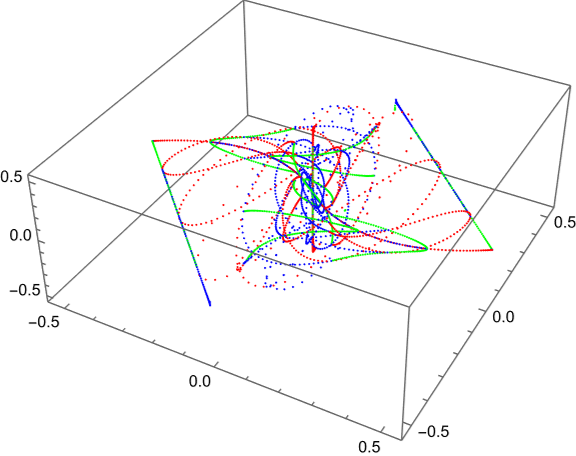}
  \includegraphics[height=.7\linewidth]{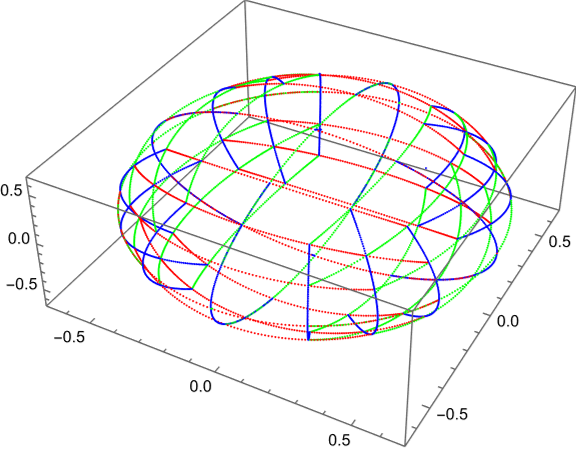}
\end{minipage}%
\begin{minipage}{.24\textwidth}
  \centering
  \includegraphics[height=.7\linewidth]{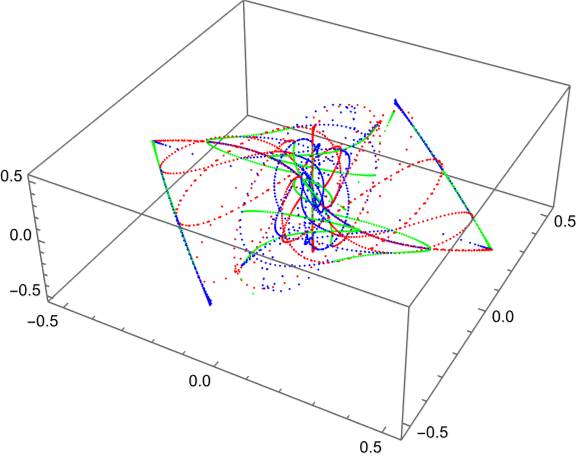}
  \includegraphics[height=.7\linewidth]{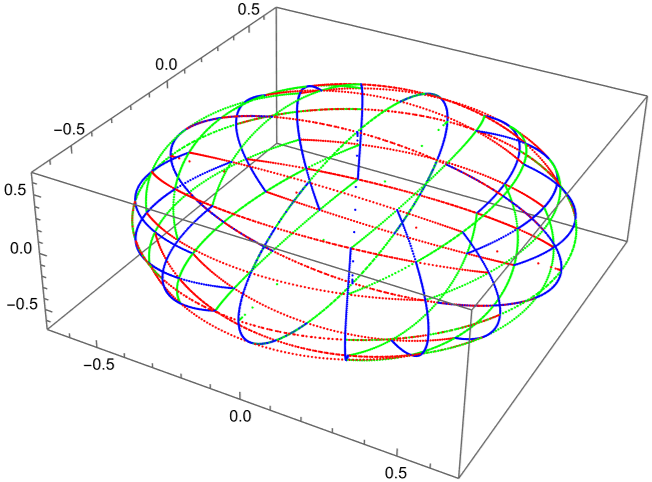}
\end{minipage}%
\begin{minipage}{.24\textwidth}
  \centering
  \includegraphics[height=.7\linewidth]{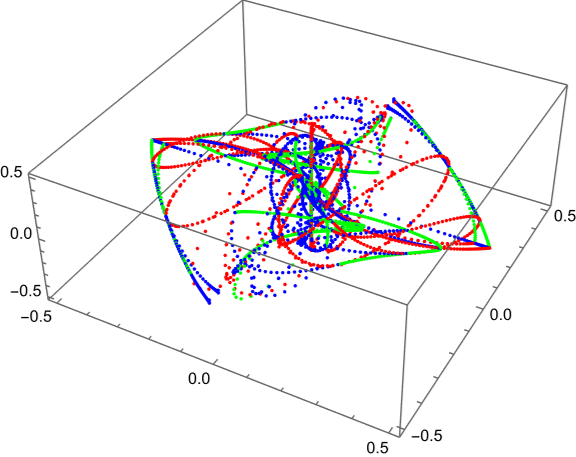}
  \includegraphics[height=.7\linewidth]{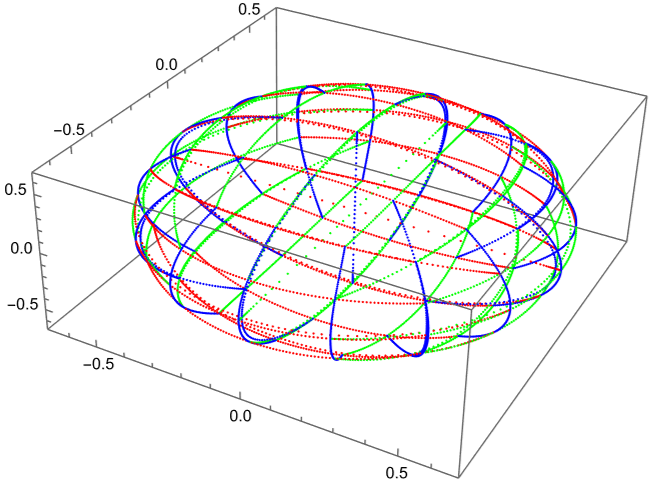}
\end{minipage}%
\begin{minipage}{.24\textwidth}
  \centering
  \includegraphics[height=.7\linewidth]{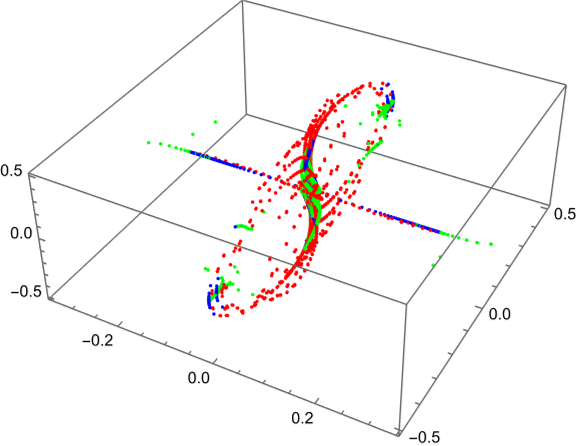}
  \includegraphics[height=.7\linewidth]{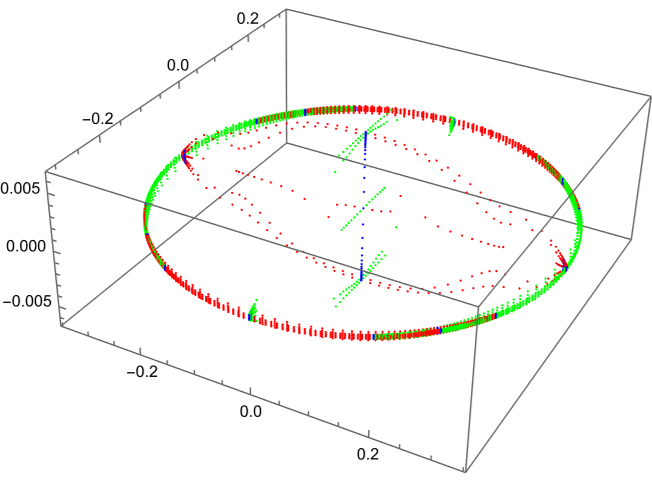}
\end{minipage}%
\caption{Large scale Cartesian coordinate lines in the directions $\hat{e}_2,\hat{e}_4,\hat{e}_6$ around a random point in the vicinity of $0$ in the squashed fuzzy $\mathbb{C}P^2$ for $n=5$. Left to right: $\alpha=1,\alpha=0.9,\alpha=0.1,\alpha_i=\text{random}$; top: projective plot of $\mathcal{M}$, bottom: projective plot of $\Tilde{\mathcal{M}}$ in the directions $\hat{e}_2,\hat{e}_4,\hat{e}_6$}
\label{fig:SU3/2B}
\end{figure}

In figure \ref{fig:SU3/2B} we see Cartesian coordinate lines in the directions $\hat{e}_2$, $\hat{e}_4$ and $\hat{e}_6$ in $\Tilde{\mathbb{R}}^D$ that reach approximately from $-1.5$ to $1.5$, a much larger sector than the one shown in figure \ref{fig:SU3/2A}. While the plots of $\mathcal{M}$ are too much entangled to give a good understanding of the geometry, the plots of $\Tilde{\mathcal{M}}$ are enlightening.\\
For $\alpha=1$ we can see a sphere, while for smaller $\alpha$ the plots look very much like what we have seen for the squashed fuzzy sphere in section \ref{sfs_results}, although, there we used different coordinate lines. Also for random $\alpha_a$, we see a similar behaviour, yet it compares to a very strong squashing.

\begin{figure}[H]
\centering
\begin{minipage}{.24\textwidth}
  \centering
  \includegraphics[height=.7\linewidth]{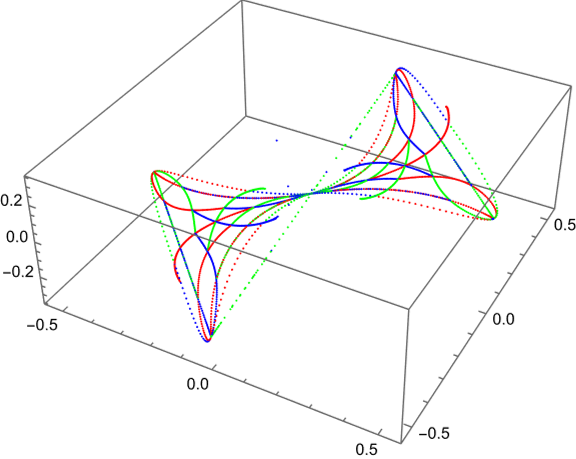}
  \includegraphics[height=.7\linewidth]{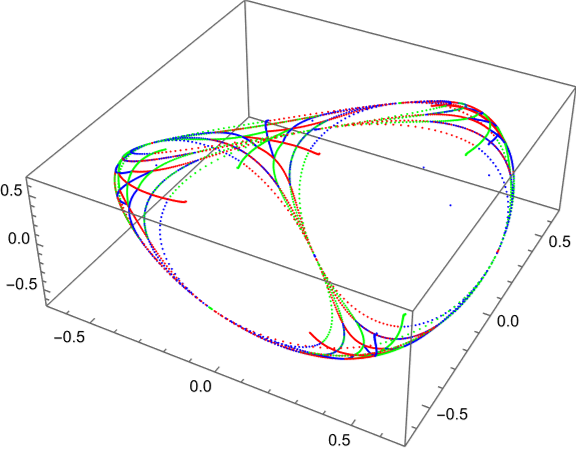}
\end{minipage}%
\begin{minipage}{.24\textwidth}
  \centering
  \includegraphics[height=.7\linewidth]{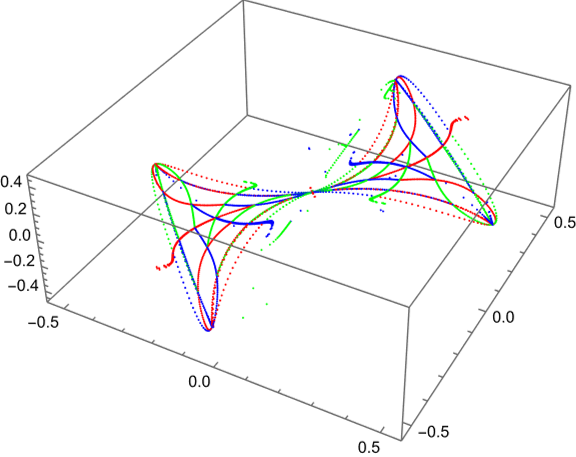}
  \includegraphics[height=.7\linewidth]{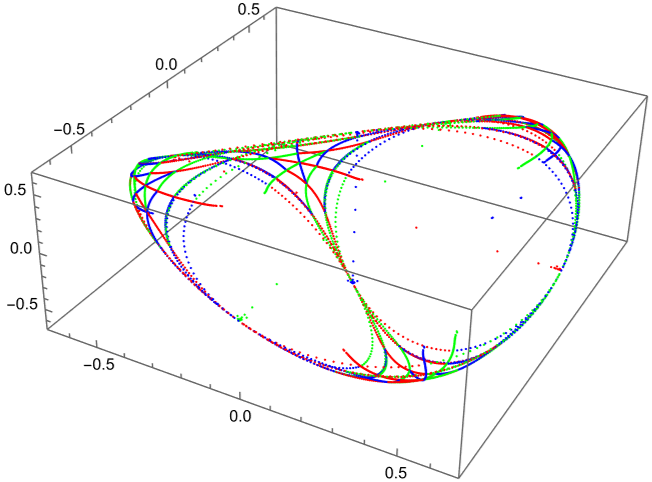}
\end{minipage}%
\begin{minipage}{.24\textwidth}
  \centering
  \includegraphics[height=.7\linewidth]{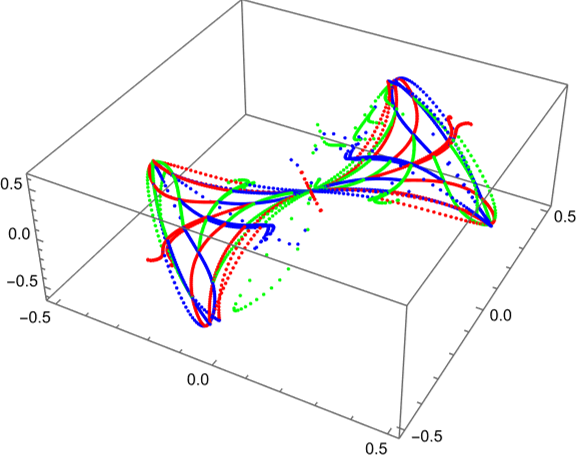}
  \includegraphics[height=.7\linewidth]{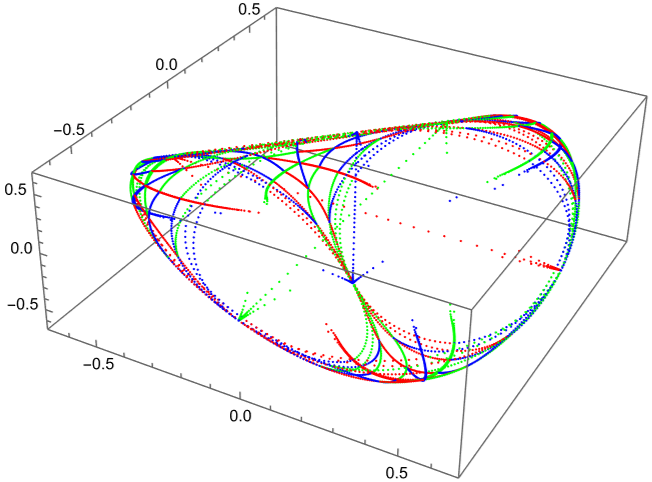}
\end{minipage}%
\begin{minipage}{.24\textwidth}
  \centering
  \includegraphics[height=.7\linewidth]{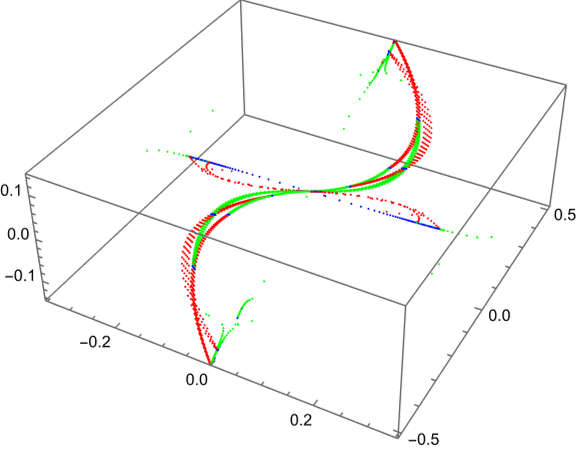}
  \includegraphics[height=.7\linewidth]{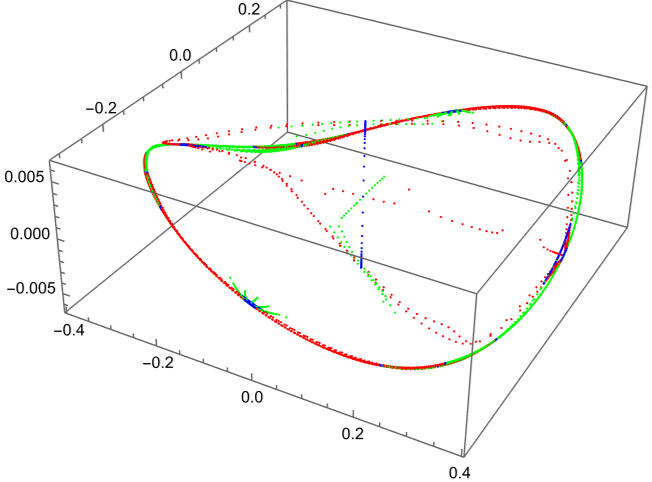}
\end{minipage}%
\caption{Large scale Cartesian coordinate lines in the directions $\hat{e}_1,\hat{e}_4,\hat{e}_6$ around a random point in the vicinity of $0$ in the squashed fuzzy $\mathbb{C}P^2$ for $n=5$. Left to right: $\alpha=1,\alpha=0.9,\alpha=0.1,\alpha_i=\text{random}$; top: projective plot of $\mathcal{M}$, bottom: projective plot of $\Tilde{\mathcal{M}}$ in the directions $\hat{e}_1,\hat{e}_4,\hat{e}_6$}
\label{fig:SU3/2C}
\end{figure}

For the visualization of $\Tilde{\mathcal{M}}$ we had to project from $\mathbb{R}^8$ to $\mathbb{R}^3$ and considered only points in a three dimensional subspace of $\Tilde{\mathbb{R}}^8$ (and correspondingly for $\mathcal{M}$), meaning we do not know how $\mathcal{M}$ and $\Tilde{\mathcal{M}}$ look in different directions.\\
Therefore, the same plots as in figure \ref{fig:SU3/2B} are shown in figure \ref{fig:SU3/2C}, yet the direction $\hat{e}_2$ has been replaced with $\hat{e}_1$.
Here, the large scale shape looks completely different, while the local structure can be very well compared to the previous perspective.

\begin{figure}[H]
\centering
\begin{minipage}{.3\textwidth}
  \centering
  \includegraphics[height=.7\linewidth]{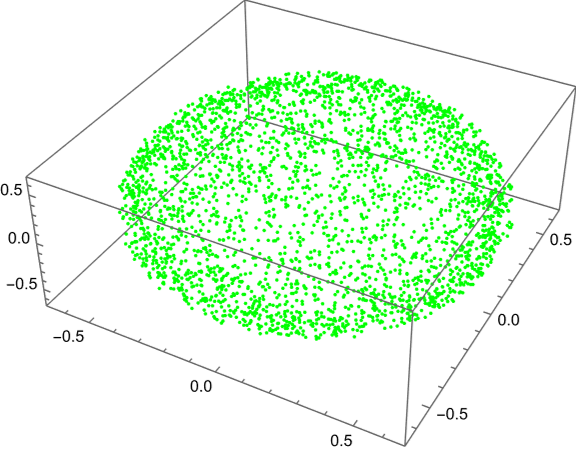}
\end{minipage}%
\begin{minipage}{.3\textwidth}
  \centering
  \includegraphics[height=.7\linewidth]{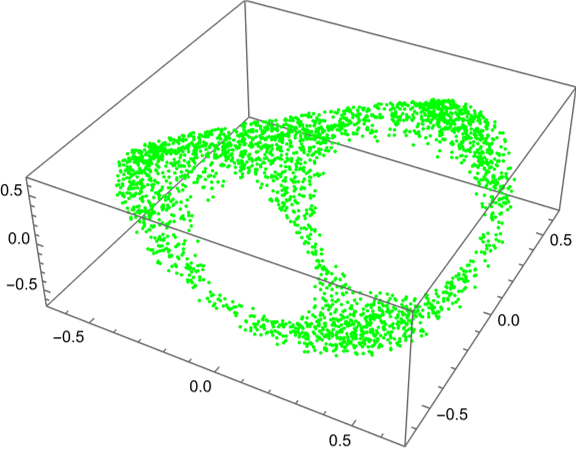}
\end{minipage}%
\begin{minipage}{.3\textwidth}
  \centering
  \includegraphics[height=.7\linewidth]{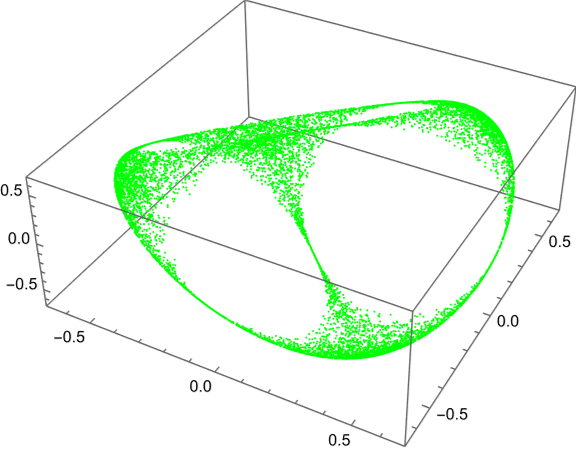}
\end{minipage}%
\caption{Plots of the squashed fuzzy $\mathbb{C}P^2$ for $n=5$ and $\alpha=1$. Left to right: sliced plot of $\Tilde{\mathcal{M}}$ in the directions $\hat{e}_2,\hat{e}_4,\hat{e}_6$ for $15000$ random points, sliced plot of $\Tilde{\mathcal{M}}$ in the directions $\hat{e}_1,\hat{e}_4,\hat{e}_6$ for $15000$ random points, projective plot of $\Tilde{\mathcal{M}}$ for $10000$ random points lying in the plane spanned by $\hat{e}_1,\hat{e}_4,\hat{e}_6$}
\label{fig:SU3/2D}
\end{figure}

The left and middle plot in figure \ref{fig:SU3/2D} show sliced plots for $\alpha=1$ in the respective directions shown in figures \ref{fig:SU3/2B} and \ref{fig:SU3/2C} for random points in the unit ball. These confirm that the apparent large scale shapes are not merely a result of considering points in a three dimensional subspace of $\Tilde{\mathbb{R}}^8$ only, but rather represent the intersection of $\Tilde{\mathcal{M}}$ with the respective planes in $\mathbb{R}^8$.\\
Yet the shapes are washed out a bit. This is due to the needed tolerance in the calculation of sliced plots as described in section \ref{Visual}.\\
The right plot in figure \ref{fig:SU3/2D} shows random points in the plane spanned by $\hat{e}_1,\hat{e}_4,\hat{e}_6$ that make the shape in figure \ref{fig:SU3/2C} for $\alpha=1$ better visible.

\subsubsection{Effective Dimension, Foliations and Integration}

Before we can do anything else, we need to determine the effective dimension $l$ of $\mathcal{M}$. For the fuzzy sphere, the only possible choice had been $l=2$, while here in principle we only know $l\in\{2,4,6,8\}$. 
In the round case we know that $l=4$ and we would readily expect that this generalizes also to the squashed cases, but this is a good opportunity to check if the methods that we have developed so far work properly.\\
In table \ref{table4} various quantities related to the different leaves are listed for $x=(1,2,1,0,0,0,0,0)$ and $n=3$ that should allow us to determine $l$ for $\alpha=1,0.9,01$ respectively for random $\alpha_a$.\\
Before we go into the discussion, we note that here it is obvious why the symplectic leaf cannot be a good choice for the squashed fuzzy $\mathbb{C}P^2$: In general, the rank of $\omega_{ab}$ is eight, meaning that there will only be one leaf that fills the whole $\mathcal{M}$.\\
We start with the Pfaffian method. In section \ref{pfaff}, we described how to choose $l$. There we said that we should calculate $v^s_{x,max}(x)$ for all possible $s$ and then choose $l$ as large as possible so that $v^s_{x,max}(x)\gg0$. Looking at the table this means in all cases that $l=4$, just as we expected.\\
The Kähler method works in a rather similar way, but here we calculate $v^s_{x,min}(x)$ and choose $l$ as the largest possible $s$ such that $v^s_{x,max}(x)\approx0$. Here we can see once again how unreliable the method is, in fact $v^s_{x,max}(x)$ should always increase with $s$ and not decrease. Still there is a tendency to $l=4$, while the matter is much less clear compared to the Pfaffian method.\\
Also for the hybrid leaves we described a way to determine $l$: We choose it as the amount of eigenvalues $\lambda_i$ of $\theta^{ab}$ respectively $\omega_{ab}$ with $\vert\lambda_i\vert\gg0$. In all cases, this means $l=4$.\\
We conclude that $l=4$, becoming less and less obvious with decreasing $\alpha$.

\begin{table}[H]
\centering
\begin{tabular}{ll|l|l|l|l}
 method &   & $\alpha=1$ & $\alpha=0.9$ & $\alpha=0.1$ & Random $\alpha_a$ \\\hline
\multicolumn{1}{l|}{Pfaffian}           & $s=2$   & $0.600$ & $0.599$ & $0.599$ &  $0.033$  \\\cline{2-6} 
\multicolumn{1}{l|}{$v^s_{x,max}(x)$} &    $s=4$  & $0.720$ & $0.659$  & $0.146$ & $0.004$    \\\cline{2-6} 
\multicolumn{1}{l|}{} &    $s=6$                  & $0.000$ & $0.000$ & $0.000$ &  $0.000$   \\\cline{2-6} 
\multicolumn{1}{l|}{} &    $s=8$                  & $0.000$ & $0.000$ & $0.000$ &  $0.000$   \\\hline
\multicolumn{1}{l|}{Kähler} &  $s=2$              & $0.000$ & $0.002$ & $0.009$ & $0.012$   \\\cline{2-6} 
\multicolumn{1}{l|}{$v^s_{x,min}(x)$}  &    $s=4$ & $0.000$ & $0.000$ & $0.006$ &  $0.006$   \\\cline{2-6} 
\multicolumn{1}{l|}{}  &    $s=6$                 & $1.414$ & $0.002$ & $0.005$ & $1.412$    \\\cline{2-6} 
\multicolumn{1}{l|}{}  &    $s=8$                 & $2.000$ & $1.414$ & $1.414$ &  $1.414$   \\\hline
$\operatorname{Eig}(\theta^{ab}(x))$    &         &
$ \begin{pmatrix}\pm 0.300i\\\pm 0.300i\\\pm0.000\\\pm 0.000\\\end{pmatrix}$ &
$ \begin{pmatrix}\pm 0.300i\\\pm 0.275i\\\pm0.000\\\pm 0.000\\\end{pmatrix}$ &
$ \begin{pmatrix}\pm 0.300i\\\pm 0.034i\\\pm0.000\\\pm 0.000\\\end{pmatrix}$ &
$ \begin{pmatrix}\pm 0.072i\\\pm 0.031i\\\pm0.000\\\pm 0.000\\\end{pmatrix}$   \\\hline
$\operatorname{Eig}(\omega_{ab}(x))$   & & $\begin{pmatrix}\pm0.333i\\\pm0.333i\\\pm0.000\\\pm0.000\\\end{pmatrix}$  &
$ \begin{pmatrix}\pm 0.328i\\\pm 0.080 i\\\pm0.000\\\pm 0.000\\\end{pmatrix}$  &
$ \begin{pmatrix}\pm 0.315i\\\pm 0.013i\\\pm0.000\\\pm 0.000\\\end{pmatrix}$ &
$ \begin{pmatrix}\pm 0.358i\\\pm 0.037i\\\pm0.000\\\pm 0.000\\\end{pmatrix}$     
\end{tabular}
\caption{Determination of the effective dimension for $n=3$ and $x=(1,2,1,0,0,0,0,0)$. Note that the eigenvalues $\pm 0.000$ are only identically zero for $\alpha=1$}
\label{table4}
\end{table}

\begin{figure}[H]
\centering
\begin{minipage}{.3\textwidth}
  \centering
  \includegraphics[height=.7\linewidth]{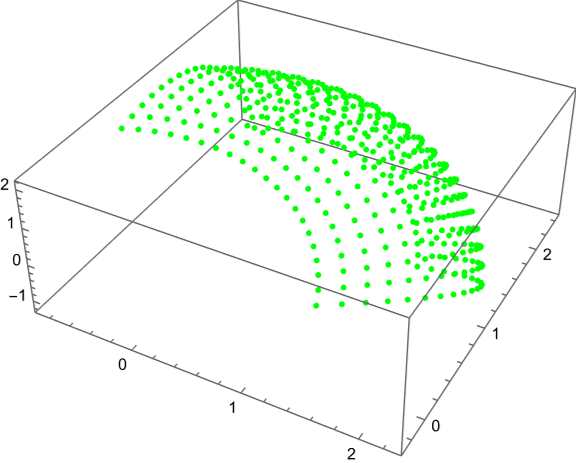}
\end{minipage}%
\begin{minipage}{.3\textwidth}
  \centering
  \includegraphics[height=.7\linewidth]{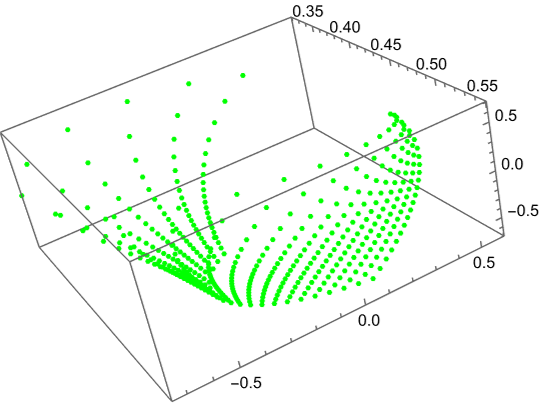}
\end{minipage}%
\caption{Two directions of local coordinates around $x=(1,2,1,0,0,0,0,0)$ in the hybrid leaf in the squashed fuzzy $\mathbb{C}P^2$ for $n=2$ and $\alpha=0.1$. Left: projective plot of $\Tilde{\mathbb{R}}^D$, right: projective plot of $\mathcal{M}$}
\label{fig:SU3/4}
\end{figure}

Now that we know that $l=4$, we can construct local coordinates\footnote{In principle it would be good to check if the leaves are integrable. In higher dimensions this is a difficult task. Still, on the one hand we can argue that the results for the fuzzy sphere were promising and on the other hand we can substantiate the assumption with our results a posteriori.} for $\mathcal{M}$. Figure \ref{fig:SU3/4} features plots of two directions of such coordinates for $n=2$ around the point $(1,2,1,0,0,0,0,0)$. We can see that these coordinates are \textit{well behaved}.

\begin{figure}[H]
\centering
\begin{minipage}{.3\textwidth}
  \centering
  \includegraphics[height=.7\linewidth]{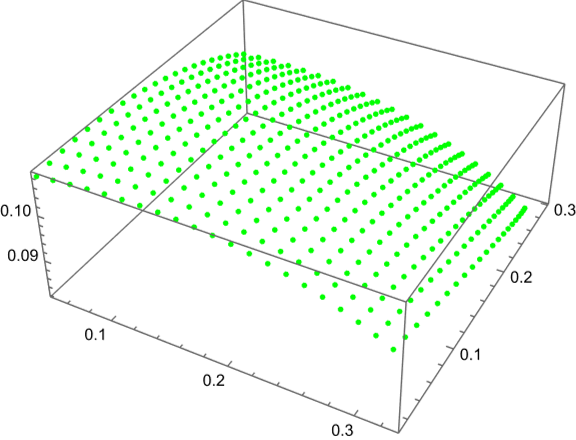}
\end{minipage}%
\begin{minipage}{.3\textwidth}
  \centering
  \includegraphics[height=.7\linewidth]{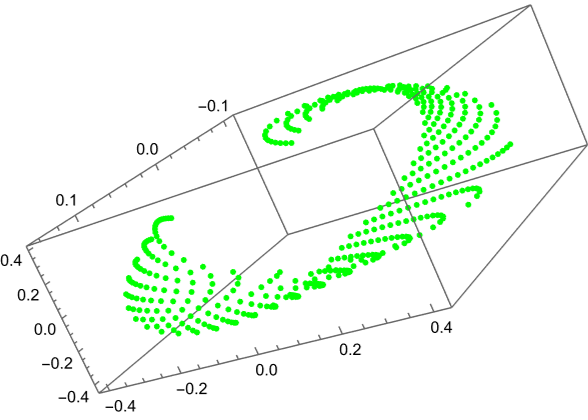}
\end{minipage}%
\caption{Two dimensions of local coordinates around a random point in the hybrid leaf in the squashed fuzzy $\mathbb{C}P^2$ for $n=5$ and $\alpha=0.1$. Left: projective plot of $\Tilde{\mathbb{R}}^D$, right: projective plot of $\mathcal{M}$}
\label{fig:SU3/4A}
\end{figure}

Figure \ref{fig:SU3/4A} shows similar coordinates for $n=5$ around a random point. Here, we note that the large scale shape winds itself strongly in the plot of $\mathcal{M}$. This is also the reason why the surface in the plot of $\Tilde{\mathbb{R}}^D$ is less curved than the one in figure \ref{fig:SU3/4}: Here, the step length has been chosen much smaller in order to produce a plot of coordinates where it is visible that they are not self-intersecting.

The next step is to integrate over $\mathcal{M}$ and to look at the usual quantities.
The tables \ref{3DepAlpha} and \ref{3DepN} show the dependence of these quantities on $\alpha$ and $N$.

Yet, there is a caveat: For $D>3$ and $l>2$ a few systematic problems arise that either make the calculations more computationally demanding or less precise.\\
For $D>3$ we first find that the visualizations of $\Tilde{\mathbb{R}}^D$ and $\Tilde{\mathcal{M}}$ cease to be faithful and it therefore is harder to estimate the quality of local coordinates what in turn makes it more difficult to select the appropriate number of points and step length. Also, if one wants to use a similar tiling as for the squashed fuzzy sphere, $2^D$ tiles are needed. This implies we have to calculate many more local coordinates.\\
If $l>2$ this means that the total number of coordinate points is given by the number of points in each direction to the power of $l$. Thus, we can either enlarge the step length or calculate significantly more points. Also, the higher dimension of the coordinates makes it harder to judge their quality.\\
In total, this means that either the computational cost is much larger or the quality suffers.

In the apparent calculations a compromise has been chosen. This means that the step length has been increased strongly with respect to the squashed fuzzy sphere and some tiles are not filled completely, still resulting in a significantly longer computational time. Thus, the results should not be taken too serious, while further computations with stronger hardware could be promising.\\
Additionally, numerical problems occurred for $\alpha=1$. Consequently we replaced $1\mapsto 0.99$. The scan has still been calculated for $\alpha=1$ since this showed better success.

\newpage
We briefly discuss the results, first for the dependence on $\alpha$.\\
Here, we see that the symplectic volume $V_\omega$ strongly depends on $\alpha$. This is in strict contrast to the squashed fuzzy sphere. Yet, it is not clear if this is only due to the fact that not all tiles are filled and thus $V_\omega$ is incorrect or if this is systematically true.\\
For the completeness relation and the quantization of the $\mathbf{x}^a$ we see that the quality for $\alpha=0.99,0.9,0.5$ is rather good, but hardly acceptable for $\alpha=0.1$ For random $\alpha_a$ the results are again a bit better. The quality for $\alpha=0.99$ is not overwhelming, which is a consequence of the earlier described numerical problems.\\
The compatibility of the different Poisson structure is rather bad for smaller $\alpha$ and terrible for random $\alpha_a$. For the Kähler properties we find a different behaviour as the quality of $V_L$ is once again better for the random $\alpha_a$.\\
We conclude that for the quantities that do not depend on the numerical integration we see a rather expected behaviour, while we cannot tell exactly for the others, although the results in principle fit to the scheme we know from the squashed fuzzy sphere.

\begin{table}[H]
\centering
\begin{tabular}{l|l|ll|ll|l|lll}
$\alpha$ & $V_\omega$ & $\sigma_{\mathbb{1}'}$ & $d_{\mathbb{1}'}$ & $d_{X'}$ & $n_{X'}$ & $d_{\{\}}$ & $c^2(V_R)$ & $c^2(V_L)$ & $c^2(V_K)$ \\\hline
$0.99^*$ & $24.513$ & $0.088$ & $0.080$ & $0.127$ & $2.481$ & $\sim 10^{-6}$ & $ 0.0053$ & $ 0.000017$ & $ 0.00081$  \\
$0.9$ & $41.698$ & $0.074$ & $0.067$ & $0.110$ & $2.495$ & $0.00030$ & $ 0.051$ & $ 0.0018$ & $\sim 10^{-6}$ \\
$0.5$ & $102.544$ & $0.095$ & $0.087$ & $0.132$ & $2.495$ & $0.000064$ & $ 0.196$ & $ 0.056$ & $\sim 10^{-7}$ \\
$0.1$ & $180.811$ & $0.617$ & $0.563$ & $0.648$ & $2.117$ & $0.064$ & $ 0.705$ & $ 0.486$ & $0.0037$ \\
random $\alpha_a$ & $92.875$ & $0.361$ & $0.329$ & $0.304$ & $2.167$ & $0.984$ & $ 0.922$ & $ 0.049$ & $0.012$
\end{tabular}
\caption{Dependence of various quantities on $\alpha$ ($^*$Here $\alpha=1$ has been used for the scan in order to improve the quality)}
\label{3DepAlpha}
\end{table}

Finally, we look at the dependence on $n$.\\
Here the symplectic volume $V_\omega$ explodes with increasing $n$, while the quality of the completeness relation and the quantization of the $\mathbf{x}^a$ decreases with $n$. On the one hand, we could repeat the argument from section \ref{sfsfoliat} that showed that stagnating values actually mean an improvement of the quality with increasing $N$, but on the other hand, the quality is already so poor that it is hard to say if the reasoning is still applicable.\\
Yet, considering the compatibility of the Poisson structures and the Kähler properties, we see a totally different picture: Here the quality is always decent and starts to improve for large $n$.\\ Under these considerations, it is at least plausible to assume that the bad behaviour in the first place is only due to the computational difficulty in the integration over the leaf. 

\begin{table}[H]
\centering
\begin{tabular}{ll|l|ll|ll|l|lll}
$n$ & $N$ & $V_\omega$ & $\sigma_{\mathbb{1}'}$ & $d_{\mathbb{1}'}$ & $d_{X'}$ & $n_{X'}$ & $d_{\{\}}$ & $c^2(V_R)$ & $c^2(V_L)$ & $c^2(V_K)$ \\\hline
$1$ & $3$ & $12.827$ & $0.071$ & $0.058$ & $0.137$ & $3.986$ & $\sim 10^{-16}$ & $0$ & $\sim 10^{-8}$ & $0$ \\
$2$ & $6$ & $41.698$ & $0.074$ & $0.067$ & $0.110$ & $2.495$ & $0.00030$ & $0.051$ & $0.0018$ & $\sim 10^{-6}$ \\
$3$ & $10$ & $96.156$ & $0.126$ & $0.120$ & $0.115$ & $1.984$ & $0.00036$ & $0.054$ & $0.0016$ & $0.00024$\\
$5$ & $21$ & $311.277$ & $0.131$ & $0.128$ & $0.149$ & $1.588$ & $0.00035$ & $0.053$ & $0.0012$ & $0.00026$ \\
$13$ & $105$ & $1992.790$ & $0.263$ & $0.262$ & $0.265$ & $1.196$ & $0.00022$ & $0.043$ & $0.00058$ & $0.00017$ 
\end{tabular}
 \caption{Dependence of various quantities on $n$}
\label{3DepN}
\end{table}

\subsection{The Completely Squashed Fuzzy \texorpdfstring{$\mathbb{C}P^2$}{CP2}}
\label{csfc}

Another matrix configuration that is derived from the fuzzy $\mathbb{C}P^2$ is the \textit{completely squashed fuzzy $\mathbb{C}P^2$}, defined via the matrices
\begin{align}
    \mathbb{C}P^2_{n,cs}:=\left(\bar{X}^1,\bar{X}^2,\bar{X}^4,\bar{X}^5,\bar{X}^6,\bar{X}^7\right)
\end{align}
(where the $\bar{X}^a$ are the matrices from the round fuzzy $\mathbb{C}P^2$).
This means we omit the two Cartan generators $\bar{X}^3,\bar{X}^8$.

\subsubsection{First Results and Dimensional Aspects}

At first, we note that also here $0\in\mathcal{K}$. In table \ref{table5} we see the dimension of $\mathcal{M}$ and the ranks of $g_{ab}$, $\omega_{ab}$ and $\theta^{ab}$ for random points, depending on $n$. We find that for $n=1$ all the latter are given by four and turn to six for $n>1$. Once again, this can be explained by the results for random matrix configurations.

\begin{table}[H]
\begin{tabular}{l|l|l|l|l}
  & $\operatorname{dim}(\mathcal{M})$ & $\operatorname{rank}(g)$ & $\operatorname{rank}(\omega)$ & $\operatorname{rank}(\theta)$ \\ \hline
$n=1$  & 4   & 4   & 4    & 4        \\ \hline 
$n>1$  & 6   & 6    & 6       & 6 
\end{tabular}
\centering
\caption{Overview of the dimensions and ranks in different scenarios}
\label{table5}
\end{table}

Yet there are special points like $(0,0,0,0,0,1)$ and $(0,1,0,0,0,0)$ where the dimension of $\mathcal{M}$ respectively the ranks of $g_{ab}$, $\omega_{ab}$ and $\theta^{ab}$ are locally reduced as shown in table \ref{table6}. These findings are not surprising, especially $(0,1,0,0,0,0)$ should be compared to $(0,1,0,0,0,0,0,0)$ for the squashed fuzzy $\mathbb{C}P^2$ where we also found a special behaviour.

\begin{table}[H]
\begin{tabular}{l|l|l|l|l}
  & $\operatorname{dim}(\mathcal{M})$ & $\operatorname{rank}(g)$ & $\operatorname{rank}(\omega)$ & $\operatorname{rank}(\theta)$ \\ \hline
$n=1$  & 3   & 3   & 2    & 2        \\ \hline 
$n>1$  & 6   & 6    & 4       & 4 
\end{tabular}
\centering
\caption{Overview of the dimensions and ranks in different scenarios for the exemplary special points $(0,0,0,0,0,1)$ and $(0,1,0,0,0,0)$}
\label{table6}
\end{table}

\subsubsection{A Global View}

As the completely squashed fuzzy $\mathbb{C}P^2$ (that has first been discussed in \cite{Steinacker_2015}) is in some sense the limit of the squashed fuzzy $\mathbb{C}P^2$ for $\alpha\mapsto 0$, we expect to find a related behaviour. This has already been manifested in the tables \ref{table5} and \ref{table6}, comparing them to the tables \ref{table2} and \ref{table3} for $\alpha=0$.

\begin{figure}[H]
\centering
\begin{minipage}{.3\textwidth}
  \centering
  \includegraphics[height=.7\linewidth]{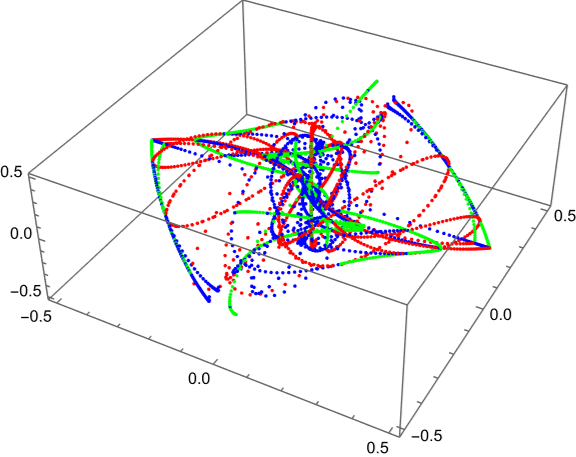}
  \includegraphics[height=.7\linewidth]{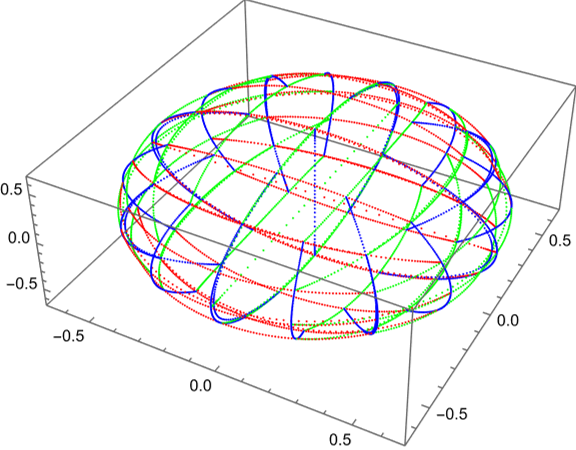}
\end{minipage}%
\begin{minipage}{.3\textwidth}
  \centering
  \includegraphics[height=.7\linewidth]{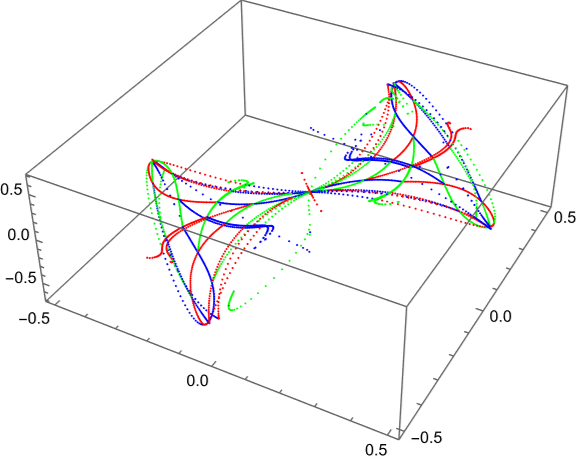}
  \includegraphics[height=.7\linewidth]{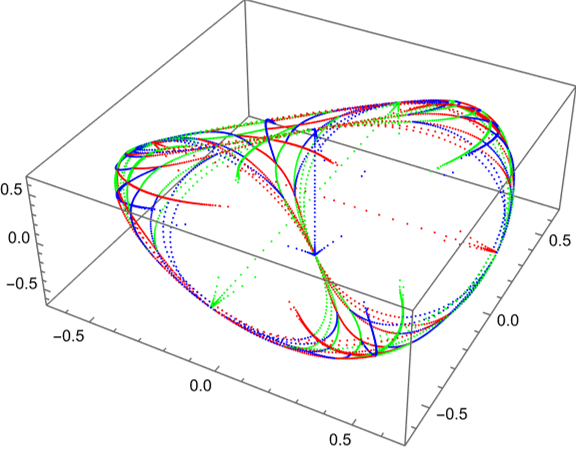}
\end{minipage}%
\caption{Large scale Cartesian coordinate lines in the directions $\hat{e}_i,\hat{e}_3,\hat{e}_5$ around a random point in the vicinity of $0$ in the completely squashed fuzzy $\mathbb{C}P^2$ for $n=5$. Top: projective plot of $\mathcal{M}$, bottom: projective plot of $\Tilde{\mathcal{M}}$ in the directions $\hat{e}_i,\hat{e}_3,\hat{e}_5$; left: $i=2$, right: $i=1$}
\label{fig:SU3cs/2BC}
\end{figure}

Looking at figure \ref{fig:SU3cs/2BC}, which is constructed analogously\footnote{Here, the directions $\hat{e}_4,\hat{e}_6$ have to be replaced by the directions $\hat{e}_3,\hat{e}_5$ as we dropped $\Bar{X}^3,\Bar{X}^8$.} to the figures \ref{fig:SU3/2B} and \ref{fig:SU3/2C} for the squashed fuzzy $\mathbb{C}P^2$, this relationship is perfectly visible.

\begin{figure}[H]
\centering
\begin{minipage}{.3\textwidth}
  \centering
  \includegraphics[height=.7\linewidth]{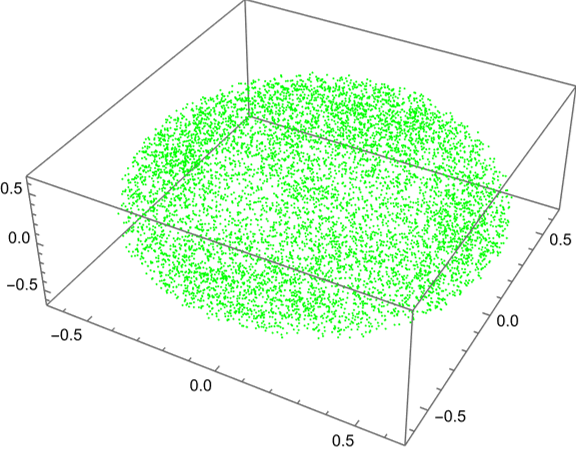}
\end{minipage}%
\begin{minipage}{.3\textwidth}
  \centering
  \includegraphics[height=.7\linewidth]{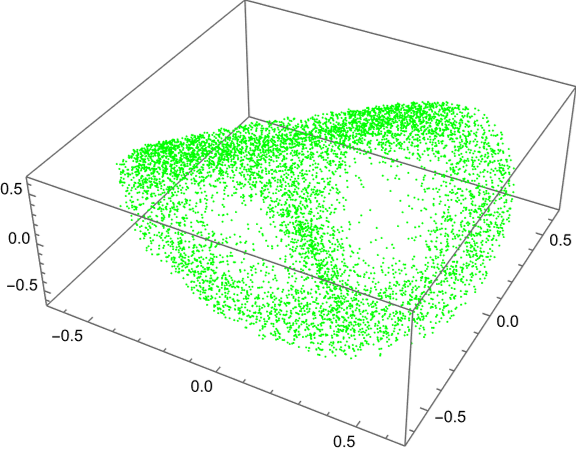}
\end{minipage}%
\caption{Plots of the completely squashed fuzzy $\mathbb{C}P^2$ for $n=5$. Left: sliced plot of $\Tilde{\mathcal{M}}$ in the directions $\hat{e}_2,\hat{e}_3,\hat{e}_5$ for $15000$ random points; right: $\Tilde{\mathcal{M}}$ in the directions $\hat{e}_1,\hat{e}_3,\hat{e}_5$ for $15000$ random points}
\label{fig:SU3cs/2D}
\end{figure}

In figure \ref{fig:SU3cs/2D} we see sliced plots similar to the ones in figure \ref{fig:SU3cs/2D} for the squashed fuzzy $\mathbb{C}P^2$ that confirm that the shape is not only caused by the restriction in $\Tilde{\mathbb{R}}^D$.

\begin{figure}[H]
\centering
\begin{minipage}{.3\textwidth}
  \centering
  \includegraphics[height=.7\linewidth]{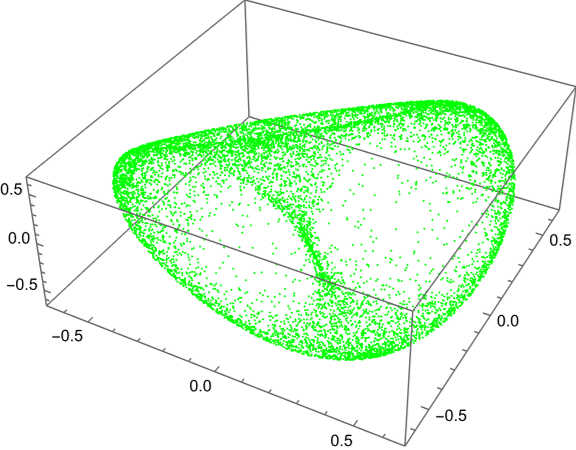}
\end{minipage}%
\begin{minipage}{.3\textwidth}
  \centering
  \includegraphics[height=.7\linewidth]{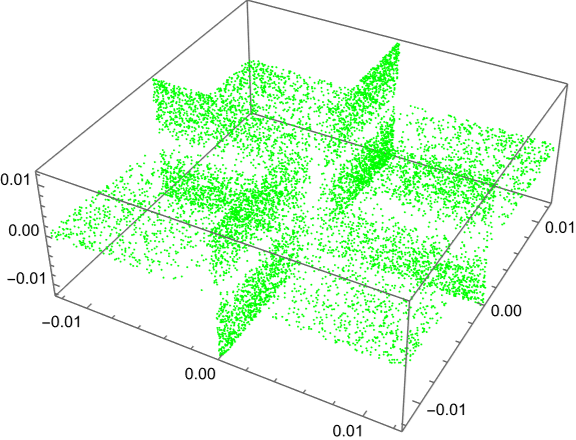}
\end{minipage}%
\caption{Plots of the completely squashed fuzzy $\mathbb{C}P^2$ for $n=5$. Left: sliced plot of $\Tilde{\mathcal{M}}$ in the directions $\hat{e}_1,\hat{e}_3,\hat{e}_5$ for $10000$ random points lying in the plane spanned by $\hat{e}_1,\hat{e}_4,\hat{e}_6$, right: the same but the components of the points are now bounded by $0.01$}
\label{fig:SU3cs/2D2}
\end{figure}

Now, we take a closer look at the slice of $\Tilde{\mathcal{M}}$ through the plane spanned by $\hat{e}_1,\hat{e}_3,\hat{e}_5$. Figure \ref{fig:SU3cs/2D2} shows plots of random points from the same plane in $\Tilde{\mathbb{R}}^3$. On the left hand side, the components of the points are bounded by $1$ and we recover the shape from figure \ref{fig:SU3cs/2BC}. Here, we might wonder how the inside looks like. This question is answered by the plot on the right hand side, where the components of the points are bounded by $0.01$. Now we can see that asymptotically there are three planes that intersect orthogonally in the origin, preventing $\Tilde{\mathcal{M}}$ from being a well defined manifold, while $\mathcal{M}$ remains a smooth manifold. 
\\
This phenomenon has been well known and was first described in \cite{Steinacker_2015} and then in \cite{Schneiderbauer_2016}.

\subsubsection{Effective Dimension, Foliations and Integration}

For the completely squashed fuzzy $\mathbb{C}P^2$ the effective dimension can be found to be $l=4$ for $n>1$, continuing the result for the squashed fuzzy $\mathbb{C}P^2$.\\
In table \ref{3csDepN} we see the most relevant quantities calculated for the hybrid leaf through $x=(1,2,1,0,0,0)$ depending on $n$. Yet, the completely squashed fuzzy $\mathbb{C}P^2$ and squashed fuzzy $\mathbb{C}P^2$ share the same problems concerning the integration over the leaf and we should not be too trustworthy with the quantities coming from integration.\\
Also here, we see that $V_\omega$ depends on $n$. The quality of the completeness relation and the quantization of the $\mathbf{x}^a$ is extremely bad already for $n=2$, while the compatibility of the different Poisson structures is not very good but acceptable. The Kähler properties hardly depend on $n$, while we see that the hybrid subspace is not very well adapted compared to the optimal Kähler subspace.

\begin{table}[H]
\centering
\begin{tabular}{ll|l|ll|ll|l|lll}
$n$ & $N$ & $V_\omega$ & $\sigma_{\mathbb{1}'}$ & $d_{\mathbb{1}'}$ & $d_{X'}$ & $n_{X'}$ & $d_{\{\}}$ & $c^2(V_R)$ & $c^2(V_L)$ & $c^2(V_K)$ \\\hline
$2$ & $6$ & $39.418$ & $0.784$ & $0.716$ & $0.767$ & $1.983$ & $0.029$ & $0.253$ & $0.168$ & $0.031$\\
$3$ & $10$ & $120.331$ & $1.695$ & $1.608$ & $1.092$ & $1.138$ & $0.043$ & $0.203$ & $0.179$ & $0.027$ \\
$5$ & $21$ & $216.906$ & $1.657$ & $1.617$ & $1.012$ & $1.236$ & $0.054$ & $0.253$ & $0.173$ & $0.019$
\end{tabular}
 \caption{Dependence of various quantities on $n$}
\label{3csDepN}
\end{table}

\subsection{The Fuzzy Torus}
\label{ft}

Our final example is the so called \textit{fuzzy torus}, defined via the two unitary \textit{clock and shift matrices} $U$ and $V$ for a given $N>0$. Their exact definition is given in appendix \ref{tor2rep}.\\
Then, we set
\begin{align}
    T^2_N:=\left(\operatorname{Re}(U),\operatorname{Im}(U),\operatorname{Re}(V),\operatorname{Im}(V)\right).
\end{align}
In principle, a discussion (that we do not repeat here) completely analogous to our first construction of the fuzzy sphere in section \ref{FuzzySphere0} is possible (where the relation to the ordinary Clifford torus is manifest), see for example \cite{Schneiderbauer_2016}.
Still, there are fundamental differences to the fuzzy sphere, since the matrices do not come from a semisimple Lie algebra.

\subsubsection{First Results and Dimensional Aspects}

As usual, we begin by calculating the dimension of $\mathcal{M}$ and the ranks of $g_{ab}$, $\omega_{ab}$ and $\theta^{ab}$ depending on $N$.
\\
Table \ref{table7} shows the results, where we find that $\mathcal{M}$ is one dimensional for $N=2$ (this can also be shown analytically \cite{Steinacker_2021}) and three dimensional for $N>2$, while the ranks follow the usual scheme.

\begin{table}[H]
\begin{tabular}{l|l|l|l|l}
  & $\operatorname{dim}(\mathcal{M})$ & $\operatorname{rank}(g)$ & $\operatorname{rank}(\omega)$ & $\operatorname{rank}(\theta)$ \\ \hline
$N=2$  & 2   & 1   & 0    & 0        \\ \hline 
$N>2$  & 3   & 3    & 2  & 2 
\end{tabular}
\centering
\caption{Overview of the dimensions and ranks in different scenarios}
\label{table7}
\end{table}

Since the Clifford torus $T^2$ is two dimensional, this is not exactly what we would expect, but since we do not deal with coadjoint orbits, it is also not completely surprising that we find a drawback.

For the fuzzy torus, there is a better way to plot points in $\mathbb{R}^4$ than our usual method from section \ref{Visual}, using the generalized stereographic map
\begin{align}
    \left(x^1,x^2,x^3,x^4\right)\mapsto \left(\frac{\sqrt{(x^1)^2+(x^2)^2}+x^3}{\sqrt{(x^1)^2+(x^2)^2}}x^1,\frac{\sqrt{(x^1)^2+(x^2)^2}+x^3}{\sqrt{(x^1)^2+(x^2)^2}}x^2,x^4\right)
\end{align}
that maps the Clifford torus $T^2$ to a better known embedding of the topological torus into $\mathbb{R}^3$ (the \textit{doughnut}) \cite{Lee_2003}.

\begin{figure}[H]
\centering
\begin{minipage}{.19\textwidth}
  \centering
  \includegraphics[height=.7\linewidth]{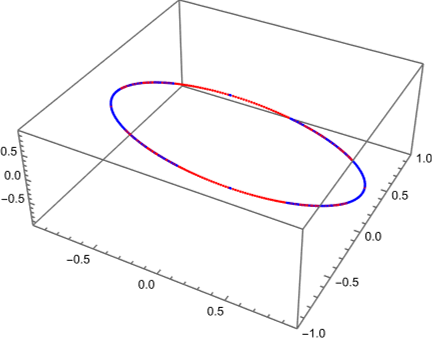}
  \includegraphics[height=.7\linewidth]{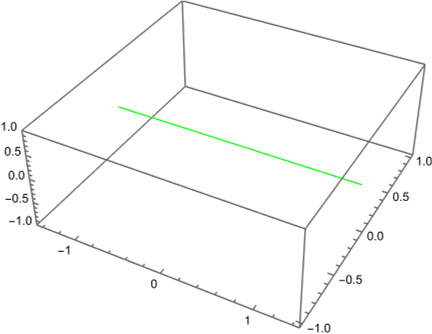}
\end{minipage}%
\begin{minipage}{.19\textwidth}
  \centering
  \includegraphics[height=.7\linewidth]{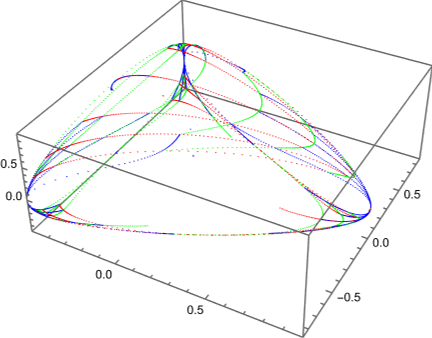}
  \includegraphics[height=.7\linewidth]{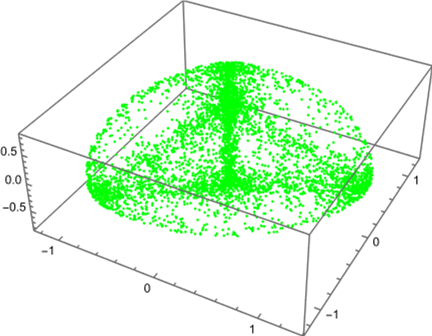}
\end{minipage}%
\begin{minipage}{.19\textwidth}
  \centering
  \includegraphics[height=.7\linewidth]{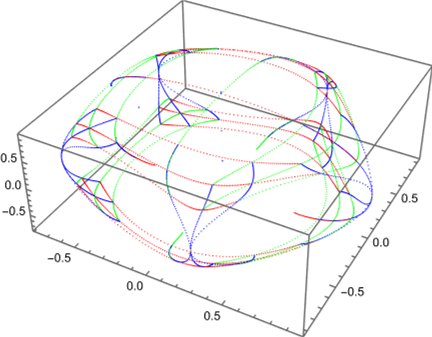}
  \includegraphics[height=.7\linewidth]{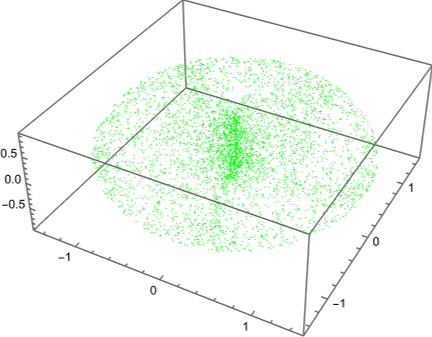}
\end{minipage}%
\begin{minipage}{.19\textwidth}
  \centering
  \includegraphics[height=.7\linewidth]{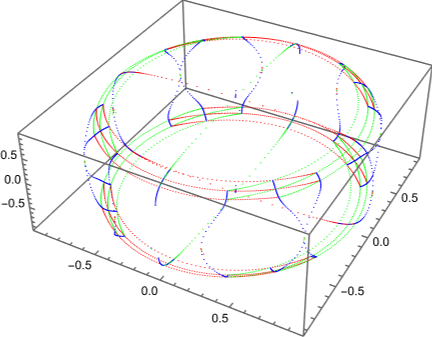}
  \includegraphics[height=.7\linewidth]{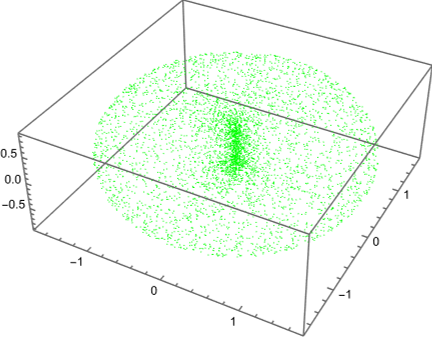}
\end{minipage}%
\begin{minipage}{.19\textwidth}
  \centering
  \includegraphics[height=.7\linewidth]{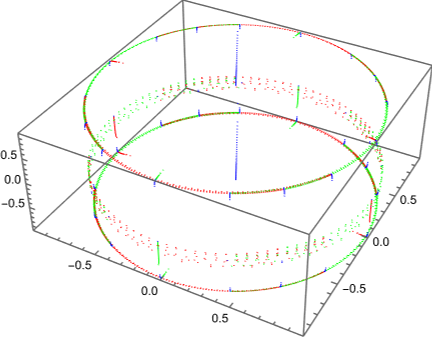}
  \includegraphics[height=.7\linewidth]{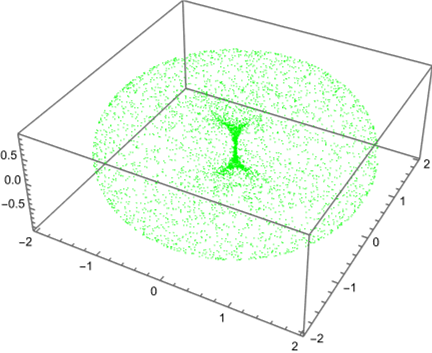}
\end{minipage}%
\caption{Plots of the fuzzy torus. Top: projective plot of $\Tilde{\mathcal{M}}$ for Cartesian coordinate lines, bottom: stereographic plot of $\Tilde{\mathcal{M}}$ for random points; left to right: $N=2,3,5,10,100$}
\label{fig:CS/2}
\end{figure}

In figure \ref{fig:CS/2} we can see projective plots of Cartesian coordinate lines and stereographic plots for random points.\\
First, we note that we truly find $\Tilde{\mathcal{M}}\cong S^1$ for $N=2$. So far, we can not easily recognize a torus in the plots of the Cartesian coordinate lines, but for $N=100$ the plot partially looks like what we would expect for a projective plot of $T^2$.\\
Considering the lower plots, we get confident that we might find $\Tilde{\mathcal{M}}\to T^2$ as $N\to\infty$.

This can easily be verified quantitatively: If we had $\Tilde{\mathcal{M}}=T^2$, we would find $(\mathbf{x}^1)^2+(\mathbf{x}^2)^2=1=(\mathbf{x}^3)^2+(\mathbf{x}^4)^2$, so we check this constraint for different $N$.
\\
In table \ref{table8} we see the average and the standard deviation of $(\mathbf{x}^1)^2+(\mathbf{x}^2)^2$ respectively $(\mathbf{x}^3)^2+(\mathbf{x}^4)^2$ for $5000$ random points in $\Tilde{\mathbb{R}}^D$, depending on $N$. The results confirm that for $N\to\infty$ we have $\mu\to1$ and $\sigma\to 0$ which shows that the claim holds.

\begin{table}[H]
\begin{tabular}{l|ll|ll}
 $N$ & $\mu((\mathbf{x}^1)^2+(\mathbf{x}^2)^2)$ & $\sigma((\mathbf{x}^1)^2+(\mathbf{x}^2)^2)$ & $\mu((\mathbf{x}^3)^2+(\mathbf{x}^4)^2)$ & $\sigma((\mathbf{x}^3)^2+(\mathbf{x}^4)^2)$ \\ \hline
$2$  & $0.502$   & $0.352$   & $0.498$    & $0.352$  \\ \hline 
$3$  & $0.441$   & $0.289$   & $0.436$    & $0.289$  \\ \hline
$5$  & $0.530$   & $0.203$   & $0.326$    & $0.204$  \\ \hline 
$10$  & $0.711$   & $0.131$   & $0.709$    & $0.133$  \\ \hline 
$100$  & $0.966$   & $0.017$   & $0.966$    & $0.017$ 
\end{tabular}
\centering
\caption{Mean and standard deviation of the toroidal constraints $(x^1)^2+(x^2)^2=1=(x^3)^2+(x^4)^2$ for different $N$ over $5000$ random points}
\label{table8}
\end{table}

\vspace{-0.4cm}
\subsubsection{Effective Dimension, Foliations and Integration}
\vspace{-0.1cm}

We now come to the most interesting part, the foliation of $\mathcal{M}$. Fittingly, we find the effective dimension $l=2$.
In table \ref{csDepN} the relevant quantities associated to the hybrid leaf through $x=(1,2,1,0)$ are listed, depending on $N$.
\\
We can see that the symplectic volume depends on $N$. Further, the quality of the completeness relation and the quantization of the $\mathbf{x}^a$ is pretty good. With the same arguments as in section \ref{sfsfoliat} it is plausible that equation (\ref{compl}) and (\ref{quant}) hold exactly for $N\to\infty$. We also see that the correction factor $n_{X'}$ goes to one in the limit. Similarly, we find that the two Poisson structures mutually approach with increasing $N$. Finally, also the Kähler properties improve with $N$.

\begin{table}[H]
\centering
\begin{tabular}{l|l|ll|ll|l|lll}
$N$ & $V_\omega$ & $\sigma_{\mathbb{1}'}$ & $d_{\mathbb{1}'}$ & $d_{X'}$ & $n_{X'}$ & $d_{\{\}}$ & $c^2(V_R)$ & $c^2(V_L)$ & $c^2(V_K)$ \\\hline
$3$ & $9.699$ & $0.057$ & $0.046$ & $0.081$ & $2.641$ & $0.185$ & $0.569$ & $0.530$ & $\sim 10^{-8}$\\
$5$ & $16.207$ & $0.044$ & $0.039$ & $0.052$ & $1.855$ & $0.734$ & $0.836$ & $0.266$ & $0.112$ \\
$10$ & $31.571$ & $0.050$ & $0.047$ & $0.056$ & $1.367$ & $0.050$ & $0.531$ & $0.122$ & $0.122$\\
$100$ & $306.157$ & $0.097$ & $0.097$ & $0.097$ & $1.027$ & $0.00017$ & $0.192$ & $0.0093$ & $0.0093$
\end{tabular}
 \caption{Dependence of various quantities on $N$}
\label{csDepN}
\end{table}

\vspace{-0.3cm}
In figure \ref{csDepN} we see projective plots of $\Tilde{\mathcal{M}}$ showing scans and stereographic plots of $\Tilde{\mathcal{M}}$ for tilings of the hybrid leaf.
The first are very interesting since we can read off $N$ from their shape: For a given $N$ it is approximately an $N$ corner in the plane. This fits extremely well to the heuristics behind the construction of $T^2_N$ that in some sense discretizes $U(1)\times U(1)$, but here we deal with true manifolds and not with discretizations.
The plots of the tilings confirm that already for $N=5$ the foliation approximately recovers a torus in $\mathcal{M}$, what is clearly better than what we have seen in figure \ref{fig:CS/2}. For larger $N$ the quality improves further.

\begin{figure}[H]
\centering
\begin{minipage}{.24\textwidth}
  \centering
  \includegraphics[height=.7\linewidth]{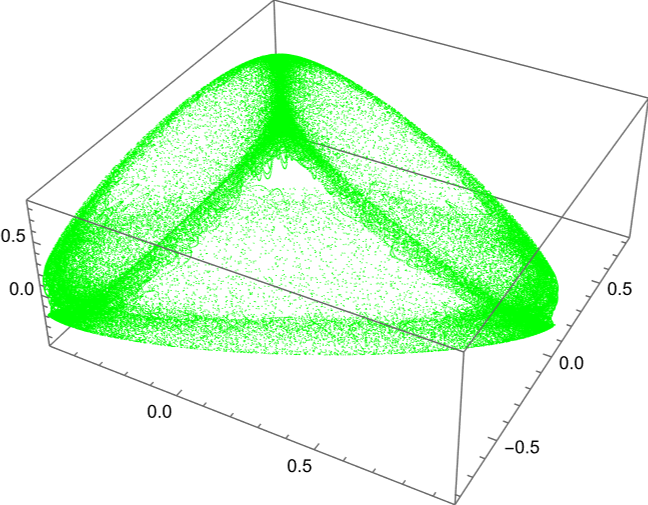}
  \includegraphics[height=.7\linewidth]{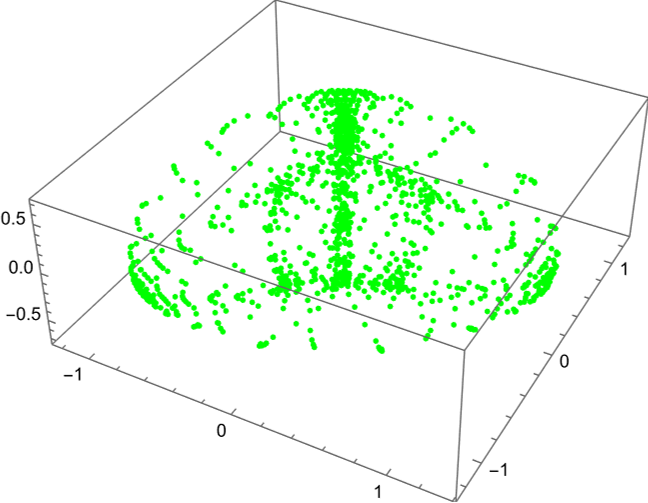}
\end{minipage}%
\begin{minipage}{.24\textwidth}
  \centering
  \includegraphics[height=.7\linewidth]{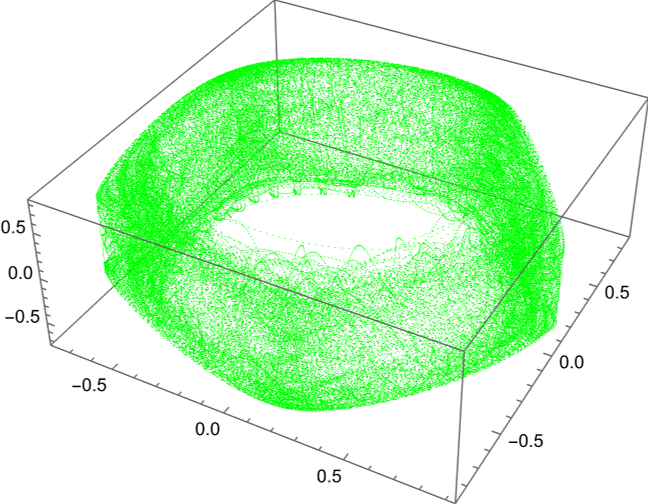}
  \includegraphics[height=.7\linewidth]{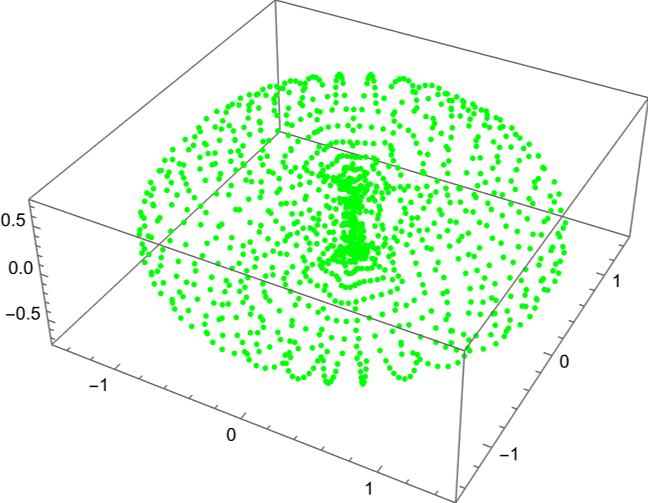}
\end{minipage}%
\begin{minipage}{.24\textwidth}
  \centering
  \includegraphics[height=.7\linewidth]{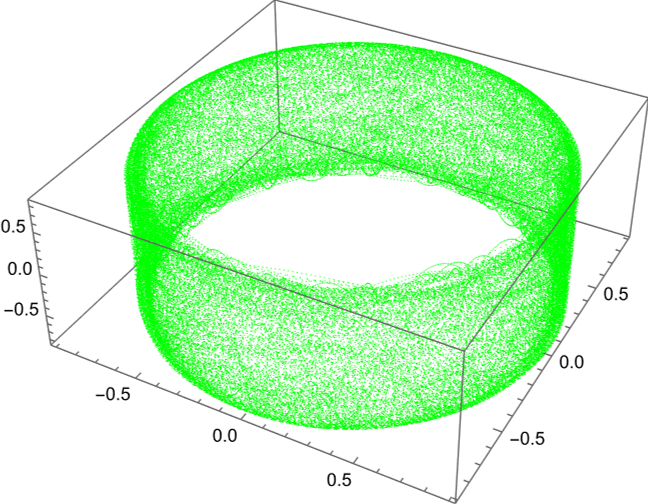}
  \includegraphics[height=.7\linewidth]{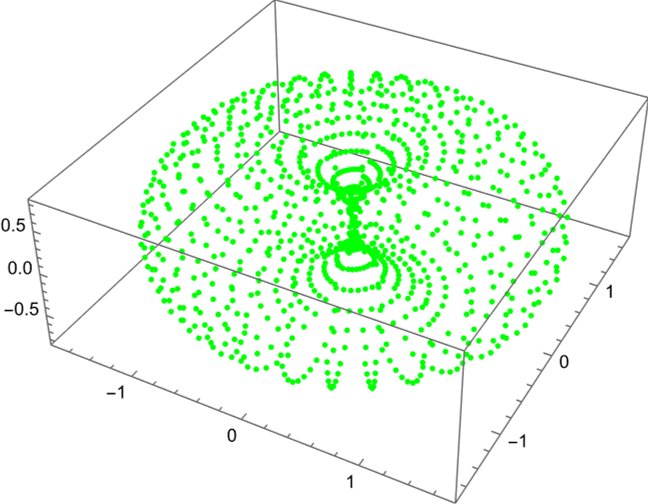}
\end{minipage}%
\begin{minipage}{.24\textwidth}
  \centering
  \phantom{\includegraphics[height=.7\linewidth]{Images2/CS/4/10/1.png}}
  \includegraphics[height=.7\linewidth]{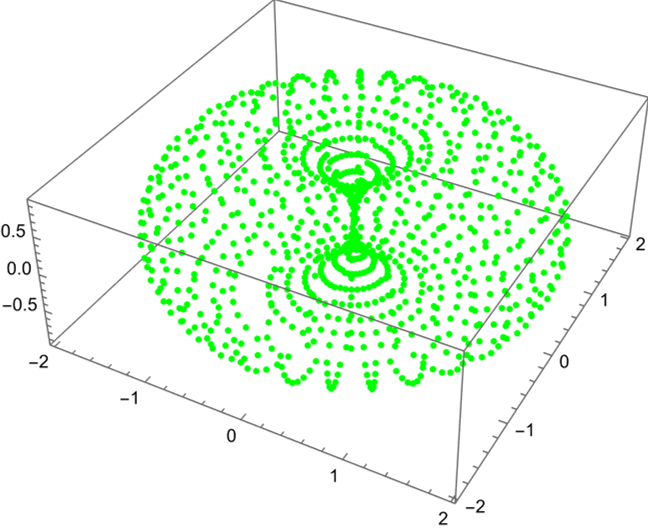}
\end{minipage}%
\caption{Plots of the fuzzy torus. Top: projective plot of $\Tilde{\mathcal{M}}$ for a scan of the hybrid leaf through $x=(1,2,1,0)$, bottom: stereographic plot of $\Tilde{\mathcal{M}}$ for a tiling of the hybrid leaf through $x=(1,2,1,0)$; left to right: $N=3,5,10,100$}
\label{fig:CS/4}
\end{figure}
\newpage
\section{Conclusion}

This work implements the framework introduced in \cite{Steinacker_2021} on a computer and makes it possible to  locally and globally visualize the so called quantum manifold that is associated to an arbitrary matrix configuration. Further it shows the necessity to work with foliations of this quantum manifold in order to maintain stability under perturbations and to produce meaningful results. For these, various approaches were discussed and compared for different examples, making it possible to construct coverings with local coordinates and to integrate over the leaves, allowing for quantitative verification.\\
Also analytical results are featured in this thesis, including the proof that the quantum manifold is a well defined smooth manifold as well as various perspectives on the Hermitian form $h_{ab}$ and
the perturbative calculations for the squashed fuzzy sphere.

We have seen in section \ref{algTricks} that $(\partial_a-iA_a)\ket{x}$ (and consequently also $h_{ab}$) can be calculated purely algebraically using the refined equation (\ref{AlgTrick}) while equation (\ref{HStraightLines}) showed that the $\mathcal{N}_x$ are convex in $\Tilde{\mathbb{R}}^D$.\\
In the sections \ref{bp} and \ref{mfp}, interwoven with appendix \ref{AppendixA}, we have seen plenty of different perspectives on the Hermitian form $h_{ab}$ and consequently the quantum metric $g_{ab}$ and the would-be symplectic form $\omega_{ab}$: While the perspective on $\omega_{ab}$ as the field strength of a principal connection respectively as the pullback of the Kirillov-Kostant-Souriau symplectic form and the quantum metric as the pullback of the Fubini-Study metric have already been discussed in \cite{Steinacker_2021}, the perspective as the pullback of a bundle metric on the bundle $T\mathcal{B}$ is new. On the other hand the convexity of the $\mathcal{N}_x$ is crucial in the proof that $\mathcal{M}$ is an immersed submanifold of $\mathbb{C}P^{N-1}$.\\
The perturbative calculations allowed for a mutual verification of themselves and later numerical computations and made it possible to analytically witness the phenomenon called \textit{oxidation}, making the use of foliations inevitable.
\\
We have seen four different approaches to foliations of $\mathcal{M}$ in section \ref{fol}, where the hybrid leaf (coming in two different flavors) showed to be the most robust and easiest method to calculate.
Concerning deformed quantum geometries, it allows one to calculate the effective dimension of $\mathcal{M}$ that always agreed with the unperturbed case. It is further the basis for many numerical results.

In section \ref{implementation} algorithms for numerical computations are described. This includes methods for the calculation and analyzation of the quasi-coherent states, the quantum metric $g_{ab}$ and the would-be symplectic form $\omega_{ab}$ as well as the visualization of $\mathcal{M}$ and $\Tilde{\mathcal{M}}$.\\
Further the integration of curves in the leaves and the construction of local coordinates and coverings with coordinates for a given leaf -- finally allowing numerical integration over the latter -- were implemented.

The explicit results in section \ref{RESULTS} show that the framework introduced in \cite{Steinacker_2021}, refined with foliations of $\mathcal{M}$, generates a meaningful semiclassical limit for matrix configurations that are not too far away from well known examples, where the perturbations may still be significant. Further, we have seen visually and quantitatively how robust these methods are against such perturbations.
\\
For the squashed fuzzy sphere we could observe the oxidation of $\mathcal{M}$ both graphically and via the computation of various ranks. We further saw how well the numerical methods work, especially to generate a covering with coordinates of the leaf $\mathcal{L}$, which turned out to be (at least approximately) well defined. The verification of the completeness relation and similar properties showed a good quality for small and medium perturbations with a tendency to reach the quality of the round case in the large $N$ limit.
\\
For the (with random matrices) perturbed fuzzy sphere we got comparable results, accompanied by intuitive pictures that come with a nice interpretation via gauge fields. However, some questions remain concerning the large $N$ limit.
\\
The (completely) squashed fuzzy $\mathbb{C}P^2$ showed to behave very similar in principle, while the computational demand increased significantly. The visual results are well comparable to the results in \cite{Schneiderbauer_2015,Schneiderbauer_2016}. Due to the lacking computational power, the quantitative results that depend on the integration over the leaf showed to be of too bad numerical quality in order to be of great significance.
\\
The fuzzy torus turned out to be an example where the methods apply extraordinarily well. Due to the foliations, we could recognize the torus in $\mathcal{L}$ already for $N=5$. Without foliations this was only possible in the large $N$ limit.
\\
In general, the results suggest that the completeness relation (\ref{compl}), the recovery of the matrix configuration $X^a$ as the quantization of the $\mathbf{x}^a$ (\ref{quant}) and the compatibility of the two different Poisson structures (\ref{CompPoisson}) hold approximately for perturbed quantum spaces, with a tendency to improve for large $N$.
Further, the results suggest that equation (\ref{CompPoisson}) reformulates results from \cite{Steinacker_2021} (discussed in section \ref{QuantMap}), such that they also hold away from local minima of $\lambda$. 
\\
This supports the assumption that the quantization map (\ref{quantmapCS}) fulfills the axioms from section \ref{QG}, at least in the large $N$ limit.

Yet, a few open tasks and questions remain. 
\\
First, some analytic results on the integrability of the distributions that define the leaves would be desirable.
Further, the suggested method to select a preferred leaf $\mathcal{L}$ did not prove very successful, thus the definition of such a choice remains an open task.
\\
On the other hand, it could be beneficial to look at more examples, including both higher $D$ and $N$. For that, clearly, more computational power is needed. Concerning the perturbed fuzzy sphere, it would be interesting to investigate the consequences when restricting the random matrices to modes of order smaller or equal $\mathcal{O}(\sqrt{N})$ as discussed in section \ref{fsr}.
\\
Finally, one could also think of a generalization of the framework and the implementation to more general target spaces, for example Minkowski space.

This thesis refined the construction of a semiclassical limit from \cite{Steinacker_2021} and showed that produced results are meaningful, stable and numerical accessible for arbitrary matrix configurations.

\newpage

\phantomsection
\addcontentsline{toc}{section}{References}
\printbibliography

\appendix

\newpage
\section{The \texorpdfstring{$U(1)$}{U(1)} Bundle and the Quantum Manifold}
\label{AppendixA}

In this appendix the details behind the bundle and manifold structure (that come with the quasi-coherent states), for which there was no room in section \ref{QMGsDescription}, are discussed.
\\
As the smooth dependence of the eigensystem of the Hamiltonian $H_x$ on the point $x$ is essential, an important result on the analytic parameter dependence of the eigensystem is reviewed in section \ref{Appendix:smoothdependence}.
\\
Based on that, in section \ref{Appendix:principalbundle} the naturally emerging $U(1)$ bundle $\mathcal{B}$ together with its natural principal connection is discussed,
followed by an alternative view on the quantum metric and the would-be symplectic form in section \ref{Appendix:mom}.\\
Finally, in section \ref{Appendix:manifold} it is shown that $\mathcal{M}$ is a smooth manifold.

\subsection{Analytic Parameter Dependence of the Eigensystem}
\label{Appendix:smoothdependence}

In the following, we need an important result on the analyticity of the eigensystem that follows from \cite{Kurdyka_2006}, noting that we only consider neighborhoods where the eigenvalue of interest is strictly separated from the others.

\textbf{Theorem:} Let $A:U\subset \mathbb{R}^n\to\operatorname{Herm}(\mathcal{H})$ be an analytic function from an open subset $U$ of $\mathbb{R}^n$ into the set of hermitian operators on a finite dimensional Hilbert space $\mathcal{H}$. Let $x\in U$ and $\lambda_x$ be an eigenvalue of $A(x)$ of multiplicity $1$ with corresponding eigenvector $v_x$.\\
Then, there exists an open neighborhood $V\subset U$ of $x$ and analytic functions $\lambda:V\to \mathbb{R}$ and $v:V\to\mathcal{H}$ such that $\lambda(x)=\lambda_x$, $v(x)=v_x$ with $\lambda(y)$ being an eigenvalue of $A(y)$ of multiplicity $1$ with corresponding eigenvector $v(y)$ $\forall y\in V$.

In our case, we deal with the analytic\footnote{The analyticity is obvious as the Hamiltonian is a polynomial in $x$.} Hamiltonian $H:\mathbb{R}^D\to\operatorname{Herm}(\mathcal{H})$ introduced in equation (\ref{Hamiltonian}), while we restrict ourselves to the subset $\Tilde{\mathbb{R}}^D:=\{x\in\mathbb{R}^D\vert \operatorname{dim}(E_x)=1\}$ (where $E_x$ is defined as the eigenspace of $H_x$ corresponding to the lowest eigenvalue $\lambda_{1,x}$).
The theorem then implies first that $\Tilde{\mathbb{R}}^D$ is open in $\mathbb{R}^D$ and second that locally around $x$ (in the notation from section \ref{IntroHamiltonian}) we can write\footnote{At first, it is not clear if the function $v(x)$ in the theorem is normalized, but, as normalizing a vector is a smooth operation, we can always choose a smooth $\Tilde{v}(x)$.} $\lambda_{1,x}=\lambda(y)$ and $\ket{1,y}=\kets{y}$ for smooth $\lambda$ and $\kets{\cdot}$.

\newpage
\subsection{The \texorpdfstring{$U(1)$}{U(1)} Bundle}
\label{Appendix:principalbundle}

We start by proofing that any matrix configuration with finite dimensional Hilbert space $\mathcal{H}$ and nonempty $\Tilde{\mathbb{R}}^D$ defines a unique $U(1)$ bundle.

\textbf{Proposition:} For a matrix configuration as defined in section \ref{QMGsDescription} (with $N$ finite), there exists a unique principal fiber bundle $p:\mathcal{B}\to\Tilde{\mathbb{R}}^D$ with fiber $U(1)$, where $\mathcal{B}:=\{(x,\ket{\psi})\in\Tilde{\mathbb{R}}^D\times\mathcal{H}\vert\;\ket{\psi}\in E_x,\braket{\psi\vert \psi}=1\}$ and $p$ is the restriction of $\operatorname{pr}_1:\Tilde{\mathbb{R}}^D\times\mathcal{H}\to\Tilde{\mathbb{R}}^D$ to $\mathcal{B}$.

\textbf{Proof:}
In the last section we have noted that $\Tilde{\mathbb{R}}^D$ is open and thus a $D$ dimensional submanifold of $\mathbb{R}^D$. There, we also saw that for any $x\in\Tilde{\mathbb{R}}^D$ exists a neighborhood $U\subset\Tilde{\mathbb{R}}^D$ and a map $\kets{\cdot}:U\to\mathcal{H}$ such that $\kets{y}\in E_y$ and $\brakets{y\vert y}=1$ for every $y\in U$.
Then, for each $\kets{\cdot}$ the map $y\mapsto(y,\kets{y})$ exactly is a candidate for a local smooth section of $\mathcal{B}$.\\
Consequently, we define the smooth map $\phi:U\times U(1)\to p^{-1}(U)\subset\mathcal{B}$ via $(y,e^{i\lambda})\mapsto (y,e^{i\lambda}\kets{y})$ (obviously being bijective) and we declare $(U,\phi^{-1})$ as a bundle chart for $\mathcal{B}$.\\
Let now $(V,\psi^{-1})$ be a chart of the same form coming from another $\ketsp{\cdot}:V\to\mathcal{H}$ with nonempty $W:=U\cap V$, providing us with the transition function $\omega:=\phi^{-1}\circ\psi:W\times U(1)\to W\times U(1)$.\\
For each $y\in W$ we find $\ketsp{y}=\braketsp{y\vert y}\kets{y}$ where $\braketsp{y\vert y}$ computes the necessary $U(1)$-phase\footnote{This for example follows from the spectral theorem in one dimension.}, thus $\omega(y,e^{i\lambda})=(y,\braketsp{y\vert y} e^{i\lambda})$ what clearly is a diffeomorphism.\\
By lemma 2.2 in \cite{Cap_2021}, $\mathcal{B}$ is a fiber bundle with standard fiber $U(1)$ and by the explicit form of the chart change we constructed a principal bundle atlas, making $\mathcal{B}$ into a principal fiber bundle. $\square$

Using the metric structure of the standard fiber, we now want to fix a natural connection on the bundle.\\
Since $\mathcal{B}\subset \Tilde{\mathbb{R}}^D\times\mathcal{H}$, we get a natural inclusion $T\mathcal{B}\hookrightarrow T\Tilde{\mathbb{R}}^D\times T\mathcal{H}\cong T\Tilde{\mathbb{R}}^D\times \mathcal{H}\times\mathcal{H}$.
In terms of local curves, we get an intuitive description of the tangent bundle of $\mathcal{B}$.
Therefore, for any manifold $\mathcal{M}$ we introduce the set $\Gamma_{x}(\mathcal{M})$ that consists of all smooth curves $c:[-\epsilon,\epsilon']\to\mathcal{M}$ for some $\epsilon,\epsilon'>0$ with $c(0)=x$.
Then we can write any tangent space of $\mathcal{B}$ as
\begin{align*}
    T_{(x,\ket{\psi})}\mathcal{B}\cong\{\at{\frac{d}{dt}}{t=0}(c(t),\ket{\rho(t)})\vert\; c\in\Gamma_x(\Tilde{\mathbb{R}}^D),\; \ket{\rho}\in\Gamma_{\ket{\psi}}(\mathcal{H}),\; \braket{\rho(t)\vert \rho(t)}=1,\; \ket{\rho(t)}\in E_{c(t)} \}.
\end{align*}
Considering the identification $T\Tilde{\mathbb{R}}^D\times T\mathcal{H}\cong T\Tilde{\mathbb{R}}^D\times \mathcal{H}\times\mathcal{H}$, we write $\at{\frac{d}{dt}}{t=0}(c(t),\ket{\rho(t)})\cong (\xi_x,\ket{\psi},\ket{v})$, where $\xi_x:=(x,\at{\frac{d}{dt}}{t=0}c(t))\in T_x\Tilde{\mathbb{R}}^D$ and $\ket{v}:=\at{\frac{d}{dt}}{t=0}\ket{\rho(t)}\in\mathcal{H}$.\\
A small side remark:
Fixing a local section $\kets{\cdot}$ such that $\kets{x}=\ket{\psi}$, we find the parameterization $\ket{\rho(t)}=e^{i\phi(t)}\kets{c(t)}$ for a smooth function $\phi:[-\epsilon,\epsilon']\to\mathbb{R}$, thus $\at{\frac{d}{dt}}{t=0}(c(t),e^{i\phi(t)}\kets{c(t)})\cong(\xi_x,\ket{\psi},(d\ket{x}_s)(\xi_x)+i\alpha\ket{\psi})$, where $\alpha:=\at{\frac{d}{dt}}{t=0}\phi(t)$. Mapping $(\xi_,\ket{\psi},(d\kets{x})(\xi_x)+i\alpha\ket{\psi})\mapsto(\xi_x,i\alpha)$ thus locally trivializes $T\mathcal{B}$. This corresponds to the local trivialization of $\mathcal{B}$ induced by any local section $\kets{\cdot}$.
\\
Based on this result, we get a simple definition of a principal connection on $\mathcal{B}$.

\textbf{Proposition:} On $\mathcal{B}$ there is a natural principal connection, given by $H:=\{(\xi_x,\ket{\psi},\ket{v})\in T\mathcal{B}\subset T\Tilde{\mathbb{R}}^D\times \mathcal{H}\times\mathcal{H}\vert\; \braket{\psi \vert v}=0 \}$. Then the associated connection 1-form is given by $\gamma(x,\ket{\psi})(\xi_x,\ket{\psi},\ket{v})=-\braket{\psi\vert v}$, while the pullback of $\gamma$ to $\Tilde{\mathbb{R}}^D$ along a local section $\kets{\cdot}$ is given by $-iA(x):=(\ket{\cdot}_s^*\gamma)(x)=-\bra{x}d\ket{x}$, where $A$ is real.

\newpage
\textbf{Proof:}
We consider $\gamma(x,\ket{\psi})(\xi_x,\ket{\psi},\ket{v}):=-\braket{\psi\vert v}$ (if $\gamma$ shows to be a principal connection 1-from, then $H$ is exactly its kernel and thus a principal connection by theorem 3.3 in \cite{Cap_2021}).\\
Obviously, $\gamma$ is smooth and $\mathcal{C}^\infty$-linear, thus a 1-from.
For every normalized $\ket{\rho}\in\Gamma_{\ket{\psi}}(\mathcal{H})$ we have $0=\at{\frac{d}{dt}}{t=0}(\braket{\rho(t)\vert \rho(t)})=\at{\bra{\rho(t)}\frac{d}{dt}\ket{\rho(t)}}{t=0}+\at{(\bra{\rho(t)}\frac{d}{dt}\ket{\rho(t)})^*}{t=0}$, implying $\gamma$ is Lie algebra valued\footnote{The Lie group of $U(1)$ is given by $\mathfrak{u}(1)\cong i\mathbb{R}$.}.\\
For $i\alpha\in \mathfrak{u}(1)$, consider the fundamental vector field  $\zeta_{i\alpha}(x,\ket{\psi}):=\at{\frac{d}{dt}}{t=0} r^{\exp(-i\alpha t)}(x,\ket{\psi})$\\$=\at{\frac{d}{dt}}{t=0} (x,\exp(-i\alpha t)\ket{\psi})\cong(0,\ket{\psi}, -i\alpha \ket{\psi})$, thus $\gamma(x,\ket{\psi})(\zeta_{i\alpha}(x,\ket{\psi}))=i\alpha$, meaning $\gamma$ is vertical.\\
Now, $((r^g)^*\gamma)(x,\ket{\psi})(\xi_x,\ket{\psi},\ket{v})=\gamma(x,g\ket{\psi})(\xi_x,g\ket{\psi},g\ket{v})$ since $r^g$ is linear. Using the $U(1)$ invariance of the inner product, we get\\ $((r^g)^*\gamma)(x,\ket{\psi})(\xi_x,\ket{\psi},\ket{v})=\gamma(x,\ket{\psi})(\xi_x,\ket{\psi},\ket{v})=\operatorname{Ad}(g^{-1})(\gamma(x,\ket{\psi})(\xi_x,\ket{\psi},\ket{v}))$, while the second step is a consequence of the fact that the adjoint representation of any abelian Lie group is trivial. So we conclude that gamma is $U(1)$ equivariant and thus a principal connection 1-form.\\
Finally, we consider $-iA(\xi_x):=(\kets{\cdot}^*\gamma)(x)(\xi_x)=\gamma(x,\ket{x}_s)(\xi_x,T_x\kets{\cdot}\cdot \xi_x)$.
The tangent map can be calculated in terms of curves: $T_x\kets{\cdot}\cdot \xi_x=\at{\frac{d}{dt}}{t=0}\kets{c(t)}=(d\kets{x})(\xi_x)$, giving us the result $-iA(\xi_x)=\gamma(x,\kets{x})(\xi_x,\kets{x},T_x\kets{x}\cdot \xi_x)=-\bras{x}(d\kets{x})(\xi_x)=-(\bras{x}d\kets{x})(\xi_x)$. Of course, $iA$ lies in the Lie algebra too, so $A$ is real. $\square$

Now we are in possession of a principal $U(1)$ bundle, equipped with a principal connection. On the other hand, $\mathcal{H}$ has a natural $U(1)$ action, inducing an associated vector bundle $\pi: E:=P\times_{U(1)}\mathcal{H}\to \Tilde{\mathbb{R}}^D$, coming with an induced linear connection on $E$.\\
For a local section $\kets{\cdot}$ of $\mathcal{B}$ on $U\subset\Tilde{\mathbb{R}}^D$, we directly get a local trivialization of both $P$ and $E$. Here, we can view local sections $\sigma$ of $E$ as smooth functions $f:\mathcal{M}\to\mathcal{H}$.\\
Through these trivializations, we can view $\kets{\cdot}$ both as a section of $P$ as well as of $E$.
Using theorem 3.4 in \cite{Cap_2021}, one finds the simple description of the induced linear connection:
\begin{align*}
    \sigma\sim f,\quad \nabla_\xi \sigma \sim \xi[f]+(\kets{\cdot}^*\gamma)(\xi)\cdot f=df(\xi)-iA(\xi)\cdot f,
\end{align*}
where $\xi$ is a smooth local vector field, and $\cdot$ indicates the $U(1)$ action.

In Cartesian coordinates, viewing $\ket{\cdot}_s$ as a local section\footnote{In the above, this exactly means $f=\ket{\cdot}_s$.} of $E$, this reads
\begin{align*}
    iA_a(x)=\bras{x}\partial_a\kets{x},\quad D_a \kets{x}=(\mathbb{1}-\kets{x}\bras{x})\partial_a\kets{x}=(\partial_a-iA_a)\kets{x},
\end{align*}
where it is conventional to write $D_a$ (what we call \textit{gauge covariant derivative}) instead of $\nabla_{\partial_a}$ in this context.
\\
Of course, we can look at the curvature 2-form $\Omega=d\gamma+[\gamma,\gamma]=d\gamma$ (since $U(1)$ is abelian), descending to the 2-form $2\omega:=-i\kets{\cdot}^*\Omega=dA$. We call $\omega$ \textit{field strength} and $A$ \textit{gauge field}.

As a final remark, we observe that $E$ inherits a complex valued bundle metric from the inner product on $\mathcal{H}$ as the latter is invariant under $U(1)$.

\newpage
\subsection{The Quantum Metric and the Would-Be Symplectic Form}
\label{Appendix:mom}

As we have seen in the last section, we can equip $E$ with a natural $U(1)$ invariant bundle metric. Using a similar construction, we can equip $T\mathcal{B}$ with a horizontal complex valued metric $\Tilde{h}:T\mathcal{B}\times_\mathcal{B}T\mathcal{B}\to\mathcal{C}^\infty_\mathbb{C}(T\mathcal{B})$, $((\xi_x,\ket{\psi},\ket{v}),(\xi'_x,\ket{\psi'},\ket{v'}))\mapsto \braket{v^H\vert v'^H}$, where $(\xi_x,\ket{\psi},\ket{v^H})$ for $\ket{v^H}:=(\mathbb{1}-\ket{\psi}\bra{\psi})\ket{v}$ is the projection on the horizontal subbundle\footnote{That this gives the horizontal projection can be seen in various ways. The easiest is to insert what we already know in the definition $\Pi_H(\Tilde{\xi}_p)=\Tilde{\xi}_p-\zeta_{\gamma(p)(\Tilde{\xi}_p)}(p)$ for $\Tilde{\xi}_p\in T_p\mathcal{B}$, while noting that $\Pi_H$ acts trivially on the first component.}.\\
Then, the pullback along a local section $\ket{\cdot}_s$ immediately gives\\ $h(\xi,\eta):=(\kets{\cdot}^*\Tilde{h})(\xi,\eta)=(D_\xi\kets{\cdot})^\dagger D_\eta\kets{\cdot}$ as $(\kets{\cdot}_*\xi)^H=(d\kets{\cdot}(\xi))^H=(d\kets{\cdot}-\kets{\cdot}\bras{\cdot}d\kets{\cdot})(\xi)=(d\kets{\cdot}-iA\kets{\cdot})(\xi)=D_\xi \kets{\cdot}$.
In Cartesian coordinates this reads
$h_{ab}(x)=((\partial_a-iA_a)\ket{x})^\dagger(\partial_b-iA_b)\ket{x}$. By construction it is clear that $h$ is independent of the chosen section.\\
Since $\Tilde{h}$ is hermitian, the same holds for $h$ and we may decompose it into its real and imaginary part $h=g+i\omega$, where $g$ is symmetric and $\omega$ is antisymmetric. Carrying this out, one immediately verifies that the so defined $\omega$ coincides with the field strength from the last section
\begin{align*}
    h_{ab}&=\partial_a\bra{x}\partial_b\ket{x}-A_a A_b,\\
    g_{ab}&=\frac{1}{2}\left(\partial_a\bra{x}\partial_b\ket{x}+\partial_b\bra{x}\partial_a\ket{x}-2A_a A_b\right),\\
    \omega_{ab}&=\frac{1}{2i}\left(\partial_a\bra{x}\partial_b\ket{x}-\partial_b\bra{x}\partial_a\ket{x}\right)=\frac{1}{2}\left(\partial_a A_b-\partial_b A_a\right)=(\frac{1}{2}dA)_{ab}.
\end{align*}
This directly implies that $\omega$ is closed (and if further nondegenerate: symplectic). However, from this construction we can not determine the rank of neither $g$ nor $\omega$.

\subsection{The Quantum Manifold}
\label{Appendix:manifold}

Recall the definition of $k$, $\mathcal{M}$ and $q$ from section \ref{mfp}.
By construction we deal with a smooth surjection $q:\hat{\mathbb{R}}^D\to\mathcal{M}\subset\mathbb{C}P^{N-1}$ of constant rank $k$.\\
Therefore, we recall the constant rank theorem:
\begin{quote}
\textbf{Theorem:} \textit{``Suppose $M$ and $N$ are smooth manifolds of dimensions $m$ and $n$, respectively, and $F:M\to N$ is a smooth map with constant rank $r$. For each $p\in M$ there exist smooth charts $(U,\phi)$ for $M$ centered at $p$ and $(V,\psi)$ for $N$ centered at $F(p)$ such that $F(U)\subseteq V$, in which $F$ has a coordinate representation} [$\hat{F}=\psi\circ F\circ \phi^{-1}:\phi(U)\to \psi(V)$] \textit{of the form $\hat{F}(x^1,\dots,x^r,x^{r+1},\dots,x^m)=(x^1,\dots,x^r,0,\dots,0)$.''} -- theorem 4.12 in \cite{Lee_2003}.
\end{quote}

For any $x\in\hat{\mathbb{R}}^D$ the theorem guarantees the existence of corresponding charts $(U,\phi)$ for $\hat{\mathbb{R}}^D$ and $(V,\psi)$ for $\mathcal{M}$.
We will use these to define a chart for $\mathcal{M}$, but before we can do so we need two technical lemmas.

In the first lemma we construct a \textit{prototypical chart} $(\mathcal{U},\bar{\beta}^{-1})$ around $q(x)\in\mathcal{M}$, using the constant rank theorem.

\textbf{Lemma 1:}
Consider $q:\hat{\mathbb{R}}^D\to\mathcal{M}\subset\mathbb{C}P^{N-1}$ as in section \ref{mfp}. Then for each $x\in\hat{\mathbb{R}}^D$ there exists an open subset $\Bar{U}\subset \mathbb{R}^k$ and a subset $\mathcal{U}\subset \mathcal{M}$ around $0$ respectively $q(x)$ and a smooth bijection $\Bar{\beta}:\Bar{U}\to\mathcal{U}$ between them.

\textbf{Proof:}
We consider the charts and the map $q':=\psi\circ q\circ \phi^{-1}: \phi(U)\to\psi(V)$, coming from the constant rank theorem.\\
Now, we define the sets $U':=\phi(U)$, $\mathcal{U}:=q(U)$ and the smooth map $\beta':=q\circ \phi^{-1}:U'\to \mathcal{U}$, thus $\beta'$ is bijective by construction.\\
Further, $\beta'=\psi^{-1}\circ q'$ and since $\psi$ is a diffeomorphism, $\beta'$ still has constant rank $k$ and depends only on the first $k$ coordinates $\chi^1,\dots,\chi^k$ in $U'\subset\mathbb{R}^D$.\\
We define\footnote{Here, we implicitly identified $\mathbb{R}^k\cong \mathbb{R}^k\times\{0\}\subset\mathbb{R}^D$ and similarly for $\mathbb{R}^{D-k}$.} $\Bar{U}:=U'\cap \mathbb{R}^k$ and $\Bar{\beta}:=\beta'\vert_{\Bar{U}}$.\\
By shrinking\footnote{Especially such that $\Bar{U}\times \mathbb{R}^{D-k}$ contains $U'$.} $U$ we can make $\Bar{\beta}$ surjective. Since $\Bar{\beta}$ has full rank it is an immersion, thus locally injective and we can shrink $U$ once again in order to make $\Bar{\beta}$ bijective. $\square$

A little comment on the notation may be helpful. All quantities marked with a \textit{comma} are related to the space $\mathbb{R}^D$ that is the target of the coordinates $\phi$, while the quantities marked with a \textit{bar} live on the $\mathbb{R}^k\cong\mathbb{R}^k\times \{0\}\subset\mathbb{R}^D$.

Now we note that for all $y\in \bar{U}$ the restriction of $\phi^{-1}$ to $\mathcal{N}'_y:=(\{y\}\times\mathbb{R}^{D-k})\cap U'$ is a diffeomorphism from $\mathcal{N}'_y$ to $\mathcal{N}_{\phi^{-1}(y,0)}\cap U$.\\
To see this we note that $\phi^{-1}(\mathcal{N}'_y)$ lies within $\mathcal{N}_{\phi^{-1}(y,0)}\cap U$ as $\beta'$ is constant here. Further $q'$ is injective on $U'\cap\mathbb{R}^k$, thus only points in $\mathcal{N}'_y$ are mapped to $\mathcal{N}_{\phi^{-1}(y,0)}\cap U$.

For any $z\in\hat{\mathbb{R}}^D$ this immediately implies $\dim(T_z\mathcal{N}_z)=D-k$, while obviously $T_z\mathcal{N}_z\subset \operatorname{ket}(T_zq)$. Thus, $T_z\mathcal{N}_z=\operatorname{ket}(T_zq)$ what in turn implies that the whole $\mathcal{N}_z$ lies within $\hat{\mathbb{R}}^D$.

This leads us to the following lemma that shows that $\mathcal{N}_W:=\cup_{x\in W}\mathcal{N}_x$ is open in $\hat{\mathbb{R}}^D$ if $W$ is. $\mathcal{N}_W$ is called \textit{saturation} of $W$ under the equivalence relation $\sim$ from section \ref{mfp}.

\textbf{Lemma 2:}
Let $W\subset\hat{\mathbb{R}}^D$. If $W$ is open in $\hat{\mathbb{R}}^D$, then also $\mathcal{N}_W$ is open in $\hat{\mathbb{R}}^D$.

\textbf{Proof:} Recall the setup from the proof of lemma 1. Since $\mathcal{N}_W\cup\mathcal{N}_X=\mathcal{N}_{W\cup X}$, it suffices to show the claim for $W\subset U$.
We define $W':=\phi(W)\subset U'$ what is clearly open. By the above we find $\mathcal{N}_{W}\cap U=\phi^{-1}((W'+\{0\}\times \mathbb{R}^{D-k})\cap U')$. Since $\phi$ is a diffeomorphism, this shows that $\mathcal{N}_{W}\cap U$ is open.\\
Consider now a point $y\in \mathcal{N}_W$. By construction there is a $y'\in W$ such that $y\in\mathcal{N}_{y'}$. Since $\mathcal{N}_{y'}$ is convex, it contains the straight line segment that joins $y$ with $y'$.\\
Then, we can pick \textit{ordered} points $y_\alpha$ for $\alpha=0,\dots,n$ on this line segment with $y_0=y'$ and $y_n=y$ such that the corresponding $U_\alpha$ (which we get in the proof of lemma 1 for the point $y_\alpha$) cover the line segment (w.l.o.g. we have $U_\alpha\cap U_\beta=\emptyset$ if $\vert \alpha-\beta\vert>1$).
For $\alpha>0$ we inductively define $W_{\alpha}:=\mathcal{N}_{W_{\alpha-1}}\cap U_{\alpha-1}\cap U_{\alpha}\subset \mathcal{N}_W$ with $W_0:=W$.\\
Repeating the above discussion, all $W_\alpha$ are open. By construction, all $W_\alpha$ contain $\mathcal{N}_{y'}\cap U_{\alpha-1}\cap U_{\alpha}$ and are thus nonempty. So finally, $y\in \mathcal{N}_{ W_n}\cap U_n \subset\mathcal{N}_{W}$.\\
This means that every point $y\in\mathcal{N}_{W}$ has an open neighborhood in $\mathcal{N}_{W}$. $\square$

This result is crucial and allows us to prove that $\mathcal{M}$ is a smooth manifold.

\textbf{Proposition:}
$\mathcal{M}$ as defined in section \ref{mfp} is a smooth manifold of dimension $k$.

\newpage
\textbf{Proof:}
By lemma 1 we can cover $\mathcal{M}$ with prototypical charts $(\mathcal{U}_\alpha,\Bar{\beta}_\alpha^{-1})$ around $q(x_\alpha)$ for appropriate points $x_\alpha$, defining a \textit{prototypical atlas}.\\
Consider now two charts with indices $\beta,\gamma$ such that $\mathcal{U}_{\beta \gamma}:=\mathcal{U}_\beta\cap \mathcal{U}_\gamma\neq \emptyset$.
That the corresponding chart changes are smooth is obvious, but it remains to show that the charts consistently define a topology on $\mathcal{M}$. Then by lemma 1.35 in \cite{Lee_2003} $\mathcal{M}$ is a smooth manifold of dimension $k$.\\
Especially, this means that we have to show that (w.l.o.g.) $\bar{U}_{\beta \gamma}:=\Bar{\beta}_\beta^{-1}(\mathcal{U}_{\beta\gamma})$ is open in $\mathbb{R}^k$.\\
Since $\bar{U}_{\beta\gamma}\subset \bar{U}_\beta$, we only have to show that $\bar{U}_{\beta\gamma}$ is open in the latter. This is equivalent to the statement that there is no sequence $\bar{U}_\beta\setminus \bar{U}_{\beta\gamma}$ that converges against a point in $\bar{U}_{\beta\gamma}$.\\
Assume to the contrary that such a sequence $(y_a)$ exists. Since $\phi_\beta$ is a diffeomorphism, this defines the convergent sequence $(x_a):=(\phi_\beta^{-1}(y_a))$ in $\hat{\mathbb{R}}^D$ with $\lim_{a\to \infty}x_a=:x'_\beta$. By assumption, $q(x'_\beta)\in \mathcal{U}_{\beta\gamma}$, thus, there is an $x'_\gamma\in U_\gamma$ with $q(x'_\beta)=q(x'_\gamma)$. But then by definition, $x'_\beta\in \mathcal{N}_{x'_\gamma}$ and consequently $x'_\beta\in \mathcal{N}_{U_\gamma}$. Since the latter is open by lemma 2, this implies that some $x_{a_i}$ lie within $\mathcal{N}_{U_\gamma}$. But for these, we find $q(x_{a_i})\in \mathcal{U}_{\beta\gamma}$ and consequently $y_{a_i}\in \bar{U}_{\beta\gamma}$ what contradicts the assumption. $\square$

As the rank of $q$ equals $k$ this shows that the inclusion $\mathcal{M}\hookrightarrow \mathbb{C}P^{N-1}$ is an immersion and thus $\mathcal{M}$ is an immersed submanifold of the latter.\\
Further, the topology that we defined on $\mathcal{M}$ via the atlas is exactly the topology that we get when declaring the bijection $\underline{q}:\hat{\mathbb{R}}^D/\sim\to\mathcal{M}$ from section \ref{mfp} as a \textit{homeomorphism} when considering the quotient topology on $\hat{\mathbb{R}}^D/\sim$.
In fact, lemma 2 exactly shows that $q:\hat{\mathbb{R}}^D\to \mathcal{M}$ is a quotient map for this topology.

If we identify $\mathbb{C}P^{N-1}\cong S^{2N-1}/U(1)$, where $S^{2N-1}\cong\{\ket{\psi}\in\mathbb{C}^N\vert\; \braket{\psi\vert \psi}=1\}$, we also get a natural identification of the tangent spaces\footnote{The tangent space of $S^{2N-1}$ at $\ket{\psi}$ is given by all vectors $\ket{v}\in\mathbb{C}^N$ that satisfy the constraint $\operatorname{Re}(\braket{\psi\vert v})=0$. In the quotient this means that the constraint has to be satisfied for all $U(1)\ket{\psi}$, meaning $\braket{\psi\vert v}=0$.} $T_{U(1)\ket{\psi}} \mathbb{C}P^{N-1}\cong \{\ket{v}\in\mathbb{C}^N\vert\; \braket{\psi\vert v}=0\}$.\\
On the other hand, complex projective space can be viewed as the natural fiber bundle $p:\mathbb{C}^N\setminus \{0\}\to\mathbb{C}P^{N-1}$, where in our identification the bundle projection $p$ maps $\ket{\psi}$ to $U(1)\ket{\psi}/\sqrt{\braket{\psi\vert \psi}}$, coming with the tangent map $T_{\ket{\psi}}p=(\mathbb{1}-\ket{\psi}\bra{\psi})$.\\
Then, we get an induced hermitian form $\hat{h}$ on $T\mathbb{C}P^{N-1}$ via $\hat{h}(U(1)\ket{\psi})(\ket{v},\ket{w})=\braket{v\vert w}$. By construction, its real part is proportional to the Fubini-Study metric and its imaginary part to the canonical symplectic form on $\mathbb{C}P^{N-1}$, the Kirillov-Kostant-Souriau symplectic form \cite{Kostant_1982}.\\
Since the inclusion $j:\mathcal{M}\hookrightarrow\mathbb{C}P^{N-1}$ is a smooth immersion, we can pull back $\hat{h}$ to $\mathcal{M}$ what we call $h_\mathcal{M}=g_\mathcal{M}+i\omega_\mathcal{M}$, providing us with a metric and an (in general degenerate but closed) would-be symplectic form. Especially, we find $h_{ab}(x):=(q^*h_\mathcal{M})(x)(\partial_a,\partial_b)= h_\mathcal{M}(q(x))(T_xq\cdot\partial_a,T_xq\cdot\partial_b)=h_\mathcal{M}(q(x))(T_{\kets{x}}p\cdot\partial_a\kets{x},T_{\kets{x}}p\cdot\partial_b\kets{x})=h_\mathcal{M}(q(x))(D_a\kets{x},D_b\kets{x})=\hat{h}(U(1)\kets{x})(D_a\kets{x},D_b\kets{x})$\\$=(D_a\kets{x})^\dagger D_b\kets{x}$, where we described $q=p\circ\kets{\cdot}$ via a local section and used $D_a\kets{x}=(\mathbb{1}-\kets{x}\bras{x})\partial_a\kets{x}=T_{\kets{x}}p\cdot\partial_a\kets{x}$.\\
But this means that the pullback of $h_\mathcal{M}$ along $q$ agrees with the $h_{ab}$ we have defined earlier, while the same holds for $g_\mathcal{M}$ and $\omega_\mathcal{M}$.

Since we now know that $\mathcal{M}$ is a manifold, locally we can get even simpler coordinates. For each point $x_0$ in $\hat{\mathbb{R}}^D$, we can find $k$ distinct indices $\mu$ such that $\partial_\mu$ is \textit{not} in the kernel of $T_xq$. But then we can find an open neighbourhood in $\mathbb{R}^k$ where the map $(\chi^\mu)\mapsto x_0+\sum_\mu \hat{e}_\mu \chi^\mu$ is bijective, providing us with a chart for $\mathcal{M}$. The pullbacks of $g_\mathcal{M}$ and $\omega_\mathcal{M}$ are then simply given by the submatrices of $(g_{ab})$ and $(\omega_{ab})$ corresponding to the selected indices.\\
Since we know that $g$ agrees up to a scale with the pullback of the Fubini-Study metric, $(g_{\mu\nu})$ cannot be degenerate, while we have no result for $(\omega_{\mu\nu})$.

\section{The Relevant Representations}
\label{AppendixB}

In this appendix we want to discuss the construction of the $\mathfrak{su}(2)$ and $\mathfrak{su}(3)$ Lie algebra generators in a given irreducible representation as well as the clock and shift matrices, where $SU(n)$ is the $n^2-1$ dimensional Lie group of unit determinant unitary matrices acting on $\mathbb{C}^n$ and $\mathfrak{su}(n)$ is the associated Lie algebra consisting of the traceless Hermitian matrices acting on $\mathbb{C}^n$.

\subsection{The Representations of \texorpdfstring{$\mathfrak{su}(2)$}{su(2)}}
\label{su2rep}

Let $J^1,J^2,J^3$ be an orthonormal basis of the Lie algebra $\mathfrak{su}(2)$ with respect to the Killing form.
Then we know their commutation relations
\begin{align}
    [J^a,J^b]=i\epsilon^{abc}J^c.
\end{align}
$SU(2)$ has exactly one nontrivial irreducible representations of dimension $N$ for every $N>1$, where we have the quadratic Casimir
\begin{align}
    J^2:=\sum_aJ^aJ^a=\frac{N^2-1}{4}\mathbb{1}:=j(j+1)\mathbb{1}=:C_N^2\mathbb{1}.
\end{align}
We can then isometrically identify the representation with $\mathbb{C}^N$, especially such that $J^3$ acts diagonally on the standard basis.
\\
We can even do more: It turns out that $J^3$ has the eigenvalues $-j,-j+1,\dots,j-1,j$ and we can implement $J^3$ as the diagonal matrix $\operatorname{diag}(-j,-j+1,\dots,j-1,j)$, while we introduce the notation $\ket{k}:=\hat{e}_{k+(j+1)}$.
So it remains to find $J^1$ and $J^2$ explicitly.
For that reason, we define $J^\pm:=J^1\pm J2$, implying the commutation relations
\begin{align}
    [J^3,J^\pm]=\pm J^\pm.
\end{align}
Thus, the $J^\pm$ raise respectively lower the eigenvalues of $J^3$ by $1$ in the eigenbasis.
But this means nothing more than $J^\pm\ket{k}\propto \ket{k+1}$, while the proportionality factors turn out to be
\begin{align}
    J^\pm\ket{k}=\sqrt{j(j+1)-k(k\pm 1)}\ket{k\pm 1}.
\end{align}
This means that $J^+$ acts as the matrix with the only non zero components being\\ $(\sqrt{j(j+1)-k(k+1)})$ for $k=-j,\dots,j-1$ on the first diagonal below the main diagonal and\\ $J^-$ with $(\sqrt{j(j+1)-k(k-1)})$ for $k=-j+1,\dots j$ on the first diagonal above the main diagonal.\\
Finally, we recover the matrices $J^1=\frac{1}{2}(J^++J^-)$ and $J^2=\frac{1}{2i}(J^+-J^-)$ \cite{Georgi_1982}.

We also note that
\begin{align}
    j=\frac{N-1}{2}
\end{align}
respectively
\begin{align}
    N=2j+1
\end{align}
and write for the explicit matrices in a chosen representation $J_N^a$ for $a=1,2,3$.

\subsection{The Representations of \texorpdfstring{$\mathfrak{su}(3)$}{su(3)}}
\label{su3rep}

For $SU(3)$ we pick eight matrices $T^a$ that form an orthonormal basis of the Lie algebra $\mathfrak{su}(3)$, again with respect to the Killing form.
These matrices satisfy the commutation relations
\begin{align}
    [T^a,T^b]=if^{abc}T^c
\end{align}
for well known coefficients $f^{abc}$ that are completely antisymmetric.
\\
Here, we have the two Cartan generators $T^3$ and $T^8$ that mutually commute.
\\
Then for each pair $(p,q)$ where $p,q\geq 0$, there is a unique irreducible representation of dimension $N=\frac{1}{2}(p+1)(q+1)(p+q+2)$, coming with the quadratic Casimir
\begin{align}
    \sum_a T^aT^a=\frac{1}{3}\left(p^2+q^2+3p+3q+pq\right)\mathbb{1}.
\end{align}

Further, the eigenvalues of $T^3$ and $T^8$ are well known and the representation can be identified with $\mathbb{C}^N$ \cite{Georgi_1982}.
The remaining task is to explicitly calculate the components of the matrices as which the $T^a$ act in an appropriate basis.\\
An algorithm for that together with an actual implementation for Mathematica can be found in \cite{Shurtleff_2009}.

We will only need the special case where $(p,q)=(n,0)$. Here the formulae reduce to
\begin{align}
    N=\frac{1}{2}(n+1)(n+2)
\end{align}
and
\begin{align}
    \sum_a T^aT^a=\frac{1}{3}(n^2+3n)\mathbb{1}:=C_n^2\mathbb{1}
\end{align}
and we write $T^a_n$ for the explicit matrices.

\subsection{The Clock and Shift Matrices}
\label{tor2rep}

Now we come to the clock and shift matrices. For a given $N>1$ we define
\begin{align}
    q:=\exp(2\pi i\frac{1}{N}),
\end{align}
and the two matrices
\begin{align}
    U_{ij}=\delta_{i,j+1} \quad \text{and} \quad V_{ij}=\delta_{ij}q^{i-1}
\end{align}
acting on $\mathbb{C}^N$.\\
These objects satisfy the relations
\begin{align}
    q^N=1,\quad U^N=\mathbb{1}=V^N \quad \text{respectively} \quad U\cdot V=qV\cdot U.
\end{align}

This construction is called \textit{clock and shift algebra} \cite{Schneiderbauer_2016}.

\section{The Perturbative Approach for the Squashed Fuzzy Sphere}
\label{Appendix:perturbativeappr}

In this appendix the perturbative approach to the calculation of the quasi-coherent states of the squashed fuzzy sphere for arbitrary $N$ is discussed.\\
We begin by replacing the squashing parameter $\alpha$ with  $1-\epsilon$ for a small $\epsilon>0$, leading to the matrix configuration $(X^1,X^2,(1-\epsilon) X^3)$, where the $X^a=\frac{1}{C_N}J^a$ are the matrices\footnote{In section \ref{SFuzzySphere} we wrote $\bar{X}^a$ for them, but here it is advantageous to stick to the notation $X^a$.} introduced in section \ref{FuzzySphere} respectively appendix \ref{su2rep}.\\
This leads to the Hamiltonian
\begin{align}
    H(x)=H_0(x)+\epsilon V(x), \quad \text{for} \quad V(x):=\frac{\epsilon-2}{2}\left(X^3\right)^2+x^3X^3,
\end{align}
where
\begin{align}
    H_0(x)=\frac{1}{2}\sum_a (X^a-x^a\mathbb{1})^2
\end{align}
is the Hamiltonian of the round fuzzy sphere, while the different notation with respect to section \ref{QMGsDescription} has only been adapted for better readability.
Here, we might drop the $\epsilon$ contribution to $V(x)$ directly.\\
Our task is to calculate the eigensystem of $H(x)$, especially $H(x)\ket{k,x}=\lambda_{k,x}\ket{k,x}$ for\footnote{Note that also here the notation is different from section \ref{QMGsDescription}. The labeling $k$ of the eigensystem is not such that $\lambda_{k,x}\leq\lambda_{k',x}$ for $k<k'$ but rather corresponds to the eigenvalues of $J^3\ket{k}=k\ket{k}$.} $k=-j,\dots,j$.
Thus, we expand the eigenvectors and eigenvalues in powers of $\epsilon$, meaning $\ket{k,x}=\ket{k,x,0}+\epsilon\ket{k,x,1}+\mathcal{O}(\epsilon^2)$ and $\lambda_{k,x}=\lambda_{k,x,0}+\epsilon\lambda_{k,x,1}+\mathcal{O}(\epsilon^2)$, where $H_0(x)\ket{k,x,0}=\lambda_{k,x,0}\ket{k,x,0}$.

From perturbation theory we know that we find
\begin{align}
\label{pertform}
    \lambda_{k,x,1}=\bra{k,x,0}V(x)\ket{k,x,0} \text{ and } \ket{k,x,1}=\sum_{l\neq k}\frac{\bra{l,x,0}V(x)\ket{k,x,0}}{\lambda_{k,x,0}-\lambda_{l,x,0}}\ket{l,x,1}
\end{align}
(see for example \cite{Dick_2012}),
thus we start with the calculation of the unperturbed eigensystem $\lambda_{k,x,0}$ and $\ket{k,x,0}$.

\subsection{The Unperturbed Eigensystem}

Any given $y\neq 0$ we can write as $y^a=rn^a$ where $r=\vert y\vert\in\mathbb{R}^+_0$ and $n=y/r\in S^2\subset\mathbb{R}^3$. Then
the $\mathfrak{su}(2)$ commutation relations imply $\operatorname{ad}(rn^aJ^a)(J^b)=rn^a[J^a,J^b]=irn^a\epsilon^{abc}J^c=(\operatorname{ad}(rn^a))^{bc}J^c$, thus the matrix form in the basis $J^a$ is
\begin{align}
    \operatorname{ad}(rn^aJ^a)=ir
\begin{pmatrix}
0 & n^2 & -n^2 \\ 
-n^3 & 0 & n^1 \\ 
n^2 & -n^1 & 0 \\ 
\end{pmatrix}=:irV_n.
\end{align}
One directly verifies that $(V_n)^3=-V_n$.

This can be used to calculate
\begin{align}
\label{AdAct1}
    e^{irn^bJb}J^ae^{-irn^bJ^b}=&\operatorname{Ad}\left(e^{irn^bJ^b}\right)(J^a)=e^{\operatorname{ad}(irn^bJ^b)}J^a=\left(\sum_{k=0}^\infty\frac{1}{k!}(-rV_n)^k\right)^{ab}J^b\\\nonumber
    =&\left(\mathbb{1}-\left(r-\frac{1}{3!}r^3+\frac{1}{5!}r^5\mp\ldots\right)V_n-\left(-1+1-\frac{1}{2}r^2+\frac{1}{4^1}r^4\mp\ldots\right)V_n^2\right)^{ab}J^b\\\nonumber
    =&\left(\mathbb{1}-\sin(r)V_n+(1-\cos(r))V_n^2\right)^{ab}J^b\\\nonumber
    =&\left(\delta^{ab}-\sin(r)n^c\epsilon^{cab}+(1-\cos(r))\left(n^an^b-\delta^{ab}\right)\right)J^b\\\nonumber
    =&J^a\cos(r)+\epsilon^{abc}n^bJ^c\sin(r)+n^an^bJ^b(1-\cos(r)).
\end{align}
This is very parallel to the derivation of the Rodrigues rotation formula (see for example \cite{Dai_2015}).
Once again, one directly verifies that for an $(m^a)\in S^2$
\begin{align}
    r_m:=\arccos(m^3) \text{ and } \left(n_m^a\right):=\frac{1}{\sqrt{(m^1)^2+(m^2)^2}}\begin{pmatrix}
m^2 \\ 
-m^1 \\ 
0 \\ 
\end{pmatrix}
\end{align}
rotates $J^3$ to $m^aJ^a$, meaning $e^{ir_mn_m^bJ^b}J^3e^{-ir_mn_m^bJ^b}=m^aJ^a$ (special care has to be taken at the poles).

We write this as $U_mJ^3U_m^\dagger=m^aJ^a$ for $U_m:=e^{ir_mn_m^bJb}$. In the eigensystem of $J^3$ (defined as $J^3\ket{k}=k\ket{k}$ for $k=-j,\dots,j$) we thus find
\begin{align}
    H_0(x)U_m\ket{k}=&\frac{1}{2}(1+\vert x\vert^2)U_m\ket{k}-\vert x\vert /\sqrt{C_N^2}m^aJ^aU_m\ket{k}\\\nonumber
    =&\frac{1}{2}(1+\vert x\vert^2)U_m\ket{k}-\vert x\vert /\sqrt{C_N^2}U_mJ^3\ket{k}=\left(\frac{1}{2}(1+\vert x\vert^2)-\vert x\vert k/\sqrt{C_N^2}\right)U_m\ket{k},
\end{align}
so $\ket{k,x,0}=U_m\ket{k}$ (where $m=x/\vert x\vert$) with the corresponding eigenvalue $\lambda_{k,x,0}=\frac{1}{2}(1+\vert x\vert^2)-\vert x\vert k/\sqrt{C_N^2}$.\\
We further conclude that we find the lowest eigenvalue for $k=j$ and thus $\lambda(x)=\lambda_{j,x}$ and $\ket{x}=\ket{j,x}$.

\subsection{The First Correction}

The next step is to calculate the first correction to the eigensystem, meaning $\lambda_{k,x,1}$ and $\ket{k,x,1}$.

For $(n^a)\in \mathbb{R}^3$ we define $n^\pm:=\frac{1}{2}(n^1\pm in^2)$. Recalling the results from section \ref{su2rep}, this allows us to rewrite
\begin{align}
\label{splitaction}
    n^aJ^a\ket{k}=\left(n^3 J^3+n^-J^++n^+J^-\right)\ket{k}=n^3C^3_k\ket{k}+n^-C^+_k\ket{k+1}+n^+C^-_k\ket{k-1},
\end{align}
where we defined $C^3_k:=k$ and $C^\pm_k:=\sqrt{j(j+1)-k(k\pm1)}$.\\
Using these observations, we can explicitly calculate the transition amplitudes
\begin{align}
    \bra{l}n^a J^a\ket{k}=&n^3C^3_k\delta^{lk}+n^-C^+_k\delta^{l-1,k}+n^+C^-_k\delta^{l+1,k},\\\nonumber
    \bra{l}n^a J^am^bJ^b\ket{k}=&\left(m^3C^3_kn^3C^3_k+m^-C^+_km^+C^-_{k+1}+m^+C^-_kn^-C^+_{k-1}\right)\delta^{lk}+\\\nonumber
    &+\left(m^3C^3_kn^-C^+_k+m^-C^+_kn^3C^3_{k+1}\right)\delta^{l-1,k}+\left(m^3C^3_kn^+C^-_k+m^+C^-_kn^3C^3_{k-1}\right)\delta^{l+1,k}+\\\nonumber
    &+m^-C^+_kn^-C^+_{k+1}\delta^{l-2,k}+m^+C^-_kn^+C^-_{k-1}\delta^{l+2,k}.
\end{align}

Reconsidering equation (\ref{AdAct1}), we note that $U_m^\dagger J^3U_m=\underline{m}^aJ^a$, where we defined
$\underline{m}^{1,2}:=-m^{1,2}$ and $\underline{m}^3:=m^3$ (this follows directly from $U_{\underline{m}}=U_m^\dagger$, what can be checked easily). Of course this also implies $\underline{m}^\pm=-m^\pm$.

Based on that, we find
\begin{align}
    \bra{k,x,0}V(x)\ket{k,x,0}=&\bra{k}U_M^\dagger\left(\frac{\epsilon-2}{2C_N^2}(J^3)^2+\frac{x^3}{\sqrt{C_N^2}}J^3\right)U_m\ket{k}\\\nonumber
    =&\frac{\epsilon-2}{2C_N^2}\bra{k}\underline{m}^aJ^a\underline{m}^bJ^b\ket{k}+\frac{x^3}{\sqrt{C_N^2}}\bra{k}\underline{m}^aJ^a\ket{k}\\\nonumber
    =&\frac{\epsilon-2}{2C_N^2}(\underline{m}^3C^3_k\underline{m}^3C^3_k+\underline{m}^-C^+_k\underline{m}^+C^-_{k+1}+\underline{m}^+C^-_k\underline{m}^-C^+_{k-1})+\frac{x^3}{\sqrt{C_N^2}}\underline{m}^3C^3_k\\\nonumber
    =&\frac{\epsilon-2}{2C_N^2}((m^3C^3_k)^2+m^+m^-(C^+_kC^-_{k+1}+C^-_kC^+_{k-1})+\frac{x^3}{\sqrt{C_N^2}}m^3C^3_k\\\nonumber
    =&\frac{\epsilon-2}{2C_N^2}(\underline{m}^3C^3_k\underline{m}^3C^3_k+\underline{m}^-C^+_k\underline{m}^+C^-_{k+1}+\underline{m}^+C^-_k\underline{m}^-C^+_{k-1})+\frac{x^3}{\sqrt{C_N^2}}\underline{m}^3C^3_k\\\nonumber
    =&\frac{\epsilon-2}{2C_N^2}\left(m^3k^2+\frac{(m^1)^2+(m^2)^2}{2}\right)(j(j+1)-k^2)+\frac{x^3}{\sqrt{C_N^2}}m^3k
\end{align}
and especially
\begin{align}
    \bra{j,x,0}V(x)\ket{j,x,0}=\frac{\epsilon-2}{2C_N^2}\left(m^3j^2+\frac{(m^1)^2+(m^2)^2}{2}\right)j+\frac{x^3}{\sqrt{C_N}}m^3j.
\end{align}
In the same way, one derives
\begin{align}
    \bra{j-1,x,0}V(x)\ket{j,x,0}=&\frac{\epsilon-2}{2C_N^2}m^3m^+(2j-1)\sqrt{2j}-\frac{x^3}{\sqrt{C_N^2}}m^+\sqrt{2j},\\\nonumber
    \bra{j-2,x,0}V(x)\ket{j,x,0}=&\frac{\epsilon-2}{2C_N^2}(m^+)^2\sqrt{2j}\sqrt{4j-2},\\\nonumber
    \bra{j-k,x,0}V(x)\ket{j,x,0}=&0\quad k>2.
\end{align}

We further have $\lambda_{j,x,0}-\lambda_{j-1,x,0}=-\vert x\vert/\sqrt{C_N^2}$ and $\lambda_{j,x,0}-\lambda_{j-2,x,0}=-2\vert x\vert/\sqrt{C_N^2}$.

Plugging all this into equation (\ref{pertform}), we get
\begin{align}
    \lambda_{j,x,1}=&\frac{\epsilon-2}{2C_N^2}\left(m^3j^2+\frac{(m^1)^2+(m^2)^2}{2}\right)j+\frac{x^3}{\sqrt{C_N}}m^3j,\\\nonumber
    \ket{j,x,1}=&\frac{\sqrt{C_N^2}}{\vert x\vert}\left(\frac{\epsilon-2}{2C_N^2}m^3m^+(2j-1)\sqrt{2j}-\frac{x^3}{\sqrt{C_N^2}}m^+\sqrt{2j}\right)\ket{j-1,x,0}-\\\nonumber
    &-\frac{\sqrt{C_N^2}}{2\vert x\vert}\frac{\epsilon-2}{2C_N^2}(m^+)^2\sqrt{2j}\sqrt{4j-2}\ket{j-2,x,0}.
\end{align}
Certainly, we could repeat the calculation for any $k$, but here we are only interested in $k=j$ as $\lambda(x)=\lambda_{j,x}$ and $\ket{x}=\ket{j,x}$.

With a little rewriting (we also drop the $j$ from the notation from now on) we can summarize
\begin{align}
    \lambda_{x,0}=&\frac{1}{2}(1+\vert x\vert^2)-\Vert x\Vert j/\sqrt{C_N^2},\\\nonumber
    \ket{x,0}=&U_m\ket{j},\\\nonumber
     \lambda_{x,1}=&\frac{\epsilon-2}{2C_N^2\vert x\vert^2}\left(x^3j^2+\frac{(x^1)^2+(x^2)^2}{2}\right)j+\frac{1}{\sqrt{C_N}\vert x\vert}(x^3)^2j,\\\nonumber
     \ket{x,1}=&\left(\frac{\epsilon-2}{2\sqrt{C_N^2}\vert x\vert^3}x^3x^+(2j-1)\sqrt{2j}-\frac{1}{C_N^2}x^3x^+\sqrt{2j}\right)U_m\ket{j-1}-\\\nonumber
    &-\frac{\epsilon-2}{2\sqrt{C_N^2}\vert x\vert^3}(x^+)^2\sqrt{2j}\sqrt{4j-2}U_m\ket{j-2},
\end{align}
where $m=x/\vert x\vert$, $U_mJ^3U_m^\dagger=m^aJ^a$, $r_m:=\arccos(m^3)$ and $(n_m^a)=(m^2,-m^1,0)^\dagger/\sqrt{(m^1)^2+(m^2)^2}$.

\subsection{The Derivatives of the Unperturbed Eigenstates}

Now, our aim is to calculate $\frac{\partial}{\partial x^a}\ket{k,x}$ and therefore especially $\frac{\partial}{\partial x^a}\ket{k,x,0}=(\frac{\partial}{\partial x^a}U_m)\ket{k}$. Therefore, we need the derivative of the exponential map.

For the latter exists the explicit formula
\begin{align}
    \frac{d}{dt}e^{X(t)}=e^{X(t)}\sum_{k=0}^\infty\frac{(-1)^k}{(k+1)!}(\operatorname{ad}(X(t)))^k\frac{dX(t)}{dt}=:e^{X(t)}\frac{1-e^{-\operatorname{ad}(X(t))}}{\operatorname{ad}(X(t))}\frac{dX(t)}{dt},
\end{align}
where the only assumption is that $X(t)$ is a smooth curve in the Lie algebra \cite{Rossmann_2002}. Therefore,
\begin{align}
    \frac{\partial}{\partial (rn^e)}e^{irn^aJ^a}=&e^{irn^aJ^a}\left(\sum_{k=0}^\infty\frac{(-1)^k}{(k+1)!}(-rV_n)^k\right)^{ef}iJ^f\\\nonumber
    =&e^{irn^aJ^a}i\left(\mathbb{1}+\frac{1}{2}rV_n+\frac{1}{3!}r^2V_n^2+\sum_{k=3}^\infty\frac{1}{(k+1)!}(rV_n)^k\right)^{ef}J^f.
\end{align}
Using $V_n^3=-V_n$, this amounts to
\begin{align}
    \frac{1}{3!}&r^2V_n^2+\sum_{k=3}^\infty\frac{1}{(k+1)!}(rV_n)^k=-\frac{1}{r}V_n\left(\sum_{k=3}^\infty\frac{1}{(k+1)!}(rV_n)^{k+1}\right)\\\nonumber
    &=-\frac{1}{r}V_n\left(e^{rV_n}-\mathbb{1}-rV_n-\frac{1}{2}r^2V_n^2-\frac{1}{3!}r^3V_n^3\right)=\left(-\frac{1}{r}e^{rV_n}+\frac{1}{r}-\frac{1}{2}r\right)V_n+\left(1-\frac{1}{3!}r^2\right)V_n^2.
\end{align}
Based on the result $e^{rV_n}=\mathbb{1}+\sin(r)V_n+(1-\cos(r))V_n^2$ and the definition of $V_n$ we find
\begin{align}
    \frac{\partial}{\partial (rn^e)}e^{irn^aJ^a}=&e^{irn^aJ^a}i\left(\mathbb{1}+\left(-\frac{1}{r}e^{rV_n}+\frac{1}{r}\right)V_n+V_n^2\right)^{ef}J^f\\\nonumber
    =&e^{irn^aJ^a}i\left(\mathbb{1}+\frac{1}{r}\left(1-\cos(r)\right)V_n+\left(1-\frac{1}{r}\sin(r)\right)V_n^2\right)^{ef}J^f\\\nonumber
    =&e^{irn^aJ^a}i\left(\delta^{ef}+\frac{1}{r}\left(1-\cos(r)\right)n^d\epsilon^{def}+\left(1-\frac{1}{r}\sin(r)\right)V_n^2\left(n^en^f-\delta^{ef}\right)\right)J^f\\\nonumber
    =&e^{irn^aJ^a}i\left(\frac{1}{r}\sin(r)\delta^{ef}-\frac{1}{r}\left(1-\cos(r)\right)n^d\epsilon^{edf}+\left(1-\frac{1}{r}\sin(r)\right)V_n^2n^en^f\right)J^f\\\nonumber
    =&:e^{irn^aJ^a}\mathcal{M}^{ef}_{rn}J^f.
\end{align}

Thus, for any given vector $V^e$ (that is not related to the matrices $V_n$) we get
\begin{align}
    V^e\frac{\partial}{\partial x^e}\ket{k,x,0}=&V^e\frac{\partial}{\partial x^e}e^{ir_mn_m^aJ^a}\ket{k}=V^e\frac{\partial(r_mn_m)^f}{\partial x^e}\frac{\partial}{\partial(r_mn_m)^f}e^{ir_mn_m^aJ^a}\ket{k}\\\nonumber
    =&e^{ir_mn_m^aJ^a}V^e\frac{\partial(r_mn_m)^f}{\partial x^e}\mathcal{M}^{fg}_{r_mn_m}J^g\ket{k}=:e^{ir_mn_m^aJ^a}\mathcal{V}_m^gJ^g\ket{k},
\end{align}
where we defined
\begin{align}
    \mathcal{V}_m^g:=&V^e\frac{\partial(r_mn_m)^f}{\partial x^e}\mathcal{M}^{fg}_{r_mn_m}\\\nonumber
    \mathcal{M}^{ef}_{rn}:=&i\left(\frac{1}{r}\sin(r)\delta^{ef}-\frac{1}{r}\left(1-\cos(r)\right)n^d\epsilon^{edf}+\left(1-\frac{1}{r}\sin(r)\right)V_n^2n^en^f\right).
\end{align}
Consequently, using equation (\ref{splitaction}), we arrive at the result
\begin{align}
    \label{dercoh}
    V^e\frac{\partial}{\partial x^e}\ket{k,x,0}=\mathcal{V}_m^3C^3_k\ket{k,x,0}+\mathcal{V}_m^-C^+_k\ket{k+1,k,0}+\mathcal{V}_m^+C^-_k\ket{k-1,x,0}.
\end{align}

\subsection{Calculation of the Quantum Metric and the Would-Be Symplectic Form}

Our final step is the calculation of $h_{ab}$, $g_{ab}$ and $\omega_{ab}$ as defined in equation (\ref{hermitianform}).
Accordingly, we expand $h_{ab}=h_{0,ab}+h'_{ab}$ (and similarly for the others).

In the round case $(\epsilon=0)$ we have $h_{ab}=h_{0,ab}=(\partial_a+iA_a)\bra{j,x,0}(\partial_b-iA_b)\ket{j,x,0}$ where $(\partial_a-iA_a)\ket{j,x,0}=(\mathbb{1}-\ket{j,x,0}\bra{j,x,0})\partial_a\ket{j,x,0}=\mathcal{V}_m^+C^-_j\ket{j-1,x,0}$ for $V^b=\delta^{ab}$.
Since $(C^-_j)^2=2j$, this implies
\begin{align}
    h_{0,ab}=2j(\mathcal{V}_m^+)^*_{V^c=\delta^{ac}}(\mathcal{V}_m^+)_{V^c=\delta^{bc}}.
\end{align}
Using Mathematica, we find the explicit result
\begin{align}
    (h_{0,ab})=\frac{j}{2\vert x\vert^4}
    \begin{pmatrix}
    (x^2)^2+(x^3)^2 & \frac{(x^2x^3+ix^1\vert x\vert)(x^1x^3+ix^2\vert x\vert)}{(x^1)^2+(x^2)^2} & -(x^1x^3+ix^2\vert x\vert) \\
   \frac{ (x^2x^3-ix^1\vert x\vert)(x^1x^3-ix^2\vert x\vert)}{(x^1)^2+(x^2)^2} & (x^1)^2+(x^3)^2 & -(x^2x^3+ix^1\vert x\vert) \\
   -(x^1x^3-ix^2\vert x\vert) & -(x^2x^3-ix^1\vert x\vert) & (x^1)^2+(x^2)^2 \\
    \end{pmatrix}
\end{align}
and consequently
\begin{align}
    (g_{0,ab})=\operatorname{Re}(h_{0,ab})=\frac{j}{2\vert x\vert^4}
        \begin{pmatrix}
    (x^2)^2+(x^3)^2 & -x^1x^2 & -x^1x^3 \\
  -x^1x^2 & (x^1)^2+(x^3)^2 & -x^2x^3 \\
   -x^1x^3 & -x^2x^3 & (x^1)^2+(x^2)^2 \\
    \end{pmatrix}
\end{align}
as well as
\begin{align}
     (\omega_{0,ab})=\operatorname{Re}(h_{0,ab})=\frac{j}{2\vert x\vert^3}
        \begin{pmatrix}
   0 & x^3 & -x^2\\
  -x^3 & 0 & x^1 \\
   x^2 & -x^1 & 0 \\
    \end{pmatrix}.
\end{align}

For $\epsilon>0$ the calculations get more involved.
We recall
\begin{align}
    \ket{j,x}=&\ket{j,x,0}+\epsilon\left(\frac{\epsilon-2}{2\sqrt{C_N^2}\vert x\vert^3}x^3x^+(2j-1)\sqrt{2j}-\frac{1}{C_N^2}x^3x^+\sqrt{2j}\right)\ket{j-1,x,0}-\\\nonumber
    &-\frac{\epsilon-2}{2\sqrt{C_N^2}\vert x\vert^3}(x^+)^2\sqrt{2j}\sqrt{4j-2}\ket{j-2,x,0}+\mathcal{O}(\epsilon^2)\\\nonumber
    =&:\ket{j,x,0}+\epsilon\left(\mathcal{W}_x^1\ket{j-1,x,0}+\mathcal{W}_x^2\ket{j-2,x,0}\right)+\mathcal{O}(\epsilon^2).
\end{align}

Using equation (\ref{dercoh}), this implies
\begin{align}
    \partial_a\ket{j,x}=&\partial_a\ket{j,x,0}+\epsilon\left(\partial_a\mathcal{W}_x^1\ket{j-1,x,0}+\partial_a\mathcal{W}_x^2\ket{j-2,x,0}+\mathcal{W}_x^1\partial_a\ket{j-1,x,0}+\right.\\\nonumber
    &\left.+\mathcal{W}_x^2\partial_a\ket{j-2,x,0}\right)+\mathcal{O}(\epsilon^2).
\end{align}
Further, we can calculate
\begin{align}
    \ket{j,x}\bra{j,x}=&\ket{j,x,0}\bra{j,x,0}+\epsilon\left(\mathcal{W}_x^1\ket{j-1,x,0}+\mathcal{W}_x^2\ket{j-2,x,0}\right)\bra{j,x,0}+\\\nonumber
    &+\epsilon\ket{j,x,0}\left((\mathcal{W}_x^1)^*\bra{j-1,x,0}+(\mathcal{W}_x^2)^*\bra{j-2,x,0}\right)+\mathcal{O}(\epsilon^2),
\end{align}
and thus
\begin{align}
    (\partial_a-iA_a)\ket{j,x}=&\left(\mathbb{1}-\ket{j,x,0}\bra{j,x,0}-\epsilon\left(\mathcal{W}_x^1\ket{j-1,x,0}+\mathcal{W}_x^2\ket{j-2,x,0}\right)\cdot\right.\\\nonumber
    &\left.\cdot\bra{j,x,0}-\epsilon\ket{j,x,0}\left((\mathcal{W}_x^1)^*\bra{j-1,x,0}+(\mathcal{W}_x^2)^*\bra{j-2,x,0}\right)\right)\partial_a\ket{j,x,0}+\\\nonumber
    &+\epsilon\left(\mathbb{1}-\ket{j,x,0}\bra{j,x,0}\right)\left(\partial_a\mathcal{W}_x^1\ket{j-1,x,0}+\partial_a\mathcal{W}_x^2\ket{j-2,x,0}+\right.\\\nonumber
    &\left.+\mathcal{W}_x^1\partial_a\ket{j-1,x,0}+\mathcal{W}_x^2\partial_a\ket{j-2,x,0}\right)+\mathcal{O}(\epsilon^2).
\end{align}
Now, we insert $V^e\frac{\partial}{\partial x^e}\ket{k,x,0}=\mathcal{V}_m^3C^3_k\ket{k,x,0}+\mathcal{V}_m^-C^+_k\ket{k+1,k,0}+\mathcal{V}_m^+C^-_k\ket{k-1,x,0}$ for $V^e=\delta^{ea}$, resulting in
\begin{align}
    &(\partial_a-iA_a)\ket{j,x}=\left(\mathbb{1}-\ket{j,x,0}\bra{j,x,0}-\epsilon\left(\mathcal{W}_x^1\ket{j-1,x,0}+\mathcal{W}_x^2\ket{j-2,x,0}\right)\bra{j,x,0}-\right.\\\nonumber
    &\phantom{+}\left.-\epsilon\ket{j,x,0}\left((\mathcal{W}_x^1)^*\bra{j-1,x,0}+(\mathcal{W}_x^2)^*\bra{j-2,x,0}\right)\right)\left(\mathcal{V}_m^3C^3_j\ket{j,x,0}+\mathcal{V}_m^+C^-_j\ket{j-1,x,0}\right)+\\\nonumber
    &\phantom{+}+\epsilon\left(\mathbb{1}-\ket{j,x,0}\bra{j,x,0}\right)\left(\partial_a\mathcal{W}_x^1\ket{j-1,x,0}+\partial_a\mathcal{W}_x^2\ket{j-2,x,0}+\right.\\\nonumber
    &\phantom{+}+\mathcal{W}_x^1\left(\mathcal{V}_m^3C^3_{j-1}\ket{j-1,x,0}+\mathcal{V}_m^-C^+_{j-1}\ket{j,k,0}+\mathcal{V}_m^+C^-_{j-1}\ket{j-2,x,0}\right)+\\\nonumber
    &\phantom{+}\left.+\mathcal{W}_x^2\left(\mathcal{V}_m^3C^3_{j-2}\ket{j-2,x,0}+\mathcal{V}_m^-C^+_{j-2}\ket{j-1,x,0}+\mathcal{V}_m^+C^-_{j-2}\ket{j-3,x,0}\right)\right)+\mathcal{O}(\epsilon^2).
\end{align}
This expression can then be rewritten as
\begin{align}
     &(\partial_a-iA_a)\ket{j,x}=\mathcal{V}^+_mC^-_j\ket{j-1,x,0}-\epsilon\mathcal{V}^3_mC^3_j\left(\mathcal{W}_x^1\ket{j-1,x,0}+\mathcal{W}_x^2\ket{j-2,x,0}\right)-\\\nonumber
&\phantom{+}-\epsilon\mathcal{V}^+_mC^-_j(\mathcal{W}^1_x)^*\ket{j,x}+\epsilon\left(\partial_a\mathcal{W}_x^1\ket{j-1,x,0}+\partial_a\mathcal{W}_x^2\ket{j-2,x,0}+\mathcal{W}_x^1\left(\mathcal{V}_m^3C^3_{j-1}\ket{j-1,x,0}+\right.\right.\\\nonumber
&\hspace{-0.08cm}\phantom{+}\left.\left.+\mathcal{V}_m^+C^-_{j-1}\ket{j-2,x,0}\right)+\mathcal{W}_x^2\left(\mathcal{V}_m^3C^3_{j-2}\ket{j-2,x,0}+\mathcal{V}_m^-C^+_{j-2}\ket{j-1,k,0}+\mathcal{V}_m^+C^-_{j-2}\ket{j-3,x,0}\right)\right)+\\\nonumber
&\phantom{+}+\mathcal{O}(\epsilon^2)\\\nonumber
&=\mathcal{V}^+_mC^-_j\ket{j-1,x,0}+\\\nonumber
&\phantom{+}+\epsilon\left(-\mathcal{V}^3_mC^3_j\mathcal{W}_x^1\ket{j,x,0}+\left(-\mathcal{V}_m^3C^3_j\mathcal{W}_x^1+\partial_a\mathcal{W}_x^1+\mathcal{W}_x^1\mathcal{V}_m^3C^3_{j-1}+\mathcal{W}_x^2\mathcal{V}_m^-C^+_{j-2}\right)\ket{j-1,x,0}+\right.\\\nonumber
&\hspace{-0.03cm}\phantom{+}\left.+\left(-\mathcal{V}_m^3C^3_j\mathcal{W}_x^2+\partial_a\mathcal{W}_x^2+\mathcal{W}^1_x\mathcal{V}_m^+C^-_{j-1}+\mathcal{W}^2_x\mathcal{V}_m^3C^3_{j-2}\right)\ket{j-2,x,0}+\mathcal{W}^2_x\mathcal{V}_m^+C^-_{j-2}\ket{j-3,x,0}\right)+\\\nonumber
&\phantom{+}+\mathcal{O}(\epsilon^2).
\end{align}
From that, we can now calculate
\begin{align}
\label{hresult}
    h_{ab}=&h_{0,ab}+\epsilon\left[\left(C^-_j\mathcal{V}^+_m\right)^*_{V^c=\delta^{ca}}\left(-\mathcal{V}_m^3C^3_j\mathcal{W}_x^1+\partial_a\mathcal{W}_x^1+\mathcal{W}_x^1\mathcal{V}_m^3C^3_{j-1}+\mathcal{W}_x^2\mathcal{V}_m^-C^+_{j-2}\right)_{V^c=\delta^{cb}}+\right.\\\nonumber
    &\left.+\left(-\mathcal{V}_m^3C^3_j\mathcal{W}_x^1+\partial_a\mathcal{W}_x^1+\mathcal{W}_x^1\mathcal{V}_m^3C^3_{j-1}+\mathcal{W}_x^2\mathcal{V}_m^-C^+_{j-2}\right)^*_{V^c=\delta^{ca}}\left(C^-_j\mathcal{V}^+_m\right)_{V^c=\delta^{cb}}\right]+\mathcal{O}(\epsilon^2).
\end{align}

As a side remark we note that we can write $\partial_a\ket{j,x}$ as
\begin{align}
    \partial_a\ket{j,x}=&\partial_a\ket{j,x,0}+\epsilon\left(\partial_a\mathcal{W}_x^1\ket{j-1,x,0}+\partial_a\mathcal{W}_x^2\ket{j-2,x,0}+\mathcal{W}_x^1\partial_a\ket{j-1,x,0}+\right.\\\nonumber
    &\left.+\mathcal{W}_x^2\partial_a\ket{j-2,x,0}\right)+\mathcal{O}(\epsilon^2)\\\nonumber
    =&\mathcal{V}^3_mC^3_j\ket{j,x,0}+\mathcal{V}^+_mC^-_j\ket{j-1,x,0}\\\nonumber
    &+\epsilon\left[\mathcal{W}^1_x\mathcal{V}^-_mC^+_{j-1}\ket{j,x,0}+\left(\partial_a\mathcal{W}^1_x+\mathcal{W}^1_x\mathcal{V}^3_mC^3_{j-1}+\mathcal{W}^2_x\mathcal{V}^-_mC^+_{j-2}\right)\ket{j-1,x,0}+\right.\\\nonumber
    &\left.+\left(\partial_a\mathcal{W}^2_x+\mathcal{W}^1_x\mathcal{V}^+_mC^-_{j-1}+\mathcal{W}^2_x\mathcal{V}^3_mC^3_{j-2}\right)\ket{j-2,x,0}+\mathcal{W}^2_x\mathcal{V}^+_mC^-_{j-2}\ket{j-3,x,0}\right]+\\\nonumber
    &+\mathcal{O}(\epsilon^2).
\end{align}

Using Mathematica, we can explicitly calculate $h'_{ab}$ defined in equation (\ref{hresult}) and consequently $g'_{ab}$ and $\omega'_{ab}$. The results are given on the next page.\\
In section \ref{BasicQuantities} we verified the results by comparing them to numerical calculations for $N=4$.

\newpage

\begin{sideways}
\resizebox{1.6\linewidth}{!}{%
  \begin{minipage}{2\linewidth}

\begin{align}
    g'_{11}=&\frac{C_3}{8 C_1 \left(x_1^2+x_2^2\right) X^8}\left[C_2^2 X \left(x_3 x_2^4 (4 C_3 X+x_3)+4 C_3 x_3^2 x_2^2 X (x_3-X)+2 x_1^2 \left(2 C_3 x_3 x_2^2 X+x_2^4\right)+x_2^6+x_1^4 \left(x_2^2-x_3^2\right)\right)+\right.\\\nonumber&\left.+4 C_3 C_2 x_3 \left(-\left(x_1^4 \left(x_2^2 (C_1 (X-x_3)+1)-2 x_3 X\right)\right)-x_1^2 \left(2 x_2^4 (C_1 (X-x_3)+1)+2 x_3 x_2^2 (C_1 x_3-1) (X-x_3)+x_3^3X\right)-x_2^2 \left(x_2^2+x_3^2\right)^2 (C_1 (X-x_3)+1)\right)+\right.\\\nonumber&\left.+4 C_1 C_3 x_3 X^2 \left(x_2^2 X^3-x_1^2 x_3 \left(x_1^2+x_2^2-x_3^2\right)\right)\right]\\\nonumber
   g'_{12}=&g'_{21}=-\frac{C_3 x_1 x_2}{8 C_1 \left(x_1^2+x_2^2\right) X^8}\left[C_2^2 X \left(2 x_1^2 \left(x_3 (2 C_3 X+x_3)+x_2^2\right)+4 C_3 x_3^2 X (x_3-X)+2 x_2^2 x_3 (2 C_3X+x_3)+x_1^4+x_2^4\right)+\right.\\\nonumber&\left.+4 C_3 C_2 x_3 \left(x_1^4 (C_1 (x_3-X)-1)-2 x_1^2 \left(x_2^2 (-C_1 x_3+C_1 X+1)+x_3 \left(C_1 x_3 X-C_1 x_3^2+x_3+X\right)\right)-x_3^3 (C_1 x_3-1) (X-x_3)+x_2^4 (C_1 x_3-C_1 X-1)-2 x_2^2 x_3 \left(C_1 x_3 X-C_1 x_3^2+x_3+X\right)\right)+\right.\\\nonumber&\left.+4 C_1 C_3 x_3 X^2 \left(x_3\left(x_1^2+x_2^2-x_3^2\right)+X^3\right)\right]\\\nonumber
   g'_{13}=&g'_{31}=\frac{C_3 x_1 x_3}{8 C_1 X^7}\left[4 C_1 C_3 \left(x_1^2+x_2^2-x_3^2\right) X+C_2^2 \left(x_1^2+x_2^2\right)-6 C_3 C_2 \left(x_1^2+x_2^2-x_3^2\right)\right]\\\nonumber
   g'_{22}=&\frac{C_3}{8 C_1 \left(x_1^2+x_2^2\right) X^8}\left[C_2^2 X \left(x_1^4 \left(x_3 (4 C_3 X+x_3)+2 x_2^2\right)+x_1^2 \left(4 C_3 x_3 x_2^2 X+4 C_3 x_3^2 X (x_3-X)+x_2^4\right)+x_1^6-x_2^4 x_3^2\right)+\right.\\\nonumber&\left.+4 C_3 C_2 x_3 \left(x_1^6 (C_1 (x_3-X)-1)-2 \left(x_2^2+x_3^2\right) x_1^4 (C_1 (X-x_3)+1)-x_1^2 \left(x_2^4 (-C_1 x_3+C_1 X+1)+2 x_3 x_2^2 (C_1 x_3-1) (X-x_3)+x_3^4 (-C_1 x_3+C_1 X+1)\right)+x_2^2 \left(2 x_2^2 x_3-x_3^3\right) X\right)+\right.\\\nonumber&\left.+4 C_1 C_3 x_3 X^2 \left(x_1^2 X^3-x_2^2 x_3\left(x_1^2+x_2^2-x_3^2\right)\right)\right]\\\nonumber
   g'_{23}=&g'_{32}=\frac{C_3 x_2 x_3}{8 C_1 X^7}\left[4 C_1 C_3 \left(x_1^2+x_2^2-x_3^2\right) X+C_2^2 \left(x_1^2+x_2^2\right)-6 C_3 C_2 \left(x_1^2+x_2^2-x_3^2\right)\right]\\\nonumber
   g'_{33}=&-\frac{C_3\left(x_1^2+x_2^2\right)}{8 C_1 X^8}\left[C_2^2 \left(x_1^2+x_2^2\right) X-4 C_3 C_2 \left(x_1^2+x_2^2-2 x_3^2\right) X+4 C_1 C_3 \left(x_1^4+2 x_2^2 x_1^2+x_2^4-x_3^4\right)\right]
\end{align}

\begin{align}
    \omega'_{11}=& 0\\\nonumber
    \omega'_{12}=&-\omega'_{21}=-\frac{C_3^2 x_3}{4 X^8 C_1}\left[X^3 C_1 \left(-x_3 (C_2 x_3+X (-C_2)+x_3+X)+x_1^2+x_2^2\right)+C_2 \left(X^2 C_2 x_3 (x_3-X)-x_1^2 \left(x_3(x_3-X)+4 x_2^2\right)+x_2^2 x_3 (X-x_3)+x_3^3 (x_3+X)-2 x_1^4-2 x_2^4\right)\right]\\\nonumber
   \omega'_{13}=&-\omega'_{31}=\frac{C_3^2 x_2}{4 X^8 C_1}\left[X^3 C_1\left(-x_3 (C_2 x_3+X (-C_2)+x_3+X)+x_1^2+x_2^2\right)+C_2 \left(X^2 C_2 x_3 (x_3-X)+x_1^2 \left(x_3 (x_3+X)-2 x_2^2\right)+x_2^2 x_3(x_3+X)+x_3^3 (2 x_3+X)-x_1^4-x_2^4\right)\right]\\\nonumber
   \omega'_{22}=& 0\\\nonumber
   \omega'_{23}=&-\omega'_{32}=-\frac{C_3^2 x_1}{4 X^8 C_1}\left[X^3C_1 \left(-x_3 (C_2 x_3+X (-C_2)+x_3+X)+x_1^2+x_2^2\right)+C_2 \left(X^2 C_2 x_3 (x_3-X)+x_1^2 \left(x_3 (x_3+X)-2 x_2^2\right)+x_2^2x_3 (x_3+X)+x_3^3 (2 x_3+X)-x_1^4-x_2^4\right)\right]\\\nonumber
   \omega'_{33}=& 0
\end{align}

\begin{align}
    \label{Ccoeffs}
    X:=\vert x\vert,\quad C_1:=\sqrt{j(j+1)},\quad C_2:=(2j-1), \quad C_3:=\sqrt{2j}
\end{align}

  \end{minipage}
}%

\end{sideways}

\end{document}